\documentclass[twocolumn,aps,rmp]{revtex4}

\usepackage{hyperref}
\usepackage{amssymb}
\usepackage{graphicx}
\usepackage{xypic}

\begin{document}

\draft
\font\Bbb =msbm10
\font\eufm =eufm10
\def\Real{{\hbox{\Bbb R}}} \def\C{{\hbox {\Bbb C}}}
\def\eea{\end{array}}
\def\bea{\begin{array}}
\newcommand{\mac}[1]{\mathsf{#1}}
\newcommand{\be}{\begin{equation}}
\newcommand{\ee}{\end{equation}}
\newcommand{\emx}{\end{array}}
\newcommand{\bex}{\begin{eqnarray}}
\newcommand{\enx}{\end{eqnarray}}
\newcommand{\ben}{\begin{enumerate}}
\newcommand{\enn}{\end{enumerate}}
\newcommand{\bei}{\begin{itemize}}
\newcommand{\eei}{\end{itemize}}
\newcommand{\enxn}{\nonumber\end{eqnarray}}
\font\Bbb =msbm10  scaled \magstephalf
\def\id{{\hbox{\Bbb I}}}
\def\hcal{{\cal H}}
\def\acal{{\cal A}}
\def\ra{\rangle}
\def\la{\langle}
\def\dim{\rm dim}
\def\lra{\longrightarrow}
\def\llra{\longleftrightarrow}

\def\Real{{\hbox{\Bbb R}}} \def\C{{\hbox {\Bbb C}}}
\def\spec{R_{\alpha\beta}}
\def\kvec{{\bbox{k}}}
\def\lvec{{\bbox{l}}}
\def\duzomniejsze{<\kern-.7mm<}
\def\duzowieksze{>\kern-.7mm>}
\def\intlarge{\mathop{\int}\limits}
\def\textbf#1{{\bf #1}}
\def\beq{\begin{equation}}
\def\eeq{\end{equation}}
\def\be{\begin{equation}}
\def\ee{\end{equation}}
\def\ben{\begin{eqnarray}}
\def\een{\end{eqnarray}}
\def\beqa{\begin{eqnarray}}
\def\eeqa{\end{eqnarray}}
\def\eea{\end{array}}
\def\bea{\begin{array}}
\newcommand{\bee}{\begin{enumerate}}
\newcommand{\eee}{\end{enumerate}}

\def\rhoiso{\rho_F^d}
\def\esq{E_{sq}}
\def\ed{E_D}
\def\er{E_R}
\def\eg{E_g}
\def\ef{E_F}
\def\ec{E_C}
\def\pptcal{{\cal PPT}}
\def\ndcal{{\cal ND}}
\def\becal{{\cal BE}}
\def\fecal{{\cal FE}}

\def\Ebar{\overline{E}}

\def\Phiplus{{\Phi^+}}
\def\Phiplusd{{\Phi^+_d}}
\def\Phiplusdwan{{\Phi^+_{2^n}}}
\def\Pplus{P^+}
\def\Pplusd{P^+_d}

\def\phiplus{\phi^+}
\def\psiplus{\psi^+}
\def\phiminus{\phi^-}
\def\psiminus{\psi^-}

\def\phiplus{{\phi^+}}
\def\singlet{singlet}
\def\singlets{singlets}
\def\ghzstate{GHZ state}
\def\eprstates{EPR states}
\def\eprpair{EPR pair}
\def\eprpairs{EPR pairs}
\def\eprstate{EPR state}
\def\bcal{{\cal B}}
\def\ccal{{\cal C}}
\def\ncal{{\cal N}}
\def\scal{{\cal S}}
\def\xcal{{\cal X}}
\def\dcal{{\cal D}}
\def\dist{\dcal}
\def\ecal{{\cal E}}
\def\pcal{{\cal P}}
\def\hcal{{\cal H}}
\def\tr{{\rm Tr}}
\def\id{{\rm I}}
\def\ra{\rangle}
\def\la{\langle}
\def\>{\rangle}
\def\<{\langle}
\def\blacksquare{\vrule height 4pt width 3pt depth2pt}
\def\ic{I_{coh}}
\def\ot{\otimes}
\def\rhoab{\varrho_{AB}}
\def\sigmaab{\sigma_{AB}}
\def\rhoa{\varrho_{A}}
\def\sigmaa{\sigma_{A}}
\def\rhob{\varrho_{B}}
\def\sigmab{\sigma_{B}}
\def\dt#1{{{\kern -.0mm\rm d}}#1\,}

\def\aaa{{A_1\ldots A_N}}
\def\aaacolon{{A_1:\ldots :A_N}}
\def\epr{{\it EPR}}
\def\ghz{{\it GHZ}}

\def\eqsq{E_{sq}^{q}}
\def\ecsq{E_{sq}^{c}}
\def\chv{C_{HV}}

\def\cka{CKA}
\def\secure{secure}
\def\security{security}
\def\private{private}
\def\privacy{privacy}
\def\bent{BE}

\newtheorem{lemma}{Lemma}
\newtheorem{corollary}{Corollary}
\newtheorem{theorem}{Theorem}
\newtheorem{proposition}{Proposition}
\newtheorem{definition}{Definition}

\newtheorem{Def}{Def.}[section]
\newtheorem{Thm}{Theorem}[section]
\newtheorem{Lem}{Lemma}[section]
\newtheorem{Prop}{Prop.}[section]
\newtheorem{Conj}{Conjecture}[section]
\newtheorem{Obs}{Observation}[section]
\newtheorem{Fact}{Fact}[section]

\newtheorem{prop}{Proposition}
\newtheorem{obs}{Observation}
\def\hcal{{\cal H}}
\def\pcal{{\cal P}}
\def\ccal{{\cal C}}
\def\ep{{\epsilon}}
\def\tr{{\rm Tr}}
\def\id{{\rm I}}
\def\>{\rangle}
\def\<{\langle}
\def\blacksquare{\vrule height 4pt width 3pt depth2pt}
\def\ic{I_{coh}}
\def\ot{\otimes}
\def\ra{\rangle}
\def\la{\langle}
\def\blacksquare{\vrule height 4pt width 3pt depth2pt}
\def\dcal{{\cal D}}
\def\pcal{{\cal P}}
\def\hcal{{\cal H}}
\def\trace{\mbox{Tr}}

\def\rhoccq{\rho^{(ccq)}}
\def\rhoccqabe{\rho^{(ccq)}_{ABE}}
\def\rhoaabb{\rho_{AA'BB'}}
\def\sigmaccqabe{\sigma^{(ccq)}_{ABE}}
\def\sigmaccq{\sigma^{(ccq)}}
\def\sigmaab{\sigma_{AB}}
\def\sigmaabtwirl{\sigma_{AB}^{twirl}}
\def\sigmae{\sigma_E}
\def\sigmaetwirl{\sigma_E^{twirl}}
\def\pmp{P\&M }
\def\mcs{entanglement distillation-like scenario}
\def\singlet{$|\psi^-\>={1\over\sqrt{2}}(|01\>-|10\>)$\,}
\def\wsinglet{singlet}

\def\PD{private distribution }
\def\intr{I_{intr}}
\def\ef{E_F}

\def\ab{X }
\def\eve{Eve }
\newtheorem{conjecture}{Conjecture}

\def\keyC{K_{LOPC}}
\def\keyQ{K_{LOCC}}

\title{Quantum entanglement}

\author{Ryszard Horodecki $^{1}$
Pawe\l{} Horodecki $^{3}$
Micha\l{} Horodecki $^{1}$,
Karol Horodecki $^{1,2}$}

\address{$^1$ Institute of Theoretical Physics and Astrophysics
 University of Gda\'nsk, 80--952 Gda\'nsk, Poland}

\address{$^2$ Faculty of Mathematics, Physics and Computer Science
University of Gda\'nsk, 80--952 Gda\'nsk, Poland}

\address{$^3$ Faculty of Applied Physics and Mathematics,
Technical University of Gda\'nsk, 80--952 Gda\'nsk, Poland
}

\def\hcal{{\cal H}}
\def\ecal{{\cal E}}
\def\ccal{{\cal C}}
\def\qcal{{\cal Q}}
\def\pcal{{\cal P}}

\def\dim{{\rm dim}}

\begin{abstract}
All our former experience with application of quantum theory seems
to say: {\it what is predicted by quantum formalism must  occur in
laboratory}. But the essence of quantum formalism --- entanglement,
recognized by Einstein, Podolsky, Rosen and Schr{\"o}dinger --- waited
over 70 years to enter to laboratories as a new resource as real as
energy.

This holistic property of compound quantum systems, which involves
nonclassical correlations between subsystems, is a potential for many
quantum processes, including ``canonical'' ones: quantum cryptography,
quantum teleportation and dense coding. However, it appeared that this
new resource is very complex and difficult to detect. Being usually
fragile to environment, it is robust against conceptual and
mathematical tools, the task of which is to decipher its rich
structure.

This article reviews basic aspects of entanglement including its
characterization, detection, distillation and quantifying. In
particular, the authors discuss various manifestations of entanglement
via Bell inequalities, entropic inequalities, entanglement witnesses,
quantum cryptography and point out some interrelations. They also
discuss a basic role of entanglement in quantum communication within
distant labs paradigm and stress some peculiarities such as
irreversibility of entanglement manipulations including its extremal
form --- bound entanglement phenomenon. A basic role of entanglement
witnesses in detection of entanglement is emphasized.

\end{abstract}
\maketitle
\tableofcontents

\section{Introduction}

Although in 1932 von Neumann had completed basic elements of
nonrelativistic quantum description of the world, it were Einstein,
Podolsky and Rosen (EPR) and Schr{\"o}dinger who first recognized a
``spooky'' feature of quantum machinery which lies at center of
interest of physics of XXI century \cite {EPR,Neumann}. {\it This
  feature implies the existence of global states of composite system
  which cannot be written as a product of the states of individual
  subsystems}. This phenomenon, known as ``entanglement'', was
originally called by Schr{\"o}dinger ``Verschr{\"a}nkung'', which
underlines an intrinsic order of statistical relations between
subsystems of compound quantum system \cite {Schrodinger}.

Paradoxically, entanglement, which is considered to be the most
nonclassical manifestations of quantum formalism, was used by Einstein
Podolsky and Rosen in their attempt to ascribe values to physical
quantities prior to measurement. It was Bell who showed the opposite:
it is just entanglement which irrevocably rules out such a
possibility.

In 1964 Bell accepted the EPR conclusion --- that quantum description
of physical reality is not complete --- as a working hypothesis and
formalized the EPR deterministic world idea in terms of local hidden
variable model (LHVM) \cite {Bell}. The latter assumes that (i)
measurement results are determined by properties the particles carry
prior to, and independent of, the measurement (``realism''), (ii)
results obtained at one location are independent of any actions
performed at spacelike separation (``locality'') (iii) the setting of
local apparatus are independent of the hidden variables which
determine the local results (``free will''). Bell proved that the
above assumptions impose constraints on statistical correlations in
experiments involving bipartite systems in the form of the Bell
inequalities. He then showed that the probabilities for the outcomes
obtained when suitably measuring some entangled quantum state violate
the Bell inequality. In this way entanglement is that feature of
quantum formalism which makes impossible to simulate the quantum
correlations within any classical formalism.

Greenberger, Horne and Zeilinger (GHZ) went beyond Bell inequalities
by showing that entanglement of more than two particles leads to a
contradiction with LHVM for nonstatistical predictions of quantum
formalism \cite {GHZk}. Surprisingly, only in the beginning of 90's
theoretical general results concerning violation of Bell inequalities
have been obtained \cite{Gisin, Popescu}.

Transition of entanglement from gedanken experiment to laboratory
reality began in the mid-60s \cite {Kocher,Freedman}. However it were
Aspect \emph{et al.}, who performed first a convincing test of
violation of the Bell inequalities \cite {AspectD,AspectGR81}. Since
then many kinds of beautiful and precise experimental tests of quantum
formalism against the LHVM have been performed in laboratories \cite
{KwiatMWZSS,OuM,Bovino2,Rowe,HasegawaLBMH} and outsides \cite
{TittelBGZ1,TittelBGZ2,WeihsRWZ,144km}. All these experiments strongly
confirmed the predictions of the quantum description\footnote {However
  so far all the above experiments suffer from loophole, see \cite
  {Gill,BrunnerGSS2007}.}.

In fact, a fundamental nonclassical aspect of entanglement was
recognized already in 1935. Inspired by EPR paper, Shr{\"o}dinger
analyzed some physical consequences of quantum formalism and he
noticed that the two-particle EPR state does not admit ascribing
individual states to the subsystems implying ``entanglement of
predictions'' for the subsystems. Then he concluded: {``\it Thus one
  disposes provisionally (until the entanglement is resolved by actual
  observation) of only a {\bf common} description of the two in that space
  of higher dimension. This is the reason that knowledge of the
  individual systems can decline to the scantiest, even to zero, while
  that of the combined system remains continually maximal. Best
  possible knowledge of a whole does {\bf not} include best possible
  knowledge of its parts --- and this is what keeps coming back to
  haunt us''}\cite{Schrodinger}\footnote{an English
translation appears in Quantum Theory and Measurement,
edited by J. A. Wheeler and W.H. Zurek Princeton University
Press, Princeton, 1983, p.167.}.

Unfortunately this curious aspect of entanglement was long
unintelligible, as it was related to the notion of ``knowledge'' in
quantum context. Only in half of 90's it was formalized in terms of
entropic inequalities based on Von Neumann entropy \cite
{RPH1994,alpha,cerfadami}\footnote {The other formalization was
  proposed in terms of majorization relations \cite
  {NielsenKempe}.}. The violation of these inequalities by entangled
states is a signature of entanglement of quantum states, however
physical meaning of this was unclear. An interesting attempt to solve
this puzzle is due to Cerf and Adami (1997)
in terms of conditional entropy. Soon afterwards it turned out that
the latter with minus sign, called {\it coherent information} is a
fundamental quantity responsible for capabilities of transmission of
quantum information \cite{SchumacherN-1996-pra,Lloyd-cap}. The
transmission is possible exactly in those situations in which
``Schr{\"o}dinger's demon'' is ``coming to haunt us'' --- i.e. when
entropy of output system exceeds the entropy of the total system.  Let
us mention, that in 2005 this story has given a new twist in terms of
quantum counterpart of the Slepian-Wolf theorem in classical
communication \cite {SW-nature,sw-long}. In this approach the
violation of entropic inequalities implies the existence of {\it
  negative quantum information}, which is ``extra'' resource for
quantum communication. Interestingly, only recently a direct violation
of the entropic inequalities was experimentally demonstrated
confirming the breaking of classical statistical order in compound
quantum systems \cite{NonlinExp}.

The present-day entanglement theory has its roots in the key
discoveries: quantum cryptography with Bell theorem \cite{E91},
quantum dense coding \cite{BennettW} and quantum teleportation
\cite{Teleportation}\footnote{Quantum teleportation with continuous
  variables in an infinite dimensional Hilbert space was first
  proposed by Vaidman \cite{Vaidman} and further investigated
  theoretically by Braunstein and Kimble \cite{BraunsteinKt}.}
including teleportation of entanglement of EPR pairs (so called
entanglement swapping)
\cite{Yurke,YurkeStoller_ent_swapping,ent_swapping,BoseWKt}.  All such
effects are based on entanglement and all of them have been
demonstrated in pioneering experiments (see \cite
{Kimble-teleportation,Rzym-teleportation,Wieden-teleportation,JenSWWZ-exp,NaikPWBK,TittelBZG,MattleWKZ1996,exp_ent_swapping}.
In fact, the above results including pioneering paper on quantum
cryptography \cite{BB84} and idea of quantum computation
\cite{Feynman,DDeutsch,Shor,Steane-codes-prl} were a basis for a new
interdisciplinary domain called quantum information \cite
{Nielsen-Chuang,Alber2001,BouwmeesterEZ-book,BraunsteinP,LoSPbook,DagBook}
which incorporated entanglement as a central notion.

It has become clear that entanglement is not only subject of
philosophical debates, but it is a new quantum resource for tasks
which can not be performed by means of classical resources \cite
{Bennett98}. It can be manipulated \cite
{BBPSSW1996,BDSW1996,Popescu2,Gisin96,RaimondBH}, broadcasted
\cite{Buzek-local-cloning}, controlled and distributed
\cite{CiracZ,MandilaraAKK2007,BeigeBBHKPV2000}.

Remarkably, entanglement is a resource which, though does not carry
information itself, can help in such tasks as the reduction of
classical communication complexity
\cite{CBcom_cplx_1,BuhrmanCD1997-complexity,BruknerZPZ2002-Bell-complexity},
entanglement assisted orientation in space
\cite{BrukerPRV2005-orientation,BovinoGSV2006-orientation-exp}, quantum estimation of a damping constant \cite{VenzlF2007},
frequency standards improvement
\cite{Wineland,GiovannettiLM2004-sci-metrology,HuelgaFreqSt} (see in
this context \cite{BotoLithography}) and clock synchronization
\cite{JozsaADW2000-synchr}. The entanglement plays fundamental role in
quantum communication between parties separated by macroscopic
distances \cite {BDSW1996}.

Though the role of entanglement in quantum computational speed-up is
not clear \cite {KendonM2006}, it has played important role in development of quantum
computing, including measurement based schemes, one-way quantum
computing \cite{RaussendorfBriegelOneWay}\footnote{For comprehensive
  review see \cite {BrowneB}.}, linear optics quantum computing
\cite{KnillLM2001-lin-opt}\footnote{It has been shown that
  linear optics quantum computing can viewed as a measurement based
  one-way computing \cite{popescu-2006}.}. Entanglement has also given
us new insights for understanding many physical phenomena including
super-radiance \cite {Lambert}, superconductivity \cite{Vedral},
disordered systems \cite {DurHHLB} and emerging of classicality
\cite{Zurek03}. In particular, understanding the role of entanglement
in the existing methods of simulations of quantum spin systems allowed
for significant improvement of the methods, as well as understanding
their limitations \cite{Vidal,Vidal1,Verstraete,AndreasPD}. The role
of entanglement in quantum phase transitions
\cite{LarssonJ,OsborneN,Osterloh,Latorre,Rico,VerstraetePC2003-locent}
was intensively studied. Divergence of correlations at critical points
is always accompanied by divergence of a suitably defined entanglement
length \cite{VerstraetePC2003-locent}. The concept of entanglement
length originates form earlier paper by D. Aharonov, where a critical
phenomenon has been studied in context of fault-tolerant quantum
computing \cite{dorit-phase}.

Unfortunately quantum entanglement has three disagreeable but
interesting features: It has in general very complex structure, it is
very fragile to environment and it can not be increased on average
when systems are not in direct contact but distributed in spatially
separated regions. The theory of entanglement tries to give answers to
fundamental questions as: i) How to detect optimally entanglement
theoretically and in laboratory; ii) How to reverse an inevitable
process of degradation of entanglement; iii) How to characterize,
control and quantify entanglement.

History of the questions i) and ii) has its origin in seminal papers
by Werner and Popescu \cite {Werner1989,Popescu2}. Werner not only gave
accurate definition of separable states (those mixed states that are
not entangled), but also noted that there exist entangled states that
similarly to separable states, admit LHV model, hence do not violate
Bell inequalities. Popescu showed \cite{Popescu2} that having system
in such state, by means of local operations one can get a new state
whose entanglement can be detected by Bell inequalities. This idea was
developed by Gisin who used so-called filters to enhance violating of
Bell inequalities \cite {Gisin96}. In fact, this idea turned out to be
a trigger for a theory of entanglement manipulations
\cite{BBPSSW1996}.

Soon afterwards, Peres showed that if the state was
separable\footnote{More formal definitions of entangled and separable
  states are given in the next section.} then after {\it partial
  transpose}\footnote{The positive partial transpose condition is
  called the Peres criterion or the PPT criterion of separability.} of
density matrix on one of the subsystems of a compound bipartite system
it is still a legitimate state \cite {Peres}. Surprisingly Peres
condition appeared to be strong test for entanglement\footnote{For low
  dimensional systems it turned out to be necessary and sufficient
  condition of separability what is called Peres-Horodecki criterion
  \cite{sep1996}.}. As the partial transpose is a {\it positive map}
it was realized that the positive maps can serve as the strong
detectors of entanglement. However they cannot be implemented directly
in laboratory, because they are unphysical\footnote {See however \cite
  {PHAE}.}. Fortunately there is ``footbridge''-Jamio\l{}kowski
isomorphism \cite {Jamiolkowski} which allowed to go to physical
measurable quantities --- Hermitian's operators. This constitutes a
necessary and sufficient condition separability on the both, physical
level of observables and nonphysical one engaging positive maps \cite
{sep1996}. This characterization of entanglement although
nonoperational, provides a basis for general theory of detection of
entanglement.  A historical note is here in order.  Namely it turned
out that both the general link between separability and positive maps
as well as the Peres-Horodecki theorem were first known and expressed
in slightly different language as early as in 70s \cite{Choi1972}
(see also \cite{Osaka,Woronowicz,Stoermer}).  The rediscovery by Peres
and Horodecki's brought therefore powerful methods to entanglement
theory as well as caused revival of research on positive maps,
especially the so-called non-decomposable ones.

Terhal was the first
to construct a family of indecomposable positive linear maps based on
entangled quantum states \cite {Terhal2000-laa}. She also pointed out,
that a violation of a Bell inequality can formally be expressed as a
witness for entanglement \cite {TerhalB}\footnote {The term
  ``entanglement witness'' for operators detecting entanglement was
  introduced in \cite {TerhalB}.}.  Since then theory of entanglement
witnesses was intensively developed \cite {Lewenstein00a,TothG_s,
  GuehneHBE_s,Kiesel_s,BrussReflections,BrandaoV,Brandao2005-witent},
including a nonlinear generalization \cite {GuehneL1,GuehneLutkenhaus}
and the study of indistinguishable systems \cite
{SchliemannCKML-fermion,EckertSBL-fermion}.

The concept of entanglement witness was applied to different problems in statistical
systems \cite {Cavalcanti05,WiesniakVB-Magnetic,LidarWBS,BruknerVZ2006}, quantum
cryptography \cite {CurtyGLL} quantum optics \cite
{StobinskaW-teor-s,stobinska-2006-20}, condensed-matter nanophysics
\cite {BlaauboerD}, bound entanglement \cite {HyllusABM}, experimental realization of cluster states
\cite {VallonePMMB2007}, hidden nonlocality \cite {MasanesLD2007}. About this
time in precision experiments a detection of multipartite entanglement
using entanglement witness operators was performed \cite
{Bourennane,Roos,Knill,Blatt,BarbieriMNMAM,Zeilinger-witness,Mikami-witness,Altepeter,LuZGGZSGYP2007}.

As one knows, the main virtue of entanglement witnesses is that they
provide an economic way of detection of entanglement, that does not
need full (tomographic) information about the state. It arises a
natural question: how to estimate optimally the amount of entanglement
of compound system in an unknown state if only incomplete data in the
form of averages values of some operators detecting entanglement are
accessible? This question posed in 1998 lead to the new inference
scheme for all the processes where entanglement is measured. It
involves a {\it principle of minimization of entanglement under a
  chosen measure of entanglement with constrains in the form of
  incomplete set of data from experiment} \cite {Jaynes}. In
particular, minimization of entanglement measures (entanglement of
formation and relative entropy of entanglement) under fixed Bell-type
witness constraint was obtained. Subsequently  the inference scheme based
of the minimization of entanglement was successfully applied \cite
{Reimpell,EisertBA,PlenioA} to estimate entanglement measures from the
results of recent experiments measuring entanglement witnesses. This
results show clearly that the entanglement witnesses are not only
economic indicators of entanglement but they are also helpful in
estimating the entanglement content.

In parallel to entanglement witnesses the theory of positive maps was
developed which provides, in particular strong tools for the detection
of entanglement \cite
{Terhal2000-laa,Kossakowski,BenattiFPquan_dyn,MPiani,PianiM,ChruscinskiK,Hall2006,Breuer2006-prl,CavesDFS,reduction,CerfAG,Majewski}.
Strong inseparability criteria beyond the positive maps approach were
also found
\cite{Rudolph,HHH02-permut,MintertKB2004-multi,OGuehne,HofmannT,ChenWu,ClarisseW2005-permut,GuhneHGE2007,DeviPR2007}.

Separability criteria for continuous variables were also proposed (see
Sec. \ref{subsec:sep_cont_var}). The necessary and sufficient
condition for separability of Gaussian states of a bipartite system of
two harmonic oscillators has been found independently by Simon and Duan \emph{et al.}
\cite{Simon,DuanGCZ1999-criterion}.
 Both approaches are equivalent. Simon's
criterion is direct generalization of partial transpose to CV systems,
while Duan \emph{et al.} started with local uncertainty principles.

Soon afterwards \cite {WernerWolf} found bound entangled Gaussian
states. Since then theory of entanglement continuous variables have
been developed in many directions, especially for Gaussian states
which are accessible at the current stage of technology (see
\cite{BraunsteinP} and references therein). For the latter states the
problem of entanglement versus separability was solved completely:
operational necessary and sufficient condition was provided in
\cite{GaussianAlgorithm}. This criterion therefore detects all
bipartite bound entangled Gaussian states. Interestingly McHugh \emph{et al.} constructed
a large class of non-Gaussian two-mode continuous variable states for which separability criterion for Gaussian states can be employed \cite {McHughBZ2007}.

Various criteria for continuous variables were obtained \cite
{ManciniGVT2001-criterion,RaymerFSG-cont-var,AgarwalB-cont-var,HilleryZ-cont-var}.
A powerful separability criterion of bipartite harmonic quantum states
based on partial transpose was derived which includes all the above
criteria as a special cases \cite {ShchukinW, MiranowiczP}.

The present day entanglement theory owes its form in great measure to
the discovery of the concept of entanglement manipulation
\cite{Popescu2,BBPSSW1996}. It was realized \cite{BBPSSW1996} that a
natural class of operations suitable for manipulating entanglement is
that of local operations and classical communication (LOCC), both of
which cannot bring in entanglement for free. So established {\it
  distant lab} (or LOCC) paradigm plays a fundamental role in
entanglement theory. Within the paradigm many important results have
been obtained. In particular, the framework for pure state
manipulations have been established, including reversibility in
asymptotic transitions of pure bipartite states \cite{BBPS1996},
connection between LOCC pure state transitions and majorization theory
\cite{SchumacherN-1996-pra} as well as a surprising effect of catalysis
\cite{JonathanP,VidalC,VidalC-irre}. Moreover, inequivalent types of
multipartite entanglement have been identified
\cite{DurVC00-wstate,BPRST-mregs}.

Since in laboratory one usually meets mixed states representing {\it
  noisy entanglement}, not much useful for quantum information
processing, there was a big challenge: to reverse the process of
degradation of entanglement by means of some active manipulations.
Remarkably in \cite{BBPSSW1996} it was shown that it is possible to
distill pure entanglement from noisy one in asymptotic regime. It
should be noted that in parallel, there was intense research aiming at
protection of quantum information in quantum computers against
decoherence. As a result error correcting codes have been discovered
\cite{Shor,Steane-codes-prl}. Very soon it was realized that quantum
error correction and distillation of entanglement are in fact
inherently interrelated \cite{BDSW1996}.

A question fundamental for quantum information processing has then
immediately arisen: ``Can noisy entanglement be always purified?''. A
promising result was obtained in \cite{HHH1997-distill}, where all
two-qubit noisy entanglement were showed to be distillable. However,
soon afterwards a no-go type result \cite{bound} has revealed dramatic
difference between pure and noisy entanglement: namely, there exists
{\it bound entanglement}. It destroyed the hope that noisy
entanglement can have more or less uniform structure: instead we
encounter peculiarity in the structure of entanglement. Namely there
is {\it free} entanglement that can be distilled and the bound one -
very weak form of entanglement. Passivity of the latter provoked
intensive research towards identifying any tasks that would reveal its
quantum features. More precisely any entangled state that has so
called {\it positive partial transpose} cannot be distilled. First
explicit examples of such states were provided in
\cite{Pawel97}. Further examples (called bound entangled) were found
in \cite{BrussPeres2000, BennettUPBI1999,WernerWolf}; see also
\cite{DagBook,ClarissePhd} and references therein.

Existence of bound entanglement has provoked many questions including
relations to local variable model and role as well as long standing
and still open question of existence of bound entangled states
violating PPT criterion, which would have severe consequences for
communication theory.

Due to bound entanglement, the question {\it can every entanglement be
used for some useful quantum task?} had stayed open for a long
time. Only quite recently positive answer both for bipartite as well
as multipartite states was given by Masanes
\cite{Masanes1_activation,Masanes-multiactiv} in terms of so-called
activation \cite{activation}. This remarkable result allows to define
entanglement not only in negative terms of Werner's definition ({\it a
  state is entangled if it is not mixture of product states}) but also
in positive terms: {\it a state is entangled, if it is a resource for
  a nonclassical task}.

One of the most difficult and at the same time fundamental questions
in entanglement theory, is quantifying entanglement. Remarkably, two
fundamental measures, {\it entanglement distillation}
\cite{BBPS1996,BBPSSW1996,BDSW1996}, and what is now called {\it
  entanglement cost} \cite{BDSW1996,cost}, appeared in the context of
manipulating entanglement, and have an operational meaning. Their
definitions brings in mind thermodynamical analogy
\cite{popescu-rohrlich,PlenioVedral1998,termo}, since they describe
two opposite processes --- creation and distillation, which ideally
should be reverse of each other. Indeed, reversibility holds for pure
bipartite states, but fails for noisy as well as multipartite
entanglement. The generic gap between these two measures shows a
fundamental irreversibility
\cite{bound,YangHHS2005-cost,VidalC-irre}. This phenomenon has its
origin deeply in the nature of noisy entanglement, and its immediate
consequence is the non-existence of a unique measure of entanglement.

Vedral \emph{et al.} \cite{VPRK1997,PlenioVedral1998} have proposed an
axiomatic approach to quantifying entanglement, in which a ``good''
measure of entanglement is any function that satisfy some
postulates. The leading idea \cite{BDSW1996} is that entanglement
should not increase under local operations and classical communication
so called monotonicity condition. The authors proposed a scheme of
obtaining based on the concept on distance from separable states, and
introduced one of the most important measures of entanglement, the so
called {\it relative entropy of entanglement}
\cite{VPRK1997,PlenioVedral1998} (for more comprehensive review see
\cite{Vedral2002-rmp}). Subsequently, a general mathematical framework
for entanglement measures was worked out by Vidal
\cite{Vidal-mon2000}, who concentrated on the axiom of monotonicity
(hence the often used term ``entanglement monotone'').

At first sight it could seem that measures of entanglement do not
exhibit any ordered behavior. Eisert and Plenio have showed
numerically that entanglement measures do not necessarily imply the
same ordering of states \cite{EisertPlenio_order}. It was further
shown analytically in \cite{GrudkaMiranowicz_order}
However, it turns out that there are constraints for measures that
satisfy suitable postulates relevant in asymptotic regime. Namely, any
such measure must lie between two extreme measures that are just two
basic operational measures {\it distillable entanglement} $E_D$ and
{\it entanglement cost} $E_C$ \cite{limits}. This can be seen as
reflection of a more general fact, that {\it abstractly} defined
measures, provide bounds for {\it operational} measures which are of
interest as they quantify how well some specific tasks can be
performed.

The world of entanglement measures, even for bipartite states, still
exhibits puzzles. As an example may serve an amazing phenomenon of
locking of entanglement \cite{lock-ent}. For most of known bipartite
measures we observe a kind of collapse: after removing a single qubit,
for some states, entanglement dramatically goes down.

Concerning multipartite states, some bipartite entanglement measures
such as relative entropy of entanglement or robustness of entanglement
\cite{VidalT1998-robustness} easily generalize to multipartite
states. See \cite{EisertB2000-Schmidt,BarnumL2001-multi} for other
early candidates for multipartite entanglement measures.  In
multipartite case a new ingredient comes in: namely, one tries to
single out and quantify ``truly multipartite'' entanglement.  First
measure that reports genuinely multipartite properties is the
celebrated ``residual tangle'' of \cite{CoffmanKW-tangle}.  It is
clear, that in multipartite case even quantifying entanglement of pure
states is a great challenge. Interesting new schemes to construct
measures for pure states have been proposed
\cite{VerstraeteDM2001-normal,Miyake2002-hyper}.

Entanglement measures allowed for analysis of dynamical aspects of
entanglement\footnote{See \cite {AmicoFOV2007,Yu-Eberly05} and references therein.}, including entanglement decay under interaction with environment
\cite{ZyczkowskiHHH2000-dynamics,FicekT2006,Miranowicz2,EberlyY1,JakobczykJ2004,CarvalhoMPB2005_ent_decoh,DoddH2004,WangBPS2006,Montangero2004,Ban2006,ShrestaADH2005,MintertCKB2005-review,KimLAK2002,YiS1999,MaloyerK2006}
entanglement production, in the course of quantum computation
\cite{ParkerPlenio,KendonMunro} or due to interaction between
subsystems. The latter problem gave rise to a notion of ``entangling
power'' \cite{ZanardiZF2000-entpower,LindenSW2005-entpower} of a
unitary transformation, which can be seen as a higher-level
entanglement theory dealing with entanglement of operations, rather
than entanglement of states
\cite{EJPP02,LindenSW2005-entpower,CLP00,HarrowShor2005}.

Interestingly even two-qubit systems can reveal not trivial features of entanglement decay. Namely, the state of the two entangled qubits, treated with unitary dynamics and subjected to weak noise can reach the set of separable states (see Sec. \ref{sec:ent_as_prop}) in finite time while coherence vanishes asymptotically \cite{ZyczkowskiHHH2000-dynamics}. Such an effect was investigated in detail in more realistic scenarios \cite {FicekT2006,EberlyY1,JakobczykJ2004,DoddH2004,WangBPS2006,YuE2007,TolkunovPA2005,LastraRLSR2007,VaglicaV2007} (see also \cite{JordanSS-ent-change2007}). It was  called "sudden death of entanglement" \cite {EberlyY2} and demonstrated in the ingenious experiment \cite {AlmeidaMHSWRD2007} (see in this context \cite {SantosMDZ2006}).

Note that apart from entanglement decay
a positive effects of environment was investigated: entanglement generated by interference in the measurement process
\cite {CabrilloCGZ1999,BoseKPV1999}, cavity-loss-induced generation of entangled atoms \cite {PlenioHBK1999}, atom-photon entanglement conditioned by photon detection  \cite {Horodecki2001}, generation of entanglement from white noise \cite{PlenioH2002}, entanglement generated by interaction with a common heat bath \cite{Braun2002}, noise-assisted preparation of entangled atoms inside a cavity \cite{YuYZS2003}, environment induced entanglement in markovian dissipative dynamics \cite {BenattiFP2003}, entanglement induced by the spin chain \cite {YiCW2006}. It has been demonstrated \cite{LamataGC2007}, that
it is possible to achieve an arbitrary amount of entanglement between two atoms by using spontaneous emitted photons, linear optics and projective measurements.

One of the pillars of the theory of entanglement is its connection
with quantum cryptography, in particular, with its subdomain ---
theory of privacy, as well as a closely related classical privacy
theory \cite{GisinWolf_QKAvsCKA,GisinWolf_linking,Collins-Popescu}. It
seems that the most successful application of quantum entanglement is
that it provides basic framework for quantum cryptography (despite the
fact that the basic key distribution protocol BB84 does not use
entanglement directly.). This is not just a coincidence. It appears,
that entanglement is a quantum equivalent of what is meant by privacy.
Indeed the main resource for privacy is a secret cryptographic key:
correlations shared by interested persons but not known by any other
person. Now, in the {\it single} notion of entanglement two
fundamental features of privacy are encompassed in an ingenuous
way. If systems are in pure entangled state then at the same time (i)
systems are correlated and (ii) no other system is correlated with
them. The interrelations between entanglement and privacy theory are
so strong, that in many cases cryptographic terminology seems to be
the most accurate language to describe entanglement (see e.g.
\cite{DevetakWinter-hash}). An example of a back-reaction --- from
entanglement to privacy --- is the question of existence of bound
information as a counterpart of bound entanglement
\cite{GisinWolf_linking}. In fact, the existing of such phenomenon
conjectured by Gisin and Wolf was founded by \cite
{AcinCM-MultiBoundInfo} for multipartite systems.

The fact that entanglement represents correlations that cannot be
shared by third parties, is deeply connected with {\it monogamy} - the
basic feature of entanglement. In 1999 Coffman, Kundu and Wootters
first formalized monogamy in quantitative terms, observing that there
is inevitable trade-off between the amount of entanglement that a
qubit $A$ can share with a qubit $B_1$ and entanglement which same
qubit $A$ shares with some other qubit $B_2$
\cite{CoffmanKW-tangle}. In fact, already in 1996 the issue of
monogamy was touched in \cite{BDSW1996}, where it was pointed out that
no system can be EPR correlated with two systems at the same time,
which had direct consequences for entanglement distillation.  Monogamy
expresses unshareability of entanglement \cite {Terhal1}, being not
only central to cryptographic applications, but allows to shed new
light to physical phenomena in many body systems such as frustration
effect leading to highly correlated ground state (see
e.g. \cite{DawsonN2004-frustration}).

The entanglement was also investigated within the framework of special
relativity and quantum field theory \cite
{SummersW,AlsingM,CabanR,Czachor1,PeresST,Terno,JordanSS} (see
comprehensive review and references there in \cite {PeresTerno}). In
particular, entanglement for indistinguishable many body systems was
investigated in two complementary directions. The first (canonical)
approach bases on the tensor product structure and the canonical
decomposition for the states \cite
{SchliemannCKML-fermion,EckertSBL-fermion,PaskauskasY,LiZLL}, while
the second one is based on the occupation-number representation \cite
{Zanardi,ZanardiW}. Moreover Cirac and Verstraete \cite{VerCir03}
considered notion of entanglement in the context of superselection
rules more generally.  However it seems that there is still
controversy concerning what is entanglement for indistinguishable
particles \cite{WisVac03}. Recently, the group-theory approach to
entanglement has been developed in \cite{KobriczL_grupent2006} and its
connection with non-commutativity was found.

In general, the structure of quantum entanglement appears to be
complex (see in \cite {Gurvits1}) and many various parameters,
measures, inequalities are needed to characterize its different
aspects (see
\cite{BengtssonZyczkowski-book,Alber2001,BrussReview,BrussReflections,Eckert,TerhalReview,DagBook}).

Finally, the list of experiments dealing with
entanglement\footnote{Historically the first experimental realization
  of a two-qubit entangling quantum gate (CNOT) is due to
  \cite{WinelandFirstGate}.} quickly grows: "Entanglement over long
distances \cite {TittelBGZ1,TittelBGZ2,13,WeihsRWZ,33}, entanglement
between many photons \cite{34} and many ions \cite{35}, entanglement
of an ion and a photon \cite {36,37}, entanglement of mesoscopic
systems (more precisely entanglement between a few collective modes
carried by many particles) (\cite {38,39,40G}), entanglement swapping
\cite {exp_ent_swapping,42}, the transfer of entanglement between
different carriers (\cite {44}), etc." \cite {NG}. Let us add few
recent experiments: multiphoton path entanglement \cite {EisenbergHKB},
photon entanglement from semiconductor quantum dots
\cite{Akopian2006}\footnote{There was a controversy
whether in previous experiment of \cite{Stevenson-nature2005} entanglement was  actually detected  \cite{Avron-comment}}.
teleportation between a photonic pulse and an atomic ensemble
\cite{Polzik2006-foton-atom}, violation of the CHSH inequality
measured by two observers separated by 144~km \cite {144km}, purification of two-atom entanglement \cite {ReichleLKBBJLOSW2006},  increasing entanglement between Gaussian states
\cite{OurjoumtsevDTG2007}, creation of entangled six photon graph states \cite {LuZGGZSGYP2007}.

\section{ Entanglement as a quantum property of compound systems}
\label{sec:ent_as_prop}
We accustomed to the statement that on the fundamental level Nature
required quantum description rather than classical one. However the
full meaning of this and its all possible experimental and theoretical
implications are far from triviality \cite{Jozsa} In particular the
``effect'' of replacement of classical concept of phase space by
abstract Hilbert space makes a gap in the description of composite
systems. To see this consider multipartite system consisting of $n$
subsystems. According to classical description the total state space
of the system is the {\it Cartesian} product of the $n$ subsystem
spaces implying, that the total state is always a product state of the
$n$ separate systems. In contrast according to the quantum formalism the
total Hilbert space $H$ is a {\it tensor} product of the subsystem
spaces $H=\ot ^n_{l=1}H _l$. Then the superposition principle allows
us to write the total state of the system in the form:
\begin{equation}
  | \psi \>  = \sum_{{ \bf i}_n} c_{{ \bf i}_n}|{{ \bf i}_n}\>,
  \label{jeden}
\end{equation}
where ${{ \bf i}_n} = i_1,i_2...i_n$ is the multiindex and $|{{ \bf
    i}_n}\>= |{i_1}\> {\ot} |{i_2}\>...{\ot}|{i_n}\>$, which cannot
be, in general, described as a product of states of individual
subsystems\footnote{Sometimes instead of notation $|\psi\rangle \ot
  |\phi\rangle$ we use simply $|\psi\rangle |\phi\rangle$ and for
  $|i\rangle \ot |j\>$ even shorter $|ij\rangle$.} $| \psi
\>\neq|\psi_{1}\>\ot|\psi_{2}\>\ot \cdots \ot |\psi_{n}\>$.

This means, that it is in general not possible, to assign a single
state vector to any of $n$ subsystems. It express formally a
phenomenon of entanglement, which in contrast to classical
superposition, allows to construct exponentially large superposition
with only linear amount of physical resources. It is just what allows
to perform nonclassical tasks. The states on left hand side (LHS) of
(\ref{jeden}) appear usually as a result of direct physical
interactions. However, the entanglement can be also generated
indirectly by application of the projection postulate (entanglement
swapping).

In practice we encounter mixed states rather than pure. Entanglement
of mixed states is not longer equivalent to being non-product, as in
the case of pure states. Instead, one calls a mixed state of $n$
systems entangled if it cannot be written as a convex combination of
product states\footnote{Note, that classical probability distributions
  can \emph{always} be written as mixtures of product distributions.}
\cite{Werner1989}:
\begin{equation}
\rho\not= \sum_i p_i \rho^i_1 \ot \ldots \ot \rho^i_n.
\end{equation}

The states which are {\it not} entangled in the light of the above
definition are called {\it separable}. In practice it is hard to
decide if a given states is separable or entangled basing on the
definition itself. Therefore one of the fundamental problems
concerning entanglement is the so called {\it separability} problem
(see Secs. \ref{sec:BipartiteEntanglement}-\ref{sec:ent_geom}).

It should be noted in the above context that the active definition of entanglement states was proposed recently.  Namely entangled states
{\it are the ones that cannot be simulated by classical correlations}
\cite{MasanesLD2007}.
This interpretation defines entanglement in terms of the behavior of the states rather than in terms of preparation of the states.

{\it Example. }for bipartite systems the Hilbert space
$\textit{H}=\textit{H}_1 \otimes\textit{H}_2$ with $\dim H_1=\dim
H_2=2$ is spanned by the four Bell-state entangled basis
\begin{equation}
\left|\psi^\pm\right\rangle=\frac{1}{\sqrt{2}}(\left|0\right\rangle \left|1\right\rangle\pm \left|1\right\rangle \left|0\right\rangle) \quad
\left|\phi^\pm\right\rangle=\frac{1}{\sqrt{2}}(\left|0\right\rangle \left|0\right\rangle\pm \left|1\right\rangle \left|1\right\rangle).
\label{basis}
\end{equation}

These states (called sometimes \eprstates) have remarkable properties
namely if one measures only at one of the subsystems one finds it with
equal probability in state $\left|0\right\rangle$ or state
$\left|1\right\rangle$.  However, the result of the measurements for
both subsystems are perfectly correlated. This is just feature which
was recognized by Schr{\"o}dinger: we know nothing at all about the
subsystems, although we have maximal knowledge of the whole system
because the state is pure (see Sec. \ref{sec:entropic}).  There is
another holistic feature, that unitary operation of only one of the
two subsystems suffices to transform from any Bell states to any of
the other three states. Moreover, Braunstein \emph{et al.}  showed
that the Bell states are eigenstates of the Bell operator (\ref{B})
and they maximally violate Bell-CHSH inequality (\ref{Tr}) (see
Sec. \ref{sec:Bell}) \cite {Braunstein2}.

The Bell states are special cases of bipartite maximally entangled
states on Hilbert space $C^d\ot C^d$ given by
\be
|\psi\>=U_A \ot U_B |\Phiplusd\>_{AB}
\ee
where
\be
|\Phiplusd\>={1\over \sqrt d}\sum_{i=1}^d |i\>|i\>
\ee
is the "canonical" maximally entangled state.
In this paper, a maximally entangled state will be  also
called EPR (Einstein-Podolsky-Rosen) state or singlet state, since it is equivalent to the true singlet state up to local unitary transformations (for $d = 2$ we call it also e-bit). We will also often drop the index $d$.

The question whether a mixture of Bell states is still entangled, is
quite nontrivial. Actually it is the case if and only if one of
eigenvalues is greater than $\frac{1}{2}$ (see
Sec. \ref{sec:BipartiteEntanglement}).

So far the most widely used source of entanglement are
entangled-photon states produced by nonlinear process of parametric
down-conversion of type I or of type II corresponding to weather
entangled photons of the down-conversion pair are generated with the
same polarization or orthogonal polarization respectively. In
particular using parametric down-conversion one can produce Bell-state
entangled basis (\ref {basis}).  There are also many other sources of
entangled quantum systems, for instance: entangled photon pairs from
calcium atoms \cite {Kocher}, entangled ions prepared in
electromagnetic Paul traps \cite {MeekhofMKIW-ion}, entangled atoms in
quantum electrodynamic cavities \cite {RaimondBH}, long-living
entanglement between macroscopic atomic ensembles \cite {PolzikJK,HaldSSP1999}, entangled microwave photons from quantum
dots \cite {EmaryTB-dot},
entanglement between nuclear spins within a single molecule \cite
{ChenHLL-nuc-spin} entanglement between light and an atomic ensembles
\cite {MuschikHPC-light}.

In the next section, we will present briefly pioneering
entanglement-based communications schemes using Bell entangled states.

\section{Pioneering effects based on entanglement}
\label{sec:effects}

\subsection{Quantum key distribution based on entanglement}
\label{subsec:Ekert}

The first invention of quantum information theory, which involves
entanglement, was done by A. Ekert \cite {E91}. There were two well
known facts: existence of highly correlated state\footnote{This state
  is also referred to as {\it singlet}, {\it EPR state} or {\it EPR
    pair}. If not explicitly stated, we will further use these names to
    denote any maximally entangled state, also in higher dimensions,
    see section \ref{subsubsec:Witnesses}.
    }
\begin{equation}
|\psi^-\> = {1\over \sqrt{2}}(|0\>|1\> - |1\>|0\>)
\label{eq:singlet_dense}
\end{equation}
and Bell inequalities (violated by these states).

Ekert showed that if put together, they become useful producing
private cryptographic key. In this way he discovered entanglement
based quantum key distribution as opposed to the original BB84 scheme
which uses directly quantum communication.  The essence of the
protocol is as follows: Alice and Bob can obtain from a source the EPR
pairs.  Measuring them in basis $\{|0\>,|1\>\}$, Alice and Bob obtain
string of perfectly (anti)correlated bits, i.e. the key. To verify
whether it is secure, they check Bell inequalities on selected portion
of pairs.  Roughly speaking, if Eve knew the values that Alice and Bob
obtain in measurement, this would mean that the values exist before
the measurement, hence Bell's inequalities would not be
violated. Therefore, if Bell inequalities are violated, the values do
not exist before Alice and Bob measurement, so it looks like nobody
can know them.\footnote{ In fact, the argument is more subtle. This is
  because in principle the values which did not preexist could come to
  exist in a way that is immediately available to a third party - Eve,
  i.e. the values that were not known to anybody could happen to be
  known to everybody when they come to exist.  To cope with this
  problem Ekert has used the fact that singlet state can not be
  correlated with any environment. Recently it turned out that one can
  argue basing solely on Bell inequalities by means of so called
  monogamy of nonlocal correlations
  \cite{AcinGM-bellqkd,BKP_Bellcrypto,BHK_Bell_key,MW_Bellcrypto}.}
  First implementations of Ekert's cryptography protocol has been performed
  using polarization-entangled photons from spontaneous parametric down-conversion \cite {NaikPWBK} and entangled photons in energy-time \cite {TittelBZG}.

After Ekert's idea, the research in quantum cryptography could have
taken two paths. One was to treat violation of Bell inequality merely
as a confirmation that Alice and Bob share good \eprstates, as put
forward in \cite{BBM92}, because this is sufficient for privacy: if
Alice and Bob have true \eprstate, then nobody can know results of
their measurements. This is what actually happened, for a long time
only this approach was developed. In this case the eavesdropper Eve
obeys the rules of quantum mechanics. We discuss it in
Sec. \ref{sec:Ent_in_QKD}.  The second path was to treat \eprstate\ as
the source of strange correlations that violate Bell inequality (see
Sec. \ref{sec:Bell}). This leads to new definition of security:
against eavesdropper who do not have to obey the rules of quantum
mechanics, but just the no-faster-then-light communication
principle. The main task of this approach, which is an unconditionally
secure protocol has been achieved only recently
\cite{Nonsig_theories,BHK_Bell_key,MW_Bellcrypto}.

\subsection{Quantum dense coding}
\label{subsec:Dense}

In quantum communication there holds a reasonable bound on the
possible miracles stemming from quantum formalism. This is the {\it
  Holevo bound} \cite {Holevo}. Roughly speaking it states, that {\it
  one qubit can carry at most only one bit of classical information}.
In 1992, Bennett and Wiesner have discovered a fundamental primitive,
called dense coding, which can come around the Holevo bound. Dense
coding allows to communicate {\it two} classical bits by sending one a
priori {\it entangled} qubit.

Suppose Alice wants to send one of four messages to Bob, and can send
only one qubit. To communicate {\it two} bits sending one qubit she
needs to send a qubit in one of $2^2=4$ states. Moreover the states
need to be mutually orthogonal, as otherwise Bob will have problems
with discriminating them, and hence the optimal bound $2$ will not be
reached. But there are only {\it two orthogonal states} of {\it one
  qubit}. Can entanglement help here? Let Alice and Bob instead share
a \eprstate. Now the clever idea comes: it is not that qubit which is
sent that should be in four orthogonal state, but the pair of
entangled qubits together. Let us check how it works. Suppose Alice
and Bob share a singlet state (\ref {eq:singlet_dense}). If Alice
wants to tell Bob one of the four events $k\in\{0,1,2,3\}$ she rotates
her qubit (entangled with Bob) with a corresponding transformation
$\sigma_k$:
\begin{eqnarray}
\sigma_{0}=\left[ \bea{cc}
1 & 0  \\
0 & 1 \\
\eea
\right],\,\,
\sigma_{1}=\left[
\bea{cc}
0 & 1  \\
1 & 0 \\
\eea
\right],\,\,
\nonumber \\
\sigma_{2}=\left[
\bea{cc}
1 & 0  \\
0 & -1 \\
\eea \right],\,\, -i\sigma_{3}=\left[ \bea{cc}
0 & -1  \\
1 & 0 \\
\eea
\right].
\label{eq:Pauli_rotations}
\end{eqnarray}
The singlet state (\ref{eq:singlet_dense}) rotated by $\sigma_k$ on
Alice's qubit becomes the corresponding $|\psi_k\>$ Bell
state\footnote{ In correspondence with Bell basis defined in
  Eq. (\ref{basis}) we have here
  $|\psi_0\>=|\psi^-\>,\,|\psi_1\>=|\phi^-\>,\, |\psi_2\>=|\psi^+\>,\,
  |\psi_3\>=|\phi^+\>$.}.  Hence $|\psi_k\>=[\sigma_k]_A\ot I_B
|\psi_0\>$ is {\it orthogonal } to $|\psi_{k'}\>=[\sigma_{k'}]_A\ot
I_B |\psi_0\>$ for $k\neq k'$ because Bell states are mutually
orthogonal. Now if Bob gets Alice's half of the entangled state, after
rotation he can discriminate between $4$ Bell states, and infer
$k$. In this way Alice sending one qubit has given Bob $\log 4 = 2$
bits of information.

Why does not this contradict the Holevo bound? This is because the
communicated qubit was a priori {\it entangled} with Bob's qubit. This
case is not covered by Holevo bound, leaving place for this strange
phenomenon. Note also that as a whole, two qubits have been sent: one
was needed to share the \eprstate. One can also interpret this in the
following way: sending first half of singlet state (say it is during
the night, when the channel is cheaper) corresponds to sending one bit
of {\it potential communication}. It is thus just as {\it creating the
  possibility} of communicating $1$ bit in future: at this time Alice
may not know what she will say to Bob in the future. During day, she
knows what to say, but can send only one qubit (the channel is
expensive). That is, she sent only one bit of {\it actual
  communication}. However sending the second half of singlet as in
dense coding protocol she uses both bits: the {\it actual} one and
{\it potential} one, to communicate in total $2$ classical bits. Such
an explanation assumes that Alice and Bob have a good quantum memory
for storing \eprstates, which is still out of reach of current
technology. In the original dense coding protocol, Alice and Bob share
pure maximally entangled state.  The possibility of dense coding based
on partially entangled pure and mixed states in multiparty settings
were considered in
\cite{BarencoE1995-dense,HausladenJSWW1996-cl-cap,mixed_dense_coding,not_so_super_dc,BosePV_1998dense,ZimanB2003_dense}
(see Sec. \ref{subsec:dense-cap}. The first experimental implementation of quantum dense coding was performed in Innsbruck \cite {Dense_practical}, using as a source of polarization-entangled photons (see further experiments with using nuclear magnetic resonance \cite{FangZFDM1999}, two-mode squeezed vacuum state \cite {MizunoWFS2005} and  controlled dense coding with EPR state for continuous variable \cite {JietaiJYFXCK2003}).

\subsection{Quantum teleportation}
\label{subsec:Telep}

Suppose Alice wants to communicate to Bob an {\it unknown quantum
  bit}. They have at disposal only a classical telephone, and one pair
of entangled qubits. One way for Alice would be to measure the qubit,
guess the state based on outcomes of measurement and describe it to
Bob via telephone. However, in this way, the state will be transferred
with very poor fidelity. In general an {\it unknown} qubit can not be
described by classical means, as it would become cloneable, which
would violate the main principle of quantum information theory: {\it a
  qubit in an unknown quantum state cannot be cloned}
\cite{WoottersZ-cloning,Dieks-cloning}.

However, Alice can send the qubit to Bob at the price of
simultaneously {\it erasing it} at her site. This is the essence of
{\it teleportation}: a quantum state is transferred from one place to
another: not copied to other place, but {\it moved} to that place. But
how to perform this with a pair of maximally entangled qubits?
Bennett, Brassard, Crepeau Jozsa, Peres and Wootters have found the
answer to this question in 1993 \cite{Teleportation}.

To perform teleportation, Alice needs to measure her qubit and part of
a maximally entangled state. Interestingly, this measurement is itself
entangling: it is projection onto the basis of four Bell states
(\ref{basis}). Let us follow the situation in which she wants to
communicate a qubit in state $|q\>=a|0\>+b|1\>$ on system $A$ with use
of a singlet state residing on her system $A'$ and Bob's system
$B$. The total initial state which is
\begin{equation}
|\psi_{AA'B}\> =
|q\>_A\ot{1\over \sqrt{2}}[|0\>|0\> + |1\>|1\>]_{A'B}
\end{equation}
can be written using the Bell basis (\ref {basis}) on the system $AA'$
in the following way:
\begin{eqnarray}
  |\psi_{AA'B}\> &=& {1\over 2}[|\phi^+\>_{AA'}(a|0\>_{B} + b|1\>_{B})\nonumber\\
  &&+|\phi^-\>_{AA'}(a|0\>_{B} - b|1\>_{B}) \nonumber \\
  &&+|\psi^+\>_{AA'}(a|1\>_{B} + b|0\>_{B})\nonumber \\
  &&+ |\psi^-\>_{AA'}(a|1\>_{B} - b|0\>_{B})].
\end{eqnarray}
Now when Alice measures her systems $AA'$ in this basis, she induces
equiprobably the four corresponding states on Bob's system. The
resulting states on system $B$ are very similar to the state of qubit
$|q\>$ which Alice wanted to send him. They however admix to the
initial state of system $B$. Thus Bob does not get any information
instantaneously. However the output structure revealed in the above
equation can be used: now Alice tells to Bob her result via
telephone. Accordingly to those two bits of information (which of Bell
states occurred on $AA'$) Bob rotates his qubit by one of the four
Pauli transformations (\ref{eq:Pauli_rotations}).  This is almost the
end. After each rotation, Bob gets $|q\>$ at his site. At the same
time, Alice has just one of the Bell states: the systems $A$ and $A'$
becomes entangled after measurement, and no information about the
state $|q\>$ is left with her. That is, the {\it no cloning} principle
is observed, while the state $|q\>$ was transferred to Bob.

There is a much simpler way to send qubit to Bob: Alice could just
send it directly. Then however, she has to use a quantum channel, just
at the time she wants to transmit the qubit. With teleportation, she
might have send half of EPR pair at any earlier time, so that only
classical communication is needed later.

This is how quantum teleportation works in theory. This idea has been
developed also for other communication scenarios (see
\cite{DurCirac-telep}). It became immediately an essential ingredient
of many quantum communication protocols. After pioneering experiments
\cite{Rzym-teleportation,Wieden-teleportation, Kimble-teleportation},
there were beautiful experiments performing teleportation in different
scenarios during last decade (see
e.g. \cite{13,Riebe-telep2004,Nielsen-telep1998,Barrett-telep2004,Ursin-telep2004}).
For the most recent one with mesoscopic objects, see
\cite{Polzik2006-foton-atom}.

\subsection{Entanglement swapping}
\label{subsec:Swapping}

Usually quantum entanglement originates in certain {\it direct
  interaction} between two particles placed close together. Is it
possible to get entangled (correlated in quantum way) some two
particles which have never interacted in the past? The answer is
positive \cite{YurkeStoller_ent_swapping,Teleportation,ent_swapping}.

Let Alice share a maximally entangled state $|\phiplus\> = {1\over
  {\sqrt 2}}(|00\> + |11\>)_{AC}$ with Clare, and Bob share the same state
with David:
\begin{equation}
|\phiplus\>_{AC}\ot|\phiplus\>_{BD}.
\end{equation}
Such state can be obviously designed in such a way, that particles $A$
and $D$ have never seen each other. Now, Clare and Bob perform a {\it
  joint} measurement in Bell basis. It turns out that for any outcome,
the particles $A$ and $D$ collapse to some Bell state. If Alice and
Bob will get to know the outcome, they can perform local rotation, to
obtain entangled state $\Phi^+_{AD}$.  In this way particles of Alice
and David are entangled although they never interacted directly with
each other, as they originated from different sources.

One sees that this is equivalent to teleporting one of the \eprpair\
through the second one. Since the protocol is symmetric, any of the
pairs can be chosen to be either the channel, or the teleported pair.

This idea has been adopted in order to perform {\it quantum repeaters}
\cite{repeaters}, to allow for distributing entanglement --- in
principle --- between arbitrarily distant parties. It was generalized
to a multipartite scenario in \cite{BoseWKt}.  Swapping can be used as
a tool in multipartite state distribution, which is for example useful
in quantum cryptography.

The conditions which should be met in optical implementation of
entanglement swapping (as well as teleportation) have been derived in
\cite{ent_swapping}. Along those lines entanglement swapping was
realized in lab \cite{exp_ent_swapping}.

\subsection{Beating classical communication complexity bounds with entanglement}
\label{subsec:Boundent}

Yao \cite{Yao_complexity} has asked the following question: how much
communication is needed in order to solve a given problem distributed
among specially separated computers? To be more concrete, one can
imagine Alice having got the $n$-bit string $x$ and Bob having got
$n$-bit string $y$. Their task is to infer the value of some a priori
given function $f(x,y)$ taking value in $\{0,1\}$, so that finally
both parties know the value. The minimal amount of bits needed in
order to achieve this task is called the {\it communication
  complexity} of the function $f$.

Again, one can ask if entanglement can help in this case. This
question was first asked by Cleve and Buhrman \cite{CBcom_cplx_1} and
in a different context independently by Grover \cite{Gcom_cplx_1}, who
showed the advantage of entanglement assisted distributed computation
over the classical one.

Consider the following example which is a three-party version of the
same problem \cite{BuhrmanCD1997-complexity}. Alice Bob and Clare get
two bits each: $(a_1,a_0)$ $(b_1,b_0)$ and $(c_1,c_0)$ representing a
binary two-digit numbers $a=a_1a_0$, $b=b_1b_0$ and $c=c_1c_0$. They
are promised to have
\begin{equation}
  a_0\oplus b_0 \oplus c_0 = 0.
 \label{eq:promise}
\end{equation}
The function which they are to compute is given by
\begin{equation}
f(a,b,c) = a_1\oplus b_1\oplus c_1\oplus (a_0 \vee b_0 \vee c_0).
\end{equation}
It is easy to see, that announcing of four bits is sufficient so that
all three parties could compute $f$. One party announces both bits
(say it is Alice) $a_1a_0$. Now if $a_0=1$, then the other parties
announces their first bits $b_1$ and $c_1$.  If $a_0=0$, then one of
the other parties, (say Bob) announces $b_1\oplus b_0$ while Clare
announces $c_1$. In both cases all parties compute the function by
adding announced bits modulo 2. Thus four bits are enough. A bit more
tricky is to show that four bits are {\it necessary}, so that the
classical communication complexity of the above function equals four
\cite{BuhrmanCD1997-complexity}.

Suppose now, that at the beginning of the protocol Alice Bob and Clare
share a quantum three-partite entangled state:
\begin{equation}
|\psi_{ABC}\> = {1\over 2}[|000\> - |011\> - |101\> - |110\>]_{ABC},
\label{eq:state_cplx}
\end{equation}
such that each party holds a corresponding qubit. It is enough to
consider the action of Alice as the other parties will do the same
with their classical and quantum data respectively.

Alice checks her second bit. If $a_0=1$ she does a
Hadamard\footnote{Hadamard transformation is a unitary transformation
  of basis which changes $|0\>$ into ${1\over \sqrt{2}}(|0\> + |1\>)$
  and $|1\>$ into ${1\over \sqrt{2}}(|0\> - |1\>)$.} on her
qubit. Then she measures it in $\{|0\>, |1\>\}$ basis noting the
result $r_A$. She then announces a bit $a_1\oplus r_A$.

One can check, that if all three parties do the same, the $r-$bits
with ``quantum origin'' fulfills $r_A\oplus r_B\oplus r_C =a_0\vee
b_0\vee c_0$. It gives in turn
\begin{eqnarray}
(a_1\oplus r_A)\oplus(b_1\oplus r_B)\oplus(c_1\oplus r_C)= \nonumber \\
a_1\oplus b_1\oplus c_1 \oplus (r_A\oplus r_B\oplus r_C)= \nonumber \\
a_1\oplus b_1 \oplus c_1 \oplus(a_0\vee b_0\vee c_0) = f(a,b,c).
\end{eqnarray}

Thus three bits are enough. The fourth, controlled by common quantum
entanglement, got hidden in the three others, hence need not be
announced.

One can bother that the quantum state (\ref{eq:state_cplx}) should
enter the communication score, as it needs to be distributed among the
parties.  Thus in addition to $3$ bits they need to send qubits.
However similarly as in case of dense coding, the entangled state is
independent of the function, i.e. it can be prepared prior to the
knowledge of the form of the function.

It is known, that if exact evaluation of the function is required (as
in the described case), sharing additionally correlated bit-strings
(shared randomness) by the parties can not help them. Thus we see,
that entanglement which can serve itself as a source of shared
randomness, is a stronger resource. If one allows nonzero probability
of error, then classical correlations can help, but it was shown in a
bipartite scenario, that entanglement allows for smaller error.
\cite{BuhrmanCD1997-complexity}. Entanglement can beat classical
correlations even exponentially as shown by R. Raz
\cite{raz99exponential}. Other schemes with exponential advantage are
provided in \cite{GaKaKeWolf-class-cmplx-gap}. They are closely
related to the phenomenon called {\it quantum fingerprinting}
\cite{Qfprint,GaKeWo-fingerprint}.

Interestingly, though the effect is of practical importance, in its
roots there is purely philosophical question of type: ``Is the Moon
there, if nobody looks?'' \cite{MerminMoon}. Namely, if the outcomes
of the Alice, Bob and Clare measurements would exist prior to
measurement, then they could not lead to reduction of complexity,
because, the parties could have these results written on paper and
they would offer three bit strategy, which is not possible. As a
matter of fact, the discoveries of reduction of communication
complexity used GHZ paradox in Mermin version, which just says that
the outcomes of the four possible measurements given by values of
$a_0$, $b_0$ and $c_0$ (recall the constraint $a_0 \oplus b_0 \oplus
c_0$) performed on the state (\ref{eq:state_cplx}) cannot preexist.

In \cite{BruknerZPZ2002-Bell-complexity} it was shown that this is
quite generic: for {\it any} correlation Bell inequality
for $n$ systems,  a state violating the inequality allows to reduce
communication complexity of some problem. For recent experiment see
\cite{Exper_cplx}.

\section{Correlation manifestations of entanglement: Bell inequalities.}
\label{sec:Bell}

\subsection {Bell theorem: CHSH inequality.}

The physical consequences of the existence of entangled (inseparable)
states are continuously subject of intensive investigations in the
context of both EPR paradox and quantum information theory. They
manifest itself, in particular, in correlation experiments via Bell
theorem which states that the probabilities for the outcomes obtained
when suitably measuring some quantum states cannot be generated from
classical correlations. As a matter of fact, Bell in his proof assumed
perfect correlations exhibited by the singlet state. However, in real
experiments such correlations are practically impossible. Inspired by
Bell's paper, Clauser, Horne, Shimony and Holt (CHSH) \cite {CHSH}
derived a correlation inequality, which provides a way of experimental
testing of the LHVM as independent hypothesis separated from quantum
formalism. Consider correlation experiment in which the variables
$({A}_1,{A}_2)$ are measured on the one of the subsystems of the whole
system while $({ B}_1,{B}_2)$ on the other one, and that the
subsystems are spatially separated. Then the LHVM imposes the
following constraints on the statistics of the measurements on the
sufficiently large ensemble of systems\footnote{It is assumed here
  that the variables $A_i$ and $B_i$ for $i=1,2$ are dichotomic
  i.e. have values $\pm 1$.}
\begin{equation}
|E(A_1,B_1)+E(A_1,B_2)+E(A_2,B_1)-E(A_2,B_2)|\ \leq 2, \label{trzy}
\end{equation}
where $E(A_i,B_j)$ is the expectation value of the correlation
experiment $A_iB_j$.

This is the CHSH inequality that gives a bound on any LHVM. It
involves only two-partite correlation function for two alternative
dichotomic measurements and it is complete in a sense that if full set
of such inequalities is satisfied there exists a joint probability
distribution for the outcomes of the four observables, which returns
the measured correlation probabilities as marginals \cite {Fine}.

In quantum case the variables convert into operators and one can
introduce the CHSH operator
\begin{equation}
\bcal_{CHSH}= \textbf{A}_1 \otimes
(\textbf{B}_1+ \textbf{B}_2) + \textbf{A}_2\otimes (\textbf{B}_1 -
\textbf{B}_2), \label{B}
\end{equation}
where $\textbf{A}_1=\textbf{a_1}\cdot\textbf{\sigma}$,
$\textbf{A}_2=\textbf{a_2}\cdot\textbf{\sigma}$, similarly for
$\textbf{B}_1$ and $\textbf{B}_2$, $\textbf{\sigma}= (\sigma_x,
\sigma_y, \sigma_z)$ is the vector of Pauli operators, $\textbf{a}=
(a_x,a_y,a_z)$ etc., are unit vectors describing the measurements that
the parties: $A$ (Alice) and $B$ (Bob) perform. Then, the CHSH
inequality requires that the condition
\begin{equation}
|{\tr}(\bcal_{CHSH}\rho)|\leq2 \label{Tr}
\end{equation}
is satisfied for all states $\rho$ admitting a LHVM.

Quantum formalism predicts Tsirelson inequality \cite {Cirel}
\begin{equation}
 |\<\bcal_{CHSH}\>_{QM}| =|{\tr}(\bcal_{CHSH}\rho )|\leq 2\sqrt{2}
\label{eq:cztery}
\end{equation}
for all states $\rho $ and all observables ${\textbf{A}}_1,{
  \textbf{A}}_2,{\textbf{B}}_1,{\textbf{B}}_2$.  Clearly, the CHSH
inequality can be violated for some choices of observables, implying
violation of LHVM.  For singlet state $\rho = |\psi^-\>\<\psi^-|$,
there is maximal violation $|{\tr}(\bcal_{CHSH}\rho )| = 2\sqrt{2}$
which corresponds to Tsirelson bound.

\subsection {The optimal CHSH inequality for $2\times2$ systems}

At the beginning of 90's there were two basic questions: Firstly, it
was hard to say whether a given state violates the CHSH inequality,
because one has to construct a respective Bell observable for it. In
addition, given a mixed state, there was no way to ensure whether ii
satisfies the CHSH inequality for each Bell observable.  This problem
was solved completely for an arbitrary quantum states $\rho $ of two
qubits by using Hilbert-Schmidt space approach \cite
{HHH1995-bell}. Namely, $n$-qubit state can be written as
\begin{equation}
\rho
={1\over 2^n}\sum^3_{i_1...i_n=0}t_{i_1...i_n} \sigma^1_{i_1}\otimes
\cdots\otimes \sigma^n_{i_n}, \label{stan}
\end{equation}
where $\sigma _0^k$ is the identity operator in the Hilbert space of
qubit $k$, and the $\sigma_{i_k}^k$ correspond to the Pauli operators
for three orthogonal directions $i_k=1,2,3$. The set of real
coefficients $t_{i_1...i_n}= {\tr}[\rho (\sigma _{i_1}\otimes
\cdots\otimes \sigma _{i_n})]$ forms a correlation tensor $T_\rho$. In
particular for the two-qubit system the $3\times 3$ dimensional tensor
is given by $t_{ij}:={\tr}[ \rho (\sigma_{i} \otimes \sigma_{j})]$ for
$i,j=1,2,3$.

In this case one can compute the mean value of an arbitrary
$\bcal_{CHSH}$ in an arbitrary fixed state $\rho$ and then maximize
it with respect to all $\bcal_{CHSH}$ observables. In result we
have: ${\max}_{\bcal_{CHSH}}|{\tr}(\bcal_{CHSH}\rho )|=
2\sqrt{M(\rho)}=2\sqrt{t^2_{11} +t^2_{22}}$, where  $t^2_{11}$ and
$t^2_{22}$ are the two largest eigenvalues of $T^T_\rho T_\rho $,
$T^T_\rho $ is the transposed of $T_\rho $.

It follows that for $2\times2$ systems the necessary and sufficient
criterion for the violation of the CHSH inequality can be written in
the form
\begin{equation}
 M(\rho )> 1. \label{M}
\end{equation}
The quantity $M(\rho )$ depends only on the state parameters and
contains all the information that is needed to decide whether a state
violates a CHSH inequality. Note that using $M(\rho )$ one can define
a measure of violation of the CHSH inequality $B(\rho )=\sqrt
{\max\{0,\sqrt{M(\rho )}-1\}}$ \cite {Miranowicz}, which for an
arbitrary two-qubit pure state equals to measures of entanglement:
{\it negativity} and {\it concurrence} (see Sec. \ref{par:concurrence}).
Clearly, in general, both $M(\rho )$ and $B(\rho )$ are not measures
of entanglement as they do not behave monotonously under LOCC.
However, they can be treated as entanglement parameters, in a sense
that they characterize a degree of entanglement. In particular the
inequality (\ref{M}) provides practical tool for the investigation of
nonlocality of the arbitrary two qubits mixed states in different
quantum-information contexts (see e.g. \cite{Z,Scarani1,Hyllus}).

\subsection {Nonlocality of quantum states and LHV model}

\subsubsection{Pure states}
\label{subsec:pure}

The second more fundamental question was: {\it Are there many quantum
  states, which do not admit LHV model?} more precisely: {\it which quantum states do not admit LHV model?}
  Even for pure states the problem is not completely solved.

Gisin proved that for the standard (ie. nonsequential) projective
measurements the only pure 2-partite states which do not violate correlation CHSH inequality
(\ref{trzy}) are product states and they are obviously local \cite
{Gisin,GisinP1992}. Then Popescu and Rohrlich showed that any n-partite pure entangled state can always be projected onto a 2-partite pure entangled state by projecting $n-2$ parties onto appropriate local pure states \cite {Popescu}. Clearly, it needs
an additional manipulations (postselection). Still the problem whether Gisin theorem, can be generalized without postselection for an arbitrary n-partite pure entangled states, remains open. In the case of three parties there {\it are} generalized GHZ states \cite {ZukowskiW,Scarani} that do not violate the MABK inequalities \cite {M,Ardehali,Belinskii} (see next subsec.). More generally it has been shown \cite {ZukowskiW} that these states do not violate any Bell inequality for n-partite correlation functions for experiments involving two dichotomic observables per site. Acin {\it et al.} and Chen {\it et al.} considered a Bell inequality that shows numerical evidence that all three-partite pure entangled state violate it \cite {AcinCGKKOZ2004,ChenWKO2004}.
 Recently a stronger Bell inequalities with more measurement settings was presented \cite {LaskowskiPZB2004,WuZ2003} which can be violated by a wide class of states, including the generalized GHZ states (see also \cite{ChenAF2006}).

\subsubsection{Mixed states}
\label{subsec:Mixed}

In the case of noisy entangled states the problem appeared to be much
more complex. A natural conjecture was that only separable mixed
states of form (\ref{eq:sep_def}) admit a LHV model. Surprisingly, Werner
constructed one parameter of family of $U \otimes U$ invariant states
(see Sec. \ref{subsec:Werner_Iso}) in $d \times d$ dimensions, where
$U$ is a unitary operator and showed that some of them can be
simulated by such model \cite {Werner1989}. In particular, two-qubit
($d=2$) Werner states are mixtures of the singlet $|\psi^-\>$ with
white noise of the form
\begin{equation}
\rho = p|\psi^-\>\<\psi^-|+ (1-p)\frac{\id}{4}.
\label {W}
\end{equation}
It has been shown, using the criterion (\ref{M}) \cite {HHH1995-bell},
that the CHSH inequality is violated when ${2^{-\frac{1}{2}} } {<p\leq
  1}$.

However Popescu noticed that Werner's model accounts only for the
correlations obtained for a reduced class of local experiments
involving projective measurements and he found that some Werner
mixtures if subjected to sequence of local generalized measurements
can violate CHSH inequality , \cite {Popescu2}. Then Gisin
demonstrated that even for the case of two qubits the ``hidden
nonlocality'' can be revealed by using local filters in the first
stage of the process (a procedure of this kind can be treated as a
preprocessing consisting of stochastic-LOCC) \cite {Gisin96}. In the
same year Peres discovered that some states admitting a LHVM for the
single copy violate Bell inequalities when jointly measuring more than
one copy with postselection procedure \cite {Peres96}.  The merging of
these two tests leads to the stronger detection of hidden nonlocality
\cite {Masanes}.

A generic structure of any LHVM for experiments with postselection was
considered and it was proved that LHVM for joint probabilities of
outcomes of sequences of measurements cannot exist in the case of
``hidden nonlocality'' \cite {Popescu2,ZH,Teufel}. Barrett constructed a LHV model
which simulates arbitrary single (nonsequential) positive-operator-valued (POV) measurements of single copies of a class of entangled Werner states (see Sec.IV) \cite {Barrett2002}.

It is interesting
that the enhancement of nonlocal correlations is possible even without
postselection by collective measurements \cite {Doherty}. The first
experimental demonstration of ``hidden nonlocality'' has been reported
by Kwiat \emph{et al.} \cite {Kwiat}, (see in this context \cite {Z}).

Remarkably, the nonlocal properties of entangled states can be
related to  Grothendieck's constant, a mathematical constant
appearing in Banach space theory \cite{Grothendieck_const}.
Namely Grothendieck's constant gives a bound on the strength of violation
of all bipartite Bell inequalities \cite{Tsirelson,AcinGT2006-grot}
Interestingly, it has been recently shown that it is no longer the case for
tripartite Bell inequalities \cite{GraziaWVJ_unbound_violBell_2007}. The situation
changes if according to Bell's idea one assumes maximal correlations in choice
of observables (see \cite{Pitowsky2007}).
It also turns out that nonlocal properties of the Werner states can be expressed
via critical values $p_c$ of the parameter $p$, for which the states
cease to be nonlocal under projective measurements and which are
related to Grothendieck's constant \cite {AcinGT2006-grot}. In this way the
existence of the LHVM for two-qubit Werner states (\ref{W}) and
projective measurements has been proved up to $p\approx 0.66$ close
to the region of the CHSH violation $2^{-\frac{1}{2}} < p \leq 1$.

Banaszek and W\'{o}dkiewicz first investigated nonlocality of the
original EPR state in the Wigner representation and showed that EPR
state is strongly nonlocal, though its Wigner function is positive
\cite {BanaszekW}. The observables producing violation of Bell-type
inequalities are displaced photon-number parities rather than
continuous variables. It has been shown, that all squeezed Gaussian
states are nonlocal despite they have positive Wigner function
\cite{LoockB2001}.

There is another algorithmic approach for the problem of nonlocality
of quantum states \cite {TerhalDS}. It is based on constructing a
symmetric quasi-extension of the quantum state via semidefinite
program. It is interesting that the method allows to construct
(analytically) LHV model for bound entangled states with two measurement
settings generated by real unextendible product basis (see in this
context \cite {KaszlikowskiZG}).

The above results allow to understand the deep nature of noisy
entanglement not only in the context of LHV model. Clearly the CHSH
inequality can be used as a tool for two nonequivalent tasks: testing
quantum formalism against LHV model (nonlocality witness) and testing for entanglement within
quantum formalism (entanglement witness). On the other hand, the idea of ``hidden
nonlocality'' leads to the concept of distillation of entanglement -
an important notion in the quantum information theory (see
Sec. \ref{sec:distil}). Finally, it turned out that hidden nonlocality
allows to reveal nonclassical  features of arbitrary entangled  state.
Namely,  Masanes {\it et al.} by use of specific entanglement witnesses characterized the states that violate CHSH inequality after local filtering operators \cite{MasanesLD2007}. Then they has proved that {\it any} bipartite entangled state $\sigma$  exhibits a hidden nonlocality which can be ``activated'' in a sense that there exist another state $\rho$ not violating CHSH such that the state $\rho \ot \sigma$ violates CHSH (see sec. \ref{subsec:activbound} in this context).

\subsection {Bell theorem beyond CHSH-setting}

The CHSH inequality is one elementary inequality, that can be viewed
as a special case of an infinite hierarchy of Bell inequalities
related to the type of correlation measurements with n-partite
systems, where each of the parties has the choice of $m$ $l$-valued
observables to be measured. For the CHSH inequality $n=m=l=2$.

As early as 1990 several generalizations of the latter were derived
for the case $n,2,2$ (MABK inequalities). The complete set of
such inequalities was constructed by Werner and Wolf \cite {Wolf} and
independently by \.Zukowski and Brukner \cite{ZB}, see also
\cite{WeinfurterZ_4fotent2001}. The inequality WWZB is given by a
linear combination of the correlation expectation values
\begin{equation}
  \sum_k f(k)E(k)\leq 2^n, \label{ogol1}
\end{equation}
where coefficients are given by $f(k)=\sum_{s} S(s)(-1)^{\<k,s\>}$,
$S(s)$ is an arbitrary function of $s=s_1...s_n\in \{-1,1\}^n$, such
that $S(s_1...s_n)=\pm 1$; $\<k,s\> = \sum^n_{j=1}{k_js_j}$ and
$E(k)=\<\Pi ^n_{j=1}A_j(k_j)\>_{av}$ is the correlation function
(average over many runs of experiment) labeled by a bit string
$k=k_1...k_n$, binary variables $k_j\in {0,1}$ indicate the choice of
the $\pm 1$-valued, observable $A_j(k_j)$ at site $j$.

There are $2^{2^n}$ different functions $S(s)$, and correspondingly
$2^{2^n}$ inequalities. In particular, putting $S(s_1...s_n)= \sqrt2 \cos
(-{\pi  \over4}+(s_1+...+s_n-n){\pi  \over4})$ one recovers the
Mermin-type inequalities and for $n=2$ the CHSH inequality
(\ref{trzy}) follows.

Fortunately the set of linear inequalities (\ref{ogol1}) is equivalent
to a single non-linear inequality
\begin{equation}
\sum_s|\sum_k(-1)^{\<k,s\>}E(k)|\leq 2^n, \label{ogol2}
\end{equation}
which characterizes the structure of the accessible classical region
for correlation function for $n$-partite systems being a
hyper-octahedron in $2^n$ dimensions as the unit sphere of the Banach
space $l^1$ \cite {Wolf}. Hence it is necessary and sufficient
condition for correlation function of $n$-partite systems to admit
LHVM.

In the quantum case the observables $A_j(k_j)$ convert into Hermitian
operators $A_j(k_j)=\textbf{a}_j(k_j)\cdot \vec {\sigma} $ with
spectrum in $\{-1,+1\}$ where $\sigma $ is the vector of Pauli
matrices and $\textbf{a}_j(k_j)$ is a normalized vector in $R^3$.
Consequently the n-qubit quantum correlation function has the form:
\ben E_q(k)={\tr}[\rho \otimes ^n_{j=1}\textbf{a}_j (k_j)\cdot\vec
{\sigma}] = \nonumber \\ \<T, \textbf{a}_1\otimes \cdots\otimes\vec
\textbf{a}_n \>. \label{eq} \een Inserting $E_q(k)$ to the expression
(\ref{ogol2}) and using (\ref{stan}) one obtains that the correlations
between the measurement on $n$ qubits admit an LHVM iff in any set of
local coordinate system and for any set of unit vectors $\textbf{c}_j=
(c^j_1,c^j_2)\in R^2$ the following inequality holds \cite{ZB}
\begin{equation}
T^{mod}_{c_1...c_n}\equiv
\sum^2_{i_1...i_n}{c^1_{i_1}...c^n_{i_n}}|t_{{i_1}...{i_n}}|\leq  1,
\label{war}
\end{equation}
where $T^{mod}_{c_1...c_n}$ is a component of moduli of correlation
tensor $T$, along directions defined by the unit vectors
$\textbf{c}_j$. In the spirit of the Peres separability criterion one
can express the above condition as follows: within LHVM $T^{mod} $ is
also a possible correlation tensor.

Note that the Bell-WWZB operator for n-partite system with two
dichotomic observables per site admits spectral decomposition into
n-qubit GHZ states (\ref {GHZ}) which lead to maximal violations \cite
{Scarani}, \cite {WW}. Thus the GHZ states play a similar role for the
WWZB inequality as the Bell states for the CHSH one.

The WWZB inequalities are important tool for the investigation of
possible connections between quantum nonlocality distillability and
entanglement for $n$-qubit systems. In particular, it has been shown
that the violation of the WWZB inequality by a multi-qubit states
implies that pure entanglement can be distilled from it. But the
protocol may require that some of the parties join into several groups
\cite {Acin1,ASW}. This result was generalized to the asymptotic
scenario \cite {Masanes}.

Other $n$-particle Bell type inequalities based on the assumption of
partial separability were found which are maximally violated by
$n$-particle GHZ-states \cite {Svetlichny}. There were different
generalizations of Bell inequalities for bipartite systems using more
than two dichotomic measurements per site (\cite
{Sliwa,CG,Zukowski06,PitowskyS_nonloc2001}. The only in 2004 Collins
and Gisin has showed that the inequality involving three dichotomic
observables per site can ``detect'' nonlocality of the two-qubit
states which escapes the detection via CHSH inequality
\cite{CG}. Collins \emph{et al.} developed a novel approach to Bell
inequalities based on logical constraints local variables models must
satisfy and introduced family of Bell inequalities for bipartite
systems of arbitrarily high dimensionality \cite{CollinsGLMP-Bell-hdim}. A very
appealing form of these inequalities has been recently presented in
\cite{ZohrenGill}. Surprisingly for the above inequalities in contrast
to the CHSH inequality the optimal state in dimension $d>2$ is not
maximally entangled.

Interestingly, it has been recognized that there is a tradeoff
between the local commutativity and Bell inequality non-violating \cite{Roy-noncomut-Bell-2007}. This idea was developed in \cite{SeevinckU-loc-com-Bell-2007}.

Note that there are Bell inequalities of other type such as: Bell
inequalities in phase space \cite{AubersonMRS2003}; and entropic Bell
inequalities, which involve classical entropy
\cite{BraunsteinC1988}. However, the entropic ones are weaker than the
CHSH inequality. In \cite{NagataLWZ2004} Bell inequalities
were considered in the context of rotational invariance.
It should be noted, that the pioneering investigation
of the Bell inequalities in the relativistic context was performed by
Werner and Summers \cite{SummersW} and Czachor \cite{Czachor1}.

\subsection{Logical versions of Bell's theorem}
\label{subsec:logicBell}

The violation of Bell inequalities as a paradigmatic test of quantum
formalism has some unsatisfactory feature as it only applies to a
statistical measurement procedure. It is intriguing, that the quantum
formalism via quantum entanglement offers even stronger departure from
classical intuition based on the logical argument which discusses only
perfectly correlated states. This kind of argument was discovered by
Greenberger, Horne and Zeilinger (GHZ) and applies for all
multipartite systems that are in the same GHZ state i.e.
$|\psi\rangle_{ABC}=\frac{1}{\sqrt{2}}(|000\rangle+|111\rangle)_{ABC}$.
The authors showed that any deterministic LHVM predicts that a certain
outcome always happen while quantum formalism predicts it never
happens.

The original GHZ argument for qubits has been subsequently developed
\cite {Mermin,CerfMP,Hardy93,Cabello1,Cabello2,ChenPZBZ2003,GHZ1,GHZ2}
and extended to continuous variables \cite {Clifton,MassarP,ChenZ} and
it is known as the ``all-versus-nothing'' proof of Bell theorem or
``Bell theorem without inequalities''. The proofs are purely logical.
However, in real experiments ideal measurements and perfect
correlations are practically impossible. To overcome the problem of a
``null experiment'' Bell-type inequalities are needed.

An important step in this direction was the reduction of the GHZ proof
to the two-particle non-maximally \cite {Hardy93} and maximally
entangled systems of high-dimensionality \cite
{Pan,Kaszlikowski,Durt,Chen2,Mandel}. Recently a two-particle
all-versus-nothing nonlocality tests were performed by using
two-photon so-called hyperentanglement \cite {Zukowski,Martini}. It is
instructive to see how the GHZ argument works in this special case,
when a single source produces a two-photon state simultaneously
maximally entangled both in the polarization and in the path degrees
of freedom. It can be written in the form:
\begin{equation}
\left|\Psi\right\rangle = {1\over 2}({\left|H\right\rangle_A}{\left|V\right\rangle_B}-{\left|V\right\rangle_A}{\left|H\right\rangle_B})({\left|R\right\rangle_A}{\left|L\right\rangle_B}-{\left|L\right\rangle_A}{\left|R\right\rangle_B}).
\label{psi}
\end{equation}

Here photon-$A$ and photon-$B$ pertain to spatially separated
observers Alice and Bob, $\left|H\right\rangle$ and
$\left|V\right\rangle$ stand for photons with horizontal and
vertical polarization and $R$ , $L$ are the spatial (path) modes.

Now, one can define the set of the nine local observables
corresponding to $\left|\Psi\right\rangle$ to be measured by Alice
and Bob as follows: $a_i = ({{z'_i}},{x_i},{x_i}\cdot{z'_i})$, $b_i
= (z_i,x'_i,z_i\cdot{x'_i})$,  $c_i =(z_iz'_i,x_ix'_i,z_iz'_i
\cdot{x_ix'_i})$, where $i=A,B$ and $z_i = \sigma_{z_i}$, $x_i =
\sigma_{x_i}$, $z'_i = \sigma'_{z_i}$, $x'_i = \sigma'_{x_i}$ are
Pauli-type operators for the polarization and the path degree of
freedom respectively. The observables are arranged in three groups
$a$, $b$, and $c$, where the observables of each group are measured
by one and the same apparatus.

Then quantum formalism makes the following predictions for perfect
correlations:
\begin{eqnarray}
z_A\cdot z_B \left|\Psi\right\rangle = - \left|\Psi\right\rangle,  z'_A\cdot z'_B \left|\Psi\right\rangle = - \left|\Psi\right\rangle, \nonumber \\
x_A\cdot x_B \left|\Psi\right\rangle = - \left|\Psi\right\rangle,  x'_A\cdot x'_B \left|\Psi\right\rangle = - \left|\Psi\right\rangle,  \nonumber\\
z_A z'_A\cdot z_Bz'_B\left|\Psi\right\rangle = \left|\Psi\right\rangle,  x_Ax'_A\cdot x_Bx'_B \left|\Psi\right\rangle = \left|\Psi\right\rangle, \nonumber \\
z_A\cdot x'_A\cdot z_Bx'_B\left|\Psi\right\rangle = \left|\Psi\right\rangle, x_A\cdot z'_A\cdot x_Bz'_B \left|\Psi\right\rangle = \left|\Psi\right\rangle, \nonumber \\
z_A z'_A\cdot x_A x'_A \cdot z_Bx'_B \cdot x_Bz'_B\left|\Psi\right\rangle = - \left|\Psi\right\rangle. \nonumber
\end{eqnarray}

According to EPR the perfect correlations allow assign the
pre-existing measurement values $v(X)$ to the above observables $X$.
Then the consistency between quantum predictions and any LHVM gives
the following relations between the ``elements of reality'':
\begin{eqnarray}
v(z_A)v(z_B)=-1, v(z'_A)v(z'_B)=-1, \nonumber\\
v(x_A)v(z_B)=-1, v(x'_A)v(x'_B)=-1, \nonumber\\
v(z_Az'_A)v(z_B)v(z'_B)=1, v(x_Ax'_A)v(x_B)v(x'_B)=1, \nonumber\\
v(z_A)v(x'_A)v(z_Bx'_B)=1, v(x_A)v(z'_A)v(x_Bz'_B)=1, \nonumber\\
v(z_Az'_A)v(x_Ax'_A)v(z_Bx'_B)v(x_Bz'_B)=-1. \nonumber
\end{eqnarray}
Multiplying the above equations except the last one, one gets
$v(z_Az'_A)v(x_Ax'_A)v(z_Bx'_B)v(x_Bz'_B)=1$, that contradicts the
last one.

In real experiments the nonlocality tests are based on the estimation
of the Bell type operator $O = -z_A\cdot z_B - z'_A\cdot z'_B -
x_A\cdot x_B - x'_A\cdot x'_B + z_Az'_A\cdot z_B\cdot z'_B +
x_Ax'_A\cdot x_B\cdot x'_B + z_A\cdot x'_A\cdot z_Bx'_B +x_A\cdot
z'_A\cdot x_Bz'_B -z_Az'_A\cdot x_Ax'_A\cdot z_Bx'_B\cdot x_Bz'_B$ and
Bell-Mermin type inequality \cite {Cabello1,Cabello2,ChenPZBZ2003}
\begin{equation}
{\left\langle  O\right\rangle}_{L} \leq 7.
\label{O}
\end{equation}
It is easy to check that the operator $O$ satisfies the relation
$O\left|\Psi\right\rangle = 9 \left|\Psi\right\rangle \nonumber$,
which is in contradiction with the above inequality implied by
LHVM. In a recent experiment \cite{Zukowski} observed value for $O$
was $8.56904\pm 0.00533$. It demonstrates a contradiction of quantum
predictions with LHVM (see in this context related experiment \cite
{Martini}). A novel nonlocality test ``stronger two observer all versus nothing'' test \cite {Cabello2005} have been performed by use of a 4-qubit linear cluster state via two photons entangled
both in polarization and linear momentum \cite {VallonePMMB2007}.

Note that, the GHZ argument has been extended to the entangled
two-particle states that are produced from two independent sources
\cite {YurkeStoller_ent_swapping} by the process of ``entanglement
swapping'' \cite {ent_swapping}. The complete proof depends crucially
on the independence of the two sources and involves the factorization
of the Bell function \cite {GHZ1}.

\subsection {Violation of Bell inequalities: general remarks}

There is a huge literature concerning interpretation of the ``Bell
effect''. The most evident conclusion from those experiments is that
it is not possible to construct a LHVM simulating all correlations
observed for quantum states of composite systems. But such conclusion
is not surprising. What is crucial in the context of the Bell theorem
is just a {\it gap} between quantum and classical description of the
correlations, which gets out of hand.

Nature on its fundamental level offers us a new kind of statistical
non-message-bearing correlations, which are encoded in the quantum
description of states of compound systems via entanglement. They are
``nonlocal''\footnote {The term nonlocality is somewhat misleading.
  In fact there is a breaking of conjunction of locality and
  counterfactuality.} in the sense, that

i) They cannot be described by a LHVM.

ii) They are {\it nonsignaling} as the local measurements performed
on spatially separated systems cannot be used to transmit messages.

Generally speaking, quantum compound systems can reveal {\it holistic
  nonsignaling} effects even if their subsystems are spatially
separated by macroscopic distances. In this sense quantum formalism
offers holistic description of Nature \cite {Primas}, where in a
non-trivial way the system is more than a combination of its
subsystems.

It is intriguing that entanglement does not exhaust full potential of
nonlocality under the constraints of no-signaling. Indeed, it does not
violate Tsirelson's bound (\ref{eq:cztery}) for CHSH inequalities. On
the other hand, one can design family of probability distributions
(see \cite {NG} and ref. therein), which would violate this bound but
still do not allow for signaling. They are called Popescu-Rohrlich
nonlocal boxes, and represent extremal nonlocality admissible without
signaling \cite{PR}. We see therefore that quantum entanglement is
situated in an intermediate level between locality and maximal
non-signaling nonlocality. Needless to say, the Bell inequalities
still involve many fascinating open problems interesting for both
philosophers and physicists (see \cite{Gisin_philo2007}).

\section {Entropic manifestations of entanglement}
\label{sec:entropic}
\subsection { Entropic inequalities: classical versus quantum order}

It was mentioned in the introduction that Schr{\"o}dinger first
pointed out, that entanglement does not manifest itself only as
correlations. In fact he has recognized an other aspect of
entanglement, which involves profoundly nonclassical relation between
the information which an entangled state gives us about the whole
system and the information which it gives us about subsystems.

This new ``nonintuitive'' property of compound quantum systems,
intimately connected with entanglement, was a long-standing puzzle
from both a physical and a mathematical point of view. The main
difficulty was, that in contrast to the concept of correlation, which
has a clear operational meaning, the concept of information in quantum
theory was obscure till 1995, when Schumacher showed that the von
Neumann entropy
\begin{equation}
S(\rho) = - {\tr}\rho \log\rho \label{von}
\end{equation}
has the operational interpretation of the number of qubits needed to
transmit quantum states emitted by a statistical source \cite
{Schumacher1}. The von Neumann entropy can be viewed as the quantum
counterpart of the Shannon entropy $H(X)=-\sum_ip_i\log{p_i}$,
$\sum_ip_i=1$, which is defined operationally as the minimum number of
bits needed to communicate a message produced by a classical
statistical source associated to a random variable $X$.

Fortunately in 1994 Schr{\"o}dinger's observation, that an entangled
state gives us more information about the total system than about
subsystems was, quantified just by means of von Neumann entropy. It
was shown that the entropy of a subsystem can be greater than the
entropy of the total system only when the state is entangled \cite
{RPH1994}. In other words the subsystems of entangled system may
exhibit more disorder than the system as a whole. In the classical
world it never happens. Indeed, the Shannon entropy $H(X)$ of a single
random variable is never larger than the entropy of two variables
\begin{equation}
  H(X,Y)\geq H(X), \quad  H(X,Y)\geq H(Y). \label{H}
\end{equation}
It has been proved \cite {MRH-PRA96,alpha,VollbrechtW,TerhalReview} that analogous
$\alpha$-entropy inequalities hold for separable states
\begin{equation}
S_\alpha(\rho_{AB}) \geq S_\alpha(\rho_A),  \quad S_\alpha(\rho_{AB}) \geq
S_\alpha(\rho_B), \label{ent}
\end{equation}
where $S_\alpha(\rho)= (1-\alpha)^{-1} \log {\tr} \rho^\alpha$ is the
$\alpha$-Renyi entropy for $\alpha \ge 0$, here
$\rho_A={\tr}_B(\rho_{AB})$ and similarly for $\rho_B$. If $\alpha$
tends to 1 decreasing, one obtains the von Neumann entropy
$S_1(\rho)\equiv S(\rho)$ as a limiting case.

The above inequalities represent scalar  separability criteria.
For pure states these inequalities are violated if and only if
the state is entangled. As separable states admit LHVM one can
expect, that the violation of the entropic inequalities is a
signature of some nonclassical features of a compound quantum system
resulting from its entanglement. The above inequalities involve a
nonlinear functionals of a state $\rho$, so they can be interpreted
as a scalar separability criteria based on a nonlinear entanglement
witness (see Sec. \ref{sec:BipartiteEntanglement}).

The $\alpha$-entropy inequalities were considered in the context of
Bell inequalities \cite {MRH-PRA96} (and quantum communication for
$\alpha =1$ \cite{cerfadami}). For $\alpha=2$ one get the inequalities
based on the 2-Renyi entropy, which are stronger detectors of
entanglement than all Bell-CHSH inequalities
\cite{alpha,habil,Santos}. The above criterion was expressed in terms
of the Tsallis entropy \cite {Abe,Tsallis2,Barranco}.

Another interesting family of separability of nonlinear criteria was
derived \cite {GuehneL2004-pra} in terms of entropic uncertainty relations
\cite{Birula,Deutsch,Kraus,MaasssenU1988}\footnote{Quite recently the
  quantum entropic uncertainty relations has been expressed in the
  form of inequalities involving the Renyi entropies
  \cite{bialynickibirula-2006-74}.} using Tsallis entropy.  In contrast
to linear functionals, such as correlation functions, the entropies,
being a nonlinear functionals of a state $\rho$, are not directly
measurable quantities. Therefore, the problem of experimental
verification of the violation of the entropic inequalities was
difficult from both a conceptual and a technical point of
view. However, recently Bovino \emph{et al.}  demonstrated
experimentally a violation of 2-entropic inequalities based on the
Renyi entropy \cite {NonlinExp}. This also was first direct
experimental measurement of a nonlinear entanglement witness (see
Sec. \ref{subsec:det_ent_col}).

\subsection {1-entropic inequalities and negativity of information}

Entropic inequalities (\ref{ent}) can be viewed as a scalar
separability criteria. This role is analogous to Bell inequalities as
entanglement witnesses. In this context a natural question arises: is
this all we should expect from violation of entropic inequalities?
Surprisingly enough, entropic inequalities are only the tip of the
iceberg which reveals dramatic differences between classical and
quantum communication due to quantum entanglement. To see it consider
again the entropic inequalities based on the von Neumann entropy
($\alpha=1$) which hold for separable states. They may be equivalently
expressed as
\begin{equation}
S(\rho_{AB}) - S(\rho _B)\geq0, \quad  S(\rho_{AB})- S(\rho _A)\geq0
\label{pur1}
\end{equation}
and interpreted as the constraints imposed on the correlations of the
bipartite system by positivity of some function
\begin{equation}
S(A|B) = S(\rho_{AB}) - S(\rho_B) \label{pur2}
\end{equation}
and similarly for $S(B|A)$. Clearly, the entropy of the subsystem
$S(\rho_A)$ can be greater than the total entropy $S(\rho_{AB})$ of
the system only when the state is entangled.  However, there are
entangled states which do not exhibit this exotic property, i.e. they
satisfy the constraints. Thus the physical meaning of the function
$S(A|B)$ ($S(B|A)$) and its peculiar behavior were an enigma for
physicists.

There was an attempt \cite {cerfadami} to overcome this difficulty by
replacing Shannon entropies with von Neumann one in the formula for
classical conditional entropy
\begin{equation}
H(X|Y)=H(X,Y)-H(X) \label{SW}
\end{equation}
to get quantum conditional entropy in form (\ref {pur2}) \cite
{Wehrl}. Unfortunately, such approach is inconsistent with the concept
of the channel capacity defined via the function $-S(A|B)$ called {\it
  coherent information} (see Sec. \ref{subsec:onewayhash})
\cite{SchumacherN-1996-pra}\footnote{The term ``coherent information''
  was originally defined to be a function of a state and a channel,
  but further its use has been extended to apply to a bipartite
  state.}. Moreover, so defined conditional entropy can be
negative.

\begin{figure}
  \centering
  \includegraphics{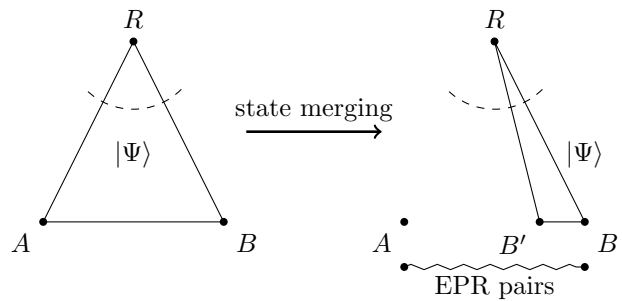}
  \caption{The concept of state merging: before and after.}
  \label{fig:merging}
\end{figure}

Recently, the solution of the problem was presented within the quantum
counterpart of the classical Slepian-Wolf theorem called ``quantum
state merging'' (\cite {SW-nature}, \cite {sw-long}). In 1971 Slepian
and Wolf considered the following problem: {\it how many bits does the
  sender (Alice) need to send to transmit a message from the source,
  provided the receiver (Bob) already has some prior information about
  the source}. The amount of bits is called the partial
information. Slepian and Wolf proved \cite {Slepian}, that the partial
information is equal to the conditional entropy (\ref {SW}),which is
always positive: $H(X|Y)\geq0$.

In the quantum state merging scenario, an unknown quantum state is
distributed to spatially separated observers Alice and Bob.  The
question is: {\it how much quantum communication is needed to transfer
  Alice's part of the state to Bob in such a way that finally Bob has
  total state} (Fig. \ref{fig:merging}). This communication measures
the { \it partial information} which Bob needs conditioned on its
prior information $S(B)$. Surprisingly, Horodecki at al. proved that
necessary and sufficient number of qubits is given by the formula
(\ref {pur2}), even if this quantity is negative.

Remarkably, this phenomenon asunders into two levels: ``classical''
and quantum depend on sign of the partial information $S(A|B)$:

1) partial information $S(A|B)$ is positive (the inequalities (\ref
{pur1}) are not violated): the optimal state merging protocol
requires sending $r\cong S(A|B)$ qubits.

2) partial information $S(A|B)$ is negative (the inequalities (\ref
{pur1}) are violated): the optimal state merging does not need
sending qubits; in addition Alice and Bob {\it gain} $r\cong
-S(A|B)$ pairs of qubits in a maximally entangled state.

It is remarkable that we have clear division into two regimes: when
the 1-entropy inequalities are not violated, in the process of quantum
state merging consumed and we have a nice analogy to the classical
Slepian-Wolf theorem where partial information (equal to conditional
entropy (\ref {SW})) is always positive; when the inequalities are
violated, apart from state merging we get additional
entanglement. Thus the quantum and classical regimes are determined by
the relations between the knowledge about the system as a whole and
about its subsystem, as considered by Schr{\"o}dinger.

Finally, let us note that the early recognized manifestations of
entanglement: ({\it nonlocality} (EPR, Bell) and what we can call {\it
  insubordination} (Schr{\"o}dinger)) had been seemingly academic
issues, of merely philosophical relevance. What is perhaps the most
surprising twist, is that both the above features qualify entanglement
as a resource to perform some concrete tasks.

Indeed, the violation of the Bell inequalities determines usefulness
of quantum states for the sake of specific nonclassical tasks, such as
e.g. reduction of communication complexity, or quantum cryptography
(see Sec.\ref{sec:effects}). Similarly, the violation of the entropic
inequalities based on the von Neumann entropy (\ref {pur1}) determines
the usefulness of states as a {\it potential} for quantum
communication. It is in agreement with the earlier results, that the
negative value of the function $S(A|B)$ is connected with the ability
of the system to perform teleportation \cite {cerfadami,MRH-PRA96} as
well as with nonzero capacity of a quantum channel
\cite{SchumacherN-1996-pra,Lloyd-cap,Devetak2003}.

\subsection {Majorization relations}
\label{subsec:majorrel}

In 2001 Nielsen and Kempe discovered a stronger version of the
``classical versus quantum order'' \cite {NielsenKempe}, which
connects the majorization concept and entanglement. Namely they
proved, that if a state is separable than the following inequalities
\begin{equation}
\lambda(\rho)\prec \lambda(\rho_A) \quad  \lambda(\rho)\prec
\lambda(\rho_B) \label{lambda}
\end{equation}
has to be fulfilled. Here $\lambda(\rho)$ is a vector of eigenvalues
of $\rho$; $\lambda(\rho_A)$ and $\lambda(\rho_B)$ are defined
similarly. The relation $x\prec y$ between d-dimensional vectors $x$
and $y$ ($x$ is majorized by $y$) means that
$\sum^k_{i=1}x^\downarrow_i\leq\sum^k_{i=1}y^\downarrow_i$, $1 \leq
k\leq d-1$ and the equality holds for $k=d$, $x^\downarrow_i(1 \leq i
\leq d)$ are components of vector $x$ rearranged in decreasing order;
$y^\downarrow_i(1 \leq i \leq d)$ are defined similarly.  Zeros are
appended to the vectors $\lambda(\rho_A)$, $\lambda(\rho_B)$ in (\ref
{lambda}), in order to make their dimension equal to the one of
$\lambda(\rho)$.

The above inequalities constitute necessary condition for separability
of bipartite states in arbitrary dimensions in terms of the local and
the global spectra of a state. This criterion is stronger than
entropic criterion (\ref{ent}) and it again supports the view that ``{\it separable states are more disordered globally
  than locally}'' \cite {NielsenKempe}. An alternative proof
of this result have been found \cite{Gurvits3}.

\section{Bipartite entanglement}
\label{sec:BipartiteEntanglement}

\subsection{Definition and basic properties \label{DefinitionBasic}}

The fundamental question in quantum entanglement theory is {\it which
  states are entangled and which are not}. Only in few cases this
question has simple answer. The simplest is the case of pure bipartite
states.  In accordance with definition of multipartite entangled
states (Sec. (\ref{sec:ent_as_prop}) any bipartite pure state
$|\Psi_{AB}\rangle \in {\cal H}_{AB}={\cal H}_{A} \otimes {\cal
  H}_{B}$ is called separable (entangled) iff it can be (can not be)
written as a product of two vectors corresponding to Hilbert spaces of
subsystems:
\begin{equation}
|\Psi_{AB}\rangle=|\phi_{A}\rangle|\psi_{B}\rangle.
\end{equation}

In general if the vector $\Psi_{AB}$ is written in any orthonormal
product basis $\{ |e_{A}^{i} \rangle \otimes |e_{B}^{j} \rangle \}$ as
follows\footnote{Here the orthonormal basis $\{ |e_{X}^{i} \rangle\}$
  spans subspace ${\cal H}_{X}$, $X=A,B$.}:
\begin{equation} |\Psi_{AB}\rangle =\sum_{i=0}^{d_{A}-1}\sum_{j=0}^{d_{B}-1}
A^{\Psi}_{ij} |e_{A}^{i} \rangle \otimes |e_{B}^{j} \rangle,
\end{equation}
then it is product {\it if and only if} the matrix of coefficients
$A^{\Psi}=\{ A^{\Psi}_{ij} \}$ is of rank one. In general the rank
$r(\Psi)\leq k\equiv min[d_{A},d_{B}]$ of this matrix is called {\it
  Schmidt rank of vector $\Psi$ } and it is equal to either of ranks
of the reduced density matrices
$\varrho_{A}^{\Psi}={\tr}_{B}|\Psi_{AB}\rangle \langle \Psi_{AB}|$,
$\varrho_{B}^{\Psi}={\tr}_{A}|\Psi_{AB}\rangle \langle \Psi_{AB}|$
(which satisfy $\varrho_{A}^{\Psi}=A^{\Psi}(A^{\Psi})^{\dagger}$ and
$\varrho_{A}^{\Psi}=[(A^{\Psi})^{\dagger}A^{\Psi}]^T$
respectively\footnote{$T$ denotes transposition.}). In particular
there always exists such a product bi-orthonormal basis $\{
|\tilde{e}_{A}^{i} \rangle \otimes |\tilde{e}_{B}^{j} \rangle \}$ in
which the vector takes the Schmidt decomposition:
\begin{equation}
|\Psi_{AB}\rangle =\sum_{i=0}^{r(\Psi)} a_{i} |\tilde{e}_{A}^{i}
\rangle \otimes |\tilde{e}_{B}^{i} \rangle,
\end{equation}
where the strictly positive numbers $a_{i}=\{ \sqrt{p_{i}} \}$
correspond to the nonzero singular eigenvalues \cite{Nielsen-Chuang}
of $A^{\Psi}$, and $ p_{i}$ are the nonzero elements of the spectrum
of either of the reduced density matrices.

Quantum entanglement is in general both quantitatively and
qualitatively considered to be a property {\it invariant} under
product unitary operations $U_{A} \otimes U_{B}$. Since in case of
pure vector and the corresponding pure state (projector)
$|\Psi_{AB}\rangle \langle \Psi_{AB}|$ the coefficients $\{a_{i}\}$
are the only parameters that are invariant under such operations, they
completely determine entanglement of bipartite pure state.

In particular, it is very easy to see that pure state
(projector)$|\Psi_{AB} \rangle \langle \Psi_{AB}|$ is separable iff
the vector $\Psi_{AB}$ is product. Equivalently the  rank of either
of reduced density matrices $\varrho_{A}$, $\varrho_{B}$ is
equal to $1$, or there is single nonzero Schmidt coefficient. Thus
for bipartite pure states it is elementary to decide whether the
state is separable or not by diagonalizing its reduced density
matrix.

So far we have considered entanglement of pure states. Due to
decoherence phenomenon, in laboratories we unavoidably deal with mixed
states rather than pure ones. However mixed state still can contain
some ``noisy'' entanglement. In accordance with general definition for
$n$-partite state (\ref{sec:ent_as_prop}) any bipartite state $
\varrho_{AB}$ defined on Hilbert space ${\cal H}_{AB}={\cal H}_{A}
\otimes {\cal H}_{B}$ is separable (see \cite{Werner1989}) if and only
if it can neither be represented nor approximated by the states of the
following form
\begin{equation}
\varrho_{AB} = \sum_{i=1}^{k} p_{i} \varrho_{A}^{i} \otimes
\varrho_{B}^{i},
\label{eq:sep_def}
\end{equation}
where $\varrho_{A}^{i}$, $\varrho_{B}^{i}$ are defined on local
Hilbert spaces ${\cal H}_{A}$, ${\cal H}_{B}$. In case of finite
dimensional systems, i.e. when $\dim{\cal H}_{AB} < \infty $ the
states $\varrho_{A}^{i}$, $\varrho_{B}^{i}$ can be chosen to be
pure. Then, from the Caratheodory theorem it follows (see
\cite{Pawel97,PlenioVedral1998}) that the number $k$ in the convex
combination can be bounded by the square of the dimension of the
global Hilbert space: $k\leq d_{AB}^{2}=(d_{A}d_{B})^{2}$ where
$d_{AB}=dim{\cal H}_{AB}$ etc. It happens that for two-qubits the
number of states (called sometimes cardinality) needed in the
separable decomposition is always four which corresponds to dimension
of the Hilbert space itself (see
\cite{SanperaTarrachVidal,Wootters-conc}). There are however $d
\otimes d$ states that for $d\ge3$ have cardinality of order of
$d^{4}/2$ (see \cite{Barely}).

We shall restrict subsequent analysis to the case of finite
dimensions unless stated otherwise.

The set ${\cal S}_{AB}$ of all separable states defined in this way is
convex, compact and invariant under the product unitary operations
$U_{A} \otimes U_{B}$. Moreover the separability property is preserved
under so called (stochastic) separable operations (see Sec.
\ref{subsec:locc}).

The problem is that given any state $\varrho_{AB}$ it is very hard to
check whether it is separable or not. In particular, its separable
decomposition may have nothing common with the eigendecomposition,
i.e. there are many separable states that have their eigenvectors
entangled, i.e. nonproduct.

It is important to repeat, what the term entanglement means on the
level of mixed states: all states that do not belong to ${\cal S}$,
i.e.  are not separable (in terms of the above definition) are called
{\it entangled}.

In general the problem of separability of mixed states appears to be
extremely complex, as we will see in the next section. The operational
criteria are known only in special cases.

\subsection{Main separability/entanglement criteria in bipartite case}

\subsubsection{Positive partial transpose (PPT) criterion}

Let us consider the characterization of the set of mixed bipartite
separable states. Some necessary separability conditions have been
provided in terms of entropic inequalities, but a much stronger
criterion has been provided by Asher Peres \cite{Peres96}, which is
called positive partial transpose (PPT) criterion. It says that if
$\varrho_{AB}$, is separable then the new matrix
$\varrho_{AB}^{T_{B}}$ with matrix elements defined in some fixed
product basis as:
\begin{equation}
\langle m| \langle \mu| \varrho_{AB}^{T_{B}} |n \rangle |\nu
\rangle \equiv \langle m| \langle \nu| \varrho_{AB} |n \rangle |\mu
\rangle
\end{equation}
is a density operator (i.e. has nonnegative spectrum), which means
automatically that $\varrho_{AB}^{T_{B}}$ is also a quantum state (It
also guarantees positivity of $\varrho_{AB}^{T_{A}}$ defined in
analogous way). The operation $T_{B}$, called partial transpose\footnote{Following
  \cite{Rains1998-def} instead of $\varrho_{AB}^{T_{B}}$ we will write
  $\varrho_{AB}^{\Gamma}$ (as the symbol $\Gamma$ is a right ``part''
  of the letter $T$).}, corresponds just to
transposition of indices corresponding to the second subsystem and has
interpretation as a partial time reversal
\cite{SanperaTarrachVidal}. If the state is represented in a block
form
\begin{equation}
\varrho_{AB}=\left(\begin{array}{cccc}
\varrho_{00} & \varrho_{01} & ... & \varrho_{0 \ d_{A}-1}\\
\varrho_{10} & \varrho_{11} & ... & \varrho_{1 \ d_{A}-1}\\
 ...  &  ...  & ...   & ...  \\
\varrho_{d_{A}-1 \ 0} & \varrho_{d_{A}-1 \ 1} & ... &
\varrho_{d_{A}-1  \ d_{A}-1}
\label{tablica}
\end{array} \right)
\end{equation}
with block entries $\varrho_{ij} \equiv \langle i |\otimes \id |
\varrho_{AB} | j \rangle \otimes I$, then one has
\begin{equation}
\varrho_{AB}^{\Gamma}=\left(\begin{array}{cccc}
\varrho_{00}^{T} & \varrho_{01}^{T} & ... & \varrho_{0 \ d_{A}-1}^{T}\\
\varrho_{10}^{T} & \varrho_{11}^{T} & ... & \varrho_{1 \ d_{A}-1}^{T}\\
 ...  &  ...  & ...   & ...  \\
\varrho_{d_{A}-1 \ 0}^{T} & \varrho_{d_{A}-1 \ 1}^{T} & ... &
\varrho_{d_{A}-1  \ d_{A}-1}^{T}
\end{array} \right).
\end{equation}

Thus the PPT condition corresponds to transposing block elements of
matrix corresponding to second subsystem. PPT condition is known to be
stronger than all entropic criteria based on Renyi $\alpha$-entropy
(\ref{sec:entropic}) for $\alpha \in [0,\infty]$ \cite{VollbrechtW}. A
fundamental fact is \cite{Peres96,sep1996} that {\it PPT condition is
  necessary and sufficient condition for separability of $2 \otimes 2$
  and $2 \otimes 3$ cases}. Thus it gives a complete characterization
of separability in those cases (for more details or further
improvements see Sec.  \ref{subsubsec:maps}).

\subsubsection{Separability {\it via} positive, but not completely positive maps}
\label{subsubsec:maps}

Peres PPT condition initiated a general analysis of the problem of the
characterization of separable (equivalently entangled) states in terms
of linear positive maps \cite{sep1996}. Namely, it can be seen that
the PPT condition is equivalent to demanding the positivity
\footnote{The operator is called positive iff it is Hermitian and has
  nonnegative spectrum.} of the operator $[I_{A} \otimes
T_{B}](\varrho_{AB})$, where $T_{B}$ is {\it the transposition map}
acting on the second subsystem. The transposition map is a positive
map (i.e. it maps any positive operator on ${\cal H}_{B}$ into a
positive one), but it is not completely positive\footnote{The map
  $\Theta$ is completely positive iff $\id \otimes \Theta$ is positive
  for identity map $\id$ on any finite-dimensional system.}. In fact,
$\id_{A} \otimes T_{B}$ is not a positive map and this is the source
of success of Peres criterion.

It has been recognized that {\it any} positive (P) but not completely
positive (CP) map $\Lambda:{\cal B}({\cal H}_{B}) \rightarrow {\cal
  B}({\cal H}_{A'})$ with codomain related to some new Hilbert space
${\cal H}_{A'}$ provides nontrivial necessary separability criterion
in the form:
\begin{equation}
  [\id_{A} \otimes \Lambda_{B}](\varrho_{AB})\geq 0.
  \label{mapscondition}
\end{equation}
This corresponds to nonnegativity of spectrum of the following matrix:
\begin{eqnarray}
&& [\id _{A} \otimes \Lambda_{B}] (\varrho_{AB}) \nonumber \\
&& =\left(\begin{array}{ccc}
\Lambda(\varrho_{00}) &  ... & \Lambda(\varrho_{0 \ d_{A}-1})\\
\Lambda(\varrho_{10}) &  ... & \Lambda(\varrho_{1 \ d_{A}-1)})\\
 ...  &   ...   & ...  \\
\Lambda(\varrho_{d_{A}-1 \ 0}) &
... & \Lambda(\varrho_{d_{A}-1  \ d_{A}-1})
\end{array} \right)
\end{eqnarray}
with $\varrho_{ij}$ defined again as in (\ref{tablica}).

It happens that using the above technique one can provide a {\it
  necessary and sufficient} condition for separability (see
\cite{sep1996}): {\it the state $\varrho_{AB}$ is separable if and
  only if the condition (\ref{mapscondition}) is satisfied for all P
  but not CP maps $\Lambda: {\cal B}({\cal H}_{B}) \rightarrow {\cal
    B}({\cal H}_{A})$ where ${\cal H}_{A}$, ${\cal H}_{B}$ describe
  the left and right subsystems of the system AB.}

Note that the set of maps can be further restricted to all P but not
CP maps that are identity-preserving \cite{Pawel2001-NATO} (the set of
witnesses can be then also restricted via the isomorphism). One could
also restrict the maps to trace preserving ones, but then one has to
enlarge the codomain \cite{HHH02-permut}.

Given characterization in terms of maps and witnesses it was natural
to ask about a more practical characterization of
separability/entanglement. The problem is that in general the set of P
but not CP maps is not characterized and it involves a hard problem in
contemporary linear algebra (for progress in this direction see
\cite{Kossakowski} and references therein).

However for very low dimensional systems there is surprisingly useful
solution \cite{sep1996}: {\it the states of $d_{A} \otimes d_{B} $
  with $d_{A}d_{B}\leq 6$ (two-qubits or qubit-qutrit systems) are
  separable if and only if they are PPT.} Recently even a simpler
condition for two-qubit systems (and only for them) has been pointed
out \cite{Augusiak} which is important in context of physical
detections (see Secs. \ref{subsubsec:CollectiveWitnesses} and
\ref{subsubsec:QuantumDataStructure}) : two-qubit state $\varrho_{AB}$
is separable iff
\begin{equation}
det(\varrho_{AB}^{\Gamma})\geq 0 \label{2QubitDeterminant}.
\end{equation}
This is the simplest two-qubit separability condition. It is a direct
consequence of two facts known earlier but never exploited in such a
way: the partial transpose of any entangled two-qubit state (i) is of
full rank and has only one negative eigenvalue.
\cite{SanperaTarrachVidal,Verstraete(i-ii)}. Note that some
generalizations of (\ref{2QubitDeterminant}) for other maps and
dimensions are also possible \cite{Augusiak}.

The sufficiency of the PPT condition for separability in low
dimensions follows from the fact \cite{Stoermer,Woronowicz} that all
positive maps $\Lambda:{\cal B}({\cal C}^{d}) \rightarrow {\cal
  B}({\cal C}^{d'}) $ where $d=2,d'=2$ and $d=2,d'=3$ are {\it
  decomposable}, i.e. are of the form:
\begin{equation}
\Lambda^{dec}=\Lambda_{CP}^{(1)} + \Lambda_{CP}^{(2)} \circ T,
\label{decomposablemap}
\end{equation}
where $\Lambda_{CP}^{(i)}$ stand for some CP maps and $T$ stands for
transposition. It can be easily shown \cite{sep1996} that among all
decomposable maps the transposition map $T$ is the ``strongest'' map
i.e. there is no decomposable map that can reveal entanglement which
is not detected by transposition.

\begin{figure}
  \centering
  \includegraphics{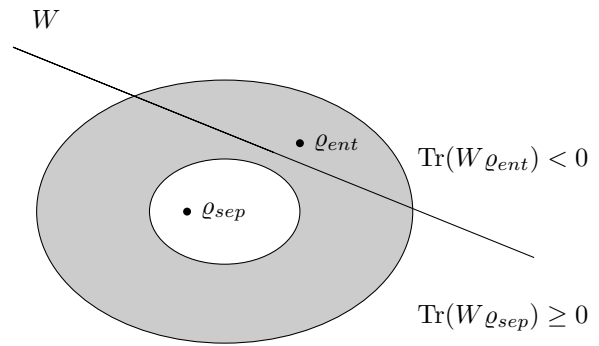}
  \caption{The line represents hyperplane corresponding to the
    entanglement witness $W$. All states located to the left of the
    hyperplane or belonging to it (in particular all separable states)
    provide nonnegative mean value of the witness,
    i.e. ${\tr}(W\varrho_{sep}) \ge 0$ while those located to the
    right are entangled states detected by the witness.}
  \label{fig:witness}
\end{figure}

\subsubsection{Separability via entanglement
witnesses\label{subsubsec:Witnesses}}

Entanglement witnesses \cite{sep1996,TerhalB} are fundamental tool in
quantum entanglement theory. They are observables that completely
characterize separable states and allow to detect entanglement
physically. Their origin stems from geometry: the convex sets can be
described by hyperplanes. This translates into the statement (see
\cite{sep1996,TerhalB}) that the state $\varrho_{AB}$ belongs to the
set of separable if it has nonnegative mean value
\begin{equation}
{\tr}(W\varrho_{AB} )\geq 0 \label{witnesscondition-v1}
\end{equation}
for all observables $W$ that (i) have at least one negative eigenvalue
and (ii) have nonnegative mean value on product states or ---
equivalently --- satisfy the nonnegativity condition
\begin{equation}
\langle \psi_{A}|\langle \phi_{B}| W
|\psi_{A}\rangle|\phi_{B}\rangle \geq 0. \label{witnesscondition-v2}
\end{equation}
for all pure product states $|\psi_{A}\rangle|\phi_{B}\rangle$ The
observables $W$ satisfying conditions (i) and (ii) above \footnote{The
  witnesses can be shown to be isomorphic to P but not CP maps, see
  Eq. (\ref{isomorphism}).} have been named {\it entanglement
  witnesses} and their physical importance as entanglement detectors
was stressed in \cite{TerhalB} in particular one says that
\emph{entanglement of $\varrho$ is detected by witness $W$} iff
${\tr}(W\varrho)<0$, see fig. \ref{fig:witness}.  (We shall discuss
physical aspects of entanglement detection in more detail
subsequently).  An example of entanglement witness for $d \otimes d$
case is (cf. \cite{Werner1989}) the Hermitian swap operator
\begin{equation}
V=\sum_{i,j=0}^{d-1}|i\rangle\langle j| \otimes |j\rangle\langle i|.
\label{swap}
\end{equation}
To see that $V$ is an entanglement witness let us note that we
have$\langle \psi_{A}|\langle \phi_{B}| V
|\psi_{A}\rangle|\phi_{B}\rangle= |\langle
\psi_{A}|\phi_{B}\rangle|^{2} \geq 0$ which ensures property (ii)
above. At the same time $V=P^{(+)} - P^{(-)}$ where $P^{(+)}={1\over 2}(\id + V)$ and
$P^{(-)}={1\over 2}(\id - V)$ correspond to projectors onto symmetric and antisymmetric
subspace of the Hilbert space ${\cal C}^{d} \otimes {\cal C}^{d}$
respectively.  Hence $V$ also satisfies (i) since it has some
eigenvalues equal to $-1$. It is interesting that $V$ is an example of
so called {\it decomposable entanglement witness} (see
(\ref{decomposablemap}) and analysis below).

The P but not CP maps and entanglement witnesses are linked by so
called Choi-Jamio\l{}kowski isomorphism \cite{Jamiolkowski,Choi82}:
\begin{equation}
W_{\Lambda}=[\id \otimes \Lambda](\Pplusd) \label{isomorphism}
\end{equation}
with pure projector
\begin{equation}
\Pplusd=|\Phiplusd \rangle \langle \Phiplusd|,
\label{MaximallyEntangledProjector}
\end{equation}
where state vector $\Phiplusd\in {\cal H}_{A} \otimes {\cal
  H}_{A}$ is defined as
\begin{equation}
|\Phiplusd\>={1\over \sqrt{d}}\sum_{i=0}^{d-1} |i\rangle \otimes |i\rangle, \
d=\mathrm{dim} {\cal H}_{A}. \label{MaximallyEntangledVector}
\end{equation}
The pure projector $\Pplusd$ is an example of maximally entangled
state\footnote{For simplicity we will drop the dimension denoting
  projector onto $\Phiplus\equiv \Phiplusd$ as $\Pplus\equiv \Pplusd$
  provided it does not lead to ambiguity.} on the space ${\cal H}_{A}
\otimes {\cal H}_{A}$.

The very important observation is that while the condition
(\ref{witnesscondition-v1}) as a whole is equivalent to
(\ref{mapscondition}), a particular witness is not equivalent to a
positive map associated via isomorphism: the map proves a stronger
condition (see discussion further in the text).

As we already said, the special class of decomposable P but not CP
maps (i.e. of the form \ref{decomposablemap}) which provide no
stronger criterion than the PPT one, is distinguished. Consequently,
all the corresponding entanglement witnesses are called {\it
  decomposable} and are of the form (see \cite{Lewenstein00a}):
\begin{equation}
W^{dec}=P+Q^{\Gamma}, \label{DecomposableWitness}
\label{eq:dec_witness}
\end{equation}
where $P$, $Q$ are some positive operators. It can be easily shown
that decomposable witnesses (equivalently --- decomposable maps)
describe the new set ${\cal S}_{PPT}$ of all states that satisfy $PPT$
criterion. Like the set ${\cal S}$ of separable states this set is
also convex, compact and invariant under product unitary operations.
It has been also found that stochastic separable operations preserve
PPT property \cite{bound}. In general we have ${\cal S} \subsetneq
{\cal S}_{PPT}$. As described in previous section the two sets are
equal for $d_{A}d_{B} \leq 6$. In all other cases, they differ
\cite{Pawel97} (see Sec. \ref{subsubsec:RangePPT} for examples),
i.e. there are entangled states that are PPT. The latter states give
rise to the so-called bound entanglement phenomenon (see Sec.
\ref{sec:distil}).

To describe ${\cal S}_{PPT}$ it is enough to consider only a subset of
decomposable witnesses where $P=0$ and $Q$ is a pure projector
corresponding to entangled vector $|\Phi\rangle$. This gives a minimal
set of entanglement witnesses that describe set of PPT states. The
required witnesses are thus of the form
\begin{equation}
W=|\Phi \rangle \langle \Phi|^{\Gamma}
\label{decomposablewitness1}
\end{equation}
with some entangled vector $|\Phi\rangle$. The swap $V$ is
proportional to a witness of this kind. Indeed, we have $V=d
P_{+}^{\Gamma}$ (hence --- as announced before --- the swap is a
decomposable witness).

For $d \otimes d$ systems there is one distinguished decomposable
witness which is not of the form (\ref{decomposablewitness1}) but is
very useful and looks very simple. This is the operator
\begin{equation}
W(\Pplus)={1\over d}\id-\Pplus.
\end{equation}
One can prove that the condition $\langle W(\Pplus)
\rangle_{\varrho_{PPT}}\geq 0$ provides immediately the restriction on
the parameter called {\it fidelity} or {\it singlet
  fraction}\footnote{ One has $0 \leq F(\varrho) \leq 1$ and
  $F(\varrho)=1$ iff $\varrho=\Pplus$.  }:
\begin{equation}
F(\varrho)={\tr}[\Pplus\varrho].
\end{equation}
Namely (see \cite{Rains2001}):
\begin{equation}
F(\varrho_{PPT}) \leq \frac{1}{d}.
\end{equation}
In particular this inequality was found first for separable states and
its violation was shown to be sufficient for entanglement distillation
\cite{reduction}.

As we have already mentioned, the {\it set} of maps conditions
(\ref{mapscondition}) is equivalent to {\it set} of witnesses
condition (\ref{witnesscondition-v1}). Nevertheless any single witness
$W_\Lambda$ condition is much {\it weaker} than the condition given by
the map $\Lambda$. This is because the first is of scalar type, while
the second represents an operator inequality condition. To see this
difference it is enough to consider the two-qubit case and compare the
transposition map $T$ (which detects all entanglement in sense of PPT
test) with the entanglement witness isomorphic to it, which is the
swap operation $V$, that does not detect entanglement of any symmetric
pure state. Indeed it is not very difficult to see (see \cite{PHAE})
that the condition based on one map $\Lambda$ is equivalent to a {\it
  continuous set} of conditions defined by all witnesses of the form
$W_{\Lambda,A}\equiv A \otimes \id W_{\Lambda}A^{\dagger} \otimes \id$
where $A$ are operators on ${\cal C}^{d}$ of rank more than one. Thus
after a bit of analysis it is not difficult to see why PPT condition
associated with {\it a single} map (transposition T) is equivalent to
set of {\it all the conditions} provided by the witnesses of the form
(\ref{decomposablewitness1}).

On the other hand one must stress condition based on a witness is
naturally directly measurable \cite{Terhal2000-laa} (and it has found
recently many experimental implementations, see introduction section)
while physical implementation of separability condition based on
(unphysical) P but not CP maps is much more complicated, though still
possible (see Sec. \ref{subsubsec:MapsDetected}).  Moreover, the power
of witnesses can be enhanced with help of nonlinear corrections (see
Sec.  \ref{subsec:NonlinearImprovement}).

For higher dimensional systems there are many nondecomposable P but
not CP maps but their construction is in general hard (see
\cite{Kossakowski} and reference therein). In particular checking that
given map is P is very hard and equivalent to checking the positivity
condition (ii) in the definition of witnesses. Note that long time ago
in literature an algorithm that corresponds to checking strict
positivity of witnesses was found (ie. (ii) condition with strict
inequality) \cite{JamiolkowskiAlgorithm}) but its complexity is very
high (it has an interesting consequences though which will be
mentioned later in more details).

The important question one can ask about entanglement witnesses
regards their optimality \cite{Lewenstein00a,Lewenstein00b}.  We say
that an entanglement witness $W_{1}$ is {\it finer} than $W_{2}$ iff
the entanglement of any $\varrho$ detected by $W_{2}$ is also detected
by $W_{1}$. A given witness $W$ is called optimal iff there is no
witness finer than it. The useful sufficient condition of optimality
\cite{Lewenstein00a} is expressed in terms of the Hilbert subspace
${\cal P}_{W}=\{ |\phi\rangle|\psi\rangle : \langle \phi|\langle\psi
|W|\phi\rangle|\psi\rangle=0\}$. Namely if ${\cal P}_{W}$ spans the
whole Hilbert space then the witness is optimal. In a sense it is then
fully ``tangent'' to set of separable states. The systematic method of
optimization of given entanglement witness have been worked out first
in \cite{Lewenstein00a,Lewenstein00b} (for alternative optimization
procedure see \cite{EisertAlgorithm}, cf.  optimization of witnesses
for continuous variables \cite{EisertCVNonlinearWitnesses}).

In some analogy to the pure bipartite case, we can define the Schmidt
rank for density matrices \cite{Terhal-Pawel-rank} as $r_{S}(\varrho)=
\min(\max_{i}[r_{S}(\psi_{i})])$ where minimum is taken over all
decompositions $\varrho=\sum_{i}p_{i}|\psi_{i}\rangle \langle
\psi_{i}|$ and $r_{S}(\psi_{i})$ are the Schmidt ranks of the
corresponding pure states (see Sec. \ref{DefinitionBasic}).  One can
easily prove that separable operations can not increase it.

Now, for any $k$ in the range $\{1,\ldots,r_{max}\}$ with
$r_{\max}=\min[d_{A},d_{B}]$, we have a set ${\cal S}_{k}$ of states
with Schmidt number not greater than $k$.  For such sets we can build
a theory similar to that for ``simple'' entanglement, with
Schmidt-number witnesses in place of usual witnesses. The family of
sets ${\cal S}_{k}$ satisfies inclusion relations: ${\cal S}_{1}
\subset {\cal S}_{2} \subset ... \subset {\cal S}_{r_{max}}$. Note
that here ${\cal S}_{1}$ corresponds just to the set of separable
states, while ${\cal S}_{r_{max}}$ --- to the set of all states. Each
set is compact, convex and again closed under separable operations.
Moreover, each such set is described by $k$-positive maps
\cite{Terhal-Pawel-rank} or by Schmidt-rank $k$ witnesses
\cite{SchmidtWitn}. A Schmidt rank $k$ witness is an observable
$W_{k}$ that satisfies two conditions (i) must have at least one
negative eigenvalue and (ii) must satisfy:
\begin{equation}
\langle \Psi_{k} |W_{k}|\Psi_{k}\rangle \geq 0,
\end{equation}
for all Schmidt rank $k$ vectors $\Psi_{k}\in {\cal H}_{AB}$. As in
case of separability problem the $k$-positive maps are related via
Choi-Jamio\l{}kowski isomorphism to special maps that are called
$k$-positive (i.e. such that $[I_{k} \otimes \Lambda_{k}]$ is positive
for $I_{k}$ being identity on ${\cal B}({\cal C}^{k})$) but not
completely positive. The isomorphism is virtually the same as the one
that links entanglement witnesses $W=W_{1}$ with $k$-positive maps
$\Lambda_{1}$. Important maps (respectively witnesses) are the ones
which discriminate between ${\cal S}_{k}$ and ${\cal S}_{k+1}$. They
are those maps, which are $k$-positive but not $k+1$ positive. A nice
example is the following family of maps $\Lambda_{p}(\varrho)=\id
{\tr}(\varrho) - p(\varrho)$ which is $k$ but not $k+1$ positive for
$\frac{1}{k+1}<p \leq \frac{1}{k}$ (see \cite{Terhal-Pawel-rank}). The
case of $p=1$ corresponds to special so called {\it reduction map}
which plays a role in entanglement distillation defining in particular
reduction separability criterion which will be discussed subsequently.
Many techniques have been generalized from separability to mixed state
Schmidt rank analysis (see \cite{SchmidtWitn}). For general review of
separability problem including especially entanglement witnesses see
\cite{BrussReview,TerhalReview,BrussReflections}. The Schmidt number
witnesses and maps description is reviewed extensively in
\cite{BrussReflections}.

\subsubsection{Witnesses and experimental detection of entanglement}
\label{subsec:wit_exp_ent_det}

As already mentioned, entanglement witnesses have been found very
important in experimental detection of entanglement
\cite{sep1996,Terhal2000-laa}: for any entangled state $\varrho_{ent}$
there are witnesses which are signatures of entanglement in a sense
that they are negative on this state ${\tr}(W\varrho_{ent})<0$. Here
we shall describe few further aspects of the detection showing both
the importance of entanglement witnesses and their possible
applications.

The first issue concerns the following question \cite{Jaynes}: Given
experimental mean values of incomplete set of observables $\langle
A_{i} \rangle=a_{i}$ what information about entanglement should be
concluded basing on those data? The idea of the paper was that {\it if
  entanglement is finally needed as a resource then the observer
  should consider the worst case scenario, i.e.  should minimize
  entanglement under experimental constraints}. In other words,
experimental entanglement should be of the form:
\begin{equation}
E(a_{1},\ldots,a_{k})={\inf}_{\{ \langle A_{i}\rangle_\varrho=a_{i}
\}_{i=1}^{k}} E(\varrho).
\end{equation}
Such minimization of entanglement of formation and relative entropy of
entanglement was performed for given mean of Bell observable on
unknown 2-qubit state. It is interesting that in general there are
many states achieving that minimum --- to get single final state the
authors also have proposed final application of maximum entropy Jaynes
principle \cite{Jaynes}.

Quite recently the idea of minimization of entanglement under
experimental constraints was applied with help of entanglement
witnesses \cite{Reimpell,EisertBA}. In \cite{Reimpell} using convex
analysis the authors have performed minimization of convex
entanglement measures for given mean values of entanglement witnesses
basing on approximation of convex function by affine functions from
below. Specific estimates have been performed for existing
experimental data. Independently a similar analysis of lower bounds
for many entanglement measures has been performed in \cite{EisertBA},
where the emphasis was put on analytical formulas for specific
examples. The derived formulas \cite{Reimpell,EisertBA} provide a
direct quantitative role for results of entanglement witnesses
measurements. Note that more refined analysis was focused on
correlations obtained in the experiment, identifying which types of
correlations measured in incomplete experiments may be already
signature of entanglement \cite{PlenioA}.

Another experimental issue, where entanglement witnesses have been
applied, is the problem of macroscopic entanglement at finite
temperature. A threshold temperature for existence of entanglement can
be identified. The relation between thermal equilibrium state and
entanglement was hidden already in 2-qubit analysis of Jaynes
principle and entanglement \cite{Jaynes}. The first explicit analysis
of entanglement in thermal state was provided by Nielsen
\cite{NielsenPhD} where first calculation of temperatures for which
entanglement is present in two-qubit Gibbs state was performed. A
fundamental observation is that entanglement witnesses theory can be
exploited to detect entanglement in general (multipartite) thermal
states including systems with large number of particles
\cite{BrucknerVedralTemperature,TothTemperature}. In the most elegant
approach, for any observable $O$ one defines entanglement witness as
follows (see \cite{TothTemperature}):
\begin{equation}
W_{O}=O-\inf_{|\Psi_{prod}\rangle} \langle
\Psi_{prod}|O|\Psi_{prod}\rangle, \label{ThermalWitness}
\end{equation}
where infimum is taken over all product pure states $\Psi_{prod}$
(note that the method can be extended to take into account partial
separability\footnote{See Sec.  \ref{MultipartiteEntanglement}.} as
well). Now if $W_{O}$ has negative eigenvalue becomes immediately an
entanglement witness by construction. In case of spin lattices one
takes $O=H$ where $H$ is a Hamiltonian of the system and calculates
$\langle W_{H}\rangle_{\varrho}$ for quantum Gibbs state
$\varrho_{Gibbs}=\exp(-H/kT)/{\tr}[\exp(-H/kT)]$. It can be
immediately seen that for $H$ with discrete spectrum the observable
$W_{H}$ has a negative eigenvalue iff the lowest energy state is
entangled and then the observable becomes entanglement witness by
construction (see (\ref{ThermalWitness})). In this one can estimate
the range of temperatures for which the mean value $\langle
W_{H}\rangle_{\varrho_{Gibbs}}$ is negative \cite{TothTemperature}.
Further improvements involve uncertainty based entanglement witnesses
\cite{AndersVedralTemperature} and applications of entanglement
measures like robustness of entanglement
\cite{MarkhamVedralTemperature} to thermal entanglement.

Finally let us recall an important issue: how to decompose given
witness into locally measurable observables
\cite{GuehneLocalMeasurementWitness}, \cite{GuehneHBE_s}. This is an
important issue since if we want to detect entanglement between
spatially separated systems we can only measure mean values of tensor
products of local observables consistent with spatial separation
(cf. Sec. \ref{sec:LOCC}). For a given witness $W$ in $d_{A}\otimes
d_{B}$ one can ask about the minimal number of local measurements on
systems $A$, $B$ that can reconstruct the mean value of the whole
witness. The problem is to find optimal decomposition
\begin{equation}
W=\sum_{k=1}^{r} \gamma_{k} X_{A}^{k} \otimes Y_{B}^{k}
\label{LocalWitness}
\end{equation}
onto a set of product of normalized (in Hilbert-Schmidt norm)
observables such that any two pairs $X_{A}^{k} \otimes Y_{B}^{k}$,
$X_{A}^{k'} \otimes Y_{B}^{k'}$ in the sum differ on at most one
side. Then the cardinality of the representation (\ref{LocalWitness})
is calculated as $s=r-r_{I}$ where $r_{I}$ is a number of those
product terms $\{ X_{A}^{k'} \otimes Y_{B}^{k'}\}_{k'=1}^{r_{I}}$ in
which at least one of local observables is proportional to identity
$\id$ (say $X_{A}^{k'}=\alpha \id$) and the second one ($Y_{B}^{k'}$)
is linearly dependent on all the other local ones (i.e. $
Y_{B}^{k'}=\sum_{k\neq k'} \alpha_k Y_{B}^{k}$). The optimal
cardinality $s_{min}$ minimize over all decompositions of the type
(\ref{LocalWitness}) gives minimal number of different measurement
settings needed to measure mean of given entanglement between
spatially separated systems. The problem of finding the optimal
decomposition (\ref{LocalWitness}) with the minimal cardinality
$s_{min}$ has been investigated in papers
\cite{GuehneLocalMeasurementWitness,GuehneHBE_s} and analytical
solutions have been found for both bipartite cases discussed above as
well as for multipartite generalizations.

\subsubsection { Entanglement witnesses and Bell inequalities}
\label{subsec:Bell_witness}

Entanglement witnesses (see Sec. \ref{sec:BipartiteEntanglement}) are
Hermitian operators that are designed directly for detection of
entanglement. In 2000 Terhal first considered a possible connection
between entanglement witnesses and Bell inequalities \cite
{TerhalB}. From a ``quantum'' point of view, Bell inequalities are
just nonoptimal entanglement witnesses. For example one can define the
CHSH-type witness which is positive on all states which admit
LHVM
\begin{equation}
  W_{CHSH}=2\cdot\id-\bcal_{CHSH}, \label{ter}
\end{equation}
where $\bcal_{CHSH}$ is the CHSH operator (\ref {B}). Such defined
witness is of course nonoptimal one since the latter is strictly
positive on {\it separable} states. However, basing on the concept of
the optimal witness \cite {Lewenstein00a} one can estimate how much
optimal witnesses have to be shifted by the identity operator to make
them positive on all states admitting a LHVM. Moreover, there exists a
natural decomposition of the CHSH witness into two optimal witnesses,
and the identity operator \cite {Hyllus}. In multipartite case, Bell
inequalities can even detect so called bound entanglement
\cite{Dur,Kaszlikowski, Sen, Augusiak} (see
Sec. \ref{subsec:bellboundent}).

Inspired by CHSH inequalities Uffink proposed inequalities
\cite{Uffink2002}, which are no longer implied by LHVM, yet constitute
a (nonlinear) entanglement witness (see
Sec. \ref{subsec:NonlinearImprovement}), see in this context \cite{UffinkS-strong-Bell-2006}.

Svetlichny and Sevnick
proposed Bell inequalities that can be used as a detector of genuinely
multipartite entanglement \cite{Svetlichny}.  Toth and G{\"u}hne
proposed a new approach to entanglement detection  \cite
{TothG_s,TothG_ent_det}based on the stabilizer theory \cite {Gottesman}. In particular,
they found interesting connections between entanglement
witnesses and Mermin-type inequalities \cite{TothG_ent_det}.

In general, the problem of the relation between Bell inequalities and
entanglement witnesses is very complex. It follows from the very large
number of degrees of freedom of the Bell inequalities.  Nevertheless
it is basic problem, as the Bell observable is a {\it double}
witness. It detects not only entanglement but also nonlocality.

It is interesting that the loophole problem in the experimental tests
of Bell inequalities (see \cite{Gill}) has found its analogy
\cite{SkwaraKKB2006} in an entanglement detection domain. In
particular the efficiency of detectors that still allows to claim that
entanglement was detected, has been analyzed and related to
cryptographic application.

\subsubsection{Distinguished maps criteria: reduction criterion and its extensions
\label{subsubsec:Reduction}}

There are two important separability criteria provided by P but not CP maps.
The first one is the so called {\it reduction criterion}
\cite{reduction,CerfAG} defined by the formula (\ref{mapscondition})
with the {\it reduction map}: $\Lambda^{red}(\varrho)= \id
{\tr}(\varrho)-\varrho$. This map is decomposable but --- as we shall
see subsequently --- plays important role in entanglement distillation
theory\cite{reduction}.  Only in case of two-dimensional Hilbert space
the map represents just a reflection in Bloch sphere representation
\cite{BengtssonZyczkowski-book} and can be easily shown to be just
equal to transpose map $T$ followed by $\sigma_y$
i.e. $\sigma_yT(\varrho)\sigma_y$. As such it provides a separability
condition completely equivalent to PPT in this special (two-qubit)
case. In general the reduction separability criterion $[\id_{A}
\otimes \Lambda^{red}_{B}](\varrho_{AB})\geq 0$ generated by
$\Lambda^{red}$ can be written as:
\begin{equation} \varrho_{A} \otimes \id
-\varrho_{AB}\geq 0
\end{equation}
and since $\Lambda^{red}$ is decomposable \cite{reduction} the
corresponding separability criterion is weaker than PPT one (see
Sec. \ref{subsubsec:maps}). On the other hand it is interesting that
this criterion is stronger \cite{Hiroshima} than mixing separability
criteria \cite{NielsenKempe} as well as some entropic criteria with
$\alpha \in [0,1]$ and $\alpha=\infty$ (for the proofs see
\cite{VollbrechtW} and \cite{reduction} respectively).

Another very important criterion is the one based on the map due to
Breuer and, independently, Hall \cite{Breuer2006-prl,Hall2006} which
is a modification of reduction map on even dimensional Hilbert space
$d=2k$. On this subspace there exist antisymmetric unitary operations
$U^{T}=-U$ (for instance the one $U=antidiag[1,-1,1,-1,\ldots,1,-1]$
\cite{Breuer2006-prl}). The corresponding antiunitary map
$U(\cdot)^{T}U$, maps any pure state to some state that is orthogonal
to it. This leads to the conclusion that the map which acts on the
state $\varrho$ as follows:
\begin{equation}
\Lambda(\varrho)=\Lambda^{red}(\varrho)-U(\varrho)^{T}U^{\dagger}
\end{equation}
is positive for any antisymmetric $U$. This map is {\it not
  decomposable} and the entanglement witness $W_{\Lambda}$ corresponding to it has an optimality property since the corresponding space ${\cal P}_{W_{\Lambda}}$
(see Sec. \ref{subsubsec:Witnesses}) is the full Hilbert space
\cite{Breuer2006-jpamg}. This nondecomposability property allows the
map to detect special class of very weak entanglement, namely PPT
entanglement mentioned already before. We shall pass to its more
detailed description now.

\subsubsection{Range criterion and its applications; PPT entanglement
\label{subsubsec:RangePPT}}

The existence of nondecomposable maps (witnesses) for the cases with
$d_{A}\cdot d_{B}>6$ implies that there are states {\it that are
  entangled but PPT} in all those cases. Thus the PPT test is no
longer a sufficient test of separability in those cases. This has
striking consequences, for quantum communication theory, including
entanglement distillation and quantum key distribution which we
discuss further in this paper. Existence of PPT entangled states was
known already in terms of cones in mathematical literature (see for
example \cite{Choi82}) sometimes expressed in direct sum language.

On physical ground, first examples of entangled states that are PPT
were provided in \cite{Pawel97}, following Woronowicz construction
\cite{Woronowicz}. Their entanglement had to be found by a criterion
that is independent on PPT one. As we already mentioned, this might be
done with properly chosen P but not CP nondecomposable map (see
Sec. \ref{subsubsec:maps}). In \cite{Pawel97} another criterion was
formulated for this purpose, which is useful for other applications
(see below). This is the {\it range criterion}: if $\varrho_{AB}$ is
separable, then there exists a set of product vectors $\{ \psi_{A}^{i}
\otimes \phi_{B}^{i}\} $, such that it spans range of $\varrho_{AB}$
while $\{ \psi_{A}^{i} \otimes (\phi_{B}^{i})^{*}\}$ spans range of
$\varrho_{AB}^{T_{B}}$, where complex conjugate is taken in the same
basis in which PPT operation on $\varrho_{AB}$ has been performed. In
particular an example of $3 \otimes 3$ PPT entangled state revealed by
range criterion (written in a standard basis) was provided:
\begin{eqnarray}
\varrho_{a}={1 \over 8a + 1}
\left[ \begin{array}{ccccccccc}
     a &0&0&0&a&0&0&0& a   \\
      0&a&0&0&0&0&0&0&0     \\
      0&0&a&0&0&0&0&0&0     \\
      0&0&0&a&0&0&0&0&0     \\
     a &0&0&0&a&0&0&0& a     \\
      0&0&0&0&0&a&0&0&0     \\
0&0&0&0&0&0&{1+a \over 2}&0&{\sqrt{1-a^2} \over 2}\\
      0&0&0&0&0&0&0&a&0     \\
     a &0&0&0&a&0&{\sqrt{1-a^2} \over 2}&0&{1+a \over 2}\\
      \end{array}
    \right ], \ \ \
\label{tran}
\end{eqnarray}
where $0 < a < 1$. Further examples of PPT states that are entangled
can be found in \cite{Alber2001}.

Especially interesting way of application of range criterion to
finding PPT states is unextendible product basis (UPB) methods
\cite{BennettUPBI1999,UPB2}. UPB is a set ${\cal S}_{UPB}$ of
orthonormal product vectors in ${\cal H}_{AB}={\cal H}_{A} \otimes
{\cal H}_{B}$ such that there is no product vector that is orthogonal
to all of them.

{\it Example .-} An example in the $3 \otimes 3$ case is
\cite{BennettUPBI1999}: $S_{UPB} \equiv \{ |0\ra(|0\ra +|1\ra),
(|0\ra+|1\ra)|2\ra, |2\ra(|1\ra+|2\ra) (|1\ra+|2\ra)|0\ra, (|0\ra-|
1\ra +|2\ra)(|0\ra-|1\ra+|2\ra) \}$.

Since there is no product vector orthogonal to the subspace ${\cal
  H}_{UPB}$ spanned by elements of ${\cal S}_{UPB}$, any vector from
the orthogonal subspace ${\cal H}_{UPB}^{\perp}$ (spanned by vectors
orthogonal to ${\cal H}_{UPB}$) is entangled. Consequently, by the
range criterion above, any mixed state with support contained in
${\cal H}_{UPB}^{\perp}$ is entangled. In particular, a special class
of states proportional to the projector $P_{{\cal
    H}_{UPB}^{\perp}}=\id - P_{{\cal H}_{UPB}}$ (here $P_{{\cal H}}$
stands for projection onto the subspace ${\cal H}$) is also entangled,
but it can be shown to be PPT because of the special way in which the
projector $P_{{\cal H}_{UPB}}$ was constructed. In this way the notion
of UPB leads to construction of PPT entangled states.  This result was
further exploited to provide new nondecomposable maps
\cite{Terhal2000-laa}. What was the mathematical origin of the
construction? This is directly related to the question: what is the
other way PPT entanglement can be detected? As we already announced
there is a rule linked directly to the positive-map separability
condition (see \cite{sep1996}): {\it for any PPT entangled state there
  is a nondecomposable P but not CP map $\Lambda$ such that the
  criterion (\ref{mapscondition}) is violated}. Thus a way to detect
PPT entanglement is to find a proper nondecomposable P but not CP
map. The alternative statement, saying that PPT state is entangled if
an only if it is detected by some nondecomposable witness (i.e. the
one that is not of the form (\ref{DecomposableWitness})) is
immediately also true.

It is rather hard to construct nondecomposable maps and witnesses
respectively. En example of P but nor CP nondecomposable map due to
Choi is \cite{Choi82} (see \cite{Kossakowski} and references therein
for generalization):
\begin{eqnarray}
&&\Lambda( \left[ \begin{array}{ccc}
a_{11} & a_{12}& a_{13}  \\
a_{21} &a_{22}&a_{23}    \\
a_{31} &a_{32}& a_{33}
       \end{array}
      \right ])=
\left[ \begin{array}{ccc}
a_{11}+a_{33}  & - a_{12}& -a_{13}  \\
-a_{21} &a_{22}+ a_{11}& -a_{23}    \\
-a_{31} &-a_{32}& a_{33}+a_{22}
       \end{array}
      \right ], \nonumber \\
\end{eqnarray}
which allows to detect PPT entangled states that help in bound
entanglement activation (see \cite{activation}).

The new technique for achieving the nondecomposable maps was worked
out by Terhal \cite{Terhal2000-laa}. Her idea was to take a projector
on UPB space $P_{{\cal H}_{UPB}}$ and observe that the following
quantity $\epsilon = min_{\Psi_{sep}} \langle \Psi_{sep}|P_{{\cal
    H}_{UPB}} | \Psi_{sep} \rangle $ is {\it strictly positive}
because of unextendibility property. Then the following operator on $d
\otimes d$ space
\begin{equation}
W_{UPB}=P_{{\cal H}_{UPB}} - d \epsilon |\Psi_{max}\rangle \langle
\Psi_{max}|, \label{TerhalMap}
\end{equation}
where $\Psi_{max}$ is a maximally entangled state such that
$|\Psi_{max} \rangle \notin {\cal H}_{UPB} $. Since one can always
show that such a vector exists any entanglement witness of the above
form detects PPT entanglement of $\varrho_{UPB}= \frac{1}{\cal N}(\id
-P_{{\cal H}_{UPB}})$ since the mean value
${\tr}(W_{UPB}\varrho_{UPB}$) will gain only the (negative)
contribution form the second term of the (\ref{TerhalMap}). At the
same time the optimization of the $\epsilon$ above guarantees that the
$W_{UPB}$ has nonnegative mean value on any product state.  Terhal has
calculated explicitly the lower bounds for the parameter $\epsilon$ in
cases of few examples of UPB-s.

This idea was further generalized to the case of {\it edge} states
\cite{Lewenstein00a}, (see also
\cite{Karnas,Ho00,LewensteinSanpera-bsa}). A state is called ``edge''
and denoted as $\delta_{edge}$ if it satisfies two properties (i) PPT
property and (ii) extremal violation of the range criterion (i. e.
there should be no $|\phi \rangle| \psi\rangle \in {\cal R}(\varrho)$
such that $|\phi\rangle| \psi^{*}\rangle \in {\cal
  R}(\varrho^{\Gamma})$). It can be seen (see below) that entanglement
of any PPT entangled state is due to some ``nonvanishing'' admixture
of edge state. An example of edge state is just any state based on UPB
construction $\varrho_{UPB}$. Another edge state is $2 \otimes 4$ PPT
entangled state from \cite{Pawel97}.

Now generalization of Terhal construction leads to a method that in
fact detects any PPT entanglement is \cite{Lewenstein00b}:
\begin{equation}
W= P + Q^{\Gamma}-\frac{\epsilon}{c}C,
\end{equation}
with $P$, $Q$ being positive operators supported on kernels of
$\delta_{edge}$ and $\delta_{edge}^{\Gamma}$ respectively while the
parameter $\epsilon=\min_{\Psi_{sep}}\langle \Psi_{sep}|P + Q^{\Gamma}
| \Psi_{sep}\rangle$ can be shown to be strictly positive (by extremal
violation of range criterion by $\delta_{edge}$ state) while $C$ is
arbitrary positive operator with ${\tr}(\delta_{edge})>0$ and
$c=max_{\Psi_{sep}}\langle \Psi_{sep}|C| \Psi_{sep}\rangle$. All the
above witnesses {\it are nondecomposable } and it is interesting that
entanglement of all PPT entangled states can be detected even by a
restricted subclass of the above, when $P$, $Q$ are just projectors on
the kernels of $\delta_{edge}$ while $C$ is the identity operator
(then $c=1$). Of course all maps isomorphic to the above witnesses are
also nondecomposable.

We would like also to mention a nice idea of construction of
nondecomposable map based on $k$-positive maps. Namely it happens that
for any map $\Lambda_{k}$ that is $k$-positive but not $k+1$ positive
(cf. application in Schmidt rank of density matrix
\cite{Terhal-Pawel-rank}) the positive (by definition) map $I_{k}
\otimes \Lambda_{k}$ is already nondecomposable \cite{PianiM}. This is
also connected to another construction of PPT entangled states
\cite{PianiM} which was inspired by very useful examples due to
\cite{Ishizaka04} in pure states interconvertibility.

Coming back to range criterion introduced here, there is an
interesting application: the so called Lewenstein-Sanpera
decomposition. Namely any bipartite state $ \varrho $ can be uniquely
decomposed (see \cite{KarnasL2000-bsa}) in the following way
\cite{LewensteinSanpera-bsa}:
\begin{equation}
\varrho=(1-p)\varrho_{sep}+ p \sigma,
\end{equation}
where $\varrho_{sep}$ (called best separable approximation BSA) is a
separable state, $\sigma$ is entangled and $p^{*}$ is a maximal
probability $p^{*}\in [0,1]$ such that the decomposition of the above
form but with $\sigma$ taken to be arbitrary state is still
true. Clearly $\varrho$ is separable iff $p^{*}=1$.  For two-qubits
the entangled part $\sigma$ is always pure
\cite{LewensteinSanpera-bsa}.  Moreover the decomposition can be then
found in a fully algebraic way without optimization procedure
\cite{WellensKus}. In particular if $\varrho_{sep}$ is of full rank,
then $\sigma$ is maximally entangled and $p^{*}$ is just equal to so
called Wootters concurrence (see \cite{WellensKus} for the proof). In
general the way of construction of the decomposition requires
technique of subtracting of product states from the range of the
matrix \cite{LewensteinSanpera-bsa} (cf.  \cite{Karnas,Ho00}). This
was the first proposal of systematic way of checking separability.

Similar technique has been used to find the decomposition of PPT
entangled states. Namely it happens that any PPT entangled state can
be decomposed into the form $\varrho=(1-p) \varrho_{sep} +
p\delta_{edge}$ where $\delta_{edge}$ is an edge state defined
above.
This is a systematic method leading to necessary and sufficient
separability test for states which are PPT i.e. in the
region where checking separability is the hardest (the state is
separable if at the end the parameter $p$ is zero). Same difficulty
like in BSA method is finding product vector in  the range of the
matrix. The problem becomes much more tractable in case of states
that satisfy low rank condition \cite{Ho00} $r(\varrho)+
r(\varrho^{\Gamma}) \leq 2d_{A}d_{B}-d_{A}-d_{B}+2$. Then
typically (in so called generic case) the state has only finite
number of product states satisfying the range criterion which can
be found by solving polynomial equations. In such cases the
separability problem can be solved in finite number of steps
\cite{Ho00}.

Let us also mention that recently in analogy to the BSA construction
above the decomposition onto the symmetrically extendible and
nonextendible parts has been found which has a very nice cryptographic
application \cite{MoroderCL2005-povmintr} (see
Sec. \ref{sec:Ent_in_QKD}). The BSA construction leads also to an
entanglement measure (see Sec. \ref{subsubsec:other-measures}).

\subsubsection{Matrix realignment criterion and linear contractions
criteria}
\label{subsubsec:realignment}

There is yet another strong class of criteria based on linear
contractions on product states. They stem from the new criterion
discovered in \cite{Rudolph2003-JPA}, \cite{ChenWu} called computable
{\it cross norm (CCN) criterion} or {\it matrix realignment criterion}
which is operational and independent on PPT test \cite{Peres96}. In
terms of matrix elements it can be stated as follows: if the state
$\varrho_{AB}$ is separable then the matrix ${\cal R}(\varrho)$ with
elements
\begin{equation}
\langle m| \langle \mu| {\cal R}(\varrho_{AB}) |n \rangle |\nu
\rangle \equiv \langle m| \langle n| \varrho |\nu \rangle |\mu
\rangle
\end{equation}
has trace norm not greater than one (there are many other variants see
\cite{HHH02-permut}.

It can be formally generalized as follows: {\it if $\Lambda$ satisfies
\begin{equation} \|\Lambda(|\phi_{A}\rangle\langle
\phi_{A}| \otimes |\phi_{B}\rangle\langle \phi_{B}|) \|_1\leq 1
\label{LinearContraction}
\end{equation}
for all pure product states $|\phi_{A}\rangle\langle \phi_{A}| \otimes
|\phi_{B}\rangle\langle \phi_{B}|$ then for any separable state
$\varrho_{AB}$ one has $\|\Lambda(\varrho_{AB})\|_1\leq 1$\footnote{
Here $\|X\|_1=\tr \sqrt{X X^\dagger}$ denotes trace norm.}}. The
matrix realignment map ${\cal R}$ which permutes matrix elements just
satisfies the above contraction on products criterion
(\ref{LinearContraction}). To find another interesting contractions of
that type that are not equivalent to realignment is an open problem.

Quite remarkably the realignment criterion has been found to detect
some of PPT entanglement \cite{ChenWu}(see also
\cite{Rudolph2003-JPA}) and to be useful for construction of some
nondecomposable maps. It also provides nice lower bound on concurrence
function (see \cite{AlbeverioLowerBoundPRL}). On the other hand it
happens that for any state that violates the realignment criterion
there is a local uncertainty relation (LUR) (see
Sec. \ref{subsec:uncertainty}) that is violated but converse statement
is not always true \cite{LURstrongerthanCCN}. On the other hand
finding LUR-s (like finding original entanglement witnesses) is not
easy in general and there is no practical characterization of LUR-s
known so far, while the realignment criterion is elementary, fast in
application and still powerful enough to detect PPT entanglement.

\subsubsection{Some classes of important quantum states: entanglement
regions of parameters}
\label{subsec:Werner_Iso}

In this section we shall recall classes of states or which PPT
property is equivalent to separability.

We shall start from Werner states that are linked to one of the most
intriguing problem of entanglement theory namely NPT bound
entanglement problem
\cite{DiVincenzoSSTT1999-nptbound,DurCLB1999-npt-bound} (see
Sec. \ref{sec:distil}).

{\it  Werner $d \otimes d$ states \cite{Werner1989} .-} Define
projectors $P^{(+)}=(\id + V)/ 2,P^{(-)}=(\id - V)/ 2$ with identity
$\id$ , and ``flip'' operation $V$ (\ref{swap}):

The following $d \otimes d$ state
\begin{equation}
W(p)=(1-p)\frac{2}{d^{2}+d}P^{(+)}+p\frac{2}{d^{2}-d}P^{(-)}, \ 0\leq p
\leq 1 \label{Werner-state}
\end{equation}
is invariant under any $U \otimes U$ operation for any unitary $U$.
$W(p)$ is separable iff it is PPT which holds for $0 \leq p \leq
\frac{1}{2}$.

{\it Isotropic states \cite{reduction} .-} They are $U \otimes U^{*}$
invariant (for any unitary $U$) $d \otimes d$ states. They are of the
form
\begin{equation}
\varrho_{F}={1-F\over d^2-1} \id +{Fd^2-1\over d^2 -1} P_+,\ \
0\leq F \leq 1 \label{ntryplet}
\end{equation}
(with $P^{+}$ defined by (\ref{MaximallyEntangledProjector})).  An
isotropic state is separable iff it is PPT which holds for $0\leq F
\leq \frac{1}{d}$.

{\it ``Low global rank class'' \cite{Ho00} .-} The general class of
$d_{A} \otimes d_{B}$ state of all states which have global rank not
greater than local ones: $r(\varrho_{AB})\leq \max [r(\varrho_{A})$,
$r(\varrho_{B})]$.  Here again PPT condition is equivalent to
separability. In particular for
$r(\varrho_{AB})=r(\varrho_{A})=r(\varrho_{B})$ the PPT property of
$\varrho_{AB}$ implies separability \cite{Ho00}. If
$r(\varrho_{AB})<\max[r(\varrho_{A})$, $r(\varrho_{B})]$ (which
corresponds to violation of entropic criterion for $\alpha=\infty$)
then PPT test is violated, because reduction criterion is stronger
than $S_{\infty}$ entropy criterion \cite{HSTT}.

\subsubsection{Characterization of bipartite separability in terms of
biconcurrence}

In this section we shall describe a quadratic function of the state
that provides necessary and sufficient condition for separability
called {\it biconcurrence}. This function was inspired by a
generalization of two-qubit Wootters' concurrence due to
\cite{RungtaBCHM2001-concurrence}, that exploited the {\it universal
  state inverter}, which in turn is actually the reduction map
$\Lambda^{red}$ (see Sec. \ref{subsubsec:Reduction}). The generalized
concurrence can be written in the form $C(\psi_{AB})= \sqrt{{1\over 2}\langle
\psi_{AB}|[\Lambda^{red}_{A} \otimes
\Lambda^{red}_{B}](|\psi_{AB}\rangle \langle \psi_{AB}|)|\psi_{AB}}
\rangle=\sqrt{2(1-{\tr}(\varrho_{B}^{2})}$, which directly reproduces
Wootters' concurrence in case of two-qubits (see Sec. \ref{par:concurrence}). In
\cite{BadziagDHHH01-conc}, a simplified form was obtained:
\begin{eqnarray}
C(\psi)&=&\sqrt{\langle \psi_{AB}|[I_{A} \otimes
\Lambda^{red}_{B}](|\psi_{AB}\rangle \langle \psi_{AB}|)|\psi_{AB}
\rangle }\nonumber\\
&=&\sqrt{1-{\tr}(\varrho_{B}^{2})}.
\end{eqnarray}
Now for any ensemble realizing mixed state $\varrho=\sum_{i=1}^{k}
p_{i}|\tilde{\psi}_{AB}^{i}\rangle \langle \tilde{\psi}_{AB}^{i}|$
with $k \leq N:=(d_{A}d_{B})^{2}$ and
$|\psi_{i}\rangle:=\sqrt{p_{i}}|\tilde{\psi}_{AB}^{i}\rangle $, the
$N \otimes N$ {\it biconcurrence matrix} $B_{m\mu,n\nu}$ is defined as:
\be
B_{m\mu,n\nu} \equiv \langle
\psi_{m}|[\id \otimes \Lambda^{red}](|\psi_{\mu}\rangle \langle
\psi_{n}|)|\psi_{\nu} \rangle
\label{biconcurrencematrix}
\ee
(we have dropped here the subsystem indices $AB$ and extended the
ensemble to $N$-element one by adding extra $N-k$ zero vectors). It
can be written equivalently as\footnote{A simple expression for biconcurrence was
  exhibited in \cite{MintertKB04-conc}.
  $B_{m,\mu,n,\nu}=\<\psi_m|\<\psi_\mu| P^{(-)}_{AA'}\ot P^{(-)}_{BB'}
  |\psi_n\>|\psi_\nu\>$.}
\begin{equation}
B_{m \mu,n \nu}=\langle \psi_m|\psi_n \rangle
\langle\psi_\mu|\psi_\nu \rangle -
 {\tr}[(A^{\psi_{m}})^{\dagger}A^{\psi_{\mu}}
(A^{\psi_{n}})^{\dagger}A^{\psi_{\nu}}]
\label{eq:bicon-latwe}
\end{equation}
with the matrix of coefficients $A^{\psi}$ defined by the relation
$\psi=\sum_{i=0}^{d_{A}-1}\sum_{j=0}^{d_{B}-1}
A^{\psi}_{ij}|i\rangle|j\rangle $. Now, there is an important
observation namely {\it the state $\varrho_{AB}$ is separable if and
only if} the scalar biconcurrence function
\begin{equation}
{\cal B}(\varrho):=\inf_{U} \sum_{m=1}^N[U \otimes U B U^{\dagger}
\otimes U^{\dagger}]_{mm,mm} \label{biconcurrencefunction}
\end{equation}
vanishes \cite{BadziagDHHH01-conc}. Here infimum (equal to minimum) is
taken over all unitary operations $U$ defined on Hilbert space ${\cal
  H}$ ($dim \hcal = d_{A}^{2}d_{B}^{2}$) while the matrix $B$
represents operator on ${\cal H} \otimes {\cal H}$.
As a matter of fact value of the biconcurrence function
(\ref{biconcurrencefunction}) is just the
square of Euclidean norm of concurrence vector introduced  in
\cite{AudenaertVM2000-concurrence} (see sec. \ref{par:concurrence}).

It is interesting that, as one can easily show, ${\cal
B}(\varrho)=\inf_{\{p_{i},\psi_{i}\}}
\sum_{i=1}^{N} p_{i}^{2}C(\psi_{i})^{2}$ where infimum is taken over
all $N$-element ensembles realizing given state $\varrho$.
Note that putting just
square root under sum provides the seminal concurrence value
for mixed states
\begin{equation}
{\cal
  C}(\varrho)={\inf}_{\{p_{i},\psi_{i}\}}
\sum_{i=1}^{k}p_{i}C(\psi_{i}),
\end{equation}
which has an advantage of being an
entanglement monotone and can be bounded analytically
\cite{MintertKB04-conc}).  Still, as in all concurrence cases
there is no direct algorithm known that finds the minimum efficiently
even for low dimensional systems.

Interestingly, the biconcurrence matrix allows to formulate separability problem in terms of a single "entanglement witness", though acting on a completely different Hilbert space
\cite{BadziagHHA-single-wit-2007}. Namely for a give state $\rho$ one constructs a special witness $W_{\rho}$ such that the state is separable iff this witness vanishes on some product state.

\subsubsection{Enhancing separability criteria by local filters}
For {\it strictly positive} density matrices one can strengthen separability criteria
by use of the following result of \cite{LeinaasMO2006}  (see also \cite{VerstraeteDM2003}).
For any such state $\rho$ there exist invertible operators $A$ and $B$ such
that
\be
\tilde \rho = A\ot B \rho A^\dagger \ot B^\dagger=
{1\over d_A d_B}(\id +\sum_{i=1}^{d_A^2-1} a_i E_i\ot F_i),
\ee
where $E_i$ ($F_i$) are traceless orthonormal hermitian operators. The operation is called filtering (see sec. \ref{subsec:locc})).
The operators $A$  and $B$ can be found constructively.
Due to invertibility  of operators $A$ and $B$, the (unnormalized) state $\tilde\rho$ is entangled if and only if the original state
is entangled. Thus the separability problem reduces to checking states of the above form.
They have, in particular, maximally mixed subsystems. Given any separability criterion,  it often proves
useful to apply it to the filtered state $\tilde \rho$ rather than to original state (see e.g. \cite{GuhneHGE2007}).

\section{Multipartite entanglement --- similarities and differences}
\label{MultipartiteEntanglement}

In multipartite case the qualitative definition of separability and
entanglement is much richer then in bipartite case. There is the
so-called full separability, which is the direct generalization of
bipartite separability. Moreover, there are many types of {\it
  partial} separability. Below we will briefly discuss the
separability criteria in this more complicated situation.

\subsection{Notion of full ($m$-partite) separability }
\label{sec:full-sep}

The definition of full multipartite separability (or just
$m$-separability) of $m$ systems $A_{1}...A_{m}$ with Hilbert space
${\cal H}_{A_{1}...A_{n}}= {\cal H}_{A_{1}} \otimes ... \otimes {\cal
  H}_{A_{m}}$ is analogous to that in bipartite case: $\varrho_{AB} =
\sum_{i=1}^{k} p_{i} \varrho_{A_1}^{i} \otimes ...\otimes
\varrho_{A_{m}}^{i}$. The Caratheodory bound is kept $k\leq dim {\cal
  H}_{A_{1}...A_{m}}^{2}$.  Such defined set of $m$-separable states
is again (i) convex and (ii) closed (with respect to trace norm)
Moreover separability is preserved under $m$-separable operations
which are immediate generalization of bipartite separable ones
\begin{equation}
\varrho_{A_1,\ldots,A_{m}} \rightarrow {\sum_{i} A_{i}^{1} \otimes \cdots
\otimes A^{n}_{i} \varrho_{A_{1},\ldots,A_{m}} (A_{i}^{1} \otimes \cdots
\otimes A^{n}_{i})^{\dagger}\over  \tr(\sum_{i} A_{i}^{1} \otimes \cdots
\otimes A^{n}_{i} \varrho_{A_{1},\ldots,A_{m}} (A_{i}^{1} \otimes \cdots
\otimes A^{n}_{i})^{\dagger})}
\end{equation}
The separability characterization in terms of positive but not
completely positive maps and witnesses generalizes in a natural way
\cite{multisep}. There is a condition analogous to
(\ref{mapscondition}) with $\id$ acting on first subsystem ${\cal
  H}_{A_{1}}$ and the map $\Lambda_{A_{2}...A_{m}}: {\cal B}({\cal
  H}_{A_{2},\ldots,A_{m}})\rightarrow {\cal B}({\cal
  H}_{A_{1}})$. Namely, in the formula (\ref{mapscondition}) we take
the maps $\Lambda_{A_{2}...A_{m}}: {\cal B}({\cal
  H}_{A_{2},\ldots,A_{m}}) \rightarrow {\cal B}({\cal H}_{A_{1}})$
that are {\it positive on product states}
ie. $\Lambda_{A_{2}...A_{m}}(|\phi_{A_{2}}\rangle \langle \phi_{A_{2}}
| \otimes \cdots \otimes |\phi_{A_{m}}\rangle \langle
\phi_{A_{m}}|)\geq 0$ (with arbitrary states $\phi_{A_{i}}\in {\cal
  H}_{A_{i}}$) but not completely positive. The corresponding
entanglement witness must have again (i) at least one negative
eigenvalue and also satisfy (ii)
\begin{equation}
\langle \phi_{A_{1}}| ...\langle \phi_{A_{m}}| W
|\phi_{A_{1}}\rangle ...|\phi_{A_{m}}\rangle \geq 0.
\label{witnesscondition(m)}
\end{equation}
Maps and witnesses are again related by the isomorphism
(\ref{isomorphism}) with maximally entangled state $P_{+}$ on
bipartite system $A_{1}A_{1}$.

The above description provides full characterization of
$m$-separability of $m$-partite system.  An example of maps positive
on product states is simply a product of positive maps. Of course
there exist maps that are positive on product states, but are not of
the latter form.  (those are in particular maps \cite{multisep}
detecting entanglement of some semiseparable states constructed in
\cite{BennettUPBI1999}, see one of examples below). Multipartite witnesses
and related maps were investigated in \cite{JafarizadehNH2006-wit}
by means of linear programming.

{\it Example .- } An elementary example of fully separable $3$-qubit
state is:
\begin{equation}
\varrho=p|0\rangle \langle 0|^{\otimes 3}+ (1-p)|1\rangle \langle
1|^{\otimes 3}.
\end{equation}

Now let us come back to the description of general mixed states
separability criteria in multipartite systems. Note that in this case
there is no simple necessary and sufficient condition for separability
like PPT characterization of $2 \otimes 2$ or $2 \otimes 3$ case. Even
for three qubits no such criterion has been found so far, in
particular checking PPT criteria with respect to all bipartite
partitions is not enough at all. However there are many criteria that
may be applied. Range criterion immediately generalizes to its many
variants since now we require range of $\varrho_{A_{1},...A_{m}}$ to
be spanned by $\{ |\phi_{A_1}\rangle...|\phi_{A_m}\rangle \}$ while
range of the state $\varrho_{A_{1}...A_{m}}^{T_{ A_{k_1}... A_{k_l}
  }}$ partially transposed with respect to subset $\{
A_{k_1},\ldots,A_{k_l} \} \subset \{ A_{1},\ldots,A_{m} \} $ is
clearly required to be spanned by the products of these vectors where
all with indices $k_1, \ldots, k_l$ are complex conjugated. Of course
if the state is separable all such partial transposes must lead to
matrices with nonnegative spectrum, more precisely all matrices of the
type $\varrho_{A_{1}...A_{m}}^{T_{ A_{k_1}... A_{k_l} }}$ should be
states themselves.

The realignment criteria are generalized to permutational criteria
\cite{HHH02-permut,ChenW2002-pla} which state that if state
$\varrho_{A_1A_2 ... A_n}$ is separable then the matrix $[{\cal
  R}_{\pi}(\varrho)]_{i_{1}j_{1},i_{2}j_{2},\ldots,i_{n}j_{n}}\equiv
{\varrho}_{\pi(i_{1}j_{1},i_{2}j_{2},\ldots,i_{n}j_{n})}$ (obtained
from the original state via permutation $\pi$ of matrix indices in
product basis) satisfies $\|{\cal R}_{\pi}(\varrho)\|_1\leq 1$.
Only some permutations give nontrivial criteria, that are also
different from partial transpose. It is possible to significantly
reduce the number of permutations to the much smaller set of those
that provide independent criteria \cite{HorodeckiWocjan}. In the case
of three particles one has a special case of partial realignment
\cite{HHH02-permut,ChenWu}).  Finally, let us recall that the
contraction criterion (\ref{LinearContraction}) generalizes
immediately.

Let us now consider the case of pure states in more detail. A pure
$m$-partite state is fully separable if and only if it is a product of
pure states describing $m$ elementary subsystems. To check it, it is
enough to compute reduced density matrices of elementary subsystems
and check whether they are pure. However, if one asks about the
possible ways this simple separability condition is violated then the
situation becomes more complicated.

The first problem is that in multipartite case (in comparison to
bipartite one) only very rarely pure states admit the generalized
Schmidt decomposition $|\Psi_{A_{1},\ldots,A_{m}}\rangle =
\sum_{i=1}^{min[d_{A_1},\ldots,d_{A_{m}}]} a_{i} |\tilde{e}_{A_{1}}^{i}
\rangle\otimes \cdots \otimes |\tilde{e}_{A_{m}}^{i}\rangle $ (see
\cite{PeresSchmidt,Thaplyial}). An example of the state admitting
Schmidt decomposition in the $d^{\otimes m}$ case is the generalized
Greenberger-Horne-Zeilinger state
\begin{equation}
|GHZ\rangle^{(m)}_{d}=\frac{1}{\sqrt{d}}
\sum_{i=0}^{d-1}(|i\rangle^{\otimes m}), \label{GHZ}
\end{equation}
which is a generalization of original GHZ state \cite{GHZk} that is
a three-qubit vector $|GHZ\rangle=
\frac{1}{\sqrt{2}}(|0\rangle|0\rangle|0\rangle+|1\rangle|1\rangle|1\rangle)$.
To give an example of state which does not admit Schmidt
decomposition, note that the latter implies that if we trace out any
subsystem, the rest is in fully separable state. One easily finds
that the following state
\begin{equation}
|W\rangle=
\frac{1}{\sqrt{3}}(|0\rangle|0\rangle|1\rangle+|0\rangle|1\rangle|0\rangle+
|1\rangle|0\rangle|0\rangle)
\end{equation}
has entangled two qubit subsystem, hence does not admit Schmidt
decomposition \cite{DurVC00-wstate}.

Thus, in general entanglement of pure state is described by spectra of
the reduced density matrices produced by all bipartite partitions. As
implied by the full separability definition it is said to be fully
$m$-partite separable iff:
\begin{equation}
|\Psi_{A_{1},\ldots,A_{m}}\rangle =|\psi_{A_{1}}\rangle \otimes \cdots
\otimes |\psi_{A_{m}} \rangle. \label{ProductPure}
\end{equation}
However violation of this condition clearly does not automatically
guarantee what can be intuitively considered as ``truly'' $m$-partite
entanglement (to understand it see for instance the $4$-qubit state
$\Psi_{A_{1}A_{2}A_{3}A_{4}}=|\Phi_{A_{1}A_{2}}\rangle \otimes
|\Phi_{A_{3}A_{4}}\rangle $ where at least one vector
$\Phi_{A_{1}A_{2}}$, $\Phi_{A_{3}A_{4}}$ is entangled).

One says that the state is {\it $m$-partite
entangled} iff
all bipartite partitions produce mixed reduce density matrices (note
that both reduced states produced in this way have the same nonzero
eigenvalues). This means that there does not exist cut, against
which the state is product. To this class belong all those pure
states that satisfy generalized Schmidt decomposition (like the GHZ
state above). But there are many others, e.g. the mentioned $W$
state. In Sec. \ref{sec:irrev} we will discuss how one can
introduce classification within the set of $m$-partite entangled
states. One can introduce a further classification by means of
stochastic LOCC (SLOCC) (see Sec. \ref{sec:irrev}),
according to which for $3$ qubits there are two classes of truly
$3$-partite entangled states, represented just by the GHZ and W
states. There are furthermore three classes of pure states which are
partially entangled and partially separable: this is the state
$|\Phiplus\rangle|0\rangle$ (where
$\Phiplus=\frac{1}{\sqrt{2}}(|0\rangle|0\rangle +
|1\rangle|1\rangle)$ and its twins produced by two cyclic
permutations of subsystems. We see that in the case of those pure
states only 2-qubit entanglement is present and explicitly
``partial'' separability can be seen. This leads us to the various
notions of partial separability which will be described in next
section. Here we will present an important family of pure entangled
states.

{\it Example: quantum graph states.} General form of graph states
has been introduced in \cite{RaussendorfEtAlGraphIntroduced2003} as
a generalization of {\it cluster states}
\cite{BriegelRaussendorfClaster} that have been shown to be a
resource for one-way quantum computer
\cite{RaussendorfBriegelOneWay}. Universality of quantum computer
based on graph states is one of the fundamental application of
quantum entanglement in a theory of quantum computer (see
\cite{GraphStates}). In general, any graph state is a pure $m$-qubit
state $|G\>$ corresponding to a {\it graph} $G(V,E)$. The graph is
described  by the set $V$ of vertices with cardinality $|V|=m$
 (corresponding to qubits of $|G\>$) and the set
$E$ of edges, i.e. pairs of vertices (corresponding to
pairs of qubits of $|G\>$)\footnote{Usually $E$ is represented by
symmetric {\it adjacency matrix} with elements
$A_{uv}=A_{vu}=1$ iff $u\neq v$ are connected by the edge
and $A_{uv}=0$ otherwise. Note, that there are no more than
$m(m-1)\over 2$ of the edges --- pairs of vertices.}. Now, the
mechanism of creating $|G\>$ is very simple. One takes as the
initial state $|+\rangle^{\otimes m}$ with
$|+\rangle=\frac{1}{\sqrt{2}}(|0\rangle + |1\rangle)$. Then
according to graph $G(V,E)$, to any of pairs of qubits corresponding
to vertices connected by an edge from $E$ one applies a controlled
phase gate: $U_{C-phase}=|0\rangle\langle 0| \otimes \id
+|1\rangle\langle 1| \otimes \sigma_{3}$. Note that since all such
operations commute even if performed according to the edges with a
common vertex, the order of applying the operations is arbitrary.
Remarkably, the set of graph states constructed is described by a
{\it polynomial} number $m(m-1)/2$ of discrete parameters, while in
general the set of all states in the m-qubit Hilbert space is
described by an exponential $2^{m}$ number of continuous parameters.
Moreover, local unitary interconvertibility under
$\otimes_{i=1}^{m}U_{i}$ of two graphs states is equivalent to
convertibility under stochastic local operations and classical
communications (SLOCC).

Any connected\footnote{A graph state $|G\>$ is called {\it connected}
  if the corresponding graph $G(V,E)$ is connected, that is when any
  two vertices from $V$ are linked by the path of subsequent edges
  from $E$.} graph state $|G\>$ is fully entangled m-particle state
and violates some Bell inequality. The latter fact has been proven
\cite{GraphBellStabiliser} using alternative (equivalent) description
of graph states in terms of stabilizer formalism \cite{GottesmanPhd}.
Moreover the idea of efficiently locally measurable entanglement
witnesses that detect entanglement of graph states has been proposed
\cite{TothG_s}. This idea can be developed, to show that entanglement
of any connected graph violates some Bell inequality that requires
only two measurements per each qubit site \cite{TothGB2006}. This is a
very useful proposal of local detection of the mean value of
corresponding entanglement witness. For of a review of many important
and very interesting applications of graph states in quantum
information, including details of their fundamental role in one-way
quantum computing, see \cite{GraphStates}.

\subsection{Partial separability \label{subsec:PartialSep}}

Here we shall consider two other very important notions of partial
separability.  The first one is just separability with respect to
partitions.  In this case, the state of $A_{1},\ldots,A_{m}$
elementary subsystems is separable with respect to a given partition
$\{ I_{1}, \ldots, I_{k} \}$, where $I_{i}$ are disjoint subsets of the
set of indices $I=\{ 1, \ldots, m\}$ ($\cup_{l=1}^{k} I_{l} =I$)iff
$\varrho=\sum_{i=1}^{N}p_{i} \varrho^{i}_{1} \otimes \cdots \otimes
\varrho^{i}_{k}$ where any state $\varrho^{i}_{l}$ is defined on
tensor product of all elementary Hilbert spaces corresponding to
indices belonging to set $I_{i}$. Now, one may combine {\it several
  separability conditions} with respect to several different
partitions. This gives many possible choices for partial separability.

Let us show an interesting example of partial separability
which requires even number of qubits in general.

{\it Example .- } Consider four-qubit Smolin states \cite{Smolin}:
\begin{equation}
\varrho^{unlock}_{ABCD}=\frac{1}{4}\sum_{i=1}^{4}
|\Psi^{i}_{AB}\rangle\langle\Psi^{i}_{AB} | \otimes
|\Psi^{i}_{CD}\rangle\langle\Psi^{i}_{CD} |, \label{SmolinS}
\end{equation}
where $|\Psi^{i}\rangle$ are four Bell states. It happens that it is
symmetrically  invariant under any permutations (to see it one can
use the symmetric Hilbert-Schmidt representation
$\varrho^{unlock}_{ABCD}=\frac{1}{4}(\id^{\otimes 4} + \sum_{i=1}^{3}
\sigma^{\otimes 4}_{i})$. Thus, the state is separable under any
partition into two two-qubit parts. Still, it is entangled under any
partition 1 versus 3 qubits since it violates PPT criterion
with respect to this partition ie.
$(\varrho^{unlock}_{ABCD})^{T_{A}}\not\geq 0$. This state has
been shown to have applications in remote concentration of quantum
information \cite{MuraoV1999-concentration}. The Smolin state has
been also shown to be useful in reduction of communication
complexity via violation of Bell inequalities \cite{AugusiakBell}.
No bound entangled states with fewer degrees of freedom
useful for communication complexity are known so far. Generalization
of the state to multipartite case is possible
\cite{LidarWBS,AugusiakGSS}.

A particularly interesting from the point of view of low partite
case systems is a special class of partially separable states called
{\it semiseparable}. They are separable under all 1-(m-1) partitions
: $\{ I_{1}=\{ k \}, I_{2}=\{ 1,\ldots,k-1,k+1,\ldots,m \} \}$, $1 \leq k
\leq m $. It allows to show a new type of entanglement: there are
semiseparable $3$-qubit states which are still entangled. To see it
consider the next example \cite{UPB2}:

{\it Examples .-} Consider the following product states: $2 \otimes 2
\otimes 2$ state composed on 3 parts $ABC$ generated by set defined as
$S_{\bf Shift}=|0\ra|0\ra|0\ra,|+\ra|1\ra|-\ra,|1\ra|-\ra|+\ra,
|-\ra|+\ra|1\ra \}$, (with $|\pm\ra=\frac{1}{\sqrt{2}}(|0\ra\pm
|1\ra)$). This set can be proved to define multipartite {\it
  unextendible product basis} in full analogy to the bipartite case
discussed in Sec. (\ref{subsubsec:RangePPT}): there is no product
state orthogonal to subspace spanned by $S_{\bf Shift}$. Thus, in
analogy to bipartite construction, the state $\varrho_{\bf
  Shift}=(\id-P_{\bf Shift})/4$ where $P_{\bf Shift}$ projects onto
the subspace spanned by $S_{\bf Shift}$ can be easily shown to be
entangled as a whole (i.e. not fully separable) but PPT under all cuts
(i.e. $A|BC$, $AB|C$, $B|AC$). However it happens that it is not only
PPT but also {\it separable} under all cuts. This means that
semiseparability is not equivalent to separability even in the most
simple multipartite case like $3$-qubit one.

Entanglement of semiseparable states shown here immediately follows
from construction by application of multipartite version of range
criterion. It is also detected by permutation criteria.  Finally, one
can use maps that are positive on all product states of two qubits but
not positive in general, described in Sec. \ref{sec:full-sep} (see
\cite{Ho00}).

Another interesting class is the set of $U \otimes U \otimes U$
invariant $d \otimes d \otimes d$ states which comprises semiseparable
and fully 3-separable subclasses of states in one $5$-parameter family
of states (see \cite{UUU}).

The moral of the story is that {\it checking bipartite separability
  with respect to all possible cuts in not enough to guarantee full
  separability}. However separability with respect to some partial
splittings gives still important generalization of separability and
have interesting applications (see
\cite{Smolin,DCpur,DurC_multi_dist2000,DurCirac_activation} and
Sec. \ref{sec:distil}). In this context we shall describe below a
particularly useful family of states.

{\it Example .-} Separability of the family of the states presented
here is fully determined by checking PPT criterion under any possible
partitions. To be more specific, PPT condition with respect to some
partition guarantees separability along that partition. The states
found some \cite{DCpur,DurC_multi_dist2000} important applications in
activation of bound entanglement \cite{DurCirac_activation},
nonadditivity of multipartite quantum channels \cite{DuHoCi04} and
multipartite bound information phenomenon
\cite{AcinCM-MultiBoundInfo}.  This is the following $m$-qubit family
\cite{DCpur}:
\begin{equation}
\varrho^{(m)}=\sum_{a=\pm}\lambda_{0}^{a} |\Psi_{0}^{a}\rangle
\langle \Psi_{0}^{a}| + \sum_{k \neq
0}\lambda_{k}(|\Psi_{k}^{+}\rangle \langle \Psi_{k}^{+}| +
|\Psi_{k}^{-}\rangle \langle \Psi_{k}^{-}|), \label{mCirac}
\end{equation}
where $|\Psi_{k}^{\pm}\rangle=\frac{1}{\sqrt{2}}(|k_{1}\rangle|k_{2}\rangle...|k_{m-1}\rangle
|0\ra \pm |\overline{k}_{1}\rangle|
\overline{k}_{2}\rangle...|\overline{k}_{m-1} \rangle |1\rangle$
with $k_{i}=0,1$, $\overline{k}_{i}=k_{i}\oplus 1 \equiv
(k_{i}+1)\, mod\, 2$ and $k$ being one of $2^{m-1}$ real numbers
defined by binary sequence $k_{1}, \ldots ,k_{m-1}$.

Let us put
$\Delta=\lambda_0^{+}-\lambda_0^{-}\geq 0$ and  define {\it
bipartite splitting} into two disjoint parts $A(k)=\{$ subset with
last (m-th) qubit $\}$, $B(k)=\{$ subset without last qubit $\}$
with help of binary sequence $k$ such that $i$-th qubit belongs to
$A(k)$ (and not to $B(k)$) if and only if the sequence $k_{1}, \ldots
,k_{m-1}$ contains $k_{i}=0$. Then one can prove \cite{DCpur} that
(i) $\varrho^{(m)}$ is separable with respect to partition $\{
A(k)$, $B(k) \}$ iff  $\lambda_{k}\geq \Delta/2$ which happens
to be equivalent to PPT condition with respect to that partition
(ie. $\varrho^{T_{B}}\geq 0$) (ii) if PPT condition is satisfied for
all bipartite splittings then $\varrho^{(m)}$ is fully separable.
Note that the condition (ii) above does not hold in general for
other mixed states  which can be seen easily on 3-qubit
semiseparable state $\varrho_{\bf Shift}$ recalled in this section.
That state is entangled but clearly satisfies PPT condition under
all bipartite splittings.

There is yet another classification that allows for stratification of
entanglement involved. For instance $m$-particle system may be
required {\it to have at most $s$-particle entanglement} which means
that it is a mixture of all states such that each of them is separable
with respect to some partition $\{ I_{1}, \ldots, I_{k} \}$ where all
sets of indices $I_{k}$ have cardinality not greater than $s$. All
m-partite states that have at most $m-1$-particle entanglement satisfy
special Bell inequalities (see \cite{Svetlichny1} for bipartite and
\cite{Svetlichny} for general case) and nonlinear separability
criteria like that of \cite{MaasssenU1988} which we shall pass to just
in next section (see Sec.  \ref{subsec:NonlinearImprovement}).

\section{Further improvements of entanglement tests: nonlinear separability criteria}

Then the nonlinear separability criteria into two different classes.
The first ones are based on functions of results of few measurements
performed in noncollective manner (on one copy of $\varrho_{AB}$ a
time). This first time comprises separability conditions in terms of
uncertainty relations that have been started to be developed first for
continuous variables (We must stress here that we do not consider
entanglement measures which are also a nonlinear functions of the
state, but belong to a very special class that has --- in a sense ---
its own philosophy.). Such conditions will be described in subsections
(\ref{subsec:uncertainty}, \ref{subsec:NonlinearImprovement}) below.

The second class of nonlinear separability conditions is based on
{\it collective} measurements on several copies and has attracted
more and more attention recently. We shall present them in
Sec. \ref{subsubsec:CollectiveWitnesses}.

\subsection{Uncertainty relation based separability tests
\label{subsec:uncertainty}}

The uncertainty relations have been first developed for continuous
variables and applied to Gaussian states \cite{DuanGCZ1999-criterion}
(see also \cite{ManciniGVT2001-criterion}). For the bipartite case
nonlinear inequalities for approximations of finite dimensional
Hilbert spaces in the limit of high dimensions have been exploited in
terms of global angular momentum-like uncertainties in \cite{Polzik0}
with further experimental application \cite{PolzikJK}\footnote {The first use of uncertainty relation to detect
entanglement (with theoretically and experimentally) can be found in \cite {HaldSSP1999} for spins of atomic ensembles.}.

General separability criteria based on uncertainty relation and valid
both for discrete and continuous variables (CV) have been introduced
in \cite{Giovannetti1}, and \cite{HofmannT} (the second one was
introduced for discrete variables but its general formulation is valid
also for the CV case). Soon further it has been shown \cite{Hoffman2}
that PPT entanglement can be detected by means of uncertainty relation
introduced in \cite{HofmannT}. This approach has been further
developed and simplified by G{\"u}hne \cite{OGuehne} and developed
also in entropic terms \cite{GuehneL2004-pra}. Another separability criterion in the two-mode continuous systems
based on uncertainty relations with the particle number and the destruction operators
was presented \cite{TothSC2003}, which may be used to detect entanglement  in light  field
 or in Bose-Einstein condensates.

Let us recall briefly the key of the approach of local uncertainty
relations (LUR) \cite{HofmannT} which has found a very nice
application in the idea of macroscopic entanglement detection {\it
via} magnetic susceptibility \cite{WiesniakVB-Magnetic}. Consider the
set of local observables $\{ A_{i} \}_{i=0}^{N}$, $\{ B_{i}
\}_{i=1}^{N}$ on Hilbert spaces ${\cal H}_{A}$, ${\cal H}_{B}$
respectively. Suppose that one has bounds on sum of {\it local}
variances ie. $\sum_{i} \delta(A_{i})^{2}\geq c_{a}$,$\sum_{i}
\delta(B_{i})^{2}\geq c_{b}$ with some nonnegative values
$c_{a},c_{b}$ and the variance definition
$\delta(M)^{2}_{\varrho}\equiv \langle M^{2} \rangle_{\varrho} -
\langle M \rangle_{\varrho}^{2}$. Then for any separable state
$\varrho_{AB}$ the following inequality holds \cite{HofmannT}:
\begin{equation}
\sum_{i} \delta(A_{i}\otimes \id + \id \otimes
B_{i})_{\varrho_{AB}}^{2}\geq c_{A} + c_{B}.
\end{equation}
Note that by induction the above inequality can be extended to the
multipartite case. Quite remarkably if the observables $A_{i}$ and
$B_{i}$ are chosen in a special asymmetric way, then the above
inequality can be shown \cite{Hoffman2} to detect entanglement of
the family of (\ref{tran}) PPT states. The LUR  approach has been
generalized in \cite{OGuehne} to separability criteria {\it via}
nonlocal uncertainty relations. That approach is based on the
observation that for any convex set ${\cal S}$ (here we choose it to
be set of separable states), any set of observables $M_{i}$ and the
state $\varrho=\sum_{i}p_i\varrho_{i}$, $\varrho_{i}\in {\cal S}$
the following inequality holds:
\begin{equation}
\sum_i \delta(M_i)^{2}_{\varrho} \geq \sum_k p_k \sum_i
\delta(M_i)^{2}_{\varrho_{k}}.
\label{uncertaintyGuehne}
\end{equation}
It happens that in many cases it is relatively easy to show that right
hand side (RHS) is separated from zero only if $\varrho$ is separable,
while at the same time LHS vanishes for some entangled states
\footnote{There is a more general form of this inequality in terms of covariance matrices, which gives rise to new separability criteria;
a simple (strong) necessary and sufficient criterion for two-qubit states
was presented which violates LUR \cite {GuhneHGE2007}.
(see also \cite{AbascalBjork2007-lur-cov} in this context)
 Other criterion for symmetric for n-qubit states
have been presented in the form of a hierarchy of inseparability condition on the intergroup covariance matrices of even order \cite{DeviPR2007}.
}.

Especially, as observed in \cite{OGuehne} if the observables $M_{i}$
have no common product eigenvector then the RHS must be strictly
greater than zero since $\delta(M)^{2}_{\Psi}$ vanishes iff $\Psi$
is an eigenvector of $M$. Consider now an arbitrary state that
violates the range criterion in such a way that it has no product
vector in its range $R(\varrho)$ (an example is just the PPT
entangled state produced by UPB meted, but discrete value PPT
entangled states from \cite{boundCV}. Now there is a simple
observation: any subspace ${\cal H}^{\perp}$ orthonormal to the
subspace ${\cal H}$ having no product vector can be spanned by
(maybe nonorthogonal) entangled vectors $\{ |\Psi_{i}\rangle
\}_{i=1}^{r}$, $r=dim{\cal H}$. The technique of \cite{OGuehne} is
now to take $M_{i}=|\Psi_i\rangle \langle \Psi_i|$, $i=1,\ldots,r$ and
$M_{r+1}=P_{R(\varrho)}$ where the last one is a projector onto the
range of the state, which --- by definition --- has no product state in
the range. Note that since all $|\Psi_i \rangle$ belong to the
kernel\footnote{By kernel of the state we mean the space spanned by
vectors corresponding to zero eigenvalues of the state.} of
$\varrho$  one immediately has LHS of (\ref{uncertaintyGuehne})
vanishing. But as one can see there is no product eigenvector that
is common to all $M_{i}$-s hence as mentioned above, RHS must be
strictly positive which gives expected violation of the inequality
(\ref{uncertaintyGuehne}).

An alternative interesting way is to consider an entropic version of
uncertainty relations, as initiated in \cite{Giovannetti2} and
developed in \cite{GuehneL2004-pra}. This technique is a next step in
application of entropies in detecting entanglement. The main idea is
(see \cite{GuehneL2004-pra}) to prove that if sum of local Klein
entropies\footnote{The Klein entropy associated with an observable and
  a state is the classical entropy of probability distribution coming
  out of measurement of this observable on the quantum state.}
satisfies uncertainty relations $S(A_{1}) + S(A_{2})\geq C$, $S(B_{1})
+ S(B_{2})\geq C$ for systems $A$, $B$ respectively then for any
separable state global Klein entropy must satisfy the same bound
$S(A_{1} \otimes B_{1}) + S(A_{2} \otimes B_{2})\geq C$. This approach
was also extended to multipartite case \cite{GuehneL2004-pra}.

\subsection{Nonlinear improvement of entanglement witnesses}
\label{subsec:NonlinearImprovement}

In general any bound on nonlinear function of means of observables
that is satisfied by separable states but violated by some entangled
states can be in broad sense considered as ``nonlinear entanglement
witnesses''. Here we shall consider those separability conditions that
had their origin in entanglement witnesses (like for instance Bell
operators) and lead to nonlinear separability test. They are still all
based on functions of mean values of noncollective measurements
(ie. single measurement per copy).

Let us come back to more general nonlinear conditions inspired by Bell
inequalities criteria. This interesting approach has been first
applied to the multipartite case \cite{MaasssenU1988}, where on the
basis of Bell inequalities nonlinear inequalities have been
constructed.  These can discriminate between full $m$-particle
entanglement of the system $A_{1}...A_{m}$ and the case when it
contains at most $m-1$-particle entanglement
(cf. Sec. \ref{subsec:PartialSep}).  It is worth to note here that
linear inequalities discriminating between those two cases have been
provided as Bell-like inequalities early in \cite{Svetlichny1} for $3$
particles and extended in \cite{Svetlichny} to $m$ particles. Though
they were not related directly to entanglement, they automatically
serve as an entanglement criteria since any separable state allows for
LHVM description of all local measurements, and as such satisfy all
Bell inequalities.

For small-dimensional systems, a nonlinear inequality inspired by Bell
inequality that is satisfied by all separable two-qubit states has
been provided in \cite{Yu}. Namely the two-qubit state $\varrho$ is
separable if and only if the following inequality holds:
\begin{eqnarray}
\sqrt{\langle A_1\otimes B_1 + A_2\otimes B_2\rangle_{\varrho}
^{2}
+ \langle A_3\otimes \id + \id \otimes B_3\rangle^{2}_\varrho} -  \nonumber \\
\langle A_3\otimes B_3\rangle_\varrho \leq 1,
\end{eqnarray}
for all sets of
dichotomic observables $\{ A_{i} \}_{i=1}^{3}$ $\{ B_{i}
\}_{i=1}^{3}$ that correspond to two local bases orthonormal vectors
$\textbf{a_{i}}$, $\textbf{b_{i}}$ that have the same
orientation in Cartesian frame (the relation
$A_{i}=\textbf{{a}_{i}}\vec{\sigma}$,$B_{i}=\textbf{{b}_{i}}\vec{\sigma}$
holds). There is also a link with PPT criterion: for any entangled
state $\varrho$ with negative eigenvalue $\lambda_{min}$ of
$\varrho^{\Gamma}$, the optimal setting of observables makes LHS
equal to $1-4\lambda_{max}$.

The above construction was inspired by a Bell-type entanglement
witness (see \cite{Yu} for details). In this context a question arises
if there is any systematic way of nonlinear improvements of
entanglement witnesses. Here we shall describe a very nice nonlinear
improvement that can be applied to any entanglement witness (see
(\ref{witnesscondition-v2})) and naturally exploits the isomorphism
(\ref{isomorphism}). Before showing that, let us recall that the whole
entanglement witnesses formalism can be translated to the level of
covariance matrix in continuous variables, and the nonlinear
corrections to such a witnesses that are equivalent to some
uncertainty relations can also be constructed
\cite{EisertCVNonlinearWitnesses}.

The above announced (quadratic) nonlinear correction worked out in
\cite{GuehneLutkenhaus} involves a set of operators
$X_{k}=X_{k}^{(h)} + i X_{k}^{(antih)}$ and its mean $\langle
X_{k}\rangle_\varrho$, which (in practice) must be collected from
mean values of their Hermitian ($X_{k}^{(h)}$) and antihermitian
$X_{k}^{(antih)}$ parts respectively. The general form of the
nonlinear improvement of the witness $W$ corresponds to the
condition
\begin{equation}
{\cal F}_\varrho=\langle W\rangle_\varrho - \sum_{k}
\alpha_{k}|\langle X_{k} \rangle_\varrho|^{2} \geq 0,
\label{nonlinearwitness}
\end{equation}
where real numbers $\alpha_{k}$ and operators $X_{k}$ are both
chosen in such a way that for all possible separable states
$\varrho_{AB}$ the condition ${\cal F}_{\varrho_{AB}}\geq 0$ is
satisfied. One can see that the second term is a quadratic
correction to the original (linear) mean value of entanglement
witness. Higher order corrections are also possible though they need
not automatically guarantee stronger condition
\cite{GuehneLutkenhaus}. Let us illustrate this by use of  example
of  partial transpose map $I_{A} \otimes
T_{B}(\cdot)=(\cdot)^{\Gamma}$ and the corresponding
decomposable witness (see Eq. (\ref{eq:dec_witness})) coming from the
minimal set of witnesses describing set of states satisfying PPT
conditions $W=|\Psi\rangle\langle \Psi|^{\Gamma}$.  In general,
virtually any entanglement witness condition can be improved in this
way, though one has to involve Hermitian conjugate of the map in
Hilbert-Schmidt space and exploit the fact (see discussion in
Sec. \ref{subsubsec:Witnesses}) that single map condition is
equivalent to continuous set of entanglement witnesses conditions.
Now, for the witness of the form $W=|\Psi\rangle\langle
\Psi|^{\Gamma}$ one of the two versions are natural. In
the first one chooses single $X=|\Phi\rangle\langle \Psi|$ (where
both vectors are normalized) and then single parameter $\alpha$ is
equal to the inverse of maximal Schmidt coefficient of $|\Psi
\rangle$. The second option is to choose $X_{i}=|\Phi\rangle\langle
\Psi_{i}|$ with $\Psi_{i}$ being an orthonormal basis in
global Hilbert space ${\cal H}_{AB}$ and then the choose
$\alpha_{i}=1$ for all parameters guarantees positivity of the value
(\ref{nonlinearwitness}).

\subsection{Detecting entanglement with collective measurements}
\label{subsec:det_ent_col}

\subsubsection{Physical implementations of entanglement criteria with
collective measurements \label{subsubsec:MapsDetected}}

The idea of direct measurement of pure states entanglement was
considered first in \cite{SH00} and involved the first
explicit application of collective measurements to entanglement
detection \cite{Acin}. In general the question here is how
to detect entanglement physically by means of a little number of
collective measurements that do not lead to complete tomography of the
state. Here we focus on the number of estimated parameters (means of
observables) and try to diminish it. The fact that the mean of an
observable may be interpreted as single binary estimated parameter
(equivalent just to one-qubit polarization) has been proven in
\cite{binpovm,PazRoncaglia}, cf. \cite{Brun}.

The power of positive maps separability criteria and entanglement
measures has motivated work on implementations of separability
criteria {\it via} collective measurements, introduced in
\cite{PHAE,PHPRL} and significantly improved in
\cite{Carteret,Noiseless}. On the other hand the entropic separability
criteria has led to the separate notion of {\it collective
  entanglement witnesses} \cite{Witnesses} which will be described in
more detail in the next section.

The  collective measurement evaluation of nonlinear state functions
\cite{Filip,Estimator} (see also \cite{Witnesses,Fiurasek}) was
implemented experimentally very recently in distance lab paradigm \cite{NonlinExp}. The
method takes an especially striking form in the two-qubit case, when
not only unambiguous entanglement detection \cite{PHAE} but also
estimation of such complicated entanglement measure as entanglement
of formation and Wootters concurrence can be achieved by measuring
only four collective observables \cite{PHPRL}, much smaller than 15
required by state estimation. The key idea of the latter scheme is
to measure four collective observables $A^{(2k)}$ on $2k$ copies of
the state that previously have been subjected to physical action of
some maps\footnote{They are so called {\it physical structural
approximations}, which we describe further in this section.}. The
mean values of these observables reproduce all four moments $\langle
A^{(2k)}\rangle  = \sum_{i}\lambda_{i}^{k}$ of spectrum $\{
\lambda_{k} \}$ of the square of the Wootters concurrence matrix
$\hat{C}(\varrho)= \sqrt{\sqrt{\varrho} \sigma_{2}\otimes
\sigma_{2}\varrho^*\sigma_{2}\otimes \sigma_{2}\sqrt{\varrho}}$.
Note that, due to the link \cite{Wootters-conc} between Wootters
concurrence and entanglement of formation, the latter can be also
inferred in such an experiment.
Recently, a collective observable very similar to those  of
\cite{Witnesses}, acting on two copies of quantum state  which
detect two-qubit concurrence has been constructed and implemented
\cite{NatureConcurrence}. The observable is much simpler, however
the method works under the promise that the state is pure.
This approach can be also generalized to multiparty case using suitable
factorisable observable corresponding to the concurrence (see
\cite{AolitaMintert}).

In methods involving positive maps criteria the main problem of how
to physically implement unphysical maps has been overcome by the
help of structural physical approximations (SPA) of unphysical maps.
In fact for any Hermitian trace nonincreasing map $L$ there is a
probability $p$ such that the new map $\tilde{L}=p{\cal D} +
(1-p)\Lambda$ can be physically implemented if ${\cal D}$ is just a
fully depolarizing map that turns any state into maximally
mixed one. Now one can apply the following procedure: (i) put $L=\id
\otimes \Lambda$ for some P but not CP map $\Lambda$, (ii) apply the
new (physical) map $\tilde{L}$ to many copies of bipartite system in
unknown state $\varrho$, (iii) estimate the spectrum of the
resulting state $\tilde{L}(\varrho)=[\id \otimes
\tilde{\Lambda}](\varrho)$. Infer the spectrum of $L(\varrho)=[\id
\otimes \Lambda](\varrho)$ (which is an easy affine transformation
of the measured spectrum of $\tilde{ L}(\varrho)$) and check in this
way whether the condition (\ref{mapscondition}) for map $\Lambda$ is
violated. Generalization to multipartite systems is immediate. This
test requires measurement of $d^{2}$ observables instead of
$d^{4}-1$ ones needed to check the condition with prior state
tomography. The corresponding quantum networks can be easily
generalized to multipartite maps criteria including realignment or
linear contractions criteria \cite{reshuff}. The implementation with
help of local measurements has also been developed (see
\cite{LOCCmapsdetection}.

However, as pointed out by Carteret \cite{Carteret}, the disadvantage
of the method is that SPA involved here requires in general
significant amount of noise added to the system. The improved method
of noiseless detection of PPT criterion, concurrence and tangle has
been worked out \cite{Carteret,Carteret2003-tangle} with help of polynomial invariants
technique which allows for very simple and elegant quantum network
designing. The problem whether noiseless networks exist for all other
positive maps (or contraction maps) have been solved quite recently
where general noiseless networks have been designed \cite{Noiseless}.

Finally let us note that the above techniques have been also developed
on the ground of continuous variables
\cite{FiurasekCerf,StobinskaW-teor-s,Pregnell}.

\subsubsection{Collective entanglement witnesses \label{subsubsec:CollectiveWitnesses}}

There is yet another technique introduced \cite{Witnesses} that
seems to be more and more important in context of experimental
implementations. This is the notion of {\it collective entanglement
witness}. Consider a bipartite system $AB$ on the space ${\cal
H}_{AB}={\cal H}_{A} \otimes {\cal H}_{B}$. One  introduces  here
the notion of collective observable $A^{(n)}$ with respect to the
single system Hilbert space ${\cal H}$ as an observable defined on
${\cal H}^{\otimes n}$ and measured on $n$ copies $\varrho^{\otimes
n}$ of given state $\varrho$. Also one defines the notion of
{\it mean of collective observable} in single copy of state
$\varrho^{\otimes n}$ as $\langle\langle A^{(n)} \rangle
\rangle:={\tr}(A^{(n)} \varrho^{\otimes n})$. Now any observable
$W^{(n)}$ defined on $({\cal H}_{AB})^{\otimes n}$ that satisfies
the following condition
\begin{equation}
\langle \langle
W^{(n)}\rangle\rangle_{\varrho_{sep}}:={\tr}(W^{(n)}\varrho_{sep}^{\otimes
n}) \geq 0,
\end{equation}
but there exists an entangled state $\varrho_{ent}$ such that
\begin{equation}
\langle \langle W^{(n)} \rangle \rangle_{\varrho_{ent}} < 0,
\end{equation}
is called {\it collective (n-copy) entanglement witness}. In
\cite{Witnesses} collective entanglement witnesses reproducing
differences of Tsallis entropies\footnote{The family of Tsallis
  entropies parametrized by $q>0$ is defined as follows:
  $S_q(\rho)={1\over {1-q}}({\tr}\rho^{q} -1)$.}, and as such
verifying entropic inequalities with help of single observable, have
been designed.

Here we shall only recall the case of $n=2$.
The following joint observables
Let us consider
\begin{eqnarray}
&&X_{AA'BB'}^{(2)}=V_{AA'} \otimes ( V_{BB'} - I_{BB'} ), \nonumber \\
&&Y_{AA'BB'}^{(2)}=(V_{AA'} - I_{AA'}) \otimes V_{BB'}.
\label{XY}
\end{eqnarray}
represent collective entanglement witnesses since they reproduce
differences of Tsallis entropies with $q=2$ : $\langle \langle X \rangle \rangle
= S_{2}(\varrho_{AB})-S_{2}(\varrho_{A})$, $\langle\langle Y \rangle
\rangle = S_{2}(\varrho_{AB})-S_{2}(\varrho_{B})$ which are always
nonnegative for separable states \cite{alpha}.

Collective (2-copy) entanglement witnesses were further applied in
continuous variables for Gaussian states
\cite{StobinskaW-teor-s}. Recently it has been observed that there is
a {\it single} 4-copy collective entanglement witness that detects all
(unknown) 2-qubit entanglement \cite{Augusiak}. On this basis a
corresponding universal quantum device, that can be interpreted as a
quantum computing device, has been designed.

It is very interesting that the following collective
entanglement witness \cite{MintertBuchleitner}:
\begin{eqnarray}
\tilde{W}_{AA'BB'}^{(2)}=2P^{(+)}_{AA'} \otimes P^{(-)}_{BB'} + 2
P^{(-)}_{AA'} \otimes P^{(-)}_{BB'} \nonumber \\- 4P^{(-)}_{AA'} \otimes
P^{(-)}_{BB'},
 \label{ConcurrenceWitness}
\end{eqnarray}
with $P^{(+)}$ and $P^{(-)}$ being projectors onto symmetric
and antisymmetric subspace respectively (see \ref{subsubsec:Witnesses}), has been shown to provide a
lower bound on bipartite concurrence  ${\cal
C}(\varrho_{AB})$ \cite{MintertKB04-conc}
\begin{eqnarray}
  {\cal C}(\varrho_{AB})\geq -\langle
 \langle \tilde{W}_{AA'BB'}^{(2)}\rangle\rangle_{\varrho_{AB}}.
 \label{con}
\end{eqnarray}
Note that $\tilde{W}_{AA'BB'}^{(2)}$ shown above is  just twice the sum
of the witnesses given in Eq. (\ref{XY}).

Further the technique leading to the above formula was applied in
the construction of   (single-copy) entanglement witnesses quantifying
concurrence as follows  \cite{Mintert2006}.
Suppose we have a given bipartite state $\sigma $ for which we know
the concurrence $C(\sigma)$ exactly.  Then one can construct the following
entanglement witness  which is parametrised by state $\sigma$ in a nonlinear
way:
$W_{AB}(\sigma_{A'B'})=-4\tr_{A'B'}(I \otimes \sigma_{A'B'}
 V_{AA'} \otimes V_{BB'})/{\cal C}(\sigma_{A'B'})$  which has the following
nice property:
\begin{equation}
{\cal C}(\varrho_{AB}) \geq - \langle W_{AB}(\sigma_{A'B'})
\rangle_{\varrho_{AB}}.
\end{equation}

\subsubsection{Detection of quantum entanglement as quantum computing
with quantum data structure \label{subsubsec:QuantumDataStructure}}

It is interesting that entanglement detection schemes involve
schemes of quantum computing. The networks detecting the moments of
the spectrum of $L(\varrho)$ that are elements of the entanglement
detection scheme \cite{PHAE,PHPRL,reshuff,Carteret} can be
considered as having fully quantum input data ({\it copies of
unknown quantum state}) and give classical output --- the moments of
the spectrum. The most striking network is a universal
quantum device detecting 2-qubit entanglement
\cite{Augusiak}\footnote{It exploits especially the idea of
measuring the mean of observable by measurement of polarization of
specially coupled qubit \cite{binpovm,PazRoncaglia}, cf.
\cite{Brun}.}. This is a quantum computing unit with an input where
(unknown) quantum data comes in. The input is represented by $4$
copies of unknown 2-qubit state. Then they are coupled by a special
unitary transformation (the network) to a single qubit polarization
of which is finally measured. The state is separable iff it is
polarization is not less than certain value.

This rises the natural question of whether in the future it will be
possible to design a quantum algorithm working on fully quantum data
(ie. copies of unknown multipartite states, entanglement of which has
to be checked) that will solve the separability/entanglement problem
much faster than any classical algorithms (cf. next section).

\section{Classical algorithms detecting entanglement}
\label{sec:alg_ent}

The first systematic methods of checking entanglement of given state
was worked out in terms of finding the decomposition onto separable
and entangled parts of the state (see \cite{LewensteinSanpera-bsa} and
generalizations to case of PPT states \cite{Karnas,Lewenstein00b}).
The methods were based on the systematic application of the range
criterion involving however the difficult analytical part of finding
product states in the range of a matrix. A further attempt to provide
an algorithm deciding entanglement was based on checking variational
problem based on concurrence vector
\cite{AudenaertVM2000-concurrence}. The problem of existence of
classical algorithm that unavoidably identifies entanglement has been
analyzed in \cite{DohertyAlgorithm1,DohertyAlgorithm2} both
theoretically and numerically and implemented by semidefinite
programming methods.  This approach is based on a theorem concerning
the symmetric extensions of bipartite quantum state
\cite{FannesEtAlSEP,RaggioWernerSEP}. It has the following
interpretation. For a given bipartite state $\varrho_{AB}$ one asks
about the existence of a hierarchy of symmetric extensions, i.e.
whether there exists a family of states $\varrho_{AB_{1}...B_{n}}$
(with $n$ arbitrary high) such that $\varrho_{AB_{i}}=\varrho_{AB}$
for all $i=1,\ldots,n$. It happens that the state $\varrho_{AB}$ is
separable if and only if such a hierarchy exists for each natural $n$
(see Sec. \ref{sec:monogamy}). However, for any fixed $n$ checking
existence of such symmetric extension is equivalent to an instance of
semidefinite programming. This leads to an algorithm consisting in
checking the above extendability for increasing $n$, which always
stops if the initial state $\varrho_{AB}$ is entangled. However the
algorithm never stops if the state is separable. Further another
hierarchy has been provided together with the corresponding algorithm
in \cite{EisertAlgorithm}, extended to involve higher order polynomial
constraints and to address multipartite entanglement question.

The idea of dual algorithm was provided in \cite{HulpkeAlgorithm},
based on the observation that in checking separability of given state
it is enough to consider countable set of product vectors spanning the
range of the state. The constructed algorithm is dual to that
described above, in the sense that its termination is guaranteed iff
the state is separable, otherwise it will not stop.  It has been
further realized that running both the algorithms (ie.  the one that
always stops if the state is entangled with the one that stops if the
state is separable) in parallel gives an algorithm that always stops
and decides entanglement definitely \cite{HulpkeAlgorithm}.

The complexity of both algorithms is exponential in the size of the
problem. It happens that it must be so. The milestone result that has
been proved in a mean time was that solving separability problem is NP
hard \cite{GurvitzNPHard,Gurvits_match_2002,Gurvits_complex_2003}.
Namely is it known \cite{YudinN_complex_1976} that if a largest ball
contained in the convex set scales properly, and moreover there exists
an efficient algorithm for deciding membership, then one can
efficiently minimize linear functionals over the convex set.  Now,
Gurvits has shown that for some entanglement witness optimization
problem was intractable. This, together with the results on radius of
the ball contained within separable states (see
Sec. \ref{sec:ent_geom}), shows that problem of separability cannot be
efficiently solved.

Recently the new algorithm via analysis of weak membership problem as
been developed together with analysis of NP hardness
(\cite{IoannouAlgorithm1,IoannouAlgorithm2,IoannouAlgorithm3}). The
goal of the algorithm is to solve what the authors call ``witness''
problem.  This is either (i) to write separable decomposition up to
given precision $\delta$ or (ii) to find an (according to slightly
modified definition) entanglement witness that separates the state
from a set of separable states by more than $\delta$ (the notion of
the separation is precisely defined). The analysis shows that one can
find more and more precisely a likely entanglement witness that
detects the entanglement of the state (or find that it is impossible)
reducing the set of ``good'' (ie. possibly detecting the state
entanglement) witnesses by each step of the algorithm. The algorithm
singles out a subroutine which in the standard picture
\cite{sep1996,Terhal2000-laa} can be, to some extent, interpreted as
an oracle calculating a ``distance'' of the given witness to the set
of separable states.

Finally, note that there are also other proposals of algorithms
deciding separability like \cite{ZapatrinAlgorithm} (for review see
\cite{IoannouAlgorithm3} and references therein).

\section{Quantum entanglement and geometry}
\label{sec:ent_geom}

Geometry of entangled and separable states is a wide branch of
entanglement theory \cite{BengtssonZyczkowski-book}. The most simple
and elementary example of geometrical representation of separable and
entangled states in three dimensions is a representation of two-qubit
state with maximally mixed subsystems \cite{MRH-PRA96}. Namely any
two-qubit state can be represented in Hilbert-Schmidt basis $\{
\sigma_{i} \otimes \sigma_{j}\}$ where $\sigma_{0}=\id$, and in this
case the correlation matrix $T$ with elements $t_{ij}={\tr}(\varrho
\sigma_{i} \otimes \sigma_{j})$, $i,j=1,2,3$ can be transformed by
local unitary operations to the diagonal form. This matrix completely
characterizes the state iff local density matrices are maximally mixed
(which corresponds to vanishing of the parameters
$r_{i}={\tr}(\sigma_{i} \otimes \id \varrho)$, $s_{j}={\tr}( \id
\otimes \sigma_{j} \varrho)$, for $i,j=1,2,3$. It happens that after
diagonalizing\footnote{Diagonalizing matrix $T$ corresponds to
  applying product $U_{A} \otimes U_{B}$ unitary operation to the
  state.}, $T$ is always a convex combination of four matrices
$T_{0}=diag[1,-1,1]$, $T_{1}=diag[-1,1,1]$, $T_{2}=diag[1,1,-1]$,
$T_{3}=diag[-1,-1,-1]$ which corresponds to maximally mixed Bell
basis. This has a simple interpretation in threedimensional real
space: a tetrahedron ${\cal T}$ with four vertices and coordinates
corresponding to the diagonals above ($(1,1,-1)$ etc.). The subset of
separable states is an octahedron that comes out from intersection of
${\cal T}$ with its reflection around the origin of the set of
coordinates. It is remarkable that all the states with maximally mixed
subsystems are equivalent (up to product unitary operations $U_{A}
\otimes U_{B}$) to Bell diagonal states (a mixture of four Bell states
(\ref{basis})).  Moreover, for all states (not only those with
maximally mixed subsystems) the singular values of the correlation
matrix $T$ are invariants under such product unitary
transformations. The Euclidean lengths of the real three-dimensional
vectors with coordinates $r_{i}$, $s_{j}$ defined above are also
similarly invariant.

Note, that a nice analogon of the tetrahedron ${\cal T}$ in the state space
for entangled two qudits was defined and investigated in the context
of geometry of separability \cite{BaumgartnerHN2007}.
It turns out that the analogon of the octahedron is no longer a polytope.

One can naturally ask about reasonable set of the
parameters or in general --- functions of the state --- that are
invariants of product unitary operations. Properly chosen invariants
allow for characterization of {\it local orbits} i.e. classes of
states that are equivalent under local unitaries.  (Note that any
given orbit contains either {\it only} separable or {\it only}
entangled states since entanglement property is preserved under local
unitary product transformations). The problem of characterizing local
orbits was analyzed in general in terms of polynomial invariants in
\cite{SchlienzMahler,Grassl}. In case of two qubits this task was
completed explicitly with $18$ invariants in which $9$ are
functionally independent \cite{MixedInvariantsI}
(cf. \cite{Grassl}). Further this result has been generalized up to
four qubits \cite{MixedInvariantsII,MixedInvariantsIII}. Other way of
characterizing entanglement in terms of local invariants was initiated
in \cite{LindenOrbitsI,LindenOrbitsII} by analysis of dimensionality
of local orbit. Full solution of this problem for mixed two-qubit
states and general bipartite pure states has been provided in
\cite{2qOrbitsKus} and \cite{SinoleckaZK2001-manifold}
respectively. For further development in this direction see
\cite{KusGeometry} and references therein. There are many other
results concerning geometry or multiqubit states
to mention only \cite{Heydari,Levay2006-4qubits,Miyake2002-hyper}.

There is another way to ask about geometrical properties of
entanglement. Namely to ask about volume of set of separable states,
its shape and the boundary of this set. The question about the volume
of separable states was first considered in \cite{ZyczkowskiHSP-vol}
and extended in \cite{volume1}. In \cite{ZyczkowskiHSP-vol} it has
been proven with help of entanglement witnesses theory that for any
finite dimensional system (bipartite or multipartite) the volume of
separable states is nonzero. In particular there exists always a ball
of separable states around maximally mixed state. An explicit bound on
the ratio of volumes of the set of all states $\scal$ and that of
separable states $\scal_{sep}$
\begin{equation}
{\mathrm{vol}(\scal)\over \mathrm{vol}(\scal_{sep})}\geq \left({1\over 1+d/2}\right)^{(d-1)(N-1)}
\end{equation}
for $N$ partite systems each of dimension $d$ was provided in
\cite{VidalT1998-robustness}.  This has inspired further discussion
which has shown that experiments in NMR quantum computing may not
correspond to real quantum computing since they are performed on
pseudopure states which are in fact separable
\cite{BraunsteinCJLS1998-nmr}. Interestingly, one can show
\cite{ZyczkowskiHSP-vol,volume2} that for any quantum system on some
Hilbert space ${\cal H}$ maximal ball inscribed into a set of mixed
states is located around maximally mixed states and is given by the
condition $R(\varrho)={\tr}(\varrho^{2})\geq \frac{1}{dim{\cal H}^{2}
  -1}$. Since this condition guarantees also the positivity of any
unit trace operator, and since, for bipartite states
${\tr}(\varrho_{AB}^{2})={\tr}[(\varrho_{AB}^{\Gamma})^{2}]$ this
means that the maximal ball contains PPT states (the same argument
works also for multipartite states \cite{volume2}). In case of $2
\otimes 2$ or $2\otimes 3$ this implies also separability giving a way
to estimate volume of separable states from below.

These estimates have been generalized to multipartite states
\cite{BraunsteinCJLS1998-nmr} and further much improved providing very
strong upper and lower bounds with help of a subtle technique
exploiting among others entanglement witnesses theory
\cite{Gurvits1,Gurvits2,Gurvits3}.  In particular it was shown that
for bipartite states, the largest separable ball around mixed
states. One of applications of the largest separable ball results is
the proof of NP-hardness of deciding weather a state is separable or
not \cite{GurvitzNPHard} (see Sec. \ref{sec:alg_ent}).

There is yet another related question: one can define the state
(bipartite or multipartite) $\varrho$ that remains separable under
action of any unitary operation $U$. Such states are called {\it
absolutely separable} \cite{2qOrbitsKus}. In full analogy one can
define what we call here {\it absolute PPT property} (ie. PPT
property that is preserved under any unitary transformation). The
question of which states are absolutely PPT has been fully solved
for $2 \otimes n$ systems \cite{AbsolutePPT}: those are all states
spectrum of which satisfies the inequality: $\lambda_{1}\leq
\lambda_{2n-1} + \sqrt{\lambda_{2n-2}\lambda_{2n}}$ where $\{
\lambda_{i} \}_{i=1}^{2n}$ are eigenvalues of $\varrho$ in
decreasing order. This immediately provides the characterization of
absolutely separable states in $d_{A} \otimes d_{B}$ systems with
$d_{A}d_{B}\leq 6$ since PPT is equivalent to separability in those
cases. Note that for $2 \otimes 2 $ states this characterization has
been proven much earlier by different methods \cite{2x2AbsoluteSep}.
In particular it follows that for those low dimensional cases the set
of absolutely separable states is strictly larger than that of
maximal ball inscribed into the set of all states. Whether it is
true in higher dimensions remains an open problem.

Speaking about geometry of separable states one can not avoid a
question about what is a boundary $\partial {\cal S}$ of set of
states? This, in general not easy question, can be answered
analytically in case of two-qubit case when it can be shown to be
smooth \cite{SmoothBoundary} relatively to set of all 2-qubit states
which is closely related to the separability characterization
\cite{Augusiak} $det(\varrho_{AB}^{\Gamma})\geq 0$.
Interestingly, it has been shown, that the set of separable state is not polytope \cite{IoannouT2006}
it has no faces \cite{GuhneL2006}.

There are many other interesting geometrical issues that can be
addressed in case of separable states. We shall recall one of them:
there is an interesting issue of how probability of finding separable
state (when the probability is measured up an a priori probability
measure $\mu$) is related to the probability (calculated by induced
measure) of finding a random boundary state to be separable. The
answer of course will depend on a choice of probability measure, which
is by no means unique.  Numerical analysis suggested
\cite{Slater1,Slater2} that in two-qubit case the ratio of those two
probabilities is equal to two if one assumes measure based the
Hilbert-Schmidt distance. Recently it has been proven that for any
$d_{A}\otimes d_{B}$ system this rate is indeed $2$ if we ask about
set of PPT states rather than separable ones
\cite{SzarekProbabilityPPTBoundary}. For $2\otimes 2$ and $2 \otimes
3$ case this reproduces the previous conjecture since PPT condition
characterizes separability there (see Sec.
\ref{subsubsec:maps}). Moreover, it has been proven (see
\cite{SzarekProbabilityPPTBoundary} for details) that bipartite PPT
states can be decomposed into the so called pyramids of constant
height.

\section{The paradigm of local operations and classical communication (LOCC)}
\label{sec:LOCC}

\subsection{Quantum channel --- the main notion}

Here we shall recall that the most general quantum operation that
transforms one quantum state into the other is a {\it probabilistic}
or {\it stochastic} physical operation of the type
\begin{equation}
\varrho \rightarrow \Lambda(\varrho)/{\tr}(\Lambda(\varrho)),
\label{PhysicalMap}
\end{equation}
with trace nonincreasing CP map, i.e. a map satisfying
${\tr}(\Lambda(\varrho))\leq 1$ for any state $\varrho$, which can be
expressed in the form
\begin{equation}
\Lambda(\varrho)=\sum_i V_{i}(\varrho)V_{i}^{\dagger},
\label{PhysicalMap2}
\end{equation}
with $\sum_{i}V_{i}^{\dagger}V_{i}\leq \id$ (domain and codomain of
operators $V_{i}$ called Kraus operators (\cite{KrausOperators}) are in general different). The operation above takes place
with the probability ${\tr}(\Lambda(\varrho))$ which depends on the
argument $\varrho$. The probability is equal to one if an only if
the CP map $\Lambda$ is tracepreserving (which corresponds to
$\sum_{i}V_{i}^{\dagger}V_{i}=\id$ in (\ref{PhysicalMap2}); in such a
case $\Lambda$ is called {\it deterministic} or {\it a quantum
channel}.

\subsection{LOCC operations}
\label{subsec:locc}

We already know that in the quantum teleportation process Alice
performs a local measurement with maximally entangled projectors
$P_{AA'}^{i}$ on her particles $AA'$ and then sends {\it classical}
information to Bob (see Sec \ref{subsec:Telep}). Bob performs accordingly a local operation
$U^{i}_{B}$ on his particle $B$. Note that the total operation acts on
$\rho$ as: $\Lambda_{AA'B}(\rho)=\sum_{i}P_{AA'}^{i} \otimes
U^{i}_{B}(\rho) P_{AA'}^{i} \otimes (U^{i}_{B})^{\dagger}$. This
operation belongs to so called one-way LOCC class which is very
important in quantum communication theory. The general LOCC paradigm
was formulated in \cite{BDSW1996}.

In this paradigm all what the distant parties (Alice and Bob) are allowed is
to perform arbitrary local quantum operations and sending classical
information. {\it No} transfer of quantum systems between the labs is
allowed. It is a natural class for considering entanglement processing
because classical bits cannot convey quantum information and cannot
create entanglement so that entanglement remains a resource that can
be only manipulated. Moreover one can easily imagine that sending
classical bits is much more cheaper than sending quantum bits, because
it is easy to amplify classical information. Sometimes it is
convenient to put some restrictions also onto classical
information. One then distinguishes in general the following
subclasses of operations described below. We assume all the operations
(except of the local operations) to be trace nonincreasing and compute
the state transformation as in (\ref{PhysicalMap}) since the
transformation may be either stochastic or deterministic (ie. quantum
channel).

{\it C1 - class of local operations .-} In this case no communication
between Alice and Bob is allowed. The mathematical structure of the
map is elementary: $\Lambda_{AB}^{ \emptyset}= \Lambda_{A} \otimes
\Lambda_{B}$ with $\Lambda_{A}$, $\Lambda_{B}$ being both quantum
channels. As we said already this operation is always deterministic.

{\it C2a - class of ``one-way'' forward LOCC operations .-} Here
classical communication from Alice to Bob is allowed. The
form of the map is: $\Lambda_{AB}^{\rightarrow} (\varrho)=\sum_{i}
V^{i}_{A} \otimes
 \id_{B}(
[\id_{A} \otimes \Lambda_{B}^{i}](\varrho))(V^{i}_{A})^{\dagger}
\otimes \id_{B} $ with  deterministic maps $\Lambda_{B}^{i}$ which
reflect the fact that Bob is not allowed to perform ``truly
stochastic'' operation since he cannot tell Alice whether it has
taken place or not (which would happen only with some probability in
general ).

{\it C2b - class of ``one-way'' backward LOCC operations .-}
Here one has $\Lambda_{AB}^{\leftarrow} (\varrho)=\sum_{i}\id_{A}
\otimes V^{i}_{B} [ \Lambda_{A}^{i} \otimes \id_{B}](\varrho)\id_{A}
\otimes (V^{i}_{B})^{\dagger} $. The situation is the same as in C2a
but with the roles of Alice and Bob interchanged.

{\it C3 - class of ``two-way'' classical communication .-} Here both
parties are allowed to send classical communication to each other.
The mathematical form of the operation is quite complicated and the
reader is referred to \cite{DonaldHR2001}. Fortunately, there are
other two larger in a sense of inclusion classes, that are much more
easy to deal with: the classes of separable and PPT
operations.

{\it C4 - Class of separable operations.-} This class was considered
in \cite{PlenioVedral1998,RainsSep}. These are operations with product
Kraus operators:
\begin{eqnarray}
\Lambda^{sep}_{AB}(\varrho)=\sum_{i} A_{i} \otimes B_{i} \varrho
A_{i}^{\dagger} \otimes B_{i}^{\dagger}, \label{Separable}
\end{eqnarray}
which satisfy $\sum_i A_{i}^{\dagger} A_{i}\ot B_{i}^{\dagger} B_{i} = \id\ot\id.$

{\it C5. PPT operations .-} Those are operations
\cite{Rains1999,Rains-erratum1999} $\Lambda^{PPT}$ such that $
(\Lambda^{PPT}[(\cdot)^{\Gamma}])^{\Gamma}$ is completely positive. We
shall see that the simplest example of such operation is $\varrho
\rightarrow \varrho \otimes \varrho_{PPT}$ i. e. the process of adding
some PPT state.

There is an order of inclusions $C1 \subset C2a,C2b \subset C3 \subset
C4 \subset C5$, where all inclusions are strict ie. are not
equalities. The most intriguing is nonequivalence $C3 \neq C4$ which
follows from so called {\it nonlocality without entanglement}
\cite{Bennett-nlwe}: there are examples of product basis which are
orthonormal (and hence perfectly distinguishable by suitable von
Neumann measurement) but are not products of two orthonormal local
ones which represent vectors that cannot be perfectly distinguished by
parties that are far apart and can use only LOCC operations. Let us
stress, that the inclusion $C3 \subset C4$ is extensively used in
context of LOCC operations. This is because they are hard to deal
with, as characterized in a difficult way. If instead one deals with
separable or PPT operations, thanks to the inclusion, one can still
conclude about LOCC ones.

Subsequently if it is not specified, the term LOCC will be referred to
the most general class of operations with help of local operations and
classical communication --- namely the class C3 above.

Below we shall provide examples of some LOCC operations

{\it Example .-} The (deterministic) ``$U \otimes U$, $U \otimes
U^{*}$ twirling'' operations:
\begin{eqnarray}
&&\tau(\varrho) =\int dU U \otimes U \varrho (U \otimes U)^{\dagger}, \nonumber \\
&&\tau'(\varrho) =\int dU U \otimes U^{*} \varrho (U \otimes
U^{*})^{\dagger}.
\end{eqnarray}
Here $dU$ is a uniform probabilistic distribution on set of unitary
matrices. They are of ``one-way'' type and can be performed in the
following manner: Alice pick up randomly the operation $U$ rotates her
subsystem with it and sends the information to Bob which $U$ she had
chosen. Bob performs on his side either $U$ or $U^{*}$ (depending on
which of the two operations they wanted to perform).  The integration
above can be made discrete which follows from Caratheodory's theorem.

The very important issue is that {\it any state can be depolarized
  with help of $\tau$, $\tau'$ to Werner (\ref{Werner-state}) and
  isotropic (\ref{ntryplet}) state respectively}. This element is
crucial for entanglement distillation recurrence protocol
\cite{BBPSSW1996} (see Sec. \ref{subsec:recur}).

Another example concerns local filtering \cite{BBPS1996,Gisin96}:

{\it Example .-} The (stochastic) local filtering operation is
\begin{eqnarray}
&&{\cal L}_{LOCCfilter}(\varrho)=\frac{A \otimes B \varrho A^\dagger
\otimes B^\dagger }{{\tr}(A \otimes B \varrho A^\dagger \otimes
B^\dagger)}, \nonumber \\
&&A^{\dagger}A \otimes B^{\dagger}B \leq \id\ot\id.
\end{eqnarray}
This operation requires two-way communication (each party must send a
single bit), however if $A=\id$ ($B=\id$) then it becomes one-way
forward (backward) filtering. This one-way filtering is a crucial
element of a universal protocol of entanglement distillation
\cite{HHH1997-distill} (see Sec. \ref{subsec:two-qubits-distil}).

The local filtering operation is special case of a more general class
operations called {\it stochastic separable operations}, which
includes class $C4$. They are defined as follows
\cite{PlenioVedral1998,RainsSep}:
\begin{eqnarray}
\Lambda^{sep}_{AB}(\varrho)={\sum_{i} A_{i} \otimes B_{i} \varrho
A_{i}^{\dagger} \otimes B_{i}^{\dagger} \over \sum_i {\tr} A_i^{\dagger}A_i\ot B_i^{\dagger}B_i\varrho}
\end{eqnarray}
where $\sum_i A_i^{\dagger}A_i\ot B_i^{\dagger}B_i\leq \id\ot\id$.
These operations can be applied probabilistically via LOCC operations.

\section{Distillation and bound entanglement}
\label{sec:distil}

Many basic effects in quantum information theory based exploit pure
maximally entangled state $|\psiplus \rangle$.  However, in
laboratories we usually have mixed states due to imperfection of
operations and decoherence. A natural question arises, how to deal
with a {\it noise} so that one could take advantages of the
interesting features of {\it pure entangled states}. This problem has
been first considered by Bennett, Brassard, Popescu, Schumacher,
Smolin, and Wootters in 1996 \cite{BBPSSW1996}. In their seminal
paper, they have established a paradigm for {\it purification} (or
distillation) of entanglement.  When two distant parties share $n$
copies of a bipartite mixed state $\rho$, which contain noisy
entanglement, they can perform some LOCC and obtain in turn some
(less) number of $k$ copies of systems in state close to a singlet
state which contains {\it pure} entanglement. A sequence of LOCC
operations achieving this task is called {\it entanglement
  purification} or {\it entanglement distillation} protocol. We are
interested in {\it optimal} entanglement distillation protocols
i.e. those which result in maximal ratio $k\over n$ in limit of large
number $n$ of input copies. This optimal ratio is called distillable
entanglement and denoted as $E_D$ (see Sec. \ref{sec:miary} for formal
definition). Having performed entanglement distillation, the parties
can use obtained nearly singlet states to perform quantum
teleportation, entanglement based quantum cryptography and other
useful pure entanglement based protocols. Therefore, entanglement
distillation is one of the fundamental concepts in dealing with
quantum information and entanglement in general. In this section we
present the most important entanglement distillation protocols. We
then discuss the possibility of entanglement distillation in general
and report the {\it bound entangled states} which being entangled can
not be distilled.

\subsection{One-way hashing distillation protocol}
\label{subsec:onewayhash}

If only one party can tell the other party her/his result during the
protocol of distillation, the protocol is called {\it one-way}, and
{\it two-way} otherwise. One-way protocols are closely connected to
error correction, as we will see below. In \cite{BDSW1996} (see also
\cite{BBPSSW1996}) there was presented a protocol for two qubit states
which originates from cryptographic privacy amplification protocol,
called {\it hashing}. Following this work we consider here the so
called Bell diagonal states which are mixtures of two qubit Bell basis
states (\ref{basis}).  Bell diagonal states $\rho_{Bdiag}$ are
naturally parametrized by the probability of mixing $\{p\}$.  For
these states, the one-way hashing protocol yields singlets at a rate
$1-H(p)$, thus proving\footnote{It is known, that if there are only
  {\it two} Bell states in mixture, then one-way hashing is optimal so
  that distillable entanglement is equal $1-H(p)$ in this case.}
$E_D(\rho_{Bdiag})\geq 1-H(p)$ In two-qubit case there are four Bell
states (\ref{basis}). The $n$ copies of the two qubit Bell diagonal
state $\rho_{Bdiag}$ can be viewed as a classical mixture of strings
of $n$ Bell states.  Typically, there are only about $2^{nH(\{p\})}$
of such strings that are likely to occur \cite{CoverThomas}. Since the
distillation procedure yields some (shorter) string of singlets
solely, there is a straightforward ``classical'' idea, that to distill
one needs to know what string of Bells occurred. This knowledge is
sufficient as one can then rotate each non-singlet Bell state into
singlet state easily as in a dense coding protocol (see sec
\ref{sec:effects}).

Let us note, that sharing $\phi^-$ instead of  $\phi^+$
can be viewed as sharing $\phi^+$ with a phase error. Similarly $\psi^+$ means bit error and $\psi^-$
--  both bit and phase error.
 The identification of
the string of all Bell states that have occurred is then equivalent to
learning which types of errors occurred in which places. Thus the
above method can be viewed as error correction
procedure\footnote{Actually this reflects a deep relation developed in
  \cite{BDSW1996} between entanglement distillation and the large
  domain of quantum error correction designed for quantum computation
  in presence of noise.}.

Now, as it is well known, to distinguish between $2^{nH(\{p\})}$
strings, one needs $\log 2^{nH(\{p\})} = nH(\{p\})$ binary
questions. Such a binary question can be the following: what is the
sum of the bit-values of the string at certain $i_1,\ldots,i_k$
positions, taken modulo 2? In other words: what is the {\it parity} of
the given subset of positions in the string. From probabilistic
considerations it follows, that after $r$ such questions about {\it
  random} subset of positions (i.e.  taking each time random $k$ with
$1\leq k\leq 2n$) the probability of distinguishing two distinct
strings is no less than $1-2^{-r}$, hence the procedure is efficient.

The trick of the ``hashing'' protocol is that it can be done in
quantum setting. There are certain local unitary operations for the
two parties, so that they are able to collect the parity of the subset
of Bell states onto a {\it single} Bell state and then get to know it
locally measuring this Bell state and comparing the results. Since
each answer to binary question consumes one Bell state, and there are
$nH(\{p\})$ questions to be asked one needs at least $H(\{p\})<1$ to
obtain nonzero amount of not measured Bell states. If this is
satisfied, after the protocol, there are $n- nH(\{p\})$ unmeasured Bell
states in a {\it known} state. The parties can then rotate them all to
a singlet form (that is correct the bit and phase errors), and hence
distill singlets at an announced rate $1-H(\{p\})$.

This protocol can be applied even if Alice and Bob share a non Bell
diagonal state, as they can twirl the state applying at random one of
the four operations: $\sigma_x\ot \sigma_x$, $\sigma_y\ot
\sigma_y$,\,$\sigma_z\ot \sigma_z$,\, $\id\ot \id$ (which can be done
one two-way).  The resulting state is a Bell diagonal state. Of course
this operation often will kill entanglement. We will see how to
improve this in Secs. \ref{subsec:recur} and
\ref{subsec:two-qubits-distil}.

The above idea has been further generalized leading to general one-way
hashing protocol which is discussed in Sec. \ref{subsec:genhash}.

\subsection{Two-way recurrence distillation protocol}
\label{subsec:recur}

The hashing protocol cannot distill all entangled Bell-diagonal
states (one easily find this, knowing that those states are
entangled if and only if some eigenvalue is greater than $1/2$). To
cover all entangled Bell diagonal states one can first launch a {\it
two-way} distillation protocol to enter the regime where one-way
hashing protocol works. The first such protocol, called {\it
recurrence}, was announced already in the very first paper on
distillation \cite{BBPSSW1996}, and developed in \cite{BDSW1996}.
It works for two qubit states satisfying
$F=\tr\rho|\phi^+\>\<\phi^+|>{1\over 2}$ with $|\phi^+\>= {1\over
\sqrt 2}(|00\>+|11\>)$.

The protocol is defined by application of certain recursive
procedure. The procedure is probabilistic so that it may fail however
with small probability. Moreover it consumes a lot of resources. In
each step the procedure uses half of initial number of states.
\begin{enumerate}
\item
\label{s:one} Divide all systems into pairs:
$\rho_{AB}\ot\rho_{A'B'}$ with AB being source system and $A'B'$ being
a target one. For each such pair do: \subitem 1.1 Apply CNOT
transformation with control at $A$($B$) system and target at
$A'$($B'$) for Alice (Bob). \subitem 1.2 Measure the target system
in computational basis on both $A'$ and $B'$ and compare the
results. \subitem 1.3 If the results are the same, keep the AB
system and discard the $A'B'$. Else remove both systems.

\item If no system survived, then stop --- the algorithm failed. Else
  twirl all the survived systems turning them into Werner states.
\subitem 2.1 For one of the system do: if its state $\rho'$ satisfies
$1-S(\rho')>0$ then stop (hashing protocol will work) else go to
step (\ref{s:one}).
\end{enumerate}

In the above procedure, one deals with Werner states, because of
twirling in step $2$. In two-qubit case, Werner states are equivalent
to isotropic states and hence are parameterized only by the singlet
fraction $F$ (See Sec. (\ref{subsec:Werner_Iso}).  In one step of the
above recurrence procedure this parameter improves with respect to the
preceding one according to the rule:
\begin{equation}
F'(F) = {{F^2+{1\over 9}(1-F)^2}\over {F^2+{2\over
3}F(1-F)+{5\over 9}(1-F)^2}}.
\end{equation}
Now, if only $F>{1\over 2}$, then the above recursive map converges to
1 for sufficiently big initial number of copies.

The idea behind the protocol is the following. Step 1 decreases bit
error (i.e. in the mixture the weight of correlated states
$|\phi^\pm\rangle$ increases). At the same time, the phase error
increases, i.e. the bias between states of type $+$ and those of type
$-$ gets smaller. Then there is twirling step, which equalizes bit and
phase error. Provided that bit error went down more than phase error went
up, the net effect is positive. Instead of twirling one can apply
deterministic transformation \cite{QPA}, which is much more efficient.

\subsection{Development of distillation protocols --- bipartite and multipartite case}

The idea of recurrence protocol was developed in different ways.  The
CNOT operation, that is made in step $1.1$ of the above original
protocol, can be viewed as a permutation. If one apply some other
permutation acting locally on $n \geq 2$ qubits and performs a testing
measurements of steps $1.2-1.3$ on $m\geq 1$ one obtains a natural
generalization of this scheme, developed in \cite{DhaeneRecur} for the
case of two-qubit states. It follows that in case of $n=4, m=1$, and a
special permutation this protocol yields higher distillation
rate. This paradigm has been further analyzed in context of so called
code based entanglement distillation protocols
\cite{Matsumoto,AmbGot2way} in \cite{DehaeneEquiv}. The original idea
of \cite{BBPSSW1996} linking entanglement distillation protocols and
error correction procedures \cite{GottesmanPhd} have been also
developed in context of quantum key distribution. See in this context
\cite{Gottesman-Lo,AbainisSY_GenEntPur} and discussion in section
\ref{subsec:Pure_proofs}.

The original recurrence protocol was generalized to higher dimensional systems
in two ways in \cite{reduction,AlberetalGXOR}.  An interesting improvement of
distillation techniques due to \cite{VollbrechtV2004-hash-rec} where a
protocol that interpolates between hashing and recurrence one was
provided. This idea has been recently developed in
\cite{DehaeneBestHashBread}. Also, distillation was considered
in the context of topological quantum memory \cite{Bombin-dist}.

The above protocols for distillation of bipartite entanglement can be
used for distillation of a multipartite entanglement, when $n$ parties
are provided many copies of a multipartite state. Namely any pair of
the parties can distill some \eprpairs\ and then, using
e.g. teleportation, the whole group can redistribute a desired
multipartite state.  Advantage of such approach is that it is
independent from the target state.  In
\cite{DCpur,DurC_multi_dist2000} a sufficient condition for
distillability of arbitrary entangled state from $n$-qubit
multipartite state has been provided, basing on this idea.

However, as it was found by Murao et. al in the first paper
\cite{MPPVKmultidist} on multipartite entanglement distillation, the
efficiency of protocol which uses bipartite entanglement distillation
is in general less then that of {\it direct} distillation. The direct
procedure is presented there which is a generalization of bipartite
recurrence protocol of $n$-partite GHZ state from its ``noisy''
version (mixed with identity). In \cite{ManevaS_multi_hash2002} the
bipartite hashing protocol has been generalized for distillation of
$n$-partite GHZ state.

In multipartite scenario, there is no distinguished state like a
singlet state, which can be a universal target state in entanglement
distillation procedures. There are however some natural classes of
interesting target states, including the commonly studied GHZ
state. An exemplary is the class of the graph states (see
Sec. \ref{sec:full-sep}), related to one-way quantum computation
model. A class of multiparticle entanglement purification protocols
that allow for distillation of these states was first studied in
\cite{DurAB_graf_distill2003} where it is shown again to outperform
bipartite entanglement purification protocols. It was further
developed in \cite{AschauerDurBriegelGraphDistPRA05} for the subclass
of graph states called two-colorable graph states. The recurrence and
breeding protocol which distills all graph states has been also
recently found \cite{KruszynskaMBD_allgrafdist2006} (for a
distillation of graph states under local noise see
\cite{KayPDB_pur_term_gs2006}).

The class of two-colorable graph states is locally equivalent to the
class of so called CSS states that stems from the quantum error
correction \cite{CS-codes,Steane-codes-prl,Steane-codes-prsl}. The
distillation of CSS states has been studied in context of multipartite
quantum cryptographic protocols in \cite{ChenLo_multi_dist} (see
Sec. \ref{subsec:othermulti}).  Recently,
the protocol which is direct
generalization of original hashing method (see
Sec. \ref{subsec:onewayhash}) has been found, that distills CSS states
\cite{HostenesDM_CSSdist2006}.
This protocol outperforms all previous versions of hashing of CSS states (or their subclasses such as Bell diagonal states)
\cite{ManevaS_multi_hash2002,DurAB_graf_distill2003,ChenLo_multi_dist,AschauerDurBriegelGraphDistPRA05}.
Distillation of the state $W$ which is not a CSS state, has been studied recently
in \cite{MiyakeB_Wdist2005}.
In \cite{GlancyKV2006} a protocol of distillation  of all stabilizer states
(a class which includes CSS states) based on stabilizer codes \cite{GottesmanPhd,Nielsen-Chuang} was proposed.
Based on this, a breeding protocol for stabilizer states was provided in \cite{HostensDM2006}.

\subsection{All two-qubit entangled states are distillable}
\label{subsec:two-qubits-distil}

The recurrence protocol, followed by hashing can distill entanglement
only from states satisfying $F>{1\over 2}$. One can then ask what is
known in the general case.  In this section we present the protocol
which allows to overcome this bound in the case of two qubits. The
idea is that with certain (perhaps small) probability, one can
conclusively transform a given state into a more desired one, so that
one knows if the transformation succeeded. There is an operation which
can increase the parameter $F$, so that one can then perform
recurrence and hashing protocol. Selecting successful cases in the
probabilistic transformation is called {\it local filtering}
\cite{Gisin96}, which gives the name of the protocol. The composition
of filtering, recurrence and hashing proves the following result
\cite{HHH1997-distill}:

\begin{itemize}
\item Any two-qubit state is distillable if and only if it is
  entangled.
\end{itemize}

Formally, in filtering protocol Alice and Bob are given $n$ copies of
state $\rho$. They then act on each copy with an operation given by
Kraus operators (see note in Sec. \ref{sec:BipartiteEntanglement})
$\{A\ot B ,\sqrt{\id-A^{\dagger}A}\ot B, A\ot\sqrt{\id-B^{\dagger}B},
\sqrt{\id-A^{\dagger}A}\ot \sqrt{\id-B^{\dagger}B} \}$. The event
corresponding to Kraus operator $A\ot B$ is the desired one, and the
complement corresponds to failure in improving properties of $\rho$.
In the case of success the state can be transformed into
\begin{equation}
\sigma = {A\ot B\rho A^{\dagger}\ot B^{\dagger}\over {\tr}(A\ot B
\rho A^{\dagger}\ot B^{\dagger})},
\end{equation}
however only with probability $p= {\tr}(A\ot B \rho A^{\dagger}\ot
B^{\dagger})$. After having performed the operation, Alice and Bob
select all good events and keep them, while removing the copies, for
which they failed.

Suppose now that Alice and Bob are given $n$ copies of entangled
state $\rho$ having $F \leq {1\over 2}$. They would like to obtain
some number $k$ of states with $F>{1\over 2}$. We show now how one
can build up a filtering operation which does this task. Namely any
two-qubit entangled state becomes not positive after
partial transpose (is NPT). Then, there is a pure state
$|\psi\>=\sum_{ij}a_{ij}|ij\>$ such, that
\begin{equation}
\<\psi|\rho^{\Gamma}_{AB}|\psi\> <0,
\end{equation}
with $\Gamma$ denoting partial transpose on Bob's subsystem. It can be
shown, that a filter $A\ot B = M_{\psi}\ot \id$ with
$[M_{\psi}]_{ij}=a_{ij}$ is the needed one. That is the state $\sigma$
given by
\begin{equation}
  \sigma={{M_{\psi}\ot \id \rho_{AB}M_{\psi}^{\dagger}\ot
      \id}\over {\tr}(M_{\psi}\ot \id \rho_{AB}M_{\psi}^{\dagger}\ot \id)},
\end{equation}
which is the result of filtering, fulfills $F(\sigma)> {1\over 2}$.

The filtering distillation protocol is just this: Alice applies to
each state a POVM defined by Kraus operators $\{M_{\psi}\ot
\id,[\sqrt{\id - M_{\psi}^{\dagger}M_{\psi}}]\ot \id\}$. She then
tells Bob when she succeeded. They select in turn on average $np$
pairs with $p={\tr}(M_{\psi}^{\dagger}M_{\psi}\ot \id
\rho_{AB})$. They then launch recurrence and hashing protocols to
distill entanglement, which yields nonzero distillable entanglement.

Similar argument along these lines, gives generalization of two-qubit
distillability to the case of NPT states acting on
$\ccal^{2}\ot\ccal^{N}$ Hilbert space \cite{DurCirac_activation,DurCLB1999-npt-bound}. It follows also, that for $N=3$
all states are also distillable if and only if they are entangled,
since any state on $\ccal^{2}\ot\ccal^3$ is entangled if and only if
it is NPT. The same equivalence has been shown for all rank two bipartite states \cite{Noiseless}

\subsection{Reduction criterion and distillability}
One can generalize the filtering protocol by use of reduction
criterion of separability (see Sec.
\ref{sec:BipartiteEntanglement}). It has been shown that any state
that violates reduction criterion is distillable \cite{reduction}.
Namely, for states which violate this criterion, there exists
filter, such that the output state has fidelity $F>{1\over d}$, where $F$
is overlap with maximally entangled state
$|\Phiplus\>={1\over \sqrt{d}}\sum_{i=0}^{d-1}|ii\>$. Such states is
distillable,  similarly, as for two-qubit states with $F>{1\over 2}$. The
simplest protocol  \cite{BraunsteinCJLS1998-nmr} is the following:
one projects such state using local rank two projectors
$P=|0\>\<0|+|1\>\<1|$, and finds that obtained two-qubit state has
$F>{1\over 2}$, hence is distillable.

The importance of this property of reduction criterion lies in the
fact, that its generalization to continuous variables allowed to
show that all two-mode Gaussian states which violate PPT criterion
are distillable.

\subsection{General one-way hashing}
\label{subsec:genhash}

One can ask what is the maximal yield of singlet as a function of a
bipartite state $\rho_{AB}$ that can be obtained by means of one way
classical communication.  In this section we will discuss protocol
which achieves this task \cite{DevetakWinter-hash,SW-nature,sw-long}.

In Sec. (\ref{subsec:onewayhash}) we learned a protocol called one-way
hashing which for Bell diagonal states with spectrum given by
distribution $\{p\}$, gives $1-H(\{p\})$ of distillable
entanglement. Since in case of these states, the von Neumann entropy
of subsystem reads $S(\rho^{B}_{Bdiag})=1$ and the total entropy of
the state is equal to $S(\rho^{AB}_{Bdiag})=H(\{p\})$, there has been
a common believe that in general there should be a one-way protocol
that yields \begin{equation} E_D(\rho_{AB})\geq S(\rho_{B})- S(\rho_{AB}).  \end{equation}
This conjecture has been proved by Devetak and Winter
\cite{DevetakWinter-hash}.

The above inequality, called {\it hashing inequality}, states that
distillable entanglement is lower bounded by {\it coherent
  information} defined as $I^{coh}_{A\>B}=S(\rho_B)-S(\rho_{AB})\equiv
-S(A|B)$.

The original proof of hashing inequality
\cite{DevetakWinter-hash,DevetakWinter-hash-prl} was based on
cryptographic techniques, where one first performs error correction
(corresponding to correcting bit) and then privacy amplification
(corresponding to correcting phase), both procedures by means of
random codes (we will discuss it in Sec. \ref{sec:Ent_in_QKD}).

Another protocol that distills the amount $I^{coh}_{A\>B}$ of singlets
from a given state is the following \cite{SW-nature,sw-long}: given
many copies of the state, Alice projects her system onto so-called
typical subspace \cite{Schumacher1995} (the probability of failure is
exponentially small with number of copies). Subsequently, she performs
incomplete measurement $\{P_i\}$ where the projectors $P_i$ project
onto blocks of size $2^{n (S_B-S_{AB})}$. If the measurement is chosen
at random according to uniform measure (Haar measure), it turns out
that for any given outcome Alice and Bob share almost maximally
entangled state, hence equivalent to $n(S_B -S_{AB})$ e-bits. Of
course, for each particular outcome $i$ the state is different,
therefore one-way communication is needed (Bob has to know the outcome
$i$).

The above bound is often too rough (e.g. because one can distill
states with negative coherent information using recurrence
protocol). Coherent information $I^{coh}_{A\>B}$
is not an entanglement monotone. It is then known, that in general
the optimal protocol of distillation of entanglement is some two-way
protocol which increases $I^{coh}$, followed by general hashing
protocol: That is we have \cite{HHH-cap2000}
\begin{equation} E_D(\rho) = \sup_{\Lambda \in LOCC} I^{coh}(\Lambda(\rho)). \end{equation}
It is however not
known how to attain the highest coherent information via two-way distillation protocol.

\subsection{Bound entanglement --- when distillability fails}
\label{subsec:bent-when-distillability-fails}

Since the seminal paper on distillation \cite{BBPSSW1996}, there was
a common expectation, that all entangled bipartite states are
distillable. Surprisingly it is not the case. It was shown in
\cite{bound}  that the PPT states cannot be distilled. It is rather
obvious, that one cannot distill from separable states.
Interestingly, the first example of {\it entangled} PPT state had
been already known from \cite{Pawel97}. Generally the states that
are entangled yet not distillable are called {\it bound entangled}.
It is not known if there are other bound entangled states
than PPT entangled states. There are quite many interesting formal
approaches allowing to obtain families of PPT entangled state. There
is however no operational, intuitive ``reason'' for existence of this
mysterious type of entanglement.

Let us first comment on the non-distillability of separable states.
The intuition for this is straightforward: separable states can be
created via LOCC from scratch. If one could distill singlets from
them, it would be creating something out of nothing.
This reasoning of course does not hold for entangled PPT states.
However, one can look from a different angle to
find relevant {\it  formal} similarities between
PPT states and separable states.   Namely, concerning separable states
one can observe, that the fidelity
$F={\tr}\sigma_{sep}|\Phiplus\>\<\Phiplus|$ is no greater then $1\over d$
for $\sigma_{sep}$ acting on $\ccal^{d}\ot \ccal^{d}$. Since LOCC operations can only transform separable state into another separable state, (i.e. the
set of separable states is closed under LOCC operations), one cannot distill singlet from separable states since one cannot increase the singlet fraction.

It turns out that also PPT states do not admit higher fidelity than
$1/d$ as well as are closed under LOCC operations. Indeed
we have ${\tr}\rho_{AB}|\Phiplus\>\<\Phiplus|={1\over d}{\tr}\rho_{AB}^{\Gamma}V $   which can not exceed $1\over d$ (here $V$ is the swap operator (\ref{swap})). Indeed, $\rho_{AB}^{\Gamma}\geq 0$, so that ${\tr}\rho_{AB}^{\Gamma}V$ can be viewed as an average value
of a random variable which can not exceed 1 since $V$ has eigenvalues $\pm 1$  \cite{Rains1999,Rains-erratum1999}. To see the second feature,
note that any LOCC operation $\Lambda$ acts on a state $\rho_{AB}$ as follows
\be
\rho_{out}=\Lambda({\rho}_{AB}) = \sum_{i}A_i\ot B_i(\rho_{AB})A^{\dagger}_i\ot
B^{\dagger}_i,
\ee
which after partial transpose on subsystem $B$ gives
\be
\rho_{out}^{\Gamma}=\sum_{i}A_i\ot
(B_i^{\dagger})^T(\rho_{AB}^{\Gamma})A^{\dagger}_i\ot B^{T}_i.
\ee
The resulting operator is positive, if only $\rho_{AB}^\Gamma$ was positive.

Since the discovery of first bound entangled states quite many further
examples of such states were found, only a few of which we have
discussed (see Sec. \ref{subsubsec:RangePPT}). The comprehensive list
of achievements in this field, as well as the introduction to the
subject can be found in \cite{ClarissePhd}.

\subsection{The problem of NPT bound entanglement}

Although it is already known that there exist entangled nondistillable
states, still we do not have a characterization of the set of such
states. The question which remains open since the discovery of bound
entanglement properties of PPT states is: are all NPT states
distillable? For two main attempts \footnote{Two recent attempts are
  unfortunately incorrect \cite{ChattoSarkar,SimonNPTbound}.} to solve
the problem see
\cite{DiVincenzoSSTT1999-nptbound,DurCLB1999-npt-bound}. In
\cite{reduction} it was shown that this holds if and only if all NPT
Werner states (equivalently entangled Werner states) are
distillable. It simply follows from the fact, that as in case of two
qubits any entangled state can be filtered, to such a state, that
after proper twirling, one obtains entangled Werner state. The
question is hard to answer because of its asymptotic nature. A
necessary and sufficient condition for entanglement distillation
\cite{bound} can be stated as follows:
\begin{itemize}
\item Bipartite state $\rho$ is distillable if and only if there
  exists $n$ such, that $\rho$ is $n$-copy distillable
  (i.e. $\rho^{\ot n}$ can be filtered to a two-qubit entangled
  state)\footnote{Equivalently, there exists pure state $|\psi\>$ of
    Schmidt rank $2$ such that $\<\psi|\left(\rho^{\otimes
        n}\right)^\Gamma|\psi\><0$
    \cite{DiVincenzoSSTT1999-nptbound,DurCLB1999-npt-bound}.}.
\end{itemize}
for some $n$.  It is however known, that there are states which are
not $n$-copy distillable but they are $n+1$-copy distillable
\cite{Watrous2003-n-dist} (see also
\cite{Bandyopadhyay2003-n-distil}), for this reason, the numerical
search concerning distillability basing on few copies may be
misleading, and demanding in computing resources. There is an
interesting characterization of $n$ copy distillable states in terms
of entanglement witnesses found in \cite{KrausLC}. Namely, a state is
separable iff the following operator
\begin{equation}
W_n= P^+_{A'B'} - [\rho_{AB}^{\ot n}]^\Gamma
\end{equation}
is not an entanglement witness (here $A'B'$ is two qubit system, $P^+$
is maximally entangled state).

One may also think, that the problem could be solved by use of simpler
class of maps - namely PPT operations. Remarkably, in \cite{WernerPPT}
it was shown that any NPT state can be distilled by PPT operations.
Recently the problem was attacked by means of positive maps, and
associated ``witnesses'' in \cite{Clarisse2004-distil-maps} (see also
\cite{Clarisse2005-distil}).

The problem of existence of NPT \bent\ states has important
consequences. If they indeed exist, then distillable entanglement is
nonadditive and nonconvex. Two states of zero $E_D$ will together give
nonzero $E_D$. The set of bipartite \bent\ states will not be closed
under tensor product, and under mixing. For the extensive review of
the problem of existence of NPT \bent\ states see \cite{ClarissePhd}.

Schematic representation of the set of all states including
hypothetical set of NPT bound entangled states is shown on
fig.~\ref{fig:states-and-witnesses}.

\begin{figure}
  \centering
  \includegraphics{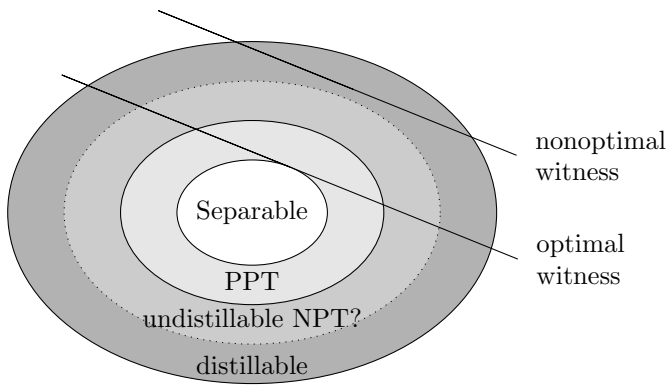}
  \caption{Schematic representation of the set of all states with
    example of entanglement witness and its optimization}
  \label{fig:states-and-witnesses}
\end{figure}

\subsection{Activation of bound entanglement}
\label{subsec:activbound}
Entanglement is always considered as a resource useful for certain
task. It is clear, that pure entanglement can be useful for many
tasks, as it was considered in Sec. \ref{sec:effects}. Since
discovery of bound entanglement a lot of effort was devoted to find
some nontrivial tasks that this type of entanglement
allows to achieve.

The first phenomenon which proves usefulness of bound entanglement,
called {\it activation} of bound entanglement, was discovered in
\cite{activation}. A parameter which is improved after activation is
the so called {\it probabilistic maximal singlet fraction}, that is
$F^{(p)}_{\max}(\rho)= max_{\Lambda}\frac{1}{{\tr}[\Lambda(\rho)]}{\tr}[\Lambda(\rho)|\Phiplus\>\<\Phiplus|]$
for $\rho$ acting on ${\cal C}^{d}\ot{\cal C}^{d}$,  where $\Lambda$ are
local filtering operations
\footnote{The superscript $(p)$ emphasizes a "probabilistic" nature of this maximal singlet fraction so that it would not  be confused with different parameter defined analogously as
$F_{\max}(\rho)=max_{\Lambda}\frac{1}{{\tr}[\Lambda(\rho)]}{\tr}[\Lambda(\rho)
|\Phiplus\>\<\Phiplus|]$ with $\Lambda$ being trace preserving LOCC operation, that is called {\it maximal
singlet fraction}.}.

Consider a state $\sigma$ with $F^{(p)}_{\max}$ bounded away
from $1$, such that no LOCC protocol can go beyond this bound.  Then a
protocol was found, involving $k$ pairs of some \bent\ state
$\rho_{be}$, which takes as an input a state $\sigma$, and outputs a
state $\sigma'$ with singlet fraction arbitrarily close to $1$ (which
depends on $k$). That is: \be \rho_{be}^{\ot k}\ot\sigma
\longrightarrow \sigma',\quad \lim_k F^{(p)}_{\max}(\sigma')=1 \ee The
protocol is actually closely related to recurrence distillation
protocol (see Sec. \ref{subsec:recur}), generalized to higher
dimension and with twirling step removed. In the above scenario the
state $\rho_{be}$ is an {\it activator} for a state $\sigma$. The
probability of success in this protocol decreases as the output
fidelity increases. To understand the activation recall that probabilistic maximal singlet fraction of every bound entangled state is by definition
bounded by ${1\over d}$, as discussed in
Sec. \ref{subsec:bent-when-distillability-fails}. It implies, that $\rho_{be}^{\ot k}$ also have $F^{(p)}_{max}$ bounded away from 1. We have then two states with probabilistic maximal fidelity bounded away from $1$, which however changes if they are put together. For this reason, the effect of activation demonstrates a sort of {\it nonadditivity} of maximal singlet fraction.

The effect of activation was further developed in various
directions. It was shown in \cite{VollbWolf_activation}, that any NPT
state, can be made one-copy distillable\footnote{A state $\rho$ is
  called one copy distillable, if there exists projectors $P,Q$ of
  rank two such that $P\ot Q \rho P\ot Q$ is NPT. In other words, from
  $\rho$ one can then obtain by LOCC a twoqubit state with $F$ greater
  than $1/2$.} by use of \bent\ states.  Moreover to this end one
needs \bent\ states which are arbitrarily close to separable states.

A remarkable result showing the power of LOCC operations supported
with arbitrarily small amount of \bent\ states is established in
\cite{Ishizaka04}. Namely, the interconversion of {\it pure} bipartite
states is ruled by entanglement measures. In particular, a pure state
with smaller measure called {\it Schmidt rank} (see Sec.
\ref{sec:miary}) cannot be turned by LOCC into a state with higher
Schmidt rank. However any transition between pure states is possible
with some probability, if assisted by arbitrarily weakly bound
entangled states. This works also for multipartite
states. Interestingly, the fact that one can increase Schmidt rank by
PPT operations (i.e. also with some probability by assistance of a PPT
state) is implicit already in \cite{AudenaertPE2002-PPT}.

Although specific examples of {\it activators} has been found, the
general question if any (bound) entangled state can be an activator
has waited until the discovery of Masanes
\cite{Masanes1_activation,Masanes-multiactiv}. He showed,
that every entangled state (even a bound entangled one) can enhance
maximal singlet fraction and in turn a fidelity of teleportation of some other entangled state, i.e. for any state $\rho$ there exists
state $\sigma$ such that
\be
F^{(p)}_{max}(\sigma) < F^{(p)}_{max}(\rho\ot\sigma).
\ee
This result\footnote{Actually, the result is even stronger: it holds also for $F_{max}$ (i.e. singlet fraction achievable with probability one).} for the first time shows that

\bei
\item {\it Every entangled state can be used to some nonclassical task}
\eei

Masanes provides an existence proof via {\it reductio ad absurdum}. It
is then still a challenge to construct for a given state some {\it
  activator}, even though one has a promise that it can be found. This
result indicates first useful task, that can be performed by all bound
entangled states.

The idea of activation for bipartite case was developed in
multipartite case in \cite{DurCirac_activation} and
\cite{Bandyo_multi_activation} where specific families of
multipartite bound entangled states were found. Interestingly, those
states were then used to find analogous phenomenon in {\it classical
key agreement} scenario (see Sec. \ref{sec:Ent_in_QKD}). It was
recently generalized by Masanes (again in existential way)
\cite{Masanes-multiactiv}.

The activation considered above is a result which concerns one copy
of the state.  An analogous result which hold in asymptotic regime,
called {\it superactivation} was found in \cite{ShorST-superactiv}.
Namely there are fourpartite states, such that no
two parties even with help of other parties can distill pure
entanglement from them:
\ben \rho_{ABCD}^{be}=\frac{1}{4}
\sum_{i=1}^{4}|\Psi_{AB}^{i}\>\<\Psi_{AB}^{i}|\ot|\Psi_{CD}^{i}\>\<\Psi_{CD}^{i}|.
\een
However the following state, consisting of five copies of the
same state, but each distributed into different parties is no longer
bound entangled \ben \rho_{free} =
\rho_{ABCD}^{be}\ot\rho_{ABCE}^{be}\ot\rho_{ABDE}^{be}\ot\rho_{ACDE}^{be}\ot\rho_{BCDE}^{be}.
\een This result is stronger than activation in two ways. First it
turns two totally non-useful states (bound entangled ones) into a
useful state (distillable one), and second the result does not
concern one copy but it has asymptotically nonvanishing rate. In
other words it shows, that there are states $\rho_1$ and $\rho_2$
such that $E(\rho_1\ot \rho_2) > E(\rho_1)+E(\rho_2)$, despite the
fact that $E(\rho_1)=E(\rho_2)=0$ where $E$ is suitable measure
describing effect of distillation (see Sec. \ref{sec:miary}).

\subsubsection{Multipartite bound entanglement}

In this case one defines bound entanglement as any entangled state
({\it i.e. violating full separability condition}) such that one can
not distill any pure entanglement between any subset of parties.
Following the fact that violation of PPT condition is necessary for
distillation of entanglement it is immediate that if the state
satisfies PPT condition under any bipartite cut then it is
impossible to distill pure entanglement at all.

The first examples of multipartite bound entanglement were
semiseparable three qubit states $\varrho^{\bf Shift}_{ABC}$ (see
Sec. \ref{subsec:PartialSep}). Immediately it is not distillable
since it is {\it separable} under any bipartite cut. Thus not only
pure entanglement, but simply {\it no} bipartite entanglement
can be obtained between any two subsystems, while any pure multipartite
state possesses entanglement with respect to some cut.

One can relax the condition about separability or PPT property and
the nice example is provided in \cite{DCpur}. This is an example of
3-qubit state $\varrho_{ABC}$ from the family (\ref{mCirac}) such
that it is PPT with respect to two partitions but not with respect
to the third one i.e. $\varrho_{ABC}^{T_{A}}\geq 0$,
$\varrho_{ABC}^{T_{B}}\geq 0$ but
$\varrho_{ABC}^{T_{C}}\not\geq 0$. Thus not all nontrivial
partitions have NPT property which, in case of this family, was
proven to be equivalent to distillability (see last paragraph of the
previous subsection). Still the state is entangled because of
violation PPT along the $AB|C$ partition, hence it is bound
entangled.

Note, however, that the bound entanglement in this state
$\varrho_{ABC}$ can be activated with partial free entanglement of
the state $|\Phi_{AB}^{+}\rangle | 0\rangle_{C} $  since then part
$A$ of bound entangled state can be teleported to $B$ and vice
versa, producing in this way two states $\sigma_{AC}=\sigma_{BC}$
between Alice and Charlie and Bob and Charlie respectively which are
$2 \otimes 4$ NPT states, but as we already know all NPT $2 \otimes
N$ states are distillable (see previous section on distillation of
bipartite entanglement). In this way one can distill entanglement
between $AC$ and $BC$ which allows to distill three particle GHZ
entanglement as already previously mentioned. Thus three qubit bound
entanglement of $\varrho_{ABC}$ can be activated with help of
two-qubit pure entanglement.

A simple, important family of bound entangled states are four-qubit
Smolin bound entangled states (\ref{SmolinS}). They have the
property, that they can be ``unlocked'': if parties $AB$ meet, then
the parties $CD$ can obtain an \eprpair. They are bound entangled,
because any cut of the form two qubits versus two other qubits is
separable. Again, if instead we could distill a pure state, the
Smolin state would be entangled with respect to some cut of this
form.

Recently a relationship between multipartite bound entangled states and the stabilizer formalism was founded \cite{WangY2007}, which allows to construct a wide class of unlockable bound entangled states in arbitrary multiqudit systems. It includes a generalized Smolin states \cite {AugusiakGSS,Bandyo_multi_activation}.

\subsection{Bell inequalities and bound entanglement}
\label{subsec:bellboundent}
Since bound entanglement is a weak resource, it was natural to ask,
whether it can violate Bell inequalities. The first paper reporting
violation of Bell inequalities was due to D\"ur \cite{Dur} who
showed  that some multiqubit bound entangled  states violate
two-settings inequalities called Mermin-Klyshko inequalities, for
$n$-partite states with $n\geq 8$. Further improvements
\cite{Kaszlikowski} ($n\geq 7$) \cite{Sen,NL} ($n\geq 6$) have been
found. In \cite{Augusiak} the lowest so far number  of particles was
established: it was shown that  the four-qubit Smolin bound
entangled states (\ref{SmolinS}) violate Bell inequality. The latter
is very simple distributed version of CHSH inequality:
\be
|a_A\ot a_B\ot a_C\ot(b_D+ b'_D)+a'_A\ot a'_B\ot a'_C\ot(b_D- b'_D)|\leq 2
\ee
We see that this is the usual CHSH inequality where one
party was split into three, and the other on ($B$) was untouched.
What is curious, the maximal violation is obtained exactly at the
same choices of observables, as for singlet state with standard
CHSH: \ben
&&a=\sigma_x,\quad a'=\sigma_y, \nonumber \\
&&b={\sigma_x+\sigma_y\over \sqrt 2},\quad b'={\sigma_x-\sigma_y\over \sqrt 2}
\een
and the value is also the same as for singlet state: $2\sqrt 2$.

Werner and Wolf analyzed multipartite states
\cite{WernerW1999-PPTBell,WernerW2001-Bell-review} and have shown that
states which are PPT in every cut (we will call them PPT states in
short) must satisfy any Bell inequality of type $N \times 2 \times 2 $
(i.e. involving N observers, two observables per site, each observable
with two outcomes).

The relation of Bell inequalities with bipartite distillability has
been also analyzed in \cite{AcinGT2006-grot} where it has been shown
that any violation of Bell inequalities from the Mermin-Klyshko class
always leads to possibility of bipartite entanglement distillation
between some subgroups of particles.

In 1999 Peres has conjectured \cite{Peres1999-PPTBell} that PPT
states do not violate Bell inequalities: \be LHVM \Leftrightarrow
PPT. \label{Peres} \ee We see that the above results support
this conjecture. The answer to this question, if positive, will
give very important insight into our understanding of classical
versus quantum behavior of states of composite systems.

Let us finally mention, that all those considerations concern
violation without postselection, and operations applied to many
copies. Recently, Masanes showed, that if we allow collective
manipulations, and postselection, no bound entangled states violate
the CHSH inequality even in asymptotic regime \cite{Masanes}.

\section{Manipulations of entanglement and irreversibility}
\label{sec:irrev}
\subsection{LOCC manipulations on pure entangled states --- exact case}
\label{subsec:puretrans}
The study on exact transformations between pure states
by LOCC was initiated by \cite{LoPopsecu1997-beyond}.
A seminal result in this area is due to Nielsen \cite{Nielsen-pure-entanglement}. It turns out that the
possible transitions can be classified in a beautiful way in terms of
squares of Schmidt coefficients $\lambda_i$ (i.e. eigenvalues of
local density matrix).
Namely, a pure state $|\psi\>=\sum_{j=1}^d\sqrt{\lambda^{(\psi)}_j}|jj\>$ can be transformed into other pure state $|\phi\>=\sum_{j=1}^d\sqrt{\lambda^{(\phi)}_j}|jj\>$ if and only if for each $k \in \{1,\ldots,d\}$ there holds
\be
\sum_{j=1}^k \lambda^{(\psi)\downarrow}_j \leq  \sum_{j=1}^k \lambda^{(\phi)\downarrow}_j,
\label{eq:major}
\ee
where $\lambda^{(\psi,\phi)\downarrow}_j$ are eigenvalues of subsystem
of $\psi$ ($\phi$) in descending order.
The above condition states that $\phi$ {\it majorizes} $\psi$ (see also Sec. (\ref{subsec:majorrel})). Thus one can transform $\psi$ into $\phi$ only when subsystems of $\psi$  are more mixed than those of $\phi$. This is compatible with Schr{\"o}dinger approach: the more mixed subsystem, the more entangled state.

Since majorization constitutes a partial order, reversible conversion
$\psi\leftrightarrow\phi$ is possible if and only if the Schmidt
coefficients of both states are equal. Moreover there exist states
either of which cannot be converted into each other. Thus generically,
LOCC transformations between pure states are {\it
  irreversible}. However as we will see, it can be lifted in
asymptotic limit.

Further important results in this area have been provided by
\cite{Vidal} who obtained optimal probability of success for
transitions between pure states (see
Sec. \ref{subsubsec:em-tr-exact}), and \cite{JonathanP}, who
considered transitions state $\to$ ensemble.

\subsubsection{Entanglement catalysis}
\label{subsec:catalysis}
The most surprising consequence of the Nielsen's laws of pure state
transitions have been discovered in \cite{JonathanP}. Namely, for
some states $\psi_1$ and $\psi_2$ for which transition $\psi_1 \to
\psi_2$  is impossible, the following process is possible: \be
\psi_1 \ot \phi \to \psi_2 \ot\phi, \ee thus we borrow state $\phi$,
run the transition, and obtain untouched $\phi$ back! The latter
state it plays exactly a role of {\it catalyst}: though not used up
in the reaction, its presence is necessary to run it.
Interestingly, it is not hard to see that the catalyst  cannot be
maximally entangled. The catalysis effect was extended to the case
of mixed states in \cite{EisertWilkens}.

\subsubsection{SLOCC classification}
\label{subsubsec:slocc}
For multipartite pure states there does not exist Schmidt
decomposition. Therefore Nielsen result cannot be easily
generalized. Moreover analysis of LOCC manipulations does not allow
to classify states into some coarse grained classes, that would give
a rough, but more transparent picture. Indeed, two pure states can
be transformed into each other by LOCC if and only if they can be
transformed by local unitary transformations so that to parameterize
classes one needs continuous labels, even in bipartite case.

To obtain a simpler, ``coarse grained'' classification, which would be
helpful to grasp important qualitative features of entanglement, it
was proposed \cite{DurVC00-wstate} to treat states as equivalent, if
with some nonzero probability they can be transformed into each
other by LOCC. This is called {\it stochastic} LOCC, and denoted by
SLOCC. It is equivalent to say that  there exists reversible
operators $A_i$ such that
\be
|\psi\>=A_1 \ot\ldots \ot   A_N |\phi\>.
\ee
For bipartite pure states of $d\otimes d$ system we obtain in this
way $d$ entangled classes of states, determined by number of nonzero
Schmidt coefficients (so called Schmidt rank). Here is example of
SLOCC equivalence: the state
\be
|\psi\>=a|00\>+b|11\>
\ee
with $a>b>0$ can
be converted (up to irrelevant phase) into $|\phiplus\>$ by filter $A\ot
\id$, with $A= \left[\bea{cc}
\frac{b}{a} & 0 \\
0 & 1 \\
\eea\right]$ with probability $p=2|b|^2$; so it is possible
to consider representative state for each class.

A surprising result is due to Ishizaka \cite{Ishizaka04} who
considered SLOCC assisted by bound entangled PPT states. He then
showed that every state can be converted into any other. This works
for both bipartite and multipartite pure states. For multipartite
states SLOCC classification was done in the case of three
\cite{DurVC00-wstate} and four qubits
\cite{VerstraeteDMV2001-4qubits}, and also two qubits and a qudit
\cite{MiyakeV2003-22n}. For three qubits there are five classes plus
fully product state, three of them being Bell states
between two qubits (i.e. states of type $\epr_{AB}\ot |0\>_C$). Two
others are GHZ state \be |GHZ\>={1\over \sqrt2} (|000\> + |111\>)
\ee and so called $W$ state \be |W\>={1\over \sqrt3}
(|100\>+|010\>+|001\>). \ee They are inequivalent, in a sense, that
none of them can be converted into the other one with nonzero
probability (unlike in bipartite state case, where one can go from
any class to any lower class  i.e. having lower Schmidt rank).

In $2\ot 2\ot d$ case \cite{MiyakeV2003-22n,Miyake2004-slocc}, there
is still discrete family of inequivalent classes, where there is
maximally entangled state --- two \eprstate\ $\phi^+_{AB_1}\ot
\phi^+_{B_2 C}$ (where the system $B$ is four-dimensional). Any
state can be produced from it simply via teleportation (Bob prepares
the needed state, and teleports its parts to Alice and to Charlie.

In four qubit case the situation is not so simple: the inequivalent
classes constitute a continuous family, which one can divide into nine
qualitatively different subfamilies.

The SLOCC classification is quite elegant generalization of local
unitary classification.  In the latter case the basic role is played
by invariants of group $SU_{d_1}\ot \ldots \ot SU_{d_N}$ for
$d_1\otimes \ldots \otimes d_N$ system, while in SLOCC, the relevant
group is $SL_{d_1,\ccal}\ot \ldots \ot SL_{d_N,\ccal}$ (one restricts
to filters of determinant $1$, because the normalization of states
does not play a role in SLOCC approach).

Finally, the SLOCC classification of pure states can be used to obtain
some classification of mixed states (see
\cite{AcinBLS2001-wghz-mixed,MiyakeV2003-22n}).

\subsection{Asymptotic entanglement manipulations and irreversibility}
\label{subsec:mixed_asym_trans}
The classifications based on exact transformations suffer for some
lack of continuity: for example in SLOCC approach $\psi$ with squares
of Schmidt coefficients $(0.5,0.49,0.01)$ is in the same class as the
state $(1/3,1/3,1/3)$, but in different class than $(0.5,0.5,0)$,
while we clearly see that the first and the last have much more in
common than the middle one. In order to neglect small differences, one
can employ some asymptotic limit. This is in spirit of Shannon's
communication theory, where one allows for some inaccuracies of
information transmission, provided they vanish in asymptotic limit of
many uses of channel. Interestingly, the first results on quantitative
approach to entanglement \cite{BBPS1996,BDSW1996,BBPSSW1996} were
based on LOCC transformations in asymptotic limit.

In asymptotic manipulations, the main question is what is the {\it
  rate} of transition between two states $\rho$ and $\sigma$. One defines the rate as follows. We assume that Alice and Bob have initially $n$ copies in state $\rho$. They apply LOCC operations, and obtain $m$ pairs\footnote{Here $m$ depends on $n$, which we do note write explicitly for brevity.} in some joint state
$\sigma_m$. If for large $n$ the latter state approaches state
$\sigma^{\ot m}$, i.e.
\begin{equation}
\|\sigma_m - \sigma^{\ot m}\|_1\to 0
\end{equation}
and the ratio $m/n$ does not vanish, then
we say that $\rho$ can be transformed into $\sigma$ with rate $R=\lim
m/n$. The largest rate of transition we denote by $R(\rho\to
\sigma)$. In particular, distillation of entanglement described in
Sec.  \ref{sec:distil} is the rate of transition to \eprstate\ \be
\ed(\rho)=R(\rho \to \psiplus). \ee The cost of creating state out of
\eprstates\ is given by \begin{equation} \ec(\rho)=1/R(\psiplus\to
  \rho) \end{equation} and it is the other basic important measure
(see Sec.  \ref{subsec:edec} for description of those measures in more
detail).

\subsubsection{Unit of bipartite entanglement}
The fundamental result in asymptotic regime is that any
bipartite pure state can be transformed into  two-qubit
singlet with rate given by entropy of entanglement $S_A=S_B$, i.e.
entropy of subsystem (either $A$ or $B$, since for pure states they
are equal). And vice versa, to create any state from  two-qubit
singlet, one needs $S_A$ singlets pair pair two-qubit
state. Thus any pure bipartite state can be reversibly transformed
into any other state. As a result, in asymptotic limit, entanglement
of these states can be described by a single parameter --- von Neumann
entropy of subsystem. Many transitions that are not allowed in exact
regime, become possible in asymptotic limit. Thus the
irreversibility implied by Nielsen result is lifted in this regime,
and \eprstate\ becomes universal unit of entanglement.

\subsubsection{Bound entanglement  and irreversibility}
However, even in asymptotic limit one cannot get rid of
irreversibility for bipartite states, due to existence of bound
entangled states.
Namely, to create such a state from pure states by LOCC one needs
entangled states, while no pure entanglement can be drawn back from
it. Thus the bound entangled state can be viewed as a sort of
black holes of entanglement theory \cite{Terhal-PhysTod}. One can
also use thermodynamical analogy \cite{termo,thermo-ent2002}. Namely,
bound entangled state is like a single heat bath: to create the heat
bath, one needs to dissipate energy, but no energy useful to perform
mechanical work can be drawn in cyclic process (counterpart of the work is here
quantum communication via teleportation). Note in this context, that
the interrelations between entanglement and energy were considered also in different
contexts (see e.g. \cite{balance,OsborneN,HughZB-ener-ent}).

One might hope to regain reversibility as follows: perhaps for many
copies of bound entangled  state $\rho$, though some pure
entanglement is needed, a {\it sublinear} amount would be
enough to create them (i.e. the rate $E_C$ vanishes). In other
words, it might be that for bound entangled states $\ec$ vanishes.
This would mean, that asymptotically, the irreversibility is lifted.

More general question is whether distillable entanglement is equal to entanglement
cost for mixed states. Already in the original papers on entanglement distillation \cite{BBPSSW1996,BDSW1996}
there was indication, that generically we would have a gap
between those quantities\footnote{Although in their operational approach
the authors meant what we now call entanglement cost, to
quantify it they used non-regularized measure entanglement of formation.
},
even though for some trivial cases we can have $\ec=\ed$ for mixed states (\cite{termo}).
Continuing thermodynamical analogy, a generic mixed state would be like
a system of two heat baths of different temperature, from which part of energy but not the whole can be transferred into mechanical work.

It was a formidable task to determine, if we have irreversibility in asymptotic setting,
as it was related to fundamental, and still unsolved problem of whether entanglement cost is equal to  entanglement of formation (see Sec. \ref{sec:miary}).
The first example of states with asymptotic irreversibility was
provided in \cite{VidalC-irre}. Subsequently more and more
examples have been revealed. In \cite{VollbrechtWW2003-irrev}
mixtures of maximally entangled states were analyzed by use of
uncertainty principle. It turns out that irreversibility for this
class of states is generic: the reversible states happen to be those
which minimize uncertainty principle, and all they turn out to be so
called pseudo-pure states (see \cite{termo}), for which
reversibility holds for trivial reasons.  Example of such state is
\be
{1\over 2} |\psi^+_{AB}\>\<\psi^+_{AB}| \ot |0\>_{A'}\<0| +
{1\over 2}|\psi^-_{AB}\>\<\psi^-_{AB}| \ot |1\>_{A'}\<1|.
\ee

The states
$|0\>,|1\>$ are local orthogonal ``flags'', which allow to return to the
pure state $\psi_+$ on system $AB$. This shows that, within mixtures
of maximally entangled  states, irreversibility is a generic
phenomenon. Last but not least, it was shown that irreversibility is
exhibited by {\it all} bound entangled states
\cite{YangHHS2005-cost} (see Sec. \ref{sec:irrev}).

One could still try to regain reversibility in some form. Recall the
mentioned thermodynamical system with difference of temperatures.
Since only part of the energy can perform useful work, we have
irreversibility. However the irreversible process occurred in the
past, while creating the system by partial dissipation of energy.
However, once the system is already created, the work can be
reversibly drawn and put back to the system in Carnot cycle. In the
case of entanglement, it would mean that entanglement can be divided
into two parts: bound entanglement and free (pure entanglement)
which then can be reversibly mixed with each other. One can test
such hypotheses by checking first, whether having bound entanglement
for free, one can regain reversibility \cite{bound}. In
\cite{thermo-ent2002} strong evidence was provided, that even this
is not the case, while conversely in \cite{AudenaertPE2002-PPT} it
was shown, that to some extent such reversibility can be reached
(see also \cite{Ishizaka04} in this context)
\footnote{It should be noted however, that in \cite{AudenaertPE2002-PPT}
PPT operations have been used, which may be a stronger resource,
than assistance of PPT states.}.

\subsubsection{Asymptotic transition rates in multipartite states}
\label{subsubsec:ghz-epr}
In multipartite case there is no such a universal unit of entanglement as the singlet state.
Even for three particles, we can have three different types of \eprstates:
the one  shared by Alice and Bob, by Alice and Charlie
and Bob and Charlie. By LOCC it is impossible to
create any of them from the others.
Clearly, one cannot create at all $\phiplus_{AB}$ from $\phiplus_{BC}$
because the latter has zero entanglement across $A:BC$ cut\footnote{By $\phiplus_{AB}$
we mean $\phiplus_{AB}\ot |0\>_C$, similarly for $\phiplus_{AC}$ and $\phiplus_{BC}$.}.
From two states $\phiplus_{AB}$ and $\phiplus_{AC}$ one can create
$\phiplus_{BC}$ via entanglement swapping, i.e. by teleportation of half
of one of pairs through the other pair. However this is {\it
irreversible}: the obtained state $\phiplus_{BC}$ does not have
entanglement across $A:BC$ cut, so that it is impossible to create
state $\phiplus_{AB}$ and for similar reason also the state $\phiplus_{AC}$.

To see why even collective operations on many copies cannot make the
transformation $\phiplus_{AB}^{\ot n} + \phiplus_{AC}^{\ot m} \to
\phiplus_{AB}^{\ot k}$ reversible it is enough to examine entropies of
subsystems. For pure states they do not increase under LOCC, so they
have to be constant during reversible process. Thus we have to have
: $S_X^{in} = S_X^{out}$  for $X=A,B,C$ which gives \be A: \,\, n +
m =k; \quad B: \,\, n = k; \quad C: \,\, m =k. \quad \ee Thus we
arrived at contradiction (to be rigorous, we should apply here
asymptotic continuity of entropy, see Sec. \ref{sec:other-post}). In the
described case irreversibility clearly enters, however the reason
for this is not deep.

Via such considerations, one can see that in four-particle case not
only EPR states are ``independent'' units, but also GHZ state cannot
be created from them, hence it constitutes another independent unit.
For three particles it is less immediate, but still true that GHZ
cannot be created reversibly from EPR states \cite{LPSW1999}. This
shows that even in asymptotic setting there is truly tripartite
entanglement, a distinct quality that cannot be reduced to bipartite
entanglements.  This shows that there is true three-particle
entanglement in asymptotic limit.

Irreversibility between $|GHZ\>$  and $\phiplus$ (much less trivial than the one between \eprstates\
themselves) can be also seen as consequence of
the fact that all two-particle reduced density matrices of $|GHZ\>$ are separable,
while for \eprstates\ two are separable and one is entangled.
In bipartite states we cannot transform separable state into entangled one by LOCC
at all. Here, the transition is possible, because the third party can help,
however this requires some measurement which introduces irreversibility.

Irreversibility was further explored in \cite{AcinVC-mregs}
and irreversibility on the level of mixed bipartite states was used
to show that even GHZ plus EPR's do not constitute what is called
{\it minimal reversible entanglement generating set}
\cite{BPRST-mregs} i.e. a minimal set of states from which any other
state can be reversibly obtained  by LOCC. Such set is still unknown
even in the case of pure states of three qubits. Recently it was
shown \cite{IshizakaP-asym} that for three qubits the set cannot be
also constituted by \epr's and W state $W={1\over \sqrt3}(|001\>
+|010\>+|100\>)$. In \cite{VidalDC2000-epr-ghz} a nontrivial class
of states was shown, which are {\it reversibly} transformable into
\epr's and \ghz:
\be |\psi\>_{ABC}=c_0|000\>+ c_1 |1\> {1\over
\sqrt2}(|11\> +|22\>). \ee

The irreversibility between pure multipartite states seems to be a
bit different than that of mixed states in that it is less
``thermodynamical''. In the former case we have two ``good'' states,
which cannot be transformed reversibly into one another (both \eprstate\
and \ghzstate\ are useful for some tasks), while bound entanglement is
clearly ``worse'' than free entanglement.

\section{Entanglement and quantum communication \label{sec:communication}}

In classical communication theory, the most important notion is that
of correlations. To send a message, means in fact to correlate
sender and receiver. Also the famous Shannon formula for channel
capacity involves mutual information, a function describing
correlations. Thus the ability to faithfully transmit a bit is
equivalent to the ability to faithfully share maximally correlated
bits. It was early recognized that in quantum communication theory
it is entanglement which will play the role of correlations. In this
way entanglement is the cornerstone of quantum communication theory.

In classical communication theory, a central task is to send some
signals. For a fixed distribution of signals emitted by a source,
there is only {\it one} ensemble of messages.  In quantum case, a
source is represented by a density matrix $\rho$, and there are {\it
many} ensembles realizing the same density matrix. What does it then
mean to send quantum information? According to \cite{Teleportation}
it is the ability of transmitting and unknown quantum state. For a
fixed source, this would mean, that all possible ensembles are
properly transmitted. For example, consider a density matrix
$\rho={1\over d} \sum_i |i\>\<i|$. Suppose that a channel decoheres
its input  in basis $\{|i\>\}$. We see that the set of states
$\{|i\>\}$ goes through the channel without any
disturbance. However complementary set consisting of states
$U|i\>$ where $U$ is discrete Fourier transform, is completely
destroyed by a channel, because the channel destroys superpositions.
(For $d=2$, example of such complementary ensemble is $|+\>,|-\>$
where $|\pm\>={1\over 2} (|0\>\pm|1\>)$).
As a matter of fact, each member of the complementary ensemble is
turned into maximally mixed state.

How to recognize whether all ensembles can go through?  Schumacher
has noted \cite{Schumacher1995} that instead of checking all
ensembles we can check
whether an {\it entangled} state $\psi_{AB}$ is preserved, if we send half
of it (the system $B$) down the channel. The state should be chosen
to be {\it purification} of $\rho$, i.e.
$\tr_A(|\psi\>\<\psi|_{AB})=\rho$. Thus sending an unknown state is
equivalent to sending faithfully entanglement.

Indeed, in our example, the state can be chosen as
$|\Phiplus\rangle={1\over \sqrt{d}}\sum_i |i\>|i\>$. One can see
that after applying our channel to one subsystem, the state becomes
classical (incoherent) mixture of states $|i\>|i\>$. This shows that
the channel cannot convey quantum information at all. It is
reflection of a mathematical fact, that if we send half of
purification of a full rank density matrix down the channel, then
the resulting state will encode all the parameters of the channel.
This heuristic statement has its mathematical form in terms of
Choi-Jamio\l{}kowski isomorphism between states and channels. Its
most standard form links the channel $\Lambda$ with a state
$\varrho_{AB}^{\Lambda}$ having maximally mixed left subsystem
${\tr}_{B}(\varrho_{AB}^{\Lambda})={1\over d_A}\id$ as follows:
\begin{equation}
\varrho_{AB}^{\Lambda}=[\id_{A} \otimes \Lambda_{B}]|\Phi^+_{AA'}\> \<\Phi^+_{AA'}|,
\label{StatesChannelsIsomorphism}
\end{equation}
where the projector onto maximally entangled state $P^{+}_{AA'}$ is defined on a
product Hilbert space ${\cal H}_{A} \otimes {\cal H}_{A'}$, with
${\cal H}_{A}\simeq{\cal H}_{A'}$\footnote{We note that this isomorphism has operational meaning, in general, only one way: given the channel,
Alice and Bob can obtain bipartite state, but usually not vice versa.
However sometimes if Alice and Bob share  mixed bipartite state, then by use of classical communication they can simulate the channel. Example is
maximally entangled state, which allows to regain
the corresponding channel via teleportation. This was first pointed out
and extensively used in \cite{BDSW1996}.}.

\subsection{Capacity of quantum channel and entanglement}
The idea of quantum capacity $Q(\Lambda)$ of quantum channel
$\Lambda$ was introduced in the seminal work \cite{BDSW1996} which
contains milestone achievements connecting quantum entanglement and
quantum data transfer. The capacity $Q$ measures the largest rate of
quantum information sent asymptotically faithfully down the channel:
\be Q=\sup{\# \mbox{ transmitted faithfully qubits} \over \# \mbox{
uses of channel}}, \ee where the  fidelity of  the transmission is
measured by minimal subspace fidelity $f(\Lambda) = \min_{\psi}
\langle \psi|\Lambda(|\psi \rangle \langle \psi|)|\psi \rangle $
\cite{BennettDS97-cap}. Also {\it average fidelity transmission} can
be used which is direct analog of average fidelity in quantum
teleportation process:
$\overline{f}(\Lambda) = \int \langle \psi|\Lambda(|\psi
\rangle \langle \psi|)|\psi \rangle d \psi$
with uniform measure $d \psi$ on unit sphere.

Other approach based on idea described above was formalized in
terms of {\it entanglement transmission}. In particular, the
quality of transmission was quantified by {\it entanglement fidelity}
\begin{equation}
F(\Lambda,\Psi_{AB})=\langle \Psi_{AB}| [I_{A} \otimes
\Lambda_{B}]|\Psi_{AB}\rangle \langle \Psi_{AB}|)| \Psi_{AB} \rangle,
\end{equation}
with respect to a given state $\Psi_{AB}$\footnote{For typical
source all three types of fidelity --- average, minimum and entanglement one
are equivalent \cite{BarnumKN-cap-equiv}.}.
The alternative
definition of quantum capacity (which has been worked out in
\cite{Schumach96PRA,SchumacherN-1996-pra,BarnSchumachNiels98PRA} and
shown to be equivalent to the original one of \cite{BDSW1996,BennettDS97-cap}
in \cite{BarnumKN-cap-equiv}) was based on counting the optimal pure
entanglement transmission under the condition of high entanglement
fidelity defined above.

A variation of the entanglement fidelity
\cite{ReimpellWerner} is when input is equal to the output
$d_{A}=d_{B}=d$ and we send half of maximally entangled state
$\Phiplusd$  down the channel. It is measured by maximal entanglement
fidelity of the channel:
\begin{equation}
F(\Lambda):=F(\Lambda,\Phiplusd),
\end{equation}
which is equal to overlap of the state $\varrho^{\Lambda}$ with
maximally entangled state namely
$F(\Lambda)=F(\varrho^{\Lambda}):={\tr}(|\Phiplusd\>\<\Phi^+_d|\varrho^{\Lambda})$. It is
interesting that one has \cite{gentele,NielsenFidelity}:
\begin{equation}
\overline{f}(\Lambda)=\frac{dF(\Lambda) + 1}{d+1}.
\label{GeneralFidelity}
\end{equation}
This formula says that possibility of sending on average faithfully
quantum information happens if and only if  it is possible to create
maximal entanglement of $\Phiplusd$ with help of the channel. The above
relation is an important element in the proof that, quite
remarkably, definition of zero-way (or --- alternatively --- one-way
forward) version of quantum capacity $Q$ (see below) remains the
same if we apply {\it any} of the fidelities recalled above (for details
of the proof see the review \cite{Tema}).

Even more interesting, LHS
of the above equality can be interpreted as an average
teleportation fidelity of the channel that results from teleporting
given state through the mixed bipartite state $\varrho_{\Lambda}$.

An impressive connection between entanglement
and quantum channels theory  has been worked out already
in (\cite{BBPSSW1996,BDSW1996}) by use of teleportation.
The authors have shown how to achieve nonzero
transmission rate by combining three elements (a) creating many
copies of $\varrho^{\Lambda}$ by sending halves of \singlets\ down the channel
$\Lambda$ (b) distilling maximal entanglement
form many copies of the created state (c) teleporting quantum
information down the (distilled) maximal entanglement. Since the
last process corresponds to ideal transmission the rate of the
quantum information transmission is here equal to
distillation rate in step (ii). In this way one can prove the
inequality linking entanglement distillation $E_{D}$ with quantum
channel capacity $Q$ as \cite{BDSW1996}:
\begin{equation}
E_{D}(\varrho^{\Lambda}) \leq Q(\Lambda), \label{DQIneq}
\end{equation}
where $\varrho^\Lambda $  is given by (\ref{StatesChannelsIsomorphism}).
The inequality holds for one-way forward an two-way scenarios of
distillation (respectively --- coding). The above
inequality is one of the central links between quantum channels and
quantum entanglement theory (see discussion below). It is not known
whether there is lower bound like $cQ(\Lambda)\leq
E_{D}(\varrho^{\Lambda})$ for some constant $c$. However there is at
least qualitative equivalence shown in \cite{PHCEJP}:
\begin{equation} E_{D}(\varrho^{\Lambda})=0
\Rightarrow   Q(\Lambda)=0. \label{DQqualitequiv}
\end{equation}
An alternative simple proof which works also for multipartite
generalization of this problem can be found
in \cite{DuHoCi04}. It uses teleportation and Choi-Jamio\l{}kowski
isomorphism to collective channels \cite{CiDuKraLe01}.

\subsection{Fidelity of teleportation via mixed states}
Before entanglement distillation was discovered,
the fundamental idea of teleporting quantum states with help of
mixed states has been put forward in the pioneering paper \cite{Popescu1}.
The fidelity of transmission that
was used there was
\begin{equation}
f_{tel}(\varrho_{AB})=\int d \phi \langle \phi|\varrho^{\phi}_{B}
|\phi \rangle,
\end{equation}
with $\varrho^{\phi}_{B}$ being the output state of the
teleportation in Bobs hands, i.e. it is average fidelity mentioned
above. Popescu used this formula to show that teleportation through
some entangled 2-qubit Werner states that satisfy some local hidden
variable models \cite{Werner1989} beats the classical
``teleportation'' fidelity threshold $2/3$. Further the formula for
$f_{tel}$ was optimized in the case of pure
\cite{GisinTeleportation} and mixed \cite{HHH1996-teleportation}
2-qubit states $\rho_{AB}$, with Alice's measurement restricted to
maximally entangled projections. For pure state $\rho_{AB}$ general
optimization was performed in $d \otimes d$ case and it was shown
to be attained just on maximally entangled Alice's projections
\cite{BanaszekTeleportation}(see \cite{BowenBoseTeleportation} for
further analysis). All that teleportation schemes that we have considered so far were deterministic. The idea of probabilistic teleportation was introduced in \cite{MorTelepovm,BrassardHorodeckiMorTeleportation}. It was employed in effect of activation of bound entanglement (see Sec. \ref{subsec:activbound}).

Finally let us note that all possible
exact teleportation and dense coding schemes were provided in
\cite{werner-alltel}. There it was proved, that they are essentially
of the same kind.  In teleportation of a state of a $d$-dimensional system, Alice measures in a maximally entangled basis\footnote{ That is performs a measurement with Kraus operators being the projects onto bipartite states $|\psi_i\>\in \ccal^d\ot\ccal^d$ which have maximally mixed subsystems and form a basis of $\ccal^d\ot\ccal^d$} $\{\psi_i\}$. Subsequently Bob decodes by use of unitary transformations related to the basis via $U_i\ot \id \, |\Psi^+\> =|\psi_i\>$ (similarly as in dense coding).
Note that for two qubits all maximally entangled bases are related by local unitary
transformation \cite{MRH-PRA96}. Thus there is in a sense a unique protocol
of teleporting a qubit.  However in higher dimension it is no longer the case  \cite{WojcikGC2003-Bell-bases} which leads to inequivalent teleportation/dense coding protocols.

\subsection{Entanglement breaking and entanglement binding channels}
The formulas (\ref{DQIneq}) and (\ref{DQqualitequiv}) naturally provoke
the question: {\it which channels have capacity zero?}.

Now, if the state $\varrho^{\Lambda}$ is separable, then the
corresponding channel is called {\it entanglement breaking}
\cite{ShorHR-break} and one cannot create entanglement at all  by
means of such channel.\footnote{Simply, its Kraus operators (in
some decomposition) are of rank one. Impossibility of creating
entanglement follows from  the fact that action of any channel of
that kind can be simulated  by classical channel: for any input
state, one measures it via some POVM, and sends the classical
outcome to the receiver. Based on this, receiver prepares some
state.} Now since it is impossible to distill entanglement from
separable state we immediately see from (\ref{DQqualitequiv}) that
the capacity of entanglement breaking channel is zero i.e. no
quantum faithful quantum transmission is possible with entanglement
breaking channel. However the converse is not true: {\it possibility
to create entanglement with help of the channel is not equivalent to
quantum communication and it is bound entanglement phenomenon which
is responsible for that}. To see it let us observe that if  the
corresponding state $\varrho^{\Lambda}$ is bound entangled, then the
channel $\Lambda$, called in this case {\it binding
entanglement channel} (introduced in \cite{bechan,UPB2}), clearly
allows for creation of entanglement. However it cannot convey
quantum information at all --- it has capacity zero.  This was first
argued in \cite{bechan} and proof was completed in \cite{PHCEJP}
just {\it via} implication (\ref{DQqualitequiv})).

Note that binding entanglement channels are not completely useless
from general communication point of view, for they can be
used to generate secure cryptographic key, see Sec.
(\ref{sec:Ent_in_QKD}).

\subsection{Quantum Shannon theorem}
\label{subsec:quantum-shannon-th}
How many qubits can be sent
per use of a given channel? It turns out that the answer to the
question is on one hand analogous to classical Shannon formula, and
on the other hand entirely different. Capacity of classical channel
is given by maximum of mutual information over all bipartite
distributions, that can be obtained by use of channel.

Quantumly, even for sending qubits, as already mentioned above, we
may have different scenarios: 1) quantum channel without help of
classical channel (capacity denoted by $Q^{ \emptyset }$) 2) with
help of classical channel: one-way forward ($Q^{ \rightarrow }$) ,
backward ($Q^{ \leftarrow }$), and two-way ($Q^{ \leftrightarrow
}$). There is elegant answer in the case 1), which is called Quantum
Shannon Theorem. The first proof, partially heuristic, was provided
in \cite{Lloyd-cap}, subsequently it was improved in
\cite{Shor-cap-notes} and finally first fully rigorous result has been obtained
in \cite{Devetak2003}. Namely, to get asymptotic capacity
formula for $Q^{ \emptyset}$ one maximizes {\it coherent
information} over all bipartite states resulting from a pure state, half of it sent down the channel (see Sec.\ref{subsec:genhash}). However, one should optimize this quantity over many uses of channel, so that the formula for capacity reads
\be
Q^{ \emptyset }(\Lambda)= \lim{1\over n} \sup_\psi I^{coh}_{A\>B}((\id\ot
\Lambda^{\ot n})|\psi\>\<\psi|).
\ee
The fact that the formula is not ``single letter'', in the sense that it involves many uses of channel was established in \cite{ShorS1996-cohinf}, see also \cite{DSSVeryNoisy}.

Coherent information can be nonzero only for entangled states.
Indeed, only entangled states can have greater entropy of subsystem
than that of total system \cite{RPH1994}. In this context, an
additional interesting qualitative link between entanglement and
channel capacity formula is the hashing inequality stating
that $E_D^\rightarrow\geq I^{coh}_{A\>B}$ (see Sec. \ref{sec:distil}).
Quite remarkably sometimes $E_{D}^{\rightarrow}(\varrho_{AB})>0$ for
$I^{coh}_{A\>B}(\varrho_{AB})=0$  which is related to the above mentioned Shor-Smolin result.

Finally, there is a remarkable result stating that {\it forward
classical communication does not increase quantum capacity} \be
Q^{\emptyset}=Q^{\rightarrow}. \ee It was argued in \cite{BDSW1996}
and  the proof was completed in \cite{BarnumKN-cap-equiv}.
Therefore, any one-way distillation scheme, provides (through
relation with $Q^{\rightarrow}$) lower bound for capacity
$Q^{\emptyset}$.

\subsection{Bell diagonal states  and related channels}
An intriguing relation  between states and channels was found
\cite{BDSW1996} --- namely there is a class of channels $\Lambda$ for
which there is equality in (\ref{DQIneq}) because of the following
equivalence: not only given $\Lambda$ one can produce
$\varrho^{\Lambda}$ (which is easy --- one sends half of maximally
entangled state  $\Phiplus$ own the channel) but also given
$\varrho^{\Lambda}$ and possibility of additional forward
(i.e. from sender to receiver) classical communication  one can
simulate again the channel $\Lambda$. There is a fundamental
observation \cite{BDSW1996}(which we recall here for qubit case but
$d \otimes d$ generalization is straightforward, see \cite{gentele})
that if the state $\varrho^{\Lambda}$ is Bell diagonal than the
operation of teleporting any state $\sigma$ through mixed state
$\varrho^{\Lambda}$ produces just $\Lambda(\sigma)$ in Bobs hands.
Thus in this special case given state isomorphic to the channel plus
possibility of forward classical communication {\it one can simulate
action of the channel}. Thus for Bell diagonal states, we have an
{\it operational} isomorphism, rather than just mathematical one
(\ref{StatesChannelsIsomorphism}). In  \cite{BDSW1996} this was
exploited to show that there is equality in (\ref{DQIneq}) i.e.
$E_{D}^{\rightarrow}(\varrho^{\Lambda})=Q^{\rightarrow}(\Lambda)$,
and $E_D(\varrho^\Lambda)=Q^\leftrightarrow(\Lambda)$. The protocol attaining
capacity of the channel is in this case sending half of \singlet\
and perform the optimal entanglement distillation protocol.

To summarize the above description consider now the example where
this analysis finds application.

{\it Example .- } Consider the quantum binary Pauli channel
acting on single qubit as follows:
\begin{equation}
\Lambda_{p}(\rho)=p(\rho) + (1-p)\sigma_{z}(\rho)\sigma_{z},
\end{equation}
where $\sigma_{z}$ is a Pauli matrix.
Applying the isomorphism (\ref{StatesChannelsIsomorphism}) we get
rank-two Bell diagonal state
\begin{equation}
\varrho^{\Lambda_{p}}=p|\phiplus\rangle\langle\phiplus|
+(1-p)|\phiminus\rangle\langle\phiminus|,
\end{equation}
with $|\phi^{\pm}\>=\frac{1}{\sqrt{2}}(|0\rangle|0\rangle \pm
|1\rangle|1\rangle)$. The formula for distillable entanglement has
been found in this case \cite{Rains1999,Rains-erratum1999}
\footnote{That $1-H(p)$ is achievable
was shown in \cite{BDSW1996}.} namely
\begin{equation}
E_{D}^{\rightarrow}(\varrho^{\Lambda_{p}})=I^{coh}_{B\>A}(\varrho^{\Lambda_{p}})=1-H(p),
\label{Rains}
\end{equation}
 where $H(p)=-p \log p-(1-p)\log (1-p)$. (Actually, for this state $E_D=E_D^{\rightarrow}$).\footnote{Note that this is equal to capacity of classical
 binary symmetric channel with error probability $p$. However,
 the classical capacity of the present channel is maximal,
 because classical information can be sent without disturbance through phase.}

Since it is
Bell diagonal state we know that capacity $Q^{\rightarrow}$
can be achieved by protocol consisting of one-way distillation of
entanglement from $\varrho^{\Lambda_{p}}$ followed by teleportation
(see discussion above). Therefore
$Q^{ \rightarrow }(\Lambda_{p})=Q^{ \emptyset
}(\Lambda_{p})=I^{coh}_{B\>A}(\varrho^{\Lambda_{p}})=1-H(p)$.

\subsection{Other capacities of quantum channels}
\label{subsec:dense-cap}
One might ask weather maximizing quantum mutual information for a
given channel (in analogy to maximizing coherent information,
Sec. \ref{subsec:quantum-shannon-th}) makes sense. Is such a maximized
quantity linked to any capacity quantity?  Recall that quantum mutual
information is defined (in analogy to its classical version) as
follows:
\begin{equation}
I_{A:B}(\varrho_{AB})=S(\varrho_{A}) + S(\varrho_{B}) -
S(\varrho_{AB}). \label{MutualInfo}
\end{equation}
First of all one can prove that the corresponding (maximized) quantity would be nonzero even
for entanglement breaking channels
\cite{AdamiC1996-neumann-cap,BelavkinO1998-mutual}. Thus it cannot
describe quantum capacity (since, as we already know, they have
quantum capacity zero). Moreover, for a qubit channel it can be
greater than 1, for example for noiseless channel it is just 2. This
reminds dense coding, where two bits are send per sent qubit, with
additional use of shared \epr\ pair. And these two things are indeed
connected. It was established in
\cite{BennettSST1999-eacap,BennettSST-eacap2001} that
the maximum of quantum mutual information taken over the states
resulting from channel is exactly the classical capacity of a channel
supported by additional \epr\ pairs (it is so called entanglement
assisted capacity).  Moreover, it is equal to twice quantum capacity:
\be
C_{ass}=2 Q_{ass}=\sup I_{A:B}\bigl((\id_A\ot \Lambda_B)(|\psi\>_{AB}\<\psi|)\bigl)
\ee
where supremum is taken over all pure bipartite states.
Note that the expression for capacity  is ``single letter'', i.e. unlike in the unassisted quantum capacity case, it is enough to optimize over one use of channel.

A dual picture is, when one has noiseless
channel, supported by noisy entangled states. Again, the formula
involves coherent information (the positive coherent information
says that we have better capacity, than without support of
entanglement) \cite{Bowen-dense,HHH-IBM-dense}.
There are also interesting developments concerning multipartite dense coding
\cite{BrussALMSS2005-multidense,HorodeckiP2006-multidense}.

\subsection{Additivity questions}
There is one type of capacity, where we do not send half of
entangled state, but restrict just to separable states. This is
classical capacity of quantum channel $C(\Lambda)$, without any
further support. However even here entanglement comes in. Namely, it
is still not resolved, whether sending signals entangled between
distinct uses of channel can increase transmission rate. This
problem is equivalent to the following one: {\it can we decrease
production of entropy by an operation $\Lambda$, by applying
entangled inputs, to $\Lambda \ot \Lambda$?} i.e.
\be 2 \inf_\rho
S(\Lambda(\rho)) \stackrel{?}{=} \inf_\rho S(\Lambda\ot
\Lambda(\rho)). \ee
One easily finds that if this  equality does not hold
it may happen only via entangled states. Quite
interestingly, these problems have even further connection with
entanglement: they are equivalent to additivity of one of the most
important measures of entanglement --- so called entanglement of
formation, see \cite{ShorAdditivity,AudenaertB2004,Matsumoto2005,KoashiW-monogamy} and references therein. We have
the following equivalent statements
\bei
\item additivity of entanglement of formation
\item superadditivity of entanglement of formation
\item additivity of minimum output entropy
\item additivity of Holevo capacity
\item additivity of Henderson-Vedral classical correlation measure  $\chv$
\eei
(see Secs. \ref{sec:miary} and \ref{sec:irrev} for definitions of $\ef$ and $\chv$).
To prove or disprove them  accounts  for one of
the fundamental problems of quantum information theory.

Discussing the links between entanglement and communication one has
to  mention about one more additivity problem inspired by
entanglement behavior. This is the problem of additivity of quantum
capacities $Q$ touched already in \cite{BDSW1996}.
Due to nonadditivity phenomenon called {\it activation of bound
entanglement} (see Sec. \ref{subsec:activbound}) it
has been even conjectured \cite{activation} that for some channels
$Q(\Lambda_{1} \otimes \Lambda_{2})>0$ even if both channels have
vanishing capacities i.e. $Q(\Lambda_{1})=Q(\Lambda_{2})=0$.
Here again, as in the question of additivity of classical capacity
$C(\Lambda)$, a possible role of entanglement in inputs of the
channel $\Lambda_{1} \otimes \Lambda_{2}$ comes into play.

Though this problem is still  open, there is its multipartite
version where nonadditivity was found \cite{DuHoCi04}. In fact, the
multiparty communication scenario can be formulated and the analog
of (\ref{DQIneq}) and (\ref{DQqualitequiv}) can be proved
\cite{DuHoCi04} together with binding entanglement channels notion
and construction. Remarkably in this case the application of
multipartite BE channels isomorphic to multipartite bound entangled
states and application of multipartite activation effect leads to
{\it nonadditivity of two-way capacity regions} for quantum
broadcast (equivalently multiply access) channel \cite{DuHoCi04}:
tensor product of three binding channels $\Lambda_{i}$ which
automatically have zero capacity leads to the channel
$\Lambda:\otimes_{i=1}^{3}\Lambda_{i}$ with nonzero capacity. In
this case bound entanglement activation (see section on activation)
leads to the proof of {\it nonadditivity effect in quantum
information transmission}. Existence of the alternative effect for
one-way or zero-way capacity is an open problem.
Finally let us note in this context that the multipartite
generalization of the seminal result $Q^{ \emptyset
}=Q^{\rightarrow}$ has been proved  \cite{DemianowiczZero-Way} exploiting the notion  of entanglement transmission.

There is an interesting application of multipartite bound
entanglement in quantum communication. As we have already mentioned,
bipartite bound entanglement, which is much harder to use than
multipartite one, is able to help in quantum information transfer
{\it via} activation effect \cite{activation}. There is a very nice
application of multipartite bound entanglement in remote quantum
information concentration \cite{MuraoV1999-concentration}. This
works as follows: consider  the 3-qubit state $\psi_{ABC}(\phi)$
being an output of quantum cloning machine
that is shared by three parties Alice, Bob and
Charlie. The initial quantum information about cloned qubit $\phi$
 has been delocalized and they cannot concentrate it back
 in this distant lab scenario. If however each of them is given in addition one
particle of the $4$-particles in a Smolin state
$\varrho^{unlock}_{ABCD}$ (\ref{SmolinS}) with remaining fourth $D$
particle handed to another party (David) then a simple LOCC action of
the three parties can ``concentrate'' the state $\phi$ back remotely
at David's site.

\subsection{Miscellanea}
{\it Negative information.} The fact that sending qubits means
sending entanglement, can be used in even more complicated
situation. Namely, suppose there is a noiseless quantum channel from
Alice to Bob, and Bob already has some partial information, i.e. now
the source $\rho$ is initially distributed between Alice and Bob.
The question is: how many qubits Alice has to send to Bob, in order
he completes full information? First of all what does it mean to
complete full information? This means that for all ensembles of
$\rho$, after transmission, they should be in Bob's hands.
Instead of checking all ensembles, one can rephrase
the above ``completing information'' in elegant
way by means of entangled state: namely, we consider purification
$\psi_{ABR}$ of $\rho_{AB}$. The transmission of the quantum
information lacking by Bob, amounts then just to {\it merging} part $A$ of
the state to Bob, without disturbing it.

In \cite{SW-nature,sw-long} it was shown that if classical
communication is allowed between Alice and Bob then the amount of
qubits needed to perform merging is equal to conditional entropy
$S(A|B)$. This holds even when $S(A|B)$ is negative, in which case
it means, that not only Alice and Bob do not need to send qubits,
but they get extra \eprpairs, which can be used for some future
communication. What is interesting in the context of channels
capacity, this result proves the Quantum Shannon Theorem and hashing
inequality.  The new proof is very simple (the protocol we have recalled in Sec. \ref{subsec:genhash}).

{\it Mother and father protocols.}
Several different phenomena of theory of entanglement
and quantum communication have been grasped by a
common formalism discovered in \cite{DevetakHW-family,DevetakHW2005-resource}.
There are two main protocols  called {\it mother} and {\it father}. In the first one,
the noisy resource is {\it state}, while in the second --- {\it channel}.
From mother protocol one can obtain other protocols, such as dense coding via
mixed states, or one-way distillation of mixed states (children).
Similarly, father protocol generates protocols achieving quantum
channel capacity as well as entangled assisted capacity of quantum
channel. In both cases  from children one can obtain parents, by
using a new primitive discovered by Harrow called {\it cobit}.

The discovery of state merging have given a new twist.
In \cite{AbeyesingheDHW2006-fullSW} a different version of
state merging (called Fully Quantum Slepian Wolf  protocol), where only quantum communication and  sharing \eprpairs  is allowed (and counted) was provided.
It turns out that the minimal number of needed
qubits is equal to ${1\over 2} I(A:R)$, and the number
${1\over 2} I(A:B)$ of \eprpairs is gained by Alice and Bob.
The protocol is based on a beautiful observation,
that a random subsystem $A'$ of $A$ of size less than ${1\over 2}I(A:B)$
is always maximally mixed and product with reference $R$. This implies that
after sending the second part of $A$ to Bob, $A'$ is maximally entangled
with Bob's system. These ideas were further developed in
\cite{DevetakY2006-cond-info}, where operational meaning of
conditional mutual information was  discovered.
It turned out that the above protocol is a powerful primitive,
allowing to unify many ideas and protocols. In particular, this
shows deep {\it operational} relations between entangled
states and quantum channels.

\section{Quantifying entanglement}
\label{sec:miary}
\subsection{Distillable entanglement and entanglement cost}
\label{subsec:edec} The initial idea to quantify entanglement was
connected with its usefulness in terms of {\it communication}
\cite{BBPSSW1996,BDSW1996}. As one knows via a two qubit maximally
entangled state {\it \singlet\ }  one can teleport one qubit.
Teleportation is a process that involves the shared singlet, and
local manipulations together with communication of classical bits.
Since qubits cannot be transmitted by use of classical communication
itself, it is clear that power of sending qubits should be
attributed to entanglement. If a state is not maximally entangled,
then it does not allow for faithful teleportation. However, in
analogy to Shannon communication theory, it turns out that when
having many copies in such  state, one can obtain asymptotically
faithful teleportation at some rate (see Sec. \ref{sec:distil}).
To find how many qubits per copy we can teleport it is enough to
determine how many e-bits we can obtain per copy, since every
$|\phi^+\>$ can then be used for teleportation. In this way we arrive
at transition rates as described in Sec. \ref{sec:irrev}, and two
basic measures of entanglement $\ed$ and $\ec$.

{\it Distillable entanglement.} Alice and Bob start from $n$ copies
of state $\rho$, and apply an LOCC operation, that ends up with a
state $\sigma_n$. We now require that for large $n$ the final state
approaches the desired state $(\Phi^+_2)^{\ot m_n}$. If it is
impossible, then $E_D=0$.  Otherwise we say that the LOCC operations
constitute a {\it distillation protocol} $\pcal$ and the rate of
distillation is given by $R_\pcal=\lim_n {m_n\over n}$. The
distillable entanglement is the supremum of such rates over all
possible distillation protocols. It can be defined concisely  (cf.
\cite{PlenioVirmani2006-review}) as follows \be
E_D(\rho)=\sup\left\{ r: \lim_{n\to \infty} \left(\inf_\Lambda \|
\Lambda(\rho^{\ot n})- \Phi^+_{2^{rn}}\|_1\right)=0\right\}, \ee
where $\|\cdot\|_1$ is the trace norm (see \cite{Rains1998-def} for
showing that other possible definitions are equivalent).
\footnote{In place of trace norm one can use fidelity,
Uhlmann fidelity $F(\rho,\sigma)=(\tr \sqrt{\sqrt \rho \sigma\sqrt
\rho})^2$ thanks to inequality proven in \cite{Fuchs-Graaf} $
1-\sqrt{F(\rho,\sigma)} \leq \frac{1}{2} \|\rho-\sigma\|_1 \leq
\sqrt{1-F(\rho,\sigma)}.
  \label{eq:fuchs}$
}

{\it Entanglement cost.} It is a measure dual to $E_D$, and it
reports how many qubits we have to communicate in order to create a
state. This, again can be translated to e-bits, so that $E_C(\rho)$
is the number of e-bits one can obtain from $\rho$ per input copy by
LOCC operations. The definition is
\be \ec(\rho)=\inf\left\{ r:
\lim_{n\to \infty} \left(\inf_\Lambda \| \rho^{\ot n}-
\Lambda(\Phi^+_{2^{rn}})\|_1\right)=0\right\}.
\ee

In \cite{cost} it was shown, that $E_C$ is equal to regularized entanglement of formation
$\ef$ --- a prototype of entanglement cost, see Sec. \ref{subsec:roof}.
Thus, if $\ef$ is additive, then the two quantities are equal.
As we have mentioned, for pure states $\ed=\ec$.

{\it Distillable key.}
There is yet another distinguished operational measure
of entanglement for bipartite states, designed in similar spirit as $\ed$ and $\ec$.
It is distillable private key $K_D$: maximum rate of bits
of private key that Alice and Bob can obtain by LOCC from state $\rho_{AB}$
where it is assumed that the rest $E$ of the total pure state $\psi_{ABE}$
is given to adversary Eve.  The distillable key satisfies obviously
$E_D\leq K_D$: indeed, a possible protocol of distilling key
is to distill \eprstates\ and then from each pair obtain one bit of
key by measuring in standard basis. A crucial property of $K_D$
is that it is equal to the rate of transition to a special class of states:
so-called private states, which are generalization of \eprstates.
We elaborate more on distillable key and the structure of private states in Sec.
\ref{sec:Ent_in_QKD}.

\subsection{Entanglement measures --- axiomatic approach}
The measures such as $E_D$ or $E_C$ are built to describe entanglement
in terms of some tasks. Thus they  arise  from optimization of some protocols performed
on quantum  states.  However one can apply axiomatic point of view, by allowing
any function of state to be a measure, provided it satisfies some postulates.
Let us now go through basic postulates.

\subsubsection{Monotonicity axiom}
The most important postulate for entanglement measures was proposed already in \cite{BDSW1996}  still in the context of operationally defined measures.

\begin{itemize}
    \item {\bf Monotonicity under LOCC}: Entanglement
cannot increase under local operations and classical communication.
\end{itemize}

\cite{VPRK1997} introduced idea of axiomatic definition of
entanglement measures and proposed that an entanglement
measure is any function that satisfies the above condition
plus some other postulates. Then
\cite{Vidal-mon2000} proposed that monotonicity
under LOCC should be the only postulate necessarily required from
entanglement measures. Other postulates would then either follow from
this basic axiom, or should be treated as optional
(see \cite{popescu-rohrlich} in this context).

Namely, for any LOCC operation $\Lambda$ we have
\be
E(\Lambda(\rho))\leq E(\rho).
\label{eq:mono}
\ee
Note that the output state  $\Lambda(\rho)$ may include some registers with stored
results of measurements of Alice and Bob (or more parties, in multipartite setting),
performed in the course
of the LOCC operation $\Lambda$. The mathematical form of $\Lambda$
is in general quite ugly (see e.g. \cite{DonaldHR2001}). A nicer mathematical expression is known as  so called ``separable operations'' \cite{Rains2001,VPRK1997}
\be
\Lambda(\rho)=\sum_i A_i \ot B_i (\rho) A_i^\dagger \ot B_i^\dagger
\ee
with obvious generalization to multipartite setting\footnote{For more extensive treatment of various classes of operations, see Sec. \ref{subsec:locc}.}.
Every LOCC operation can be written in the above form, but not vice versa, as proved
in \cite{Bennett-nlwe}. (For more extensive treatment of various classes of
operations see Sec. \ref{subsec:locc}.)

The known entanglement measures usually satisfy a stronger condition,
namely, they do not increase on average
\be
\sum_i p_i E(\sigma_i) \leq E(\rho),
\label{eq:mono-strong}
\ee
where $\{p_i, \sigma_i\}$  is ensemble obtained from the state $\rho$
by means of LOCC operations. This condition was earlier considered as mandatory,
see e.g. \cite{Michal2001,Plenio2005-logneg}, but there is now common agreement
that the condition (\ref{eq:mono}) should be considered as the only necessary requirement
\footnote{Indeed, the condition (\ref{eq:mono}) is more fundamental, as
it tells about entanglement of {\it state}, while (\ref{eq:mono-strong})
says about average entanglement of an ensemble (family $\{p_i,\rho_i\}$), which is
less operational notion than notion of state.  Indeed it is not quite clear
what does it mean to ``have an ensemble''. Ensemble can always be treated
as a state $\sum_i p_i |i\>\<i|\ot \rho_i$, where $|i\>$ are local orthogonal flags.
However it is not  clear at all  why one should require a priori that
$E(\sum_i p_i |i\>\<i|\ot \rho_i)=\sum_ip_i E(\rho_i)$.}.
However, it is often easier to prove the stronger condition.

Interestingly, for bipartite measures monotonicity also implies
that there is {\it maximal entanglement} in bipartite systems.
More precisely, if we fix Hilbert space $C^d\ot C^{d'}$,
then there exist states from which any other state can be created:
these are states $U_A\ot U_B$ equivalent to singlet. Indeed, by teleportation, Alice
and Bob  can create from singlet any pure bipartite state. Namely, Alice prepares
locally two systems in a joint state $\psi$, and one of systems teleports through singlet.
In this way Alice and Bob share the state $\psi$. Then they can  also prepare
any mixed state. Thus $E$ must take  the greatest value to the state $\phiplus$.

Sometimes, one considers monotonicity under LOCC operations
for which the output system has the same local dimension as the input system.
For example, for n-qubit states, we are interested only in output
n-qubit states. Measures that satisfy such monotonicity
can be useful in many contexts, and sometimes it is easier to
prove such monotonicity \cite{VerstraeteDM2001-normal},
(see section \ref{subsubsec:multi-pure}).

\subsubsection{Vanishing on separable states.}
If a function  $E$ satisfies the monotonicity axiom, it turns out that
it is constant on separable states. It follows from  the fact that
every separable state can be converted to any other separable  state by LOCC
\cite{Vidal-mon2000}.
Even more, $E$ must be {\it minimal} on separable states, because any separable
state can be obtained by LOCC from any other state. It is reasonable to set
this constant to zero. In this way we arrive at even more  basic axiom,
which can be formulated already on qualitative level:

\begin{itemize}
    \item  Entanglement vanishes on separable states
\end{itemize}

It is quite interesting, that the LOCC monotonicity axiom
almost imposes the latter axiom. Note also that those two axioms impose
$E$ to be a nonnegative function.

\subsubsection{Other  possible postulates.}
\label{sec:other-post}
The above two axioms are essentially the only
ones that should be necessarily required from entanglement measures. However there are other
properties, that may be useful, and are natural in some context.
{\it Normalization.}
First of all
we can require that entanglement  measure behaves in an ``information theoretic
way'' on maximally entangled states, i.e. it counts e-bits:
\be
E((\Phi^+_2)^{\ot n})= n.
\ee
A slightly stronger condition would be $E(\Phiplusd)=\log d$.
For multipartite entanglement, there is no such a natural condition,
due to nonexistence of maximally entangled state.

{\it Asymptotic continuity.}
Second, one can require some type of {\it continuity}. The asymptotic manipulations
paradigm suggest continuity of the form \cite{Vidal-mon2000,limits,DonaldHR2001}:
\be
\|\rho_n-\sigma_n\|_1\to 0\quad \Rightarrow \quad  {|E(\rho_n)-E(\sigma_n)|\over \log d_n}\to 0,
\ee
for states $\sigma_n$, $\rho_n$ acting on Hilbert space $\hcal_n$ of dimension $d_n$.
This is called {\it asymptotic continuity}. Measures which satisfy this postulate,
are very useful in estimating $E_D$, and other transition rates, via inequality
(\ref{eq:rates}) (Sec. \ref{subsec:rates}). The most prominent
example of importance of asymptotic continuity is that together with the above normalization and additivity is enough to obtain a {\it unique} measure
of entanglement for pure states (see Sec. \ref{subsubsec:extremal}).

{\it Convexity.}
Finally, entanglement measures are often convex. Convexity used to
be considered as a mandatory ingredient of the mathematical
formulation of monotonicity. At present we consider convexity as
merely a convenient mathematical property.
Most of known measures are convex, including relative entropy of
entanglement, entanglement of formation, robustness of entanglement,
negativity and  all measures constructed by means of convex
roof (see Sec. \ref{sec:measures-survey}). It is open question whether
distillable entanglement is convex \cite{ShorST2001}.
In multipartite setting it is known, that a version of distillable entanglement\footnote{E.g. maximal amount of EPR pairs between two chosen parties,
that can be distilled with help of all parties.} is not convex \cite{ShorST-superactiv}.

\subsubsection{Monotonicity for pure states.}
\label{subsubsec:monot_pur}
For many applications, it is important to know whether a given function is a good measure
of entanglement just  on pure states, i.e. if the initial state is pure,
and the final state (after applying LOCC operation) is pure.

Since by Nielsen theorem (sec. \ref{subsec:puretrans}) for any pair
of states $\phi$ and $\psi$,
one can transform $\phi$ into $\psi$ iff  $\phi$ is majorized by $\phi$,
one finds, that a function $E$ is monotone on pure states
if and only if it is Schur-concave as a function of spectrum
of subsystem. I.e. whenever $x$ majorizes $y$ we have $E(x)\leq E(y)$.

Again, it is often convenient to consider a stronger monotonicity
condition (\ref{eq:mono-strong}) involving  transitions from pure
states to ensemble of pure states. Namely, it is then enough to
require, that any {\it local} measurement will not increase
entanglement {\it on average} (cf. \cite{LPSW1999}). We obtain condition
\be
E(\psi) \geq \sum_i p_i E(\psi_i), \label{eq:mono-pure}
\ee
where
$\psi_i$ are obtained from  local operation \be \psi_i= {V_i \psi
\over \|V_i \psi\|}. \ee $V_i=A_i\ot \id $ (or $V_i = \id\ot B_i$) are
Kraus operators of local  measurement (satisfying $\sum_i
V_i^\dagger V_i=\id$) and $p_{i}$ are probabilities of
outcomes. The inequality should be satisfied for both Alice
and Bob measurements. Recall, that the notion of ``measurement''
includes also unitary operations, which are measurements with single
outcomes. The condition obviously generalizes to multipartite
states.

\subsubsection{Monotonicity for convex functions}
\label{mono-convex}

For convex entanglement measures, the strong monotonicity under LOCC
of Eq. (\ref{eq:mono-strong}) has been set in a simple form by Vidal
\cite{Vidal-mon2000}, so that the main difficulty --- lack of concise
mathematical description of LOCC operations --- has been overcome.
Namely, a convex function $f$ is LOCC monotone in the strong sense
(\ref{eq:mono-strong}) if and only if it does not increase under
\bei
\item[a)] adding local ancilla
\be
f(\rho_{AB}\ot \sigma_X)\leq f(\rho_{AB}),\quad X=A',B',
\ee
\item[b)] local partial trace
\be
f(\rho_{AB})\leq f(\rho_{ABX}),
\ee
\item[c)] local unitary transformations
\item[d)] local von Neumann measurements (not necessarily complete),
\be
f(\rho_{AB})\geq \sum_i p_i f(\sigma_{AB}^i),
\ee
where $\sigma_{AB}^i$ is state after obtaining outcome $i$,
and $p_i$ is probability of such outcome
\eei
Thus it is enough to check monotonicity under uni-local operations
(operations that are performed only on  single  site, as it was in the case of pure states).
For a convex function, all those conditions are equivalent
to a single one, so that we obtain a compact condition:

{\it For convex functions, monotonicity
(\ref{eq:mono-strong}) is equivalent to the following condition \be
\sum_ip_i E(\sigma_i) \leq E(\rho), \label{eq:mono-unilocal} \ee
where the inequality holds for $\sigma_i=\frac{1}{p_i}W_A^i \ot I_B
\rho W_A^i{}^\dagger\ot I_B$, and $p_i={\tr}(W_A^i \ot I_B \rho
W_A^i{}^\dagger\ot I_B)$, with $\sum_i W_A^i{}^\dagger W_A^i=I_A$,
and the same for $B$ (with obvious generalization to many parties).
}

Convexity allows for yet another, very simple formulation of monotonicity.
Namely, for convex functions strong monotonicity can be phrased in
terms of two {\it equalities} \cite{MH2004-mono}. Namely, a nonnegative
convex function $E$ is LOCC monotone (in the sense of inequality (\ref{eq:mono-strong})),
if and only if
\bei
\item{}[{\tt LUI}] $E$ is invariant under local unitary transformations
\item{}[{\tt FLAGS}] for any chosen party $X$,  $E$ satisfies equality
\be
E\bigl(\sum_ip_i \rho^i \ot |i\>_X\<i| \bigr)=\sum_ip_i E(\rho_i),
\ee
where $|i\>_X$ are local orthogonal flags.
\eei

\subsubsection{Invariance under local unitary transformations}
Measure which satisfies monotonicity condition,
is invariant under local unitary transformations
\be
E(\rho)=E(U_1\ot \ldots U_N \rho U_1^\dagger\ot \ldots U_N^\dagger).
\ee
Indeed these operations are particular cases of LOCC operations,
and are reversible. Thus monotonicity requires that $E$ does not change  under
those operations. This condition is usually first checked for candidate for
entanglement measures (especially if it is difficult to prove
monotonicity condition).

\subsection{Axiomatic measures --- a survey}
\label{sec:measures-survey}
Here we will review bipartite entanglement measures built on axiomatic basis.
Some of them immediately generalize to multipartite case. Multipartite
entanglement measures we will present in Sec. \ref{subsec:multipartite}

\subsubsection{Entanglement measures based on distance}
\label{subsec-distance}

A class of entanglement measures \cite{VPRK1997,PlenioVedral1998} are based on the natural
intuition, that the closer the
state is to the set of separable states, the less entangled it is. The measure is
minimum distance $\dist$\footnote{We do not require the distance to  be a metric.} between the given state and the states in $\scal$:
\be
E_{\dist, \scal}(\varrho)=\inf_{\sigma\in \scal}\dist(\varrho,\sigma).
\label{eq-distance-measure}
\ee
The set $\scal$ is chosen to be closed under LOCC operations.
Originally it was just the set of separable states $S$.
It turns out that such function is monotonous under LOCC, if
distance measure is monotonous under {\it all} operations. It
is then possible to use known, but so far unrelated, results from literature
on monotonicity under completely positive maps. Moreover, it proves that it is
not only a technical assumption to generate entanglement measures:
monotonicity is a condition for a distance to be  a measure
of {\it distinguishability} of quantum states \cite{Fuchs-Graaf,VedralPJK1997-stat}.

We thus require that \be \dist(\rho,\sigma)\geq
\dist(\Lambda(\rho),\Lambda(\sigma)) \label{eq-distance-monotone}
\ee and obviously $\dist(\rho,\sigma)=0$ for $\rho=\sigma$. This
implies nonnegativity of $\dist$ (similarly as it was in the case of
vanishing of entanglement on separable states). More
importantly, the above condition immediately implies monotonicity
(\ref{eq:mono}) of the measure $E_{\dist,\scal}$. To obtain stronger
monotonicity, one requires $\sum_ip_i\dist(\varrho_i,\sigma_i)\leq
\dist(\varrho,\sigma)$. for ensembles $\{p_i,\varrho_i\}$ and
$\{q_i,\sigma_i\}$ obtained from $\rho$ and $\sigma$ by applying an
operation.

Once good distance was chosen, one can consider different measures by
changing the sets closed under LOCC operations. In this way we
obtain $E_{\dist,PPT}$ \cite{Rains2001} or $E_{\dist,ND}$ (the distance from nondistillable states). The measure
involving set $PPT$ is much easier to evaluate. The greater the set (see fig. \ref{fig:states-and-witnesses}), the smaller the measure is, so that if we consider the set of separable states, those with positive partial transpose and the set of nondistillable states, we have
\be
E_{\dist,ND}\leq E_{\dist,PPT}\leq  E_{\dist,S}.
\ee
In Ref. \cite{PlenioVedral1998} two distances were shown to satisfy
(\ref{eq-distance-monotone}) and convexity: square of Bures metric $B^2=2-2\sqrt
{F(\varrho,\sigma)}$ where $F(\varrho,\sigma)=[\tr(\sqrt\varrho
\sigma \sqrt\varrho)^{1/2}]^2$ is fidelity
\cite{Uhlmann-fidelity,Jozsa-fidelity} and relative entropy
$S(\varrho|\sigma)=\tr\varrho(\log\varrho-\log\sigma)$.
Originally,
the set of separable states was used and the resulting measure
\be
E_R=\inf_{\sigma\in \rm SEP} \tr\varrho(\log\varrho-\log\sigma)
\label{eq:relent}
\ee
is called {\it relative entropy of entanglement}.
It  is one of the fundamental entanglement measures, as the relative
entropy is one of the most important functions in quantum
information theory (see \cite{Vedral2002-rmp,SchumacherW2000-relent}).
Its other versions -  the relative entropy distance from PPT states  \cite{Rains2001}
and from nondistillable states \cite{Vedral-distbound} -
will be denoted as $E_{R}^{PPT}$, and $E_R^{ND}$ respectively.
Relative  entropy of entanglement (all its versions)
turned out to be powerful upper bound for entanglement of
distillation \cite{Rains2001}. The distance based
on fidelity received interpretation in terms of Grover algorithm
\cite{Biham2005-ent-Grov}.

\subsubsection{Convex roof measures}
\label{subsec:roof}

Here we consider the following method of obtaining entanglement
measures: one starts by imposing a measure $E$ on pure states, and
then extends it to mixed ones by {\it convex roof}
\cite{Uhlmann-roof} \be E(\varrho)=\inf \sum_ip_iE(\psi_i),\quad
\sum_ip_i=1 ,  p_i\geq0, \ee where the infimum is taken over all
ensembles $\{p_i,\psi_i\}$ for which
$\varrho=\sum_ip_i|\psi_i\ra\la\psi_i|$. The infimum is reached on a
particular ensemble \cite{Uhlmann-roof}. Such ensemble we call {\it
optimal}. Thus $E$ is equal to average under {\it optimal ensemble}.

The first entanglement measure built in this way  was {\it entanglement of formation} $\ef$
introduced in \cite{BDSW1996}, where $E(\psi)$  is von Neumann entropy of the reduced density
matrix of $\psi$. It constituted first upper bound for distillable
entanglement. In Ref. \cite{BDSW1996} monotonicity of $\ef$ was shown. In Ref.
\cite{Vidal-mon2000} general proof for monotonicity of all possible
convex-roof measures was exhibited. We will recall here the latter  proof,
in the form  \cite{Michal2001}.

One easily checks that $E$ is convex. Actually, convex roof measures
are the largest functions that are (i) convex (ii) compatible with
given values for pure states. Now, for convex functions, there is
the very simple condition of Eq. (\ref{eq:mono-unilocal}) equivalent
to strong monotonicity. Namely, it is enough to check,
whether the measure does not increase on average under local
measurement (without coarse graining i.e. where outcomes are given
by Kraus operators).

Using this condition, we will show, that if a measure is monotone on
pure states (according to Eq. (\ref{eq:mono-pure})), then its convex
roof extension  is monotonous on mixed states. Thus the condition
(\ref{eq:mono-strong}) is reduced to monotonicity  for pure states.
To see it, consider $\varrho$ with optimal ensemble
$\{p_i,\psi_i\}$. Consider local measurement with Kraus operators
$V_k$.  It transforms initial state $\varrho$ as follows \be
\varrho\to \{q_k,\sigma_k\},\quad q_k=\tr V_k\varrho
V^\dagger_k,\quad \sigma_k={1\over q_k} V_k\varrho V^\dagger_k.
\ee
The members of the ensemble $\{p_i,\psi_i\}$ transform into
ensembles of pure states (because operation is pure) \be
\psi_i\to\{q_k^i,\psi_k^i\},\quad q_k^i=\tr (V_k|\psi_i\ra\la\psi_i|
V^\dagger_k),\quad \psi_k^i={1\over \sqrt{q_k^i}} V_k\psi_i. \ee One
finds that $\sigma_k={1\over q_k}\sum_ip_i {q_k^i} |\psi_k^i\ra \la
\psi_k^i|$.

Now we want to show that the initial entanglement $E(\varrho)$ is no
less than the final average entanglement $\Ebar=\sum_k q_k
E(\sigma_k)$, assuming, that for pure states $E$ is monotonous under
the operation. Since $\sigma_k$ is a mixture of $\psi_k^i$'s, then
due to convexity of $E$ we have \be E(\sigma_k)\leq {1\over
q_k}\sum_ip_iq_k^iE(\psi_k^i). \ee Thus $\Ebar\leq \sum_ip_i
\sum_kq_k^i E(\psi_i^k)$. Due to monotonicity on pure states
$\sum_kq_k^i E(\psi_k^i)\leq E(\psi_i)$. Thus $\Ebar\leq
\sum_ip_iE(\psi_i)$. However, the ensemble $\{p_i,\psi_i\}$ was
optimal, so that the latter term is equal simply to $E(\varrho)$.
This ends the proof.

Thus, for any function (strongly) monotonous for pure states, its
convex roof is monotonous for {\it all} states. As a result, the
problem of monotonicity of convex roof measures reduces to pure
states case. Let us emphasize that the above proof bases solely on
the convex-roof construction, hence is by no means restricted to
bipartite systems. In Sec. \ref{subsec:all-pure} we will discuss the
question of monotonicity  for pure states. In particular we will see
that any concave, expansible function of reduced density matrix of
$\psi$ satisfies (\ref{eq:mono-pure}). For example
\cite{Vidal-mon2000} one can take $E_\alpha(\psi)$ given by Renyi
entropy ${1\over 1-\alpha}\log_2\tr (\varrho^\alpha)$ of the
reduction for $0\leq \alpha \leq \infty$. For $\alpha=1$ it  gives
$\ef$, while for $\alpha=0$ --- the average logarithm of the
number of nonzero Schmidt coefficients. Finally, for
$\alpha=\infty$ we obtain a measure related to the one introduced by
Shimony \cite{Shimony1995-miara} when theory of entanglement did not
exist: \begin{equation} E(\psi)= 1 - \sup_{\psi_{prod}}|\<\psi|\psi_{prod}\>|, \end{equation}
where the supremum is taken over all product pure states
$\psi_{prod}$ (see Sec. \ref{subsubsec:multi-pure} for multipartite
generalizations of this measure).

We will consider measures for pure bipartite states
in more detail  in Sec. \ref{subsec:all-pure}, and multipartite
in Sec. \ref{subsubsec:multi-pure}.

\paragraph{Schmidt number}
The Schmidt rank can be extended to mixed states by means of convex
roof. A different extension was considered in
\cite{Terhal-Pawel-rank,SchmidtWitn} (called Schmidt number) as follows
\be
r_{S}(\varrho)= \min(\max_{i}[r_{S}(\psi_{i})]),
\ee
where minimum is
taken over all decompositions $\varrho=\sum_{i}p_{i}|\psi_{i}\rangle
\langle \psi_{i}|$ and $r_{S}(\psi_{i})$ are the Schmidt ranks of
the corresponding pure states. Thus instead of average Schmidt rank,
supremum is taken. An interesting feature of this measure is that
its logarithm is strongly nonadditive. Namely there exists a state
$\rho$ such that $r_S(\rho)=r_S(\rho\ot \rho)$.

\paragraph{Concurrence}
\label{par:concurrence}

For two qubits the measure called {\it concurrence} was introduced
for pure states in \cite{Hill-Wootters}. Wootters
\cite{Wootters-conc} provided a closed expression for its convex
roof extension and basing on it derived computable formula for $\ef$
in two-qubit case. For pure states concurrence is given by
$C=\sqrt{2(1- \tr \rho^2)}$ where $\rho$ is reduced state. For two
qubits this gives $C(\psi)=2a_1a_2$ where $a_1,a_2$ are Schmidt
coefficients. Another way of representing C for two qubits is the
following \begin{equation} C=\<\psi|\theta|\psi\>, \end{equation} where $\theta$ is
antiunitary transformation $\theta \psi=\sigma_y\ot \sigma_y
\psi^*$, with ${}^*$ being complex conjugation in standard basis,
and $\sigma_y$ is Pauli matrix. It turns out that the latter
expression for $C$ is the most useful in the context of mixed
states, for which the convex roof of $C$ can be then computed as
follows. Let us denote $\tilde\rho=\theta \rho \theta$, and consider
operator \begin{equation} \omega=\sqrt \rho \sqrt {\tilde \rho}. \end{equation} Let
$\lambda_1,\ldots,\lambda_4$ be singular values of $\omega$ in
decreasing order. Then we have \be
C(\rho)=\max\{0,\lambda_1-\lambda_2-\lambda_3-\lambda_4\}
\label{eq:concurrence} \ee Interestingly Uhlmann has shown
\cite{Uhlmann2000-conj} that for any conjugation $\Theta$, i.e.
antiunitary operator satisfying $\Theta=\Theta^{-1}$, the convex
roof of the function $\Theta$-concurrence
$C_\Theta(\psi)=\<\psi|\Theta|\psi\>$ is given by generalization of
Wootters'  formula: \be
C_\Theta(\rho)=\max\{0,\lambda_1-\sum_{i=2}^d \lambda_i\}, \ee where
$\lambda_i$ are  eigenvalues  of operator $\sqrt {\rho} \sqrt{\Theta
\rho \Theta}$ in decreasing order.

The importance of the measure stems from the fact that it allows to
compute entanglement of formation for two qubits according to
formula \cite{Wootters-conc}
\begin{equation} \ef(\rho)= H({1+\sqrt{1-C^2(\rho)}\over 2}), \end{equation} where $H$ is binary entropy $H(x)=-x
\log x -(1-x)\log (1-x)$. Another advantage of concurrence is that
it is very simple for pure states. Namely  $C^2$ is a polynomial
function of coefficients of a state written in standard basis \be
C(\psi)=2 |a_{00} a_{11} - a_{01} a_{10}|, \ee for $\psi=a_{00} |00\>
+ a_{01}|01\> + a_{10}|10\>+a_{11}|11\>$. One can extend this
quantity to higher dimensions
\cite{AudenaertVM2000-concurrence,RungtaBCHM2001-concurrence}
(see also \cite{BadziagDHHH01-conc})
by building {\it concurrence vector}. The elements $C\alpha$ of the vector are
all minors of rank two:
\be
C_\alpha(\psi)=2(A^\psi_{ij} A^\psi_{i'j'}-A^\psi_{ij'} A^\psi_{i'j})
\ee
where $\psi=\sum_{ij} A^\psi_{ij} |ij\>$, $\alpha=(ii',jj')$ with $i<i',j<j'$.
So in the case of two qubits, the matrix
reduces to a number, as there is only one such minor --- the
determinant. For $d_1 \ot d_2$ system, there are ${d_1 \choose
2}\times {d_2 \choose 2}$ minors. The norm of such concurrence
vector gives  rise to concurrence in higher dimension
\be
C(\psi)=\sqrt{\sum_\alpha
C_\alpha(\psi)^2}=\sqrt{\<\psi|\psi\>-\tr \rho^2},
\ee
where $\rho$ is density matrix of subsystem.
Strong algebraic lower bound  for $C$ was obtained in  \cite{MintertKB04-conc}
allowing to detect states which are bound entangled (see also \cite{MintertCKB2005-review}). Namely,
consider ensemble $\rho=\sum_m|\psi_m\>\<\psi_m|$
\be
C(\rho)=\inf_{\psi_m}\sum_m C(\psi_m)
\ee
One finds that
\be
C(\rho)=\inf \sum_m \sqrt{(U\ot U B U^\dagger \ot U^\dagger)_{mm,mm}}
\ee
where $B$ is biconcurrence matrix \ref{eq:bicon-latwe}.
This matrix can be written as follows $B_{m\mu,n\nu}=\<\psi^m_{AB}|\<\psi^\mu_{A'B'}| P^{(-)}_{AA'}\ot P^{(-)}_{BB'} |\psi^n_{AB}\>|\psi^\nu_{A'B'}\>$, from which
one can show that it is positive, hence it can be written in
eigendecomposition as $B=\sum_\alpha |\chi_\alpha\>\<\chi_\alpha|$,
with unnormalized, orthogonal $\chi_\alpha$.
It  was shown that
\be
C(\rho)\geq \max\{\lambda_1 - \sum_{\lambda_i>1} \lambda_i,0\}
\ee
where $\lambda_i$ are singular values of operator $\sum_\alpha z_\alpha
A^{\chi_\alpha}$  put in decreasing order, with $z_\alpha$
being arbitrary chosen complex numbers satisfying $\sum_\alpha |z_\alpha|^2$
and $\chi_\alpha=\sum_{ij} A^{\chi_\alpha}_{ij}|ij\>$. A very good computable
bound  is obtained already if the linear combination
$\sum_\alpha z_\alpha \chi_\alpha$  is reduced to a chosen single $\chi_\alpha$.
In the case of two qubits $B$ is of rank one, hence we obtain
the Wootters result.

There are other interesting measures introduced in
\cite{SinoleckaZK2001-manifold,FanMI2002-concurrence} and developed
in \cite{Gour-mon2004}, which are built by means of polynomials of
the Schmidt coefficients $\lambda$'s:
\ben
\tau_1&=& \sum_{i=1}^d \lambda_i=1, \nonumber \\
\tau_2&=& \sum_{i>j}^d \lambda_i\lambda_j, \nonumber \\
\tau_3&=& \sum_{i>j>k}^d \lambda_i\lambda_j \lambda_k, \nonumber \\
&\vdots& \nonumber \\
\label{eq:tau-miary}
\een
The above measures $\tau_p$ are well defined if the dimension
of the Hilbert space $d$ is no smaller than the degree $p$.
For convenience, one can also set $\tau_p=0$ for $p<d$,
so that each of the above measures is well defined for
all pure states. The functions are generalizations of
concurrence and can be thought as higher level concurrences.
In particular $\tau_2$ is square of concurrence.
The measures are Schur-concave (i.e. they preserve
majorization order), so that by Nielsen theorem  (see sec. \ref{subsec:puretrans})
they satisfy monotonicity.

Due to  simplicity of concurrence measure, an interesting
quantitative connection has been found with complementarity between
visibility and which-path information in interference experiments
\cite{JakobBergou2003-compl}. Interesting generalizations of
concurrence were found in multi-partite case (see Sec.
\ref{subsec:multipartite})

\subsubsection{Mixed convex roof measures}
One can consider some variation of convex roof method by allowing
decompositions of a state into arbitrary states rather than just
pure ones
\cite{multisquash}.
This allows to produce entanglement measures
from other functions that are not entanglement measures themselves.
Namely, consider function $f$ which does not increase on average
under local measurement (recall, that we mean here generalized measurement,
so that it includes unitary transformations).
I.e. $f$ satisfies the condition (\ref{eq:mono-unilocal}),
so that if only it were convex, it would be monotone.
However, what if the function is not convex?
Example of such quantity is quantum mutual information.
It is obviously not a monotone, because e.g. it takes different values
on separable states.

From such function we could obtain entanglement measure
by taking usual convex roof. However, we can also take {\it mixed
convex roof} defined by \begin{equation} E(\rho)=\inf \sum_ip_i f(\rho_i), \end{equation}
where infimum is taken over all decompositions $\rho=\sum_ip_i
\rho_i$. The new function is already convex, and keeps the feature of
nonincreasing on average under local measurements, so that it is
LOCC monotone. If we start with mutual information, we obtain
measure  interpolating between entanglement of formation and squashed entanglement
introduced in \cite{Tucci2002-squashed}
and later independently in \cite {NagelR2003-ent-mutual}. Its monotonicity under
LOCC was proved later \cite{multisquash}.

Let us mention, that if initial function $f$ is asymptotically
continuous, then both usual convex roof, as well as mixed convex
roof are asymptotically continuous  too (it was shown in
\cite{Synak05-asym}, generalizing the result of Nielsen
\cite{Nielsen-cont} for $\ef$).

\subsubsection{Other entanglement measures}
\label{subsubsec:other-measures}

\paragraph{Maximal teleportation fidelity/maximal singlet fraction.}
For a state $\rho$ one can consider fidelity of teleportation of a
qudit averaged uniformly over inputs and maximized over trace preserving LOCC operations.
Denote it by $f_{max}$. It is related to maximal fidelity $F_{max}$ with $d\ot d$
maximally entangled state $|\Phiplus\>=\sum_i{1\over \sqrt{d}}|ii\>$ optimized over trace preserving LOCC \cite{Rains1999} as follows
\cite{gentele}
\begin{equation}
f_{max}={dF_{max} +1 \over d+1}.
\end{equation}
Both quantities are
by construction LOCC monotones. They are not equal to zero on
separable states, but are constant on them: $F_{max}(\rho_{sep})=1/d$.
For two qubits, the protocol to obtain $F_{max}$ is of the following
simple form \cite{VerstraeteV2003-maxtel}: Alice applies some filter
$A_i\ot \id$ (see Sec. \ref{sec:LOCC}), and tells Bob whether  she
succeeded or not. If not, then they remove the state and produce
some separable state which has overlap $1\over d$ with singlet
state.

Another measure was constructed in \cite{Brandao2005-witent} by means
of activation concept \cite{activation}. It is connected with the maximal
fidelity with $\Psi^+$, obtainable by means of local filtering, denoted here by $F_{max}^{(p)}$ (i.e. probabilistic maximal singlet fraction, see sec.
\ref{subsec:activbound}). Namely consider a state
$\sigma$ having some value $F_{max}^{(p)}(\sigma)$. With help of some other
state $\rho$ one can obtain better $F_{max}^{(p)}$, i.e. $F_{max}^{(p)}(\rho\ot \sigma)$ may be larger than $F_{max}^{(p)}(\sigma)$. What is rather nontrivial, this
may happen even if $\rho$ is bound entangled, i.e. for which $F_{max}^{(p)}$
is the same as for separable states. The activation power of a given
state can be quantified as follows \be E^{(d)}(\rho)=\sup_\sigma
{F_{max}^{(p)}(\rho\ot \sigma)-F_{max}^{(p)}(\sigma)\over F_{max}^{(p)}(\sigma)}. \ee Since Masanes \cite{Masanes1_activation} showed that any entangled state can activate
some other state, i.e. increase its $F_{max}^{(p)}$, the quantity $E^{(d)}$ is
nonzero for all entangled states,  including all bound entangled
states.

\paragraph{Robustness measures}

Robustness of entanglement was introduced in
\cite{VidalT1998-robustness}. For a state $\rho$
consider separable state $\sigma_{sep}$. Then $R(\rho|\sigma_{sep})$
is defined as minimal $t$ such that  the state
\be
{1\over 1+t}(\rho + t\sigma_{sep})
\ee
is separable. Now robustness of entanglement is defined
as
\be
R(\rho)=\inf_{\sigma_{sep}} R(\rho|\sigma_{sep}).
\ee
It is related to the quantity
$P(\rho)$ given by minimal $p$ such that the state
\be
(1-p) \rho  +p \sigma_{sep}
\ee
is separable. We have $P=R/(1+R)$. Though $P$ is more intuitive, it turns out
that $R$ has better mathematical properties, being e.g. convex. $R$ satisfies
monotonicity (\ref{eq:mono-strong}). In \cite{Steiner2003-robust-gen,HarrowN2003-robust}
generalized robustness $R_g$ was considered, where the infimum is taken
over all states rather than just separable ones. Interestingly it was shown that for
pure states it does not make difference. $R_g$ is monotone too.
Brandao \cite{Brandao2005-witent} showed that  the generalized
robustness $R_g$ has operational interpretation: it is
just equal to the measure $E^{(d)}$ quantifying the activation power.
Moreover  \cite{Brandao2005-witent} $R_g$ gives rise to the following upper bound for $E_D$:
\be
E_D(\rho)\leq \log_2(1+R_g(\rho)).
\ee

\paragraph{Best separable approximation measure} was introduced in
\cite{KarnasL2000-bsa} by use of {\it best separable approximation} idea
 \cite{LewensteinSanpera-bsa}. The latter is defined as follows: one  decomposes
state $\rho$  as a mixture
\be
\rho=(1-p) \delta \rho + p \sigma_{sep},
\ee
where $\sigma_{sep}$ is separable state, and $\delta\rho$ is arbitrary state.
Denote by $p^*$ the maximal possible $p$.
Now, it turns out that $E_{bsa}(\rho)=1-p^*$ is an entanglement
measure, i.e. it is LOCC monotone  and vanishes on separable states.
For all pure entangled states the measure is equal to $1$.

\paragraph{Witnessed entanglement}
In \cite{BrandaoV,Brandao2005-witent} entanglement
measures are constructed using entanglement witnesses  as follows:
\begin{equation} E=-\inf_W \tr \rho W, \end{equation} where infimum is taken over some set of
entanglement witnesses. It turns out that many measures can be
recast in this form. For example random robustness is equal
to $-\inf_{W: \tr W=1} (\tr \rho W)$, $R_g=-\inf_{W\leq \id}(-\tr \rho
W)$, $E_{bsa}=-\inf_{W\geq \id} (\tr \rho W)$.

Interestingly, the singlet fraction maximized over PPT operations \cite{Rains2001} can be represented as:
\be
F_{max}^{ppt}={1\over d} -\inf_{W}\tr W \rho,
\ee
where the infimum is taken over the set $\{W: {1\over (1-d)}\id\leq W \leq {1\over d}\id,0\leq W^\Gamma
\leq 2 {1\over d}\id\}$.

\paragraph{Negativity.} A simple computable measure was introduced in
\cite{ZyczkowskiHSP-vol} and then shown in \cite{Vidal-Werner} to be
LOCC monotone. It is negativity
\be
\ncal=\sum_{\lambda<0}\lambda
\ee
where $\lambda$ are eigenvalues of $\rho^\Gamma$ (where $\Gamma$ is
partial transpose). A version of the measure called logarithmic
negativity given by
\be
E_\ncal(\rho)=\log \|\rho^\Gamma\|_1
\ee
is upper bound for distillable entanglement \cite{Vidal-Werner}. It can be also written as $E_\ncal(\rho)= \log {2\ncal(\rho) +1\over 2}$.
The measure $E_\ncal(\rho)$ is easily seen to satisfy monotonicity
(\ref{eq:mono}), because $\ncal(\rho)$ does satisfy it, and
logarithm is monotonic function. However logarithm it is
not convex, and as such might be expected not to satisfy the
stronger monotonicity condition (\ref{eq:mono-strong}). However it
was recently shown that it does satisfy it \cite{Plenio2005-logneg}.
The measure is moreover additive. For states with positive $|\rho^\Gamma|^\Gamma$
$E_\ncal$ has operational interpretation --- it is equal to
exact entanglement cost of creating state by PPT operations from \singlets\
\cite{AudenaertPE2002-PPT}.

It turns out that negativity, robustness, BSA can be also obtained from one
scheme originating form base norm \cite{Vidal-Werner,PlenioVirmani2006-review}

\paragraph{Greatest cross norm.}
In Ref. \cite{Rudolph2001-measureJMP} a measure based on the so called {\it greatest cross-norm} was
proposed. One decomposes $\rho$ into sum of product operators
\be
\rho=\sum_i A_i\ot B_i.
\label{eq:prod}
\ee
Then the measure is given by
\be
E(\rho)=\sup\sum_i \|A_i\|_1\cdot \|B_i\|_1,
\ee
where supremum is taken over all decompositions (\ref{eq:prod}).
It is not known if the function satisfies monotonicity (in \cite{Rudolph2001-measureJMP} monotonicity under local operations
and convexity was shown).
However it was shown \cite{Rudolph2002-criterion} that  if in infimum one restricts to Hermitian operators,
then it is equal to $2R+1$, where $R$ is robustness of entanglement.

\paragraph{Rains bound}
Rains \cite{Rains2001}  has combined two different concepts (relative entropy of entanglement
and negativity) to give the following quantity
\be
E_{R+\ncal}=\inf_\sigma \left(S(\varrho|\sigma) + \|\sigma^\Gamma\|_1\right),
\ee
where the infimum is taken under the set of {\it all} states.
It is the best known upper bound on distillable entanglement. However it
turns out to be an entanglement measure itself.  One can show that
the measure satisfies  monotonicity (\ref{eq:mono-strong}).
To see this, consider optimal $\sigma$ (i.e. reaching infimum),  the existence of which is assured by continuity of norm and lower semi-continuity of relative
entropy (\cite{OhyaPetz}). For such a $\sigma$ by monotonicity of relative entropy under any
quantum operation and monotonicity of $E_\ncal$ under LOCC operation we have for any LOCC map $\Lambda$
\be
 S(\varrho|\sigma) + \|\sigma^{\Gamma}\|_1 \geq
 S(\Lambda(\varrho)|\Lambda(\sigma)) + \|\Lambda(\sigma)^{\Gamma}\|_1.
\ee
Hence the infimum  can not be increased under LOCC.

\paragraph{Squashed entanglement.}
Squashed entanglement was introduced by  \cite{Tucci2002-squashed}
and then independently by Christandl and Winter
\cite{Winter-squashed-ent}, who showed that it is monotone, and
proved its additivity. It is therefore the  first additive measure
with good asymptotic properties. In the latter paper, definition of squashed
entanglement $\esq$ has been inspired by relations between
cryptography and entanglement. Namely, $\esq$ was designed on the
basis of a quantity called {\it intrinsic information}
\cite{MauWol97c-intr,GisinWolf_linking,renner-wolf-gap}, which was
monotonic under local operations and public communication. The
squashed entanglement is given by \be
\esq(\rho_{AB})=\inf_{\rho_{ABE}} {1\over 2} I(A:B|E) \ee
where $I(A:B|E)= S_{AE}+S_{BE}-S_E - S_{ABE}$ and infimum is
taken over all density matrices $\rho_{ABE}$ satisfying $\tr_E
\rho_{ABE}=\rho_{AB}$. The measure is additive on tensor product
and superadditive in general, i.e. \ben
\esq(\rho_{AB} \ot \rho_{A'B'}) = \esq(\rho_{AB})+\esq(\rho_{A'B'});\quad \nonumber\\
\esq(\rho_{AA'BB'})\geq  \esq(\rho_{AB})+\esq(\rho_{A'B'}). \een
It is not known whether it vanishes if and only if the state
is separable. (It would be true, if infimum could be turned into
minimum, this is however unknown). The measure is asymptotically
continuous \cite{Alicki-Fannes}, and therefore lies between $E_D$
and $E_C$.  Even though it was computed for just two families of
states, a clever guess for $\rho_{ABE}$ can give very good estimates
for $E_D$ in some cases.

It is an open question if the optimization in the definition of
$\esq$ can be restricted to the subsystem $E$ being a classical
register \cite{Tucci2002-squashed}. If it is so, then we could
express $\esq$ as follows: \be \esq\stackrel{?}{=}\inf \sum_i p_i
I_M(\rho_{AB}^i), \ee where the infimum is taken over all
decompositions $\sum_i p_i \rho^i_{AB} =\rho_{AB}$. (This measure was  called c-squashed entanglement, because it can be recast as a version of $\esq$ with  $E$ being a classical register). $\esq$ would be then
nothing else but a mixed convex roof measure based on mutual information mentioned above, which is an entanglement measure itself.

\paragraph{Conditioning entanglement}
It is worth to mention that there is a method to obtain a new entanglement measure from a given one as follows \cite{DongHW2007-cond-ent}:
\be
CE(\rho_{AB})=\inf_{\rho_{AA'BB'}} (E(\rho_{AA'BB'}) - E(\rho_{A'B'})),
\ee
where infimum is taken over all states $\rho_{AA'BB'}$ such that $\tr_{A'B'}\rho_{AA'BB'}=\rho_{A'B'}$.
Provided the initial measure was convex and satisfied strong monotonicity,
the same holds for the new measure. Moreover it is automatically superadditive, and is a lower bound for regularization $E^{\infty}$ of the original measure.

\subsection{All measures for pure bipartite states}
\label{subsec:all-pure}

In Ref. \cite{Vidal-mon2000} it was shown that measures for pure states
satisfying strong monotonicity (\ref{eq:mono-strong}) are in
one-to-one correspondence to functions $f$ of density matrices satisfying
\bei
\item[(i)] $f$ is symmetric, expansible function of eigenvalues of $\varrho$
\item[(ii)] $f$ is concave function of $\varrho$
\eei
(by expansibility we mean $f(x_1,\ldots,x_k,0)=$ $f(x_1,\ldots,x_k)$).
In this way all possible entanglement measures for pure  states were characterized.

More precisely, let $E_p$, defined for pure states,
satisfy $E_p(\psi)=f(\varrho_A)$, where $\varrho_A$ is reduction of $\psi$,
and $f$ satisfies (i) and (ii). Then there exists an entanglement  measure $E$
satisfying LOCC monotonicity coinciding with $E_p$ on pure states ($E$ is convex-roof
extension of $E_p$). Also conversely, if we have arbitrary measure
$E$ satisfying  (\ref{eq:mono-strong}), then  $E(\psi)=f(\varrho_A)$ for
some $f$ satisfying (i) and (ii).

We will recall the proof of the direct part. Namely, we will show
that convex-roof extension $E$ of $E_p$ satisfies
(\ref{eq:mono-strong}). As mentioned earlier, it suffices to show it
for pure states. Consider then any operation on, say, Alice side
(for Bob's one, the proof is the same) which produces ensemble
$\{p_i,\psi_i\}$ out of state $\psi$. We want to show that the final
average entanglement $\Ebar=\sum_ip_iE(\psi_i)$ does not exceed the
initial entanglement $E(\psi)$. In other words, we need to show
$\sum_ip_i f(\varrho_A^{(i)})\leq f(\varrho_A)$, where
$\varrho_A^{(i)}$ are reductions of $\psi_i$ on Alice's side.  We
note that due to Schmidt decomposition of $\psi$, reductions
$\varrho_A$ and $\varrho_B$ have the same non-zero eigenvalues.
Thus, $f(\varrho_A) =f(\varrho_B)$, due to (i). Similarly
$f(\varrho_A^{(i)})= f(\varrho_B^{(i)})$. Thus it remains to show
that $\sum_ip_i f(\varrho_B^{(i)})\leq f(\varrho_B)$. However
$\varrho_B= \sum_ip_i\varrho_B^{(i)}$ (which is algebraic fact, but
can be understood as no-superluminal-signaling condition --- no action
on Alice side can influence statistics on Bob's side, provided  no
message was transmitted from Alice to Bob). Thus our question
reduces to the inequality $\sum_ip_i f(\varrho_B^{(i)})\leq
f(\sum_ip_i\varrho_B^{(i)})$. This is however true, due to concavity
of $f$.

As we have mentioned, examples of entanglement measures for pure
states are quantum Renyi entropies of subsystem for
$0\leq\alpha\leq1$. Interestingly, the Renyi entropies for
$\alpha>1$ are not concave, but are Schur concave. Thus they are
satisfy monotonicity (\ref{eq:mono}) for pure states, but do not
satisfy the strong one (\ref{eq:mono-strong}). It is not  known
whether the convex roof construction will work, therefore it is open question
 how to extend such measure to mixed states.

Historically first measure was the von Neumann entropy of subsystem
(i.e. $\alpha=1$) which has operational interpretation --- it is equal
to distillable entanglement and entanglement cost. It is the unique
measure for pure states, if we require some additional postulates,
especially asymptotic continuity (see Sec. \ref{subsec:asym}).

\subsubsection{Entanglement measures and transition between states --- exact case}
\label{subsubsec:em-tr-exact}
Another family of entanglement measures is the following. Consider
squares of Schmidt coefficients of a pure state $\lambda_k$ in
decreasing order so that $\lambda_1 \geq \lambda_2 \geq \ldots
\lambda_d$, $\sum_{i=1}^d\lambda_i=1$. Then the sum of the $d-k$
smallest $\lambda$'s \begin{equation} E_k(\psi)=\sum_{i=k}^d \lambda_i \end{equation} is an
entanglement monotone \cite{Vidal1999-ent-pure}. Thus for a state
with $n$ nonzero Schmidt coefficients we obtain $n-1$ nontrivial
entanglement measures. It turns out that these measures constitute
in a sense a complete set of entanglement measures for bipartite
pure states.

Vidal proved the following inequality relating probability $p(\psi\to \phi)$
of obtaining state $\phi$ from $\psi$ by LOCC with entanglement measures:
\be
p(\psi\to \phi)\leq {E(\psi)\over E(\phi)}.
\label{eq:p-e}
\ee

If we fix an entangled state $\phi_0$, then $p(\psi\to \phi_0)$ is
itself a measure of entanglement, as a function of $\psi$. The
relation (\ref{eq:p-e}) gives in particular \be p(\psi\to \phi) \leq
{E_k(\psi)\over E_k(\phi)} \ee for all $k$. It turns out that these
are the only constraints for transition probabilities. Namely
\cite{Vidal1999-ent-pure} {\it
The optimal probability of transition from state $\psi$ to $\phi$ is
given by \begin{equation} p(\psi\to \phi)=\min_k {E_k(\psi)\over E_k(\phi)}. \end{equation}
} This returns, in particular, Nielsen's result
\cite{Nielsen-pure-entanglement}. Namely, $p=1$, if for all $k$,
$E_k(\psi)\geq E_k(\phi)$, which is precisely the majorization
condition (\ref{eq:major}).

Thus the considered set of abstractly defined measures determines
possible transformations between states, as well as optimal probabilities
of such transformations. We will further see generalization of such
result to asymptotic regime, where nonexact transitions are  investigated.

\subsection{Entanglement measures and transition between states --- asymptotic case}
\label{subsec:asym} In asymptotic  regime, where we tolerate small
inaccuracies, which disappear in the limit of large number of
copies, the landscape of entanglement looks more ``smooth''.
Out of many measures for pure states only one becomes
relevant: entropy of entanglement, i.e. any measure significant for
this regime reduces to entropy of entanglement for pure states.
\footnote{One can consider half-asymptotic regime, where one takes limit
of many copies, but does not allow for inaccuracies. Then other measures
can still be of use, such as logarithmic negativity which is related
to PPT-cost of entanglement \cite{AudenaertPE2002-PPT} in such regime.}
Moreover, only the measures with some properties, such as asymptotic
continuity, can be related to operational quantities such as $\ed$,
or more generally, to asymptotic transitions rates. We refer to such
measures as good asymptotic measures.

\subsubsection{$E_D$ and $E_C$ as extremal measures. Unique measure
for pure bipartite states.}
\label{subsubsec:extremal} If measures satisfy some properties, it
turns out that their regularizations are bounded by $E_D$ from one
side and by $E_C$ from the other side. By regularization of any
function $f$ we mean $f^{\infty}(\rho)=\lim_n {1\over n} f(\rho^{\ot
n})$ if such a function exists. It turns out that if a function $E$
is monotonic under LOCC, asymptotically continuous and satisfies
$E(\psiplus_d)=\log d$, then we have
\be
E_D\leq  E^\infty  \leq E_C.
\ee
In particular this implies that for pure states, there is {\it unique}
entanglement measure in the sense that regularization of any
possible entanglement measure is equal to entropy of
subsystem\footnote{Uniqueness of entanglement measure for pure states
was put forward in \cite{popescu-rohrlich}. The postulates that
lead to uniqueness were further worked out in \cite{Vidal-mon2000,limits,DonaldHR2001}.}. An
exemplary measure that fits this scheme is relative entropy of
entanglement (related either to the set of separable states or to PPT
states). Thus whenever we have reversibility, then the transition
rate is determined by relative entropy of entanglement. Some
versions of the theorem are useful to find upper bounds for $E_D$ -
one of central quantities in entanglement theory. We have for
example that any function $E$ satisfying the following conditions:
\bee
\item $E$ is weakly subadditive, i.e. $E(\rho^{\ot n})\leq nE(\rho)$,
\item For isotropic states ${E(\rhoiso)\over \log d} \to 1$
for $F\to 1$, $d\to \infty$,
\item $E$ is monotonic under LOCC (i.e. it satisfies Eq. (\ref{eq:mono})),
\eee
is an upper bound for distillable entanglement. This theorem
covers all known bounds for $\ed$.

There are not many measures which fit the asymptotic regime. apart
from operational measures such as $\ec$, $\ed$ and $K_D$ only
entanglement of formation, relative entropy of entanglement
(together with its PPT version) and squashed entanglement belong
here. For review of properties of those measures see
\cite{Christandl-PhD}.

\subsubsection{Transition rates}
\label{subsec:rates}
One can consider transitions between any two states \cite{BDSW1996} by means of LOCC:
$R(\rho\to \sigma)$ defined analogously to $E_D$,
but with $\sigma$ in place of maximally entangled state. Thus we
consider $n$ copies  of $\rho$ and want to obtain  a state $\sigma_n^{out}$
that will converge $m$ copies  of $\sigma$ for large $n$.
$R(\rho\to \sigma)$ is then defined  as maximum asymptotic rate $m/n$ that can be achieved
by LOCC operations.  One then can generalize the theorem about
extreme measures as follows:
\be
R(\rho\to \sigma) \leq {E^\infty(\rho)\over E^\infty(\sigma)}
\label{eq:rates}
\ee
for any $E$ satisfying:
\bee
\item $E$ is nonincreasing under LOCC,
\item regularizations exist for states $\rho$ and $\sigma$ and $E^\infty(\sigma)>0$,
\item $E$ is asymptotically continuous (see Sec. \ref{sec:other-post}).
\eee This result is an asymptotic counterpart of Vidal's relation
between optimal probability of success and entanglement measures
(\ref{eq:p-e}). Noting that, in particular, we have \be R(\rho\to
\psiplus_2)=E_D(\rho),\quad R(\psiplus_2 \to \rho)={1\over
E_C(\rho)}, \ee one easily arrives at extreme measures theorem. For
example to obtain bounds  for  transition rates between maximally
correlated we can choose measures $E_R$ and $\ef$, as they satisfy
the conditions  and are additive for those states.  For more
sophisticated transitions see \cite{MichalSS2002}. One notes that
$R$ gives rise to plethora entanglement measures, since the
following functions \be E_\sigma^D (\rho)=R(\rho\to \sigma),\quad
E^C_\sigma(\rho)={1\over R(\sigma \to \rho)}, \ee where $\sigma$ is
an entangled state, are nonincreasing under LOCC. (Thanks to the
fact that every entangled state have nonzero $E_C$ the above
measures are well defined.)

{\it Upper bound on distillable key.} This paradigm allows to
provide upper bound $K_D$  (see Sec. \ref{subsubsec:kd-pbits}), because
distillable key is also some rate of transition under LOCC
operations. Namely, any entanglement measure which is asymptotically
continuous, and for so-called private state $\gamma_d$ satisfies
$E(\gamma_d)\geq \log d$, then it is upper bound for $K_D$
\cite{Christandl-PhD}. Roughly speaking it follows from the fact,
that distillable key can be expressed as rate of transition \be
K_D(\rho)=\sup_{\gamma_d} R(\rho\to \gamma_d) \log d, \ee where
supremum is taken over all p-dits  $\gamma_d$ (see Sec. \ref{subsubsec:private}).

\subsection{Evaluating measures}
\label{subsec-evaluating}

It is usually not easy to evaluate measures. The only measure that
is easily computable for any state is $E_\ncal$ (logarithmic
negativity). Entanglement of formation is efficiently computable for
two-qubits \cite{Wootters-conc}. Other measures are usually
computable for states with high symmetries, such as Werner states,
isotropic state, or the family of ``iso-Werner'' states, see
\cite{BDSW1996,TerhalV2000-eof-iso,VollbrechtW2000-ent-sym,
Rains1999,Rains-erratum1999}.

An analytical lower bounds for concurrence for all states were provided
in \cite{MintertKB04-conc} (see also \cite{AudenaertVM2000-concurrence}).
The bound constitutes also a new criterion of separability. A way to bound a convex roof measure, is to provide a computable convex function, that is greater than or equal to the measure on the pure states.
For example  we have
\be
\|(|\psi\>\<\psi|)^\Gamma\|_1=\|R(|\psi\>\<\psi|)\|_1=
\left(\sum_i \sqrt{p_i}\right)^2
\ee
where $p_i$ are squares of Schmidt coefficients of $\psi$, and $R$
is realignment map \ref{subsubsec:realignment}. Comparing
this with concurrence one gets  a bound obtained in \cite{AlbeverioLowerBoundPRL1}
\be
C(\rho)\geq \sqrt{{2\over m(m-1)}}(\max (\|\rho^\Gamma\|_1,\|R(\rho)\|_1)-1).
\ee

As far as entanglement of formation is concerned, in \cite{TerhalV2000-eof-iso} a method was introduced for which
it is enough to optimize over some  restricted set rather than the set of pure
states. This was further successfully developed in
\cite{FeiX2006-R-EF,AlbeverioLowerBoundPRL,DattaFSC2006-maps-EMs}
where lower bounds for $\ef$ were obtained based on known separability criteria such as PPT, realignment or the recent Breuer's map
(see Secs. \ref{subsubsec:realignment} and  \ref{subsubsec:Reduction}).

In \cite{VollbrechtW2000-ent-sym} a surprising result was obtained, concerning possible
additivity of $E_{R}$. They have shown that $E_R$ is nonadditive
for Werner asymmetric states, and moreover for large $d$, $E_R$ of two copies is almost the same as for one copy. Thus, the
relative entropy of entanglement can be strongly nonadditive.
Therefore regularization of $E_R$ is not equal to $E_R$. In \cite{AEJPVM2001}
for the first time $E_R^\infty$ was computed  for some states.
Namely for Werner states we have:
\be
E^\infty_{R,\scal}=E_{R+\ncal}=
\left\{
\bea{ll}
1-H(p) & {1\over 2} < p\leq {1\over 2} +{1\over d}\\
\log \left({d-2\over d}\right) + p \log \left({d+2\over d-2}\right) &{1\over 2}
+{1\over d}<
p\leq 1
\eea
\right.
\ee

Concerning the operational measures, we know that
$E_C=\ef^\infty\equiv \lim_{n\to\infty} \ef(\varrho^{\ot n})$
\cite{cost}. If $\ef$ were additive (which is a long-standing
problem) then it would be equal to $E_C$. $E_D$ is bounded from
above by $\ef$ \cite{BDSW1996}. For pure states
$E_D=\ef=E_C=E_R=S(\varrho_A)$ where $\varrho_A$ is reduced density
matrix of the given pure state \cite{BBPS1996,PlenioVedral1998}. In
\cite{VidalC-irre} it was found that for some bound entangled
state (i.e. with $E_D=0$) $E_C>0$. It seems that we can have
$E_D=E_C$ only for states of the form
\be
\sum_ip_i
|\psi_i\>\<\psi_i|_{AB}\ot |i\>\<i|_{A'B'}
\ee
where $|i\>_{A'B'}$
are product states distinguishable by Alice and Bob \cite{termo} (or
some generalizations in  similar spirit, that Alice and Bob can
distinguish states, which satisfy  $\ed=\ec$ trivially).

Apart from the above trivial case of locally orthogonal mixtures
the value of the measure $E_D$ is known only for
maximally correlated states $\sum_{ij} a_{ij} |ii\>\<jj|$ for which
$E_D =S_A - S_{AB}$. It is upper bound, since it is equal to $E_R$
\cite{Rains1999}. That it can be achieved follows from general
result of \cite{DevetakWinter-hash} stating that $E_D\geq
S_A-S_{AB}$. Example is  mixture of two maximally entangled two
qubit states where we have
\be
\ed(\varrho)=1-S(\varrho).
\ee

For higher dimension  powerful tools for evaluating $E_D$ were provided
in \cite{Rains2001}. One knows several upper bounds for $E_D$
\cite{BDSW1996,PlenioVedral1998,Rains1999,irrev,Vidal-Werner,Rains2001}.
The best known bound is $E_{R+\ncal}$ provided in \cite{Rains2001}. For Werner
states it is equal  to regularization of $E_{R,\scal}$
(this is true for more general class of symmetric states  \cite{AudenaertMVW2002-asym-oo}).

Squashed entanglement has been evaluated for so called {\it flower state}
(purification of which is given by (\ref{eq:flower-purif}))
and its generalizations in \cite{ChristandlW-lock}.

Several entanglement measures have been evaluated for some multipartite
pure and mixed graphs states in \cite{MarkhamMV2007-em-graph}.
They have used results concerning distinguishing states via LOCC (\ref{eq:locc-dist-jap}), see sec. \ref{sec:rozroz}.

\subsubsection{Hashing inequality}
\label{subsec:hasineq}
Any entanglement measure $E$ that is good in asymptotic regime satisfy inequality $E\geq S(\varrho_B)-S(\varrho_{AB})\equiv I^{coh}_{A\>B}$. For $\ef$ the
inequality follows from concavity of $I^{coh}_{A\>B}$ \cite{termo}. For
$E_R$ the proof is much more involved \cite{PlenioVP2000}. In
\cite{HHH-cap2000} the {\it hashing inequality} was conjectured
$E_D^\to\geq I^{coh}_{A\>B}$, where $E_D^\to$ is one-way distillable
entanglement (classical communication only from Alice to Bob is
allowed \cite{BDSW1996}). In
\cite{DevetakWinter-hash,DevetakWinter-hash-prl} the inequality was
proven. However, from extreme measures result (see Sec.
\ref{subsubsec:extremal}) it follows that all asymptotic measures of
entanglement (in the sense that they satisfy conditions discussed
in Sec. \ref{subsec:asym}) are upper bounds  for $E_D$, hence they are also greater than $I^{coh}_{A\>B}$.

\subsubsection{Evaluating $E_C$ vs additivity problem}
\label{subsubsec:cost-addit} It is long open question of whether in
general $\ef$ is additive. Since $E_D$ is lower bound for $E_C$, it
follows that for distillable states $E_C$ is nonzero. Though for
bound entangled states (for which $E_D=0$) this does not give any
hint. In \cite{VidalC-irre} the first example of bound
entangled state with $E_C>0$ was found. Later \cite{Vidal-cost2002}
examples of states with additive $\ef$ were found.

{\it Example.} We will now show a simple proof (see \cite{lock-ent}),
that for any state of the form
\be
\rho_{AB}=\sum_{ij} a_{ij}  |ii\>\<jj|
\ee
$\ef$ is additive.  Such states are called maximally correlated
\footnote{Additivity was shown in \cite{MichalSS2002}
along the lines of Ref. \cite{Vidal-cost2002}.}.
Namely, one considers purification $\psi_{ABC}$ which can be taken
as $\sum_i |ii\>_{AB}|\tilde \psi_i\>_C$ where $\<\tilde \psi_i|\tilde \psi_j\>=a_{ij}$. The state of subsystems $B$ and $C$
is of the form  $\rho_{BC}=\sum_{i} p_i  |i\>\<i|\ot |\psi_i\>\<\psi_i|$,
where $\psi_i$ are normalized vectors $\tilde \psi_i$ and
$p_i=|a_{ii}|^2$.
Then using the fact that $\ef$ can be also defined as
infimum of average entanglement between Alice and Bob
over all measurements performed by Charlie on system $C$ one finds that
\be
\ef= S_B - I_{acc}(\{p_i,\psi_i\})
\ee
where $I_{acc}$ is so called accessible information\footnote{It is defined as follows:
$I_{acc}(\{p_i,\psi_i\})=\sup_{\{A_j\}}I(i:j)$ where
supremum is taken over all measurements, and $I(i:j)$
is mutual information between symbols $i$ and measurement outcomes $j$.}.
Now, following Wootters (see \cite{hiding-ieee}),  one can easily show  by use
of chain rule that $I_{acc}$ is additive.
Since $S_B$ is additive too, we obtain that $\ef$ is additive.
More generally we have \cite{KoashiW-monogamy}
\be
\ef(A:B)=S(B)- \chv(C\>B)
\ee
where $\chv$ is a measure of classical correlations \cite{HendersonVedral}.
It is defined as follows:
\be
\chv(A\>B)=\sup_{\{A_i\}} \bigl(S(\rho_B)-\sum_ip_iS(\rho_B^i)\bigr)
\label{eq:chv}
\ee
where supremum is taken over all measurements $\{A_i\}$ on
system $A$, and $p_i$ is probability of outcome $i$, $\rho^i_B$ is
the state of system $B$, given the outcome $i$ occurred.
In \cite{DevetakW03-common} it was shown that for separable states
$\chv$ is additive. This is related to the result by Shor
\cite{Shor-break} on additivity of classical capacity for
entanglement breaking channels (actually the original results on additivity of $\ef$
were based on Shor result).

Another result on $E_C$ was provided in \cite{YangHHS2005-cost}.
Namely, $E_C$ is lower bounded by a function $G$
which is (mixed) convex  roof of the function $\chv$.
Such a function is nonzero if and only if a state is entangled.
It is worth to present a proof, as it is quite  simple, and
again uses duality between $\ef$ and $\chv$.

We first consider a pure state $\psi_{AA'BB'}$.
For this state we have
\ben
&&\ef(\psi_{AA':BB'})=S(AA')=\ef(\rho_{AA':B})+C_{HV}(B'\>AA')\geq\nonumber \\
&&\geq \ef(\rho_{A:B})+C_{HV}(B'\>A')
\een
where $\rho_{XXX}$ denote suitable partial traces.

Now consider mixed state $\rho_{AA':BB'}$. We obtain \be
\ef(\rho_{AA':BB'})\geq \ef(\rho_{A:B})+G(B'\>A'), \ee where $G$ is
mixed convex roof of $C_{HV}$. This follows from applying left hand
side to optimal decomposition of $\rho_{AA':BB'}$ into pure states
(which gives mixed convex roof on the right hand side --- as a result,
$\ef$ remains $\ef$, but $C_{HV}$ changes into $G$).

Now we apply this last inequality to $n$ copies of the same state
$\rho$ and obtain \be \ef(\rho^{\ot (n-1)}\ot \rho) \geq
\ef(\rho^{\ot (n-1)}) + G(\rho). \ee Iterating this equation we
obtain \begin{equation} \ef(\rho^{\ot n}) \geq \ef(\rho) + (n-1) G(\rho). \end{equation}
Dividing both sides by $n$ we obtain $\ec \geq G(\rho)$. It remains
to prove that $G$ is nonzero for any entangled state. This follows
from standard arguments of Caratheodory's type.

\subsection{Entanglement imposes different orderings}
One can ask, whether different entanglement measures impose the same
ordering in set of states. The question was first posed in
\cite{VirmaniP1999-order}. Namely, suppose that $E(\rho)\geq
E(\sigma)$. Is it also the case that $E'(\rho)\geq E'(\sigma)$?
That it is not the case, we can see just on pure states. There
exist incomparable states, i.e. such states $\psi$, $\phi$, that
neither $\psi\to \phi$ nor $\phi\to\psi$ is possible by LOCC. Since
LOCC transitions is governed by entanglement measures (see
Sec.\ref{subsec:all-pure}) we see that there are two measures which
give opposite ordering on those states.

In asymptotic regime there is unique measure for pure states.
However, again it is easy to see \cite{VirmaniP1999-order}, that a unique order
would imply $\ed=\ec$ for all states, while we know that it is not the case.

On can interpret this lack of single ordering as follows:
there are many different types of entanglement,
and in one state we have more entanglement of one type, while in other state,
there is more entanglement of some other type
(See \cite{Miranowicz,Verstraete} in this context).

\subsection{Multipartite entanglement measures}
\label{subsec:multipartite}
Many of the axiomatic measures, are immediately extended to multipartite case.
For example relative entropy of entanglement is generalized, by taking a suitable
set in place of bipartite separable states. One can take the set of fully separable states
(then the measure will not distinguish between ``truly multipartite''
entanglement and several instances of bipartite entanglement such as $\phi^+_{AB}\ot \phi^+_{CD}$)\footnote{Some inequalities between so chosen version of $\er$
and bipartite entanglement were provided in \cite{PlenioV2001-rel-multi}}.
To analyze truly multipartite entanglement, one has to consider as
in \cite{VPRK1997} the set of all states containing no more than
$k$-particle entanglement (see Sec. \ref{MultipartiteEntanglement}). Similarly
one can proceed with robustness of entanglement. It is not easy
however to compute such measures even for pure states (see e.g.
\cite{PlenioV2001-rel-multi}). Moreover, for multipartite states,
much more parameters to describe entanglement is needed, therefore
many new entanglement measures have been designed, especially for
pure states. Then they can be extended to all states by convex roof
(which is however also hard to compute).

Before we present several such measures, let us make a digression on
behavior of  $k$-party entanglement (see Sec.
\ref{MultipartiteEntanglement}) under tensor product. Consider two
states, $|\phiplus\>_{AB}|0\>_C$ and $|\phiplus\>_{AC}|0\>_B$. Both are $2$
entangled: they contain no three-partite entanglement. But if Alice,
Bob and Charlie have both states,  they can create e.g. \ghz\ state.
It thus follows, that the "$k$-partyness" of entanglement is not closed under tensor
product. One of  implications is that distillable-\ghz\ entanglement
is extremely  superadditive: it jumps from zero to $1$ upon
tensoring the above states.

The above discussion concerns scenario where we tensor systems
such a way that the
number of parties does not increase: i.e. $\psi_{ABC}\ot \psi_{A'B'B'}=\psi_{AA':BB':CC}$.
This way of interpreting tensoring is characteristic for information theoretic
applications, being analogous to many uses of channel. It seems that
for analysis of physical many-body systems, it is more natural
to interpret tensoring as adding new parties: $\psi_{ABC}\ot \psi_{DEF}=\psi_{ABCDEF}$.
In such case, ``$k$-partyness'' of entanglement is preserved under tensoring.

There is even more subtle notion of "genuinely multipartite entanglement"
for pure states, which is defined recursively as follows.
An entangled bipartite pure states has always genuine
bipartite entanglement.
A m-partite pure state has genuine m-partite entanglement
if its (m-1)-particle reduced density matrices can be written as mixtures
of pure states which do not have genuinely
(m-1)-partite entanglement. For three qubits it is 3-tangle,
which is nonzero iff the state is genuinely 3-partite entangled.
It is then in GHZ class.  \footnote{Let us emphasize,
that there is a different notion of genuine multiparty entanglement
in asymptotic domain. There a pure state contains genuinely triparty entanglement
if it cannot be reversibly transformed into EPR pairs (see sec. \ref{subsubsec:ghz-epr}).
It is then not known, whether W state is of this sort or not.}

\subsubsection{Multipartite entanglement measures for pure states}
\label{subsubsec:multi-pure}

There are measures that are simple functions of sums of bipartite entanglement measures.
Example is ``global entanglement'' of  \cite{MeyerW2001-global} which is
sum of concurrences  between single qubit versus all other qubits.
Their monotonicity under LOCC is simply inherited from bipartite measures.

The first measure that is neither easy combination of bipartite measures,
nor an obvious generalization of such a measure is 3-tangle (or residual tangle)
introduced in \cite{CoffmanKW-tangle}.
It is defined as follows
\be
\tau(A:B:C)=\tau(A:BC)-\tau(AB)-\tau(AC),
\label{tau}
\ee
where $2$-tangles  on the right hand side
are squares of concurrence (\ref{eq:concurrence}). 3-tangle is permutationally invariant,
even though the definition does not suggest it.
It may be though zero  for pure states that are $3$-entangled (i.e. that
are not product with respect to any cut). Example is  so called W state. Tangle vanishes
on any states that are separable under any cut, and is nonzero for example on
$\ghz$ state.\footnote {In turns out, that if tangles in (\ref{tau}) are replaced by squares of negativities
the obtained quantity (after symmetrizing over systems permutation) gives also rise to an entanglement monotone \cite {OuF2007}.}
There are attempts to define a good generalization
of tangle for multiqubit systems  by means  of hyper-determinant \cite{Miyake2002-hyper}
(see below).  In \cite{LohmayerOSU2006-tangle} a convex roof of 3-tangle was computed for
mixture of GHZ state and W state orthogonal to it.

Shortly after introducing tangle, a concept of another measure
for tripartite states was introduced in the context of asymptotic
rate of transitions \cite{LPSW1999}:
\be
E(\psi)=\er(\rho_{AB}) + S(\rho_C)
\ee
where $\rho_{AB},\rho_C$ are reductions of $\psi_{ABC}$.
The measure allowed to detect truly tripartite
entanglement in GHZ state in asymptotic regime (see sec. \ref{subsubsec:ghz-epr}).
It is easy to see that it is monotonic under Alice and Bob actions.
Namely, the term $\er(\rho_{AB})$ represents entanglement of $\rho_{AB}$
(hence does not increase under Alice and Bob actions)
and the term $S(\rho_C)$ represents entanglement of the total
state under the cut $AB:C$ (hence does not increase under action of any party).
However the first term can be usually increased by Charlie.
Still the whole measure is monotone, because this increase is always
accompanied by a larger decrease of the term $S(\rho_C)$.
Thus to increase entanglement between $A$ and $B$,
one has to use up entanglement between $AB:C$.
Equivalently $\er(\rho_{A:B})$ goes down under mixing less than the entropy
goes up. Any measure having this feature can be put to the above formula,
to create a new entanglement measure. However, it is known that entanglement
of formation will not work here, because it is lockable, i.e. it can decrease
arbitrarily under loss of one bit, i.e. increase entropy by 1 (see section
\ref{subsec:locking}).

One of the first measures designed specifically for multipartite
states was Schmidt measure \cite{EisertB2000-Schmidt}. This
is minimum of $\log r$ where $r$ is number of terms in an expansion
of the state in product basis. For GHZ this measure is $1$, because
there are just two terms: $|000\>$ and $|111\>$. One can show, that
for W state it is impossible to write it by means of less than three
terms (otherwise it would either belong to GHZ class, or to EPR
class).
The measure is zero if and only if the state is fully product.
Therefore, it cannot distinguish truly multipartite entanglement from
bipartite entanglement. However it may be useful in many contexts,
see eg. \cite{MoraB2005-compl-ent}.

An interesting general class of multipartite entanglement measures was
obtained in the context of classification of states via
so called {\it normal forms} \cite{VerstraeteDM2001-normal}.
Namely, consider any homogeneous function of the state. Then if it is
invariant under determinant one SLOCC i.e. it satisfies
\be
f(A_1\ot \ldots A_n\psi)=f(\psi)
\ee
for $A_i$ being square matrices satisfying $\det A_i=1$,
then it is entanglement monotone in strong sense
(\ref{eq:mono-strong}), but under the restriction that the LOCC operation
produces output states on the Hilbert space of {\it the same dimension}.
The 3-tangle is example of such a measure.
Many measures designed for pure multipartite states like those
obtained in \cite{OsterlohS2005-filters,Miyake2002-hyper,
WongC2000-multi,Akhtarshenas2003-multiconc} are originally defined only for
a fixed dimension hence it is simply not possible to check the
standard monotonicity (\ref{eq:mono}).
However concurrence  though initially defined for qubits, can be  written in
terms of linear entropy of subsystems, being thus well defined for
all systems. Therefore there is a hope, that one can arrive at definition
independent of dimension for other measures. Then to obtain full
monotonicity, one will need in addition to prove, that  the measure
does not change, if the state is embedded into larger Hilbert spaces
of subsystems (equivalently, that the measure does not change under
adding local ancilla). However, for four-qubit concurrence of
\cite{WongC2000-multi} $\<\psi^*|\sigma_y^4|\psi\>$
its  natural generalization was shown to be not monotonous \cite{DemkowiczBKM2006-multi} (see below).
Of course, even the functions that are only  monotonous for
fixed dimension are useful quantities in many contexts.

{\it Measures based on hyperdeterminant.} Miyake noticed that
measures of entanglement such as concurrence and tangle are special
cases of hyperdeterminant \cite{Miyake2002-hyper}. Consider for
example qubits. For two qubits concurrence is simply modulus of
determinant, which is hyperdeterminant of first order. Tangle is
hyperdeterminant of second order --- a function of tensor with three
indices. Though computing hyperdeterminants of higher order than
tangle is rather complex, basing just on properties of
hyperdeterminant Miyake proved, that hyperdeterminants of higher
degree are also entanglement monotones \cite{Miyake2004-slocc}. They
describe truly multipartite entanglement, (in a sense, that states
such as product of \epr's have zero entanglement). The proof of
monotonicity bases on geometric-arithmetic mean, and is closely
related to the construction of entanglement measures based on
homogeneous functions  described above. Explicit formula
for hyperdeterminant for four qubits can be found in
\cite{Levay2006-4qubits}.

{\it Geometric measure.}
A family of measures have been defined by Barnum and Linden \cite{BarnumL2001-multi}.
In particular so called geometric measure is defined as
\be
\eg(\psi) = 1-\Lambda^k(\psi),
\ee
where $\Lambda^k(\psi)=\sup_{\phi\in S_k}|\<\psi|\phi\>|^2$, with $S_k$ being set of $k$-separable states.
This is generalization of Shimony measure \cite{Shimony1995-miara}, which for bipartite states was related to Renyi entropy with $\alpha=\infty$.
For relations with robustness of entanglement see \cite{Cavalcanti2006-geom-rob}.
The measure was also investigated in \cite{WeiG2003-geom-em,WeiAGM2003-em-bound}
where it was in particular computed for Smolin four qubit bound entangled
states (\ref{SmolinS}).

{\it Concurrence-like measures.} There were other attempts to generalize concurrence. Christensen and
Wong \cite{WongC2000-multi} obtained a measure for even number of
qubits by exploiting conjugation that appeared in original
definition of concurrence for two qubits. Their concurrence
works for even number of qubits  and is given by $\<\psi^*|\sigma_y^n|\psi\>$.
The measure is nonzero for a four partite states two pairs of EPR states.
This approach was generalized in \cite{OsterlohS2005-filters,OsterlohS2006-filters} who
analysed systematically possible quantities built out of antilinear
operations, also of higher order in $\psi$ than concurrence. For example,
they obtained the following nice representation for
3-tangle:
\be
\tau=\<\psi|\sigma_\mu\ot\sigma_y\ot\sigma_y|\psi^*\>
\<\psi|\sigma^\mu\ot\sigma_y\ot\sigma_y|\psi^*\>
\ee
where $\mu=0,1,2,3$ and the contraction is described by tensor
$g^{\mu,\nu}={\rm diag}[-1,1,0,1]$.
They have also designed measures that  distinguish between three different SLOCC classes of states (see sec. \ref{subsubsec:slocc}):

\ben
&&|\Phi_1\>={1\over \sqrt2}(|0000\>+|1111\>) \nonumber \\
&&|\Phi_2\>={1\over \sqrt6}(\sqrt2 |1111\>+|1000\>+|0100\>+|0010\>+|0001\>)
\nonumber \\
&&|\Phi_3\>={1\over 2}(|1111\> + |1100\> + |0010\>+ |0001\>).
\een
 An interesting proposal is due to Akhtarshenas
\cite{Akhtarshenas2003-multiconc}, which however is not proved to be
a monotone. In \cite{MintertKB2004-multi,DemkowiczBKM2006-multi}
a family of functions of the form
\be
C_{\acal}(|\psi\>)=2\sqrt{\<\psi|\ot \<\psi| {\acal} |\psi\>\ot |\psi\>}
\ee
was introduced, where $\acal=\sum_{s_1 \ldots s_n} p_{s_1 \ldots s_n}P_{s_1}
\ot \ldots P_{s_n}$,
with $s_i=\pm 1$, $P^{(\pm1)}$ is projector onto symmetric (antisymmetric)
subspace (see Sec. \ref{subsubsec:Witnesses}), and the coefficients $p_{s_1\ldots s_n}$ are nonnegative.
They have given sufficient conditions that must be satisfied by the coefficients to ensure monotonicity of $C$ (now without the restriction
of fixed dimension). On the other hand they have shown that  if $\acal = P^{(-)}\ot P^{(-)}\ot P^{(-)}\ot P^{(-)}$
then the function returns concurrence $\<\psi^*|\sigma_y^4|\psi\>$, and it is not monotonic.
The main tool was the following condition for monotonicity derived on basis of conditions {\tt LUI} and {\tt FLAGS} see sec. \ref{mono-convex}).
Namely  a function $C$ that\bei
\item is real, nonnegative and invariant under local unitaries
\item satisfies $C(a|\psi\>) = |a|^2C(|\psi\>)$
\item is defined for mixed states as a convex roof
\eei
is an entanglement monotone if and only if
$C(a|\psi\>\ot |\eta_1\> +b|\phi\>\ot |\eta_1\>) \leq |a|^2 C(a|\psi\>\ot |\eta_1\>) +bC(|\phi\>\ot |\eta_1\>)$
with equality for $a=0$ or $b=0$, where $\psi$  and $\phi$ are
arbitrary multipartite pure states, and $\eta_1$,  $\eta_2$  are local
orthogonal flags.

\noindent{\it Multipartite version of squashed entanglement.}
We can obtain multipartite versions of both squashed and c-squashed entanglement,
by putting in place of mutual information of bipartite system,
its generalization for multipartite systems \cite{multisquash}
\be
I(A_1:\ldots :A_N)=S(A_1) + \ldots + S(A_N) - S(A_1\ldots A_N),
\ee
choosing normalization factor ${1\over N}$, to get $1$ on GHZ state,
and applying the conditioning rule $S(X|Y)=S(XY)-S(Y)$.
So constructed multipartite entanglement measure
is closely related to multipartite secrecy monotones introduced in
\cite{Cerf-secr-mono}.

In the context of spin chains, another measure of entanglement was
designed: localisable entanglement \cite{VerstraetePC2003-locent}. Namely,
one chooses two spins,
and  performs LOCC operations aiming at obtaining the largest
bipartite entanglement between them (measured according to a chosen
entanglement measure for two bipartite states). Localisable
entanglement is a generalization of {\it entanglement of
assistance}, initially defined for tripartite pure states in
\cite{entass} as maximal entropy of entanglement that can be created
between Alice and Bob, if Charlie helps them by measuring his system
and telling the outcomes. For tripartite pure states entanglement of
assistance is simply a function of bipartite state $\rho_{AB}$ of
Alice and Bob (however it reflects entanglement properties of joint
state, not $\rho_{AB}$). Its formula is dual to entanglement of
formation: instead of infimum, there is supremum \be
E_{ass}(\rho)=\sup_{\{p_i,\psi_i\}}\sum_i p_i S(\rho_A^i), \ee where
$\rho_A^i$ is reduced density matrix of $\psi_i$, and supremum is
taken over all decompositions of $\rho$. For example consider \ghzstate.
Alice and Bob themselves share classically correlated states.
However, Charlie can measure in basis $|+\>,|-\>$, and if he tells
them result, they obtain \eprpair. So $E_{ass}= 1$ in this case.

Entanglement of assistance is not a monotone for tripartite states,
because it could be increased by cooperation between Alice, Bob and
Charlie \cite{GourS2005-eass}. It is also nonadditive \cite{entass}.
However in the limit of many copies, it turns out that it becomes a
monotone, namely it reduces to minimum of entropies of Alice's and
Bob's subsystem \cite{svw2005}, and for larger number of parties to
minimum entropy over all cuts that divide Alice from Bob. (This was
generalized to $N$ parties in \cite{SW-nature,sw-long}.)

\subsection{Entanglement parameters}

Some quantities even though are not monotonic under LOCC, seem still
useful in quantitative description of entanglement. For example the
parameter $M(\rho)$ \cite{HHH1995-bell} reporting maximal violation of
Bell-CHSH inequality for two-qubit states, or $N(\rho)$ reporting
maximal fidelity of teleportation for a class of protocols
\cite{HHH1996-teleportation} (see Sec. \ref{sec:Bell}). One of the
most important quantities in quantum communication theory, coherent
information, introduced in \cite{SchumacherN-1996-pra} (see also
\cite{Lloyd-cap}), can be positive only if the state is entangled
\cite{RPH1994}. One feels that the greater are such quantities, the
more entangled is the state. However those quantities can increase
under LOCC. For example coherent information can increase even under
local partial trace.

Thus they cannot describe entanglement directly, as it would imply,
that entanglement can be increased by means of LOCC. However, it is
plausible that the above parameters simply underestimate some
measures.  Let us consider coherent information in more detail.  One
can consider maximum value of coherent information attainable by
LOCC. This is already an entanglement measure.  One can show, that
this value does not exceed $\log d$, i.e. the value on singlet state.
Thus coherent information can only underestimate the value of
entanglement measure. Let us note that it is important to know the
maximal value of the entanglement measure induced by the given
parameter, so that we have a reference point.

\subsection{How much can entanglement increase under communication of one qubit?}

In \cite{LoPopescu} it was postulated that under sending $n$ qubits
entanglement shouldn't increase more than by $n$. In
\cite{ChenYang2000-ent-qubit} it was shown for entanglement of
formation.  Due to teleportation, sending qubits is equivalent to
bringing in a singlet. The question can then be recast as follows:
which entanglement measures satisfy
\begin{equation}
  E(\rho\ot |\phi^+\>\<\phi^+|) \leq E(\rho) + n.
  \label{eq:ent-incr-qubit}
\end{equation}
(Of course it is meaningful to ask such questions only for those
entanglement measures that exhibit a sort of extensive behavior.) If a
measure is subadditive, i.e. $E(\rho\ot \sigma) \leq
E(\rho)+E(\sigma)$ then the condition is satisfied. This is the case
for such measures as $E_R$, $\ef$, $E_C$, $E_N$, $E_{sq}$. More
problematic are $E_D$ and $K_D$. As far as $E_D$ is concerned, it is
easy to see that (\ref{eq:ent-incr-qubit}) is satisfied for
distillable states.  Simply, if by adding singlet, we can increase
$E_D$, then we could design protocol that would produce more singlets
than $E_D$, by using singlets obtained from distilling a first bunch
of copies to distillation of the next bunch\footnote{D. Gottesman,
private communication}. A more rigorous argument
includes continuity of $E_D$ on isotropic states (singlets with
admixture of random noise).  It is not hard to see, that for PPT
states the condition also holds (it is actually enough to check it on
two qubit singlet):
\ben
E_D(\rho\ot |\phi^+\>\<\phi^+|)\leq
E_N(\rho\ot |\phi^+\>\<\phi^+|)\nonumber \\
=E_N(\rho)+E_N(|\phi^+\>\<\phi^+|).
\een
This was later shown \footnote{P. Shor, A. Harrow and D. Leung, private communication} for all states, by
exploiting results of \cite{DiVincenzoMSST1999-UPB-huge}. The question
is still open for $K_D$.

\section{Monogamy of entanglement}
\label{sec:monogamy}

One of the most fundamental properties of entanglement is monogamy
\cite{CoffmanKW-tangle,Terhal2000-laa}. In its extremal form it can be
expressed as follows: If two qubits A and B are maximally quantumly
correlated they cannot be correlated at all with third qubit C. In
general, there is trade-off between the amount of entanglement between
qubits A and B and the same qubit A and qubit C. This property is
purely quantum: in classical world if A and B bits are perfectly
correlated, then there is no constraints on correlations between bits
A and C. For three qubits the trade-off is described by Coffman-Kundu-Wootters
monogamy inequality
\begin{equation}
 C^2_{A:B}+ C^2_{A:C}\leq C^2_{A:BC},
\label{CKW}
\end{equation}
where $C_{A:B}$ is the concurrence between $A$ and $B$, $C_{A:C}$ ---
between $A$ and $C$, while $C_{A:BC}$ --- between system $A$ and
$BC$. There was a conjecture that the above inequality can be extended
to $n$-qubits. The conjecture has been proved true only recently
\cite{OsborneV-monogamy}. The monogamy is also satisfied for Gaussian states \cite{HiroshimaAI2007,GTangle_AdeFab}
see sec. \ref{subsec:gaussian-measures}.
However, it does not hold anymore in higher dimension \cite{Ou2006-monogamy-viol}.
Namely consider Aharonov state of three qutrits
\begin{equation}
\psi_{ABC}={1\over \sqrt 6}( |012\> +|120\> + |210\> -
|021\> - |210\> - |102\>).
\end{equation}
One finds that $C^2_{A:B}=C^2_{AC}=1$, while $C^2_{A:BC}=3/4$.

Interestingly the monogamy for concurrence
implies monogamy of negativity  (see Sec.\ref{subsubsec:monot_pur}) \cite {OuF2007}
\begin{equation}
\ncal^2_{A:B}+  \ncal^2_{A:C}\leq  \ncal^2_{A:BC}.
\label{neg}
\end{equation}
It is not known if this holds in higher dimension.

More generally in terms of entanglement measures monogamy takes the
following form:

{\it For any tripartite state of systems $A_1$,  $A_2$, $B$}
\begin{equation}
E(A:B) + E(A:C) \leq E(A:BC).
\label{eq:monogamy}
\end{equation}
Note that if the above inequality holds in general (i.e. not only for
qubits), then it already itself implies (by induction) the inequality
\begin{eqnarray}
  E(A:B_1) + E(A:B_2) + \ldots + E(A:B_N) \qquad \nonumber \\
  \leq E(A:B_1\ldots B_N).
\end{eqnarray}
In \cite{KoashiW-monogamy} it was shown that squashed entanglement
satisfies this general monogamy
\begin{equation}
E_{sq}(A:B) + E_{sq}(A:C) \leq E_{sq}(A:BC).
\label{eq:monogamy-sq}
\end{equation}
This is the only known entanglement measure having this property.
$\ef$ and $E_C$ are not monogamous \cite{CoffmanKW-tangle,KoashiW-monogamy}
e.g. on the above Aharonov state. This somehow support a view, that
$E_C$ does not say about entanglement contents of the state, but
rather describes entanglement that must be ``dissipated'' while
building the state by LOCC out of pure entanglement
\cite{thermo-ent2002,SynakHH04}. (It is also supported by the locking effect, see Sec. \ref{subsec:locking}). In the context of monogamy, one can
consider other tradeoffs similar to (\ref{eq:monogamy}). For example
we have \cite{KoashiW-monogamy}
\begin{equation}
\ef(C:A) + C_{HV}(B\>A) = S_A,
\end{equation}
where $C_{HV}$ is a measure of classical \cite{HendersonVedral} given
by \ref{eq:chv}.

This says that if with one system we share too much entanglement, this
must suppress classical correlations with the other system. One might
think, that since classical correlations are suppressed, then also
quantum should, as existence of quantum correlations imply existence
of classical ones. Qualitatively it is indeed the case: no classical
correlations means that a state is product, hence cannot have
entanglement too. Quantitatively the issue is a bit more complicated,
as $\ef$ can be greater than $C_{HV}$. However, if other prominent
measures of entanglement would be smaller than $C_{HV}$, we could
treat it as an oddity of $E_C$ as mentioned above.  To our knowledge
it is not known, whether $E_R$, $E_D$, $K_D$ are smaller
than $C_{HV}$. Let us note that this latter relation would imply
monogamy for those measures (since they are all upper bounded by
$\ef$).

A beautiful monogamy was found for Bell inequalities. Namely, basing
on the earlier results concerning a link between the security of
quantum communication protocols and violation of Bell's inequalities
\cite {Scarani1} and theory of nonlocal games \cite{Cleve}, Toner
proved the CHSH inequality is monogamous \cite {Toner}. This means,
that if three parties A,B and C share quantum state $\rho$ and each
chooses to measure one of two observables then trade-off between
$AB$'s and $AC$'s violation of the CHSH inequality is given by
\begin{equation}
|{\tr}(\bcal^{AB}_{CHSH}\rho)|+ |{\tr}(\bcal^{AC}_{CHSH}\rho)|\leq 4.
\label{trad}
\end{equation}
It shows clearly that CHSH correlations are monogamous i.e. if $AB$
violate the CHSH inequality, then $AC$ cannot. This is compatible with
the present ideas of drawing cryptographic key from nonlocality, where
monogamy of nonlocal correlations is the main feature allowing to
bound the knowledge of eavesdropper.

There is a qualitative aspect of monogamy, recognized quite early
\cite{Werner1989-sym-ext,DohertyPS04}. Namely, a state $\rho_{AB}$ is
separable if and only if for any $N$ there exists its $N+1$ partite
{\it symmetric extension}, i.e. state $\rho_{A B_1\ldots B_N}$, such
that $\rho_{AB_i}=\rho_{AB}$. D. Yang in a recent paper
\cite{Yang2006-sym-ext} has provided an elegant proof of this result
and has given an explicit bound on the number of shared in terms of
quantity $G$ (see Sec.  \ref{subsubsec:cost-addit}) which is a mixed
convex roof from Henderson-Vedral measure of classical correlations,
and itself is indicator of entanglement, in a sense that it is zero if
and only if a state is entangled. Namely we have
\begin{equation}
N\leq {S_A\over G(A\>B)}.
\end{equation}
For pure states $S_A=G(A\>B)$ which shows that no symmetric extension
is possible, if only state is entangled, i.e.  $G\not=0$.

\section{Entanglement in continuous variables systems}
\subsection{Pure states}

Many properties of entanglement (separability) change when passing to
continuous variables since the infinite dimensional Hilbert space is
not compact.

The term continuous variables comes from the fact that any infinite
dimensional Hilbert space is isomorphic to any of the two spaces:
\begin{itemize}
\item[(i)] $l^{2}({\cal C})$ which is space of sequences $\Psi=\{
  c_{i} \}$ with $\sum_{i=1}^{\infty} |c_{i}|^{2} < \infty $ and
  scalar product $\langle\Psi|\Phi\rangle=\sum_{i=1}^{\infty}
  a_{i}^{*}b_{i}$) and

\item[(ii)] space $L^{2}(R)$ of all functions $\Psi:R \rightarrow
  {\cal C}$ with $\int_{R}|\Psi(x)|^{2}(x)dx < \infty$ and scalar
  product defined as $\int_{R}\Psi(x)^{*} \Phi(x) dx $ The variable
  $x$ is just a {\it continuous variable} (CV) here.
\end{itemize}

An example of entangled state from such space is two-mode squeezed
state: which has its $l^{2}\otimes l^{2}$-like representation (in so
called Fock basis considered to be a standard one):
\begin{equation}
|\Psi_{\lambda}\>=\sqrt{1-\lambda^{2}}\sum_{n=0}^{\infty}\lambda^{n}|n\rangle|,
n \rangle \label{squeezed}
\end{equation}
where index goes from zero for physical reasons ($n$ represents the
photon number). Here coefficients $a_{n}:=(1-\lambda^{2})\lambda^{n}$
are just Schmidt coefficients.

Alternatively the state has its $L^{2}\otimes L^{2}$ representation:
\begin{equation}
\Psi_{\lambda}(q_1,q_2)=\sqrt{\frac{2}{\pi}}\exp[-e^{-2r}(q_1+q_2)^{2}/2
-e^{2r}(q_1-q_2)^{2}/2],
\end{equation}
related to previous representation  by
\begin{equation}
\lambda=\tanh(r).
\end{equation}
In the case of infinite squeezing $r \rightarrow \infty$ the
$\Psi(q_1,q_2)$ becomes more and more similar to the delta function
$\delta(q_{1} - q_{2})$ while its Fourier transform representation
(changing ``positions'' $q_{i}$ into ``momenta'' $p_{i}$) becomes
almost $\delta(p_{1} + p_{2})$. This limiting case was originally
discussed in the famous EPR paper \cite{EPR}, and perfect correlations
in positions as well as momenta resemble us perfect correlations of
local measurements $\sigma_{x}$ and $\sigma_{z}$ on sides of the
two-qubit state $\Psi_{+}=\frac{1}{\sqrt{2}}(|0\rangle|0\rangle +
|1\rangle| 1\rangle)$.

The separability property are in case of bipartite CV pure states easy
--- as in discrete case we have for bipartite pure states equivalence
\begin{eqnarray}
\textrm{separability} &\Leftrightarrow& \textrm{PPT criterion} \nonumber \\
& \Leftrightarrow& \textrm{reduced state pure} \nonumber \\
&\Leftrightarrow& \textrm{Schmidt rank one}
\end{eqnarray}

The entropy of entanglement of pure states remains a good measure of
entanglement, exhibiting however some oddities. In the case of the
state (\ref{squeezed}) it is given by
\cite{GaussianEoF,GaussianEoFPRA}
\begin{equation}
E_{F}(\Psi_\lambda)= \cosh^2(r)\log_2(\cosh^2(r)) - \sinh^2(r)
\log_2(\sinh^2(r)).
\end{equation}
However in bipartite case with both subsystems of CV type typically
the entropy of entanglement is {\it infinite}.  This is a consequence
of the fact that generically quantum states on CV spaces have the
entropy infinite \cite{Wehrl} (or alternatively the set of density
matrices with finite entropy is {\it nowhere dense} i.e. contains no
ball).

As an example of such state take $\Psi_{AB}=\sum_{n} \sqrt{p_{n}}
|n\rangle_{A}|n\rangle_{B}$, with $p_{n}$ proportional (up to
normalization factor) to ${1}/{(n+2) \log(n + 2)^{4}}$. Then entropy
of entanglement is infinite, since the series $\sum_{n}p_{n} \log
p_{n}$ is not convergent.

As one can expect the very important fact connected to it is that
there is no maximally entangled state in such spaces. Simply the state
with all Schmidt coefficients equal does not mathematically exists (it
would have infinite norm).

A natural question here arises: what about usefulness of infinite
entanglement which is so common phenomenon in CV?  For example is it
possible to distill infinite amount of two-qubit entanglement from
single copy of bipartite CV quantum states with infinite entanglement?
The answer to this is negative and can be proven formally
\cite{InfiniteEntanglementKeyl}.

Also, one can easily see that even for pure states $\ef$ is not
continuous \cite{InfiniteEntanglementEisert}. Consider the sequence of
states
\begin{equation}
|\Psi_{AB}^{n}\>=\sqrt{1-\frac{1}{n\log n}} |e_{0}\rangle |f_{0}\rangle
+ \sqrt{\frac{1}{n \log n}} \sum_{i=1}^{n}| e_{i} \rangle
|f_{i}\rangle
\end{equation}
it is elementary to see that the sequence converges to a state with
zero entanglement, $\| |\Psi_{AB}\rangle \langle\Psi_{AB}| -
|e_{0}\rangle\langle e_{0}| \otimes
 |f_{0}\rangle\langle f_{0} \|_1 \rightarrow 0$,
while its entanglement diverges ($\ef(\Psi_{AB}^{n})\rightarrow
\infty$).

\subsection{Mixed states}

The definition of mixed separable states has to be changed slightly if
compared to discrete variables: the state is separable if it is a
limit (in trace norm) of finite convex combination of, in general
mixed and not pure, product states\footnote{This is actually the
  original definition of separable states \cite{Werner1989}.}:
\begin{equation}
\|\varrho^{sep}_{AB} - \sum_{i}p_{i}^{(n)} \varrho^{(n),i}_{A}
\otimes \varrho^{(n),i}_{B}\|_1 \rightarrow 0.
\end{equation}

The characterization of entanglement in terms of positive maps and
entanglement witnesses is again true since the corresponding proofs
are valid for general Banach spaces (see \cite{sep1996}).

There is a small difference here: since there is no maximally
entangled state one has to use the original version of Jamio\l{}kowski
isomorphism between positive maps and entanglement witnesses:
\begin{equation}
W^{\Lambda}_{AB}=(\id \otimes \Lambda)(V_{AA'}),
\end{equation}
with $d_{A}\geq d_{B}$ (remember that it may happen that only one of
subspaces is infinite) where $V_{AA'}$ is a swap operator on ${\cal
  H}_{AA'}={\cal H}_{A}\otimes {\cal H}_{A'}$, (${\cal H}_{A'}$ being
a copy of ${\cal H}_{A}$) and the map acting from system $A'$ to $B$.
This is because there is no maximally entangled state in infinite
dimensional space.

The PPT criterion is well defined and serves as a separability
criterion as it was in finite dimension. It has a very nice
representation in terms of moments \cite{SV}.

There exist nontrivial PPT states \cite{boundCV} that cannot be
constructed as a naive, direct sum of finite dimensional structures.
It seems that such states are generic, though the definition of
generic CV state in case of mixed state is not so natural as in case
of pure state where infinite Schmidt rank (or rank of reduced density
matrix) is a natural signature defining CV property (for discussion of
the property of being generic see \cite{BoundCon}).

The first important observation concerning CV separability is
\cite{CliftonHalvorson02} that in bipartite case the set of separable
states is nowhere dense or --- equivalently --- any state on this
space is a limit (in trace norm) of sequence of entangled state.  Thus
set of separable states contains {\it no} ball of finite radius and in
that sense is ``of zero volume'' unlike it was in finite dimensions
\cite{ZyczkowskiHSP-vol}.  This result can be extended \cite{BoundCon}
to the set of all nondistillable states (in a sense of definition
inherited from discrete variables i.e. equivalent to impossibility of
producing two-qubit \singlets) is also nowhere dense in set of all
states.  Thus CV bound entanglement like CV separability is a rare
phenomenon.

Let us pass to quantitative issues involving entanglement
measures. If one tries to extend definition of entanglement of
formation to mixed states \cite{InfiniteEntanglementEisert} then again
set of states with finite $\ef$ has the same property as the set of
separable states --- it is again nowhere dense.\footnote{This problem
  does not occur when one of the local Hilbert spaces ${\cal H}_{A}$,
  ${\cal H}_{B}$ is finite i.e. $min[d_{A},d_{B}] <\infty$ then
  entanglement of formation is well defined and restricted by
  logarithm of the dimension of the finite space
  \cite{MajewskiJPA2002}.} Also, it is not continuous (as we have
already seen in the case of pure states).

The question was how to avoid, at least partially, the above problems
with entanglement that occur when both dimensions are infinite? The
authors of \cite{InfiniteEntanglementEisert} propose then to consider
subset ${\cal S}_{M}(H) \subset {\cal S}$ (${\cal S}$ set of all
bipartite states defined as ${\cal S}_{M}(H):= \{ \varrho:
{\tr}(\varrho H)< M \}$ for some fixed constant $M$ and Hamiltonian H
(some chosen Hermitian operator with spectrum bounded from below). The
set is nowhere dense but it is defined by natural physical requirement
of bounded mean energy in a physical system.  Remarkably for fixed $M$
and all states from ${\cal S}_{M}$, the entanglement of formation
$E_{F}$ and the relative entropy of entanglement $E_R$ are continuous
in trace norm on pure states.  Moreover those measures are
asymptotically continuous on pure states of the form $\sigma^{\otimes
  n}$ with finite-dimensional support of $\sigma $.

\subsection{Gaussian entanglement}

There is a class of CV states that are very well characterized with
respect to separability. This is the class of Gaussian states.
Formally a {\it Gaussian state} of $m$ modes (oscillators) is a
mixed state on Hilbert space ${\cal L}^{2}(R)^{\otimes m}$ (of
functions of $\xi=[q_{1},\ldots,q_{m}]$ position variables) which is
{\it completely characterized} by the vector of its first moments
$d_{i}={\tr}(\varrho R_{i})$ (called {\it displacement vector}) and
second moments {\it covariance matrix}
$\gamma_{ij}={\tr}(\varrho\{R_{i}-d_{i}\id),(R_{j}-d_{j}\id\}_{+})$, where
we use anticommutator $\{ , \}_{+}$ and the cannonball observables
are position $Q_{k}=R_{2k}$ and momentum $P_{k}=R_{2k+1}$ operators
of $k$-th oscillator which satisfy the usual Heisenberg {\it
commutation relations} $[R_{k},R_{k'}]=iJ_{kk'}$ where
$J=\oplus_{i=1}^{n} J_{i}$, with one mode symplectic matrices
$J_{i}=\left(
\begin{array}{cc}
0 &-1\\
1 &0
\end{array} \right)$.
Any general matrix $S$ is called {\it symplectic} iff it satisfies
$SJS^{T}=J$. They represent all canonical transformations
$S:\xi\rightarrow \xi'=S\xi$ where
$\xi=[q_{1},p_{1};q_{2},p_{2};\ldots;q_{m}p_{m}]^{T}$ is a vector of
{\it canonical variables}. The corresponding action on the Hilbert
space is unitary. There is also a broader set of unitary operations
called {\it quasifree} or {\it linear Bogolubov} transformations
$\xi \rightarrow S \xi + d$ where $S$ is symplectic.

The canonical operators $Q_{i},P_{i}$ are real and Hermitian part of
the creation $a_{k}^{\dagger}$ and annihilation $a_{k}$ operators
that provide a natural link to $(l^{2})^{\otimes m}$ representation
since they distinguish a special $l^{2}$ {\it Fock basis} $\{
|n\rangle \}$ of each mode ($a_{k}^{\dagger}=\sum_{n=0}^{\infty}
\sqrt{n+1}|n+1 \rangle_{k} {}_{k}\langle n|$ and
$a_{k}=(a_{k}^\dagger)^\dagger $) {\it via} number operator
$N=a_{k}^{\dagger}a_{k}=\sum_{n=0}^{\infty}n|n\rangle\langle n|$
which is diagonal in that basis.

Since displacement $d$ can be easily removed by quasifree {\it
local} (i.e. on each mode separately) unitary operations
\cite{DuanGCZ1999-criterion}), only the properties of variance matrix are
relevant for entanglement tests. Before recalling them we shall
provide conditions for $\gamma$ to be physical. Let us recall that
{\it via} Williamson theorem $\gamma$ can be diagonalised with some
symplectic matrix $\gamma_{diag}=S_{V}\gamma
S_{V}^{T}=diag[\kappa_{1},\kappa_{1};\ldots;\kappa_{m},\kappa_{m}]$,
with $\kappa_{i}$ real. The physical character of the variance
matrix $\gamma$ is guaranteed by the condition:
\begin{eqnarray}
&&\gamma+iJ\geq 0\Leftrightarrow \\
\label{1}
&&\gamma\geq J^{T}\gamma^{\-1}J \Leftrightarrow \\
\label{2}
&&\gamma\geq S^{T}S \Leftrightarrow \\
\label{3}
&&\kappa_{i}\geq 1, i=1,\ldots,m.
\label{4}
\end{eqnarray}
There is a remarkable fact\cite{Simon}:
Gaussian state is pure if and only if its variance matrix is of the form
\begin{eqnarray}
\gamma=S^{T}S,
\label{Pure}
\end{eqnarray}
for some symplectic matrix $S$. Generally, given $m$ modes
can be divided into $k$ groups containing $m_{1},\ldots,m_{2},m_{k}$
modes ($m=\sum_{i}m_{i}$), belonging to different local observers
$A_{1},\ldots,A_{k}$. We say that the state is a $k$-partite {\it
Gaussian state of $m_{1} \times m_{2} \times \cdots \times m_{k}$
type}. For instance the bipartite state is  of $m_{1} \times m_{2}$
type iff the first $m_{1}$ modes are on Alice side, and the rest
$m_{2}$ on Bob one. All reduced states of the systems $A_{i}$ are
Gaussians. With each site we associate the symplectic matrix
$J_{A_{k}}$ as before. There is general necessary and sufficient
separability condition that can resemble to some extent the range
criterion (see \cite{WernerWolf}):
\begin{eqnarray}
&&\varrho \ (\text{Gaussian separable} \Leftrightarrow \\
&&\gamma_{AB}(\varrho) \geq \gamma_{A} \oplus \gamma_{B}
\label{GaussMajor}
\end{eqnarray}
for some variance matrices $\gamma_{A}$,$\gamma_{B}$ (which, as
further was shown \cite{Simon1}, can be chosen to be pure, i.e. of the
form (\ref{Pure})). Quite remarkably the above criterion can be
generalized to an arbitrary number of parties (see \cite{Eisert}).  In
general --- if the state is not Gaussian the criterion becomes only a
necessary condition of separability. The criterion is rather hard to
use (see however \cite{WernerWolf} and the discussion below).

The PPT separability criterion takes a very easy form for
Gaussians. On the level of canonical variables partial transpose with
respect to the given subsystem (say $B$) corresponds to reversal of
its conjugate variables $\xi=[\xi_{A},\xi_{B}] \rightarrow
\tilde{\xi}=\Lambda$ where $\Lambda=diag[\id_{A},\sigma^z_{B}]$.  If
we introduce transformation
\begin{equation}
  \tilde{X}=\Lambda X \Lambda,
\end{equation}
then we have PPT criterion for Gaussian states: bipartite Gaussian
state is PPT iff $\tilde{\gamma}$ is physical i.e. satisfies either of
the conditions (\ref{1})-(\ref{4}). Note that the first one satisfied
by $\tilde{\gamma}$ can be written as:
\begin{equation}
\gamma + i \tilde{J} \geq 0.
\end{equation}
There is a very important separability characterization: PPT criterion
has been shown to be {\it both} necessary and sufficient for $1 \times
1$ \cite{Simon,DuanGCZ1999-criterion} and subsequently generalized to
$1 \times n$ Gaussians \cite{WernerWolf}. Further the same result has
been proven for $m \times n$ ``bisymmetric'' \cite{Serafini,SerafiniAI2005}
(i.e. symmetric under permutations of Alice and Bob modes
respectively) Gaussian states. The equivalence of PPT condition to
separability is {\it not} true, in general, if both Alice and Bob have
more than one mode. In particular an example of $ 2 \times 2$ Gaussian
bound entanglement has been shown by proving ({\it via} technique
similar to that from \cite{LewensteinSanpera-bsa}) that the following
covariance matrix
\begin{equation}
\gamma=\left(\begin{array}{cccccccc}
2 & 0&0 & 0& 1&0 &0 &0\\
0 & 1&0 & 0& 0&0 &0 &-1\\
0 & 0&2 & 0& 0&0 &-1 &0\\
0 & 0&0 & 1& 0&-1 &0 &0\\
1 & 0&0 & 0& 2&0 &0 &0\\
0 & 0&0 & -1& 0&4 &0 &0\\
0 & 0&-1 & 0& 0&0 &2 &0\\
0 &-1&0 & 0& 0&0 &0 & 4
\end{array} \right)
\end{equation}
does not satisfy the condition (\ref{GaussMajor}).

The PPT test for $1\times 1$ Gaussian states
can be written elementary if we represent the
variance as $\gamma=\left(
\begin{array}{cc}
A &C\\
C^{T} &B
\end{array} \right)$.
Then the two elementary conditions \cite{Simon} $\det(A)det(B)+(1/8
\pm \det(C))^{2} - \tr(AJCJBJC^{T}J)/8 + (\det(A) +\det(B))/8 \geq 0$
represent the physical and PPT character of $\gamma$ respectively.
These conditions can be further simplified if the variance is driven
by local linear unitaries (corresponding to symplectic operations)
to canonical form of type I, where matrices A and B are proportional
to identities and C is diagonal \cite{Simon}. An interesting and
more fruitful for general (nongaussian) case approach is offered by
uncertainty relations approach where the so called type II is
achieved by local quasi free unitary operations and PPT condition is
represented by some {\it uncertainty relation}
(see \cite{DuanGCZ1999-criterion} and subsequent section).

In \cite{1x1x1} the separability problem for three mode Gaussian
states was also completely solved in terms of operational criterion.

{\it Operational necessary and sufficient condition.}  In
\cite{GaussianAlgorithm} operational necessary and sufficient
condition for separability for all bipartite Gaussian states have been
presented. It is so far the only operational criterion of separability
that detects all PPT entangled states within such broad class of
states. Entanglement is detected {\it via} a finite algorithm, that
transforms the initial covariance matrix into a sequence of matrices
which after finite number of steps (i) either becomes not physical
(not represents a covariance matrix) and then the algorithm detects
entanglement (ii) or its special affine transformation becomes
physical and then the initial state is recognized to be separable.

\subsection{General separability criteria for continuous variables}
\label{subsec:sep_cont_var}

One of the natural separability criteria is local projection or --- in
general LOCC transformation of CV state onto product of finite
dimensional Hilbert spaces and then application of one of separability
criteria for discrete variables. This method was used in
\cite{boundCV} where finally discrete variables range criterion was
applied.

First, it must be stressed that any Gaussian separability criterion
that refers only to well defined variances, and does not use the fact
that the variance matrix completely describes the state, is also a
separability criterion for general CV states.

Separability criteria that do not refer to discrete quantum states
usually are based on some uncertainty type relations. As an example
of such relation, consider position and momenta operators
$Q_{A_{1}}$, $Q_{A_{2}}$, $P_{A_{1}}$, $P_{A_{2}}$ for a bipartite
system $A_{1}A_{2}$, which satisfy the commutation relations
$[Q_{A_{i}},P_{A_{j}}]=i\delta_{ij}$ and define
$U=|a|Q_{A_{1}}+\frac{1}{a}Q_{A_{2}}$,
$V=|a|P_{A_{1}}+\frac{1}{a}P_{A_{2}}$ for arbitrary nonzero real
number $a$. Then any separable bipartite CV state $\varrho$
satisfies \cite{DuanGCZ1999-criterion}
\begin{equation}
\langle (\Delta U)^{2} \rangle_{\varrho} + \langle (\Delta V)^{2} \rangle_{\varrho}
\geq \frac{1}{2}(a^{2} + \frac{1}{a^{2}}).
\end{equation}
The above criterion can be modified to the form which refers to so
called type II standard form of two mode variance and becomes then
necessary and sufficient separability criterion (and hence equivalent
to PPT) in this two mode case \cite{DuanGCZ1999-criterion}. Further,
it is interesting to note that the PPT separability criterion implies
a series of uncertainty principle like relations. For any $n\times n$
mode state the physical condition $\gamma_{AB} + iJ\geq 0$ and
observables equivalent to the statement that observables
$X(d)=\vec{d}\vec{\xi}$,\,$X(d')=\vec{d'}\vec{\xi}$ obey the
uncertainty relation (see \cite{Simon1})
\begin{equation}
\langle (\Delta X(d))^{2} \rangle_{\varrho} + \langle (\Delta X(d'))^{2} \rangle_{\varrho}
\geq |\vec{d}_{A}J_{A}\vec{d}'_{A} + \vec{d}_{B}J_{B}\vec{d}'_{B}|,
\end{equation}
for any real vectors $\vec{d}=(\vec{d}_{A},\vec{d}_{B})$,
$\vec{d'}=(\vec{d}'_{A},\vec{d}'_{B})$.
The PPT condition leads to another restriction \cite{Simon1}
\begin{equation}
\langle (\Delta X(d))^{2} \rangle_{\varrho} + \langle (\Delta X(d'))^{2} \rangle_{\varrho}
\geq |\vec{d}_{A}J_{A}\vec{d}'_{A} - \vec{d}_{B}J_{B}\vec{d}'_{B}|,
\end{equation}
which may be further  written in one combined inequality with
$|\vec{d}_{A}J_{A}\vec{d}'_{A}| +|\vec{d}_{B}J_{B}\vec{d}'_{B}|$ in
its RHS. Special case of this inequality was considered by
\cite{Giovannetti1} together with relation to other criteria.
Note that the above criterion is only necessary for the state to
satisfy PPT criterion since it refers to the variance including only the first
and second moments.

The practical implementation of PPT criterion in terms of all moments
that goes beyond variance properties of CV states is Shchukin-Vogel
criterion \cite{SV,MiranowiczP}.  It turned out that their criterion
covers many known separability criteria.  The idea is that with any
state of two modes, one can associate the following matrix of moments
\begin{equation}
M_{ij}=\tr (\hat a^{\dagger q} \hat
a^{p}\hat a^{\dagger n} \hat a^{m} \ot \hat b^{\dagger l} \hat
b^{k}\hat b^{\dagger r} \hat b^{s}\,\rho_{AB}),
\end{equation}
where $i=(pqrs)$ and $j=(nmkl)$. The operators $a,b$ act on systems
$A,B$ respectively. It turns out that the above matrix is positive if
and only if the state is PPT\footnote{See \cite{VerchWerner} in this
  context.}. Positivity of matrix can be expressed in terms of
nonnegativity of subdeterminants. It then turns out that many known
separability criteria are obtained by imposing nonnegativity of a
suitably chosen subdeterminant.

An example is Simon criterion which for $1\times 1$ Gaussian states is
equivalent to PPT. One finds that it is equivalent to nonnegativity of
determinant of a $5\times 5$ main submatrix the matrix $M$. Other
criteria published in
\cite{DuanGCZ1999-criterion,ManciniGVT2001-criterion,RaymerFSG-cont-var,AgarwalB-cont-var,HilleryZ-cont-var}
can be also reduced to positivity conditions of some determinants of
matrix of moments.

The approach of Shchukin and Vogel was developed in
\cite{MiranowiczPHH2006-moments}. The authors introduced a modified
matrix of moments:
\begin{equation}
M_{kk'll'}=\tr ( (\hat a^{\dagger k_1} \hat
a^{k_2})^\dagger \hat a^{\dagger k_1'} \hat a^{k_2'} \ot (\hat
b^{\dagger l_1} \hat b^{l_2})^\dagger \hat b^{\dagger l_1'} \hat
b^{l_2'} \,\rho_{AB}),
\end{equation}
so that it is labeled with four indices, and can be treated itself (up
to normalization) as a state of compound system. Now, it turns out
that the original state is separable, then so is the matrix of
moments. In this way, one can obtain new separability criteria by
applying {\it known} separability criteria to matrix of moments. The
PPT condition applied to matrix of moments turns out to be equivalent
to the same condition applied to original state. Thus applying PPT to
matrix of moments reproduces Shchukin-Vogel result. What is however
intriguing, that so far no criterion stronger than PPT was found.

\subsection{Distillability and entanglement measures of Gaussian states}
\label{subsec:gaussian-measures}

The question of distillability of Gaussian states has attracted a lot
of effort. In analogy to the two-qubit distillability of quantum
states in finite dimensions, it has been first shown, that all
two-mode entangled Gaussian states are distillable
\cite{G2D_GieDuaCirZol}. Subsequently it was shown, that all NPT
entangled Gaussian states are distillable \cite{GallD_GieDuaCirZol}.
In other words, there is no NPT bound entanglement in Gaussian
continuous variables: any NPT Gaussian state can be transformed into
NPT two-mode one, and then distilled as described in
\cite{G2D_GieDuaCirZol}. However the protocol which achieves this task
involves operations which are not easy to implement nowadays.  The
operations feasible for present linear-optic based technology are so
called {\it Gaussian operations}. The natural question was risen then,
whether entangled Gaussian states are distillable by means of this
restricted class of operations. Unfortunately, it is not the case: one
cannot obtain pure entanglement from Gaussian states using only
Gaussian operations \cite{GnonD_GieCir} (see in this context
\cite{GnonD_Fiu} and \cite{GnonD_EisSchPle}). Although these
operations are restrictive enough to effectively ``bind''
entanglement, they are still useful for processing entanglement: by
means of them, one can distill key from entangled NPT Gaussian states
\cite{GKD_NavBaeCirLewSanAci}. Interestingly, no PPT Gaussian state
from which key can be distilled is known so far
\cite{GPPTnonD_Nav_Aci}.

Apart from question of distillability and key distillability of
Gaussian states, entanglement measures such as entanglement of
formation and negativity have been studied. It led also to new
measures of entanglement called {\it Gaussian entanglement measures}.

In \cite{GEform_GieVolKruWerCir} entanglement of formation was
calculated for symmetric Gaussian states. Interestingly, the optimal
ensemble realizing $\ef$ consists solely from Gaussian states. It is
not known to hold in general. One can however consider the so called
{\it Gaussian entanglement of formation} $E_G$ where infimum is taken
over decompositions into Gaussian states only. Gaussian entanglement
of formation was introduced and studied in
\cite{GEform_GieVolKruWerCir}. It is shown there, that $E_G$ is
monotonous under Gaussian operations. For two-mode Gaussian states its
value can be found analytically. If additionally, the state is
symmetric with respect to sites, this measure is additive. On a single
copy it is shown to be equal to $\ef$.

The idea of Gaussian entanglement of formation has been extended to
other convex-roof based entanglement measures in
\cite{GNeg_AdeFab}. The log-negativity of Gaussian states defined
already in \cite{Vidal-Werner} has also been studied in
\cite{GNeg_AdeFab}. In this case the analytic formula has been found,
in terms of symplectic spectrum $\lambda_i$ of the partially
transposed covariance matrix:
\begin{equation}
E_N = -\sum_{i=1}^{n}\log_2 [\min(1,\lambda_i)].
\end{equation}

The continuous variable analogue {\it tangle} (squared concurrence, see sec \ref{sec:miary}), called
{\it contangle} was introduced in \cite{GTangle_AdeFab} as the Gaussian convex
roof of the squared negativity. It is shown, that for three-mode
Gaussian states contangle exhibits Coffman-Kundu-Wootters monogamy.
Recently the general monogamy inequality for all N-mode Gaussian states
was established \cite{HiroshimaAI2007} (in full analogy with the qubit case \cite {OsborneV-monogamy}).
For three modes, the 3-contangle  - analogue of Coffman-Kundu-Wooters 3-tangle is monotone under Gaussian operations.

Surprisingly, there is a symmetric Gaussian state which is a
counterpart of both GHZ as well as W state \cite{GTangle_AdeFab}.
Namely, in finite dimension, when maximizing entanglement of
subsystems, one obtains $W$ state, while maximization of tangle leads
to GHZ state. For Gaussian states, such optimizations (performed for a
fixed value of mixedness or of squeezing of subsystems) leads to a
{\it single} family of pure states called $GHZ/W$ class. Thus to
maximize tripartite entanglement one has to maximize also bipartite
one.

An exemplary practical use of Gaussian states apart from the quantum
key distribution (e.g. \cite{Gcrypto_GottesmanPreskill}) is the
application for continuous quantum Byzantine agreement protocol
\cite{GByzant_NeigSan}. There are many  other theoretical and experimental issues
concerning Gaussian states and their entanglement properties,
that we do not touch here. For a recent review on this topic see
\cite{G_review05,AdessoFE2007}.

\section{Miscellanea}

\subsection{Entanglement under information
loss: locking entanglement}
\label{subsec:locking}
Manipulating a quantum state with local operations and classical
communication in non-unitary way usually decreases its entanglement
content. Given a quantum bipartite system of $2\times\log d$ qubits in
state $\rho$ one can ask how much entanglement can decrease if one
traces out a single qubit. Surprisingly, a lot of entanglement
measures can decrease by arbitrary large amount, i.e. from
$O(\log d)$ to zero.
Generally, if some quantity of $\rho$ can decrease by an arbitrarily
large amount (as a function of number of qubits) after LOCC operation on
few qubits, then it is called {\it lockable}. This is because a huge
amount of quantity can be controlled by a person who posses only a
small dimensional system which plays a role of a ``key'' to this
quantity.

The following related question was asked earlier in \cite{Wilkens}:
how entanglement behaves under classical information loss?
It was quantified by means of entropies
and for convex entanglement measures it takes the form
\be
\Delta E \leq \Delta S
\label{eq:ent-info}
\ee
where $\Delta E= \sum_i p_i E(\rho_i)-E(\sum_ip_i\rho_i$ and
$\Delta S=S(\rho)-\sum_i p_is(\rho_i)$.
It holds for relative entropy of entanglement \cite{LPSW1999}
(see also \cite{Synak05-asym}). It turns out however that this inequality
can be drastically violated, due to above locking effect\footnote{Using
the fact that  that a loss of  one qubit can be simulated by
applying one of four random Pauli matrices
to the qubit one easily arrives at the connection between locking,
and violation of the above inequality.}.

The phenomenon of drastic change of the content of state  after tracing out one qubit, was earlier recognized in \cite{DiVincenzo-locking}  in the case of
classical correlations  of quantum states (maximal mutual information of
outcomes of local measurements) (see in this context
\cite{string-commitment,bh-locking,small-locking,3basis-locking}).
Another effect of this sort was found in a classical key agreement
\cite{renner-wolf-gap} (a theory bearing some analogy to
entanglement theory, see sec. \ref{subsec:inter-ent-CKA}).

Various entanglement measures have been shown to be lockable.
In \cite{lock-ent} it has been shown that {\it
entanglement cost}, and {\it log-negativity} are lockable measures.
The family of states which reveal this property of measures (called
{\it flower states}) can be obtained via partial trace over system $E$
of the pure state
\begin{eqnarray}
|\psi_{AaBE}^{(d)}\> = {1\over \sqrt{2d}}\sum_{i=0}^{d-1} |i\>_A|0\>_a\{|i\>|0\>\}_B|i\>_E
\label{eq:flower-purif} \\+ |i\>_A|1\>_a \{|i\>|1\>\}_B U|i\>_E,
\nonumber
\end{eqnarray}
where $U=H^{\ot \log d}$ is a tensor product of Hadamard
transformations. In this case $E_C(\rho_{AaB}) = {1\over 2}\log d $
and $E_N(\rho_{AaB})= \log (\sqrt{d} +1)$.  However after tracing
out qubit $a$ one has $E_C(\rho_{AB})=E_N(\rho_{AB})=0$, as the state
is separable. The gap between initial $E_C$ and final one, can be made
even more large, using more unitaries instead of $U$ and $\id$ in the
above states and the ideas of randomization \cite{RandomQstates}.

The examples of states for which $\ef$ can change from nearly maximal
($\log d $) to nearly zero after tracing out $O(\log \log d)$ qubits
was found in \cite{AspGenEnt}. It can indicate either drastic
nonadditivity of entanglement of formation, or most probably an
extreme case of irreversibility in creation-distillation LOCC
processes\footnote{Because distillable entanglement of the states
  under consideration is shown to be small we have either $\ef >>E_C$ (reporting nonadditivity of $\ef$) or $\ef =E_C>>E_D$ (reporting extreme creation-distillation irreversibility).}. The analogous effect has been found earlier in
classical key agreement \cite{renner-wolf-gap}. It was shown that {\it
  intrinsic information} $I(A:B\downarrow Ee)$ (an upper bound on
secret key extractable from triples of random variables) can decrease
arbitrarily after erasing 1-bit random variable $e$ (see Sec. \ref{subsec:inter-ent-CKA}).

In \cite{ChristandlW-lock} it was shown that {\it squashed
entanglement} is also lockable measure, which is again exhibited by
flower states. The authors has shown also that regularized
entanglement of purification $E_p^{\infty}$ is lockable quantity, as it
can go down from maximal to zero after erasing one-qubit system $a$ of
the state
\begin{equation}
\omega_{aAB} = {d+1\over 2d}|0\>\<0|_a \ot
P_{AB}^{(+)} + {d-1\over 2d}|1\>\<1|_a \ot P_{AB}^{(-)},
\label{eq:eplocking}
\end{equation}
with $P_{AB}^{(\pm)}$ being normalized projectors onto symmetric and
antisymmetric subspace respectively (see Sec. \ref{subsubsec:Witnesses}).  Although $E_p^{\infty}$ is not
an entanglement measure, it is apart from $E_C$ another example of
lockable quantity which has operational meaning \cite{IBMHor2002}. In
case of (\ref{eq:eplocking}) $E_p = E_p^{\infty}$, so the $E_p$ itself
also can be locked.

It is known that all measures based on the convex roof method (see Sec. \ref{subsec:roof}) are lockable. There is also a general connection between lockability and
asymptotic non-continuity. Namely it was shown in \cite{lock-ent} that
any measure which is (i) subextensive i.e. bounded by $M \log d$ for
constant M, (ii) convex (iii) {\it not} asymptotically continuous, is
lockable. The only not lockable measure known up to date is the
relative entropy of entanglement \cite{lock-ent},
which can be derived with help of (\ref{eq:ent-info})

Since fragility of measure to one qubit erasure is a curious property,
for any lockable entanglement measure $E$ one could design its {\it
reduced} version, that is a non-lockable version of $E$.
One of the possible definitions of {\it reduced entanglement measure} is proposed in \cite{lock-ent}. Possible consequences of locking for multipartite entanglement
measures can be found in \cite{GroismanLindenPopescu}.

Although the phenomenon of locking shows that certain measures can
highly depend on the structure of particular states, it is not known how many
states have a structure which can contribute to locking of measures,
and what kind of structures are specific for this purpose.

It is an open question whether {\it distillable entanglement} can be
locked, although it is known that its {\it one-way} version is
lockable, which follows from monogamy of entanglement. In case of
distillable key, one can consider two versions of locking: the one
after tracing out a qubit from Eve ($E$-locking), and the one after
the qubit of Alice's and Bob's system is traced out ($AB$-locking).
It has been shown, that both classical and quantum {\it distillable
  key} is not $E$-lockable
\cite{renner-wolf-gap,EkertCHHOR2006-ABEkey}. It is however not known
if distillable key can be $AB-$lockable i.e. that if erasing a qubit
from Alice and Bob systems may diminish by far their ability to obtain
secure correlations. Analogous question remains open in classical key
agreement (see Sec. \ref{subsec:inter-ent-CKA}).

\subsection{Entanglement and distinguishing states by LOCC}
\label{sec:rozroz}

Early fundamental results in distinguishing by LOCC are the following:
there exist sets of orthogonal product states that are not perfectly
distinguishable by LOCC \cite{Bennett-nlwe} (see also \cite{WH2002})
and every two orthogonal states (even multipartite) are
distinguishable by LOCC \cite{Walgate-twoent}. In a qualitative way,
entanglement was used in the problem of distinguishability in
\cite{hiding-prl}.  To show that a given set of states cannot be
distinguished by LOCC, they considered all measurements capable to
distinguish them. Then applied the measurements to $AB$ part of the
system in state $\psi_{AA'}\ot \psi_{BB'}$ where the components are
maximally entangled states. If the state after measurement is
entangled across $AA':BB'$ cut, then one concludes that the measurement cannot be done by
use of LOCC because the state was initially product across this cut.

An interesting twist was given in \cite{BellPRL} where distillable
entanglement was used. Let us see how they argued that four Bell
states $\psi_i$ (\ref{basis}) cannot be distinguished. Consider fourpartite state
$\rho_{ABA'B'}={1\over 4}\sum_i |\psi_i\>\<\psi_i|_{AB} \ot
|\psi_i\>\<\psi_i|_{A'B'}$. Suppose that it is possible to distinguish
Bell states by LOCC. Then Alice and Bob will distinguish Bell states
of system $AB$ (perhaps destroying them). Then they will know which of
Bell states they share on system $A'B'$, obtaining then 1 e-bit of
pure entanglement (hence $E_D\geq 1$). However, one can check that the
initial state $\rho_{ABA'B'}$ is separable \cite{Smolin-unlock}, so that $E_D=0$
and we get a contradiction.  This shows that entanglement measures can
be used to prove impossibility of distinguishing of some
states. Initially entanglement measures have not been used for this
problem. It is quite nontrivial concept, as naively it could seem that
entanglement measures are here useless. Indeed, the usual argument
exploiting the measures is the following: a given task cannot be
achieved, because some entanglement measure would increase. In the
problem of distinguishing, this argument cannot be directly applied,
because while distinguishing, the state is usually destroyed, so that
final entanglement is zero.

This approach was developed in \cite{morenon} (in particular, it was shown
that any measure of entanglement can be used) and it was proved that a
full basis cannot be distinguished even probabilistically, if at least
one of states is entangled. Other results can be found in
\cite{VirmaniSPM2001-locc-discr, Fan-LOCCdist,Nathanson-LOCCdist}. In
\cite{VirmaniSPM2001-locc-discr} it was shown that for any two pure
states the famous Helstrom formula for maximal probability of distinguishing
by global measurements, holds also for LOCC distinguishing.

In \cite{Badziag-Holevo} a general quantitative bound for LOCC distinguishability
was given in terms of the maximal mutual information achievable by LOCC measurements
\footnote{It is defined as follows:
$I_{acc}^{LOCC}(\{p_i,\psi_i\})=\sup_{\{A_j\}}I(i:j)$ where
supremum is taken over measurements that can be implemented by LOCC, and $I(i:j)$
is mutual information between symbols $i$ and measurement outcomes $j$.}  $I^{LOCC}_{acc}$.
The bound can be viewed is a generalization of Holevo bound:
\begin{equation}
I_{acc}^{LOCC} \leq \log N- \overline E,
\end{equation}
where $N$ is number of states, and $\overline E$ is average
entanglement (it holds for any $E$ which is convex and is equal to
entropy of subsystems for pure states).  The usual Holevo bound would
read just $I_{acc}\leq \log N$, so we have correction coming from
entanglement. The following generalization of the above inequality
obtained in \cite{HorodeckiOSS-dist}
\begin{equation}
I_{acc}^{LOCC} \leq \log N-\overline E_{in} -\overline E_{out}
\end{equation}
shows that if some entanglement, denoted here by $\overline E_{out}$,
is to survive the process of distinguishing (which is required e.g in
process of distillation) then it reduces distinguishability
even more\footnote{The inequality was independently
conjectured and proven for one way LOCC in \cite{Ghosh2004-locc-acc}}.

Finally in \cite{HayashiMMOV-LOCCdist} the number $M$ of orthogonal
multipartite states distinguishable by LOCC was bounded in terms of
several measures of entanglement, in particular we have
\begin{equation}
M \leq D 2^{-{1\over M}\sum_i \bigl(S(\rho_i)+E_R(\rho_i)\bigr)},
\label{eq:locc-dist-jap}
\end{equation}
where $D$ denotes dimension of the total system, $S$ denotes the von Neumann entropy and $E_R$ is the relative entropy of entanglement (see Sec. \ref{subsec-distance}).
A striking phenomenon is possibility of ``hiding'' discovered in
\cite{hiding-prl,hiding-ieee}.  Namely, one can encode one bit into
two states $\rho_0$ and $\rho_1$, for which probability of error in
LOCC-distinguishing can be arbitrary close to $1$. In
\cite{WernerHide} it was shown that this can be achieved even with
$\rho_i$ being separable states.  Thus unlike for pure states, for two
mixed states, Helstrom formula does not hold any longer.

Finally, a lower bound on $I_{acc}^{LOCC}$ was provided in
\cite{SenSL2006-Ilocc-bound} and relation with entanglement
distillation was discussed.

\subsection{Entanglement and thermodynamical work}
\label{sec:delta}

In \cite{OHHH2001} an approach to composite systems was proposed based
on thermodynamics. Namely, it is known, that given a pure qubit and a
heat bath, one can draw amount $kT\ln 2$ of work\footnote{It is
  reverse of Landauer principle, which says that resetting bit/qubit
  to pure state costs the same amount of work.}. The second law is not
violated, because afterward the qubit becomes mixed serving as an
entropy sink (in place of reservoir of lower temperature) (see
\cite{Vedral1999,Scully-negentropy,silnik}).  In this way the noisy
energy from the heat bath is divided into noise, that goes to the
qubit, and pure energy that is extracted.  More generally, a $d$ level
system of entropy $S$ allows to draw amount of work equal to $kT (\log
d -S)$. If we treat $I=\log d -S$ as information about the state (it
is also called purity), then the work is proportional to
information/purity \cite{igor-deficit}.

Suppose now that Alice and Bob have local heat baths, and want to draw
work from them by means of shared bipartite state. If they do not
communicate, then they can only draw the work proportional to {\it
  local} purity. If they share two maximally classically correlated
qubits ,
local states are maximally mixed, and no work can be
drawn. However if Bob will measure his qubit, and tell Alice the
value, she can conditionally rotate her qubit, so that she obtains
pure qubit, and can draw work. For classically correlated systems, it
turns out that the full information content $\log d -S$ can be changed
into work. How much work can Alice and Bob draw from a given quantum
state, provided that they can communicate via a classical channel? It
turns out that unlike in the classical case, part of information
cannot be changed into work, because it cannot be concentrated via
classical channel. For example, for singlet state the information
content is $2$, but only $1$ bit of work (in units kT) can be drawn.

The amount of work that can be drawn in the above paradigm is
proportional to the amount of pure local qubits that can be obtained
(localisable information/purity). The difference between total
information $I$ and the localisable information represents purely
quantum information, and is called {\it quantum deficit}
\cite{OHHH2001}. It can be viewed as a measure of quantumness of correlations, which is zero for a classically correlated states\footnote {A bipartite state is properly classically correlated or shortly classically correlated, if it can be written in the form: $\rho_{AB}= \sum_{ij} p_{ij}|i\>\<i|\ot |j\>\<j|$ with coefficients $
0 \leq p _{ij}\leq 1$,\, $\sum_{ij} p_{ij} =1$;\, $\{|i\>\}$ and  $\{|j\>\}$ are local bases \cite{OHHH2001}.}. (A closely related quantity called {\it quantum  discord} was obtained earlier in \cite{Zurek-demons-02} on the basis
of purely information-theoretical considerations. It was then also
related to thermodynamics \cite{Zurek-discord}.) Recently other measures of quantumness correlations
based on distance from classically correlated states  were  introduced \cite {GroismanKM2007,SaiTohRN2007}.

Localisable information is a notion dual to distillable entanglement:
instead of singlets we want to distill product states.  Thus local
states constitute valuable resource (the class of involved operations
exclude bringing fresh local ancillas). It turns out that techniques
from entanglement distillation can be applied to this paradigm
\cite{nlocc,SynakHH04}. One of the main connections between
distillation of local purity and entanglement theory is the following:
{\it For pure bipartite states the quantum deficit is equal to entropy
  of entanglement} \cite{nlocc}. Thus one arrives at entanglement by
an opposite approach to the usual one: instead of looking at nonlocal
part, one considers local information, and subtracts it from total
information of the state. It is an open question, if for pure
multipartite states, the quantum deficit is also equal to some
entanglement measure (if so, the measure is most likely the relative
entropy of entanglement).

Another link is that the above approach allows to define cost of
erasure of entanglement in thermodynamical spirit. Namely the cost is
equal to entropy that has to be produced while resetting entangled
state to separable one, by {\it closed} local operations and classical
communication \cite{huge-delta}. (closed means that we do not trace
out systems, to count how much entropy was produced).  It was proved
that the cost is lower bounded by relative entropy of
entanglement. (some related ideas have been heuristically claimed
earlier in \cite{Vedral1999}).

Note that another way of counting cost of entanglement erasure is
given by robustness of entanglement. In that case it is a mathematical
function, while here the cost is more operational, as it is connected
with entropy production.

In a recent development the idea of drawing work from local heat baths
have been connected with a separability criterion. Namely, in
\cite{MaruyamaMV2003-thermo-ent} the following non-optimal scheme of
drawing work was considered for two qubits: Alice and Bob measure the
state in the same randomly chosen basis. If they share singlet state
$\psi^-$, irrespectively of choice of basis, they obtain perfect
anti-correlations, allowing to draw on average 1 bit (i.e.  $kT \ln2$)
of work. Consider instead classically correlated state
$\frac{1}{2}(|00\>\<00|+|11\>\<11|)$. We see that any measurement apart from
the one in bases $|0\>,|1\>$ will decrease correlations. Thus if we
average over choice of bases, the drawn work will be significantly
smaller. The authors have calculated the optimal work that can be
drawn from separable states, so that any state which allows to draw
more work, must be entangled.

\subsection{Asymmetry of entanglement}
\label{sec:asym}

To characterize entanglement one should recognize its different
qualities. Exemplary qualities are just transition rates (exact or asymptotic) between different states under classes of operations such as LOCC  (see Secs. \ref{subsec:puretrans} and \ref{subsec:mixed_asym_trans}).

Recently, in \cite{HHHH-asymmetry} a different type of transition was
studied. Namely, transition of a bipartite state into its own swapped
version: $\rho \longrightarrow V\rho V$ with $V$ being a swap operation
which exchanges subsystems of $\rho$ (see Sec. (\ref{swap})). In other
words this is transition from $\rho_{AB}$ into
$\rho_{BA}$ under LOCC. If such a transition is possible for a given state, one
can say, that entanglement contents of this state is symmetric.
Otherwise it is asymmetric, and the {\it asymmetry of entanglement} can be quantified as follows
\begin{equation}
A(\rho)=\sup_\Lambda \|\Lambda(\rho) - V\rho V \|_1,
\end{equation}
where supremum is taken over all LOCC operations $\Lambda$.
Note that if a state is symmetric under swap, then also its entanglement
contents is symmetric, however the converse does not hold (a trivial example
is a pure nonsymmetric state: one can produce its swapped version by
local unitaries).

It was shown that there exist states entanglement of which is asymmetric
in the above sense. Namely, one can prove if the entanglement measure called G-concurrence
\cite{SinoleckaZK2001-manifold,FanMI2002-concurrence,Gour-mon2004} is nonzero,
then the only LOCC operation that could swap the state is just local
unitary operation. Thus any state which has nonzero G-concurrence, and
has different spectra of subsystems cannot be swapped by LOCC and
hence any it contains {\it asymmetric entanglement}. It is an
open question if there exist states which maintain this asymmetry even
in asymptotic limit of many copies.

One should note here, that the asymmetry in context of quantum states
has been already noted in literature. In particular the asymmetric
correlation functions $f$ of a quantum state are known, such as
e.g. one-way distillable entanglement $E_D^{\rightarrow}$
\cite{BBPSSW96} or {\it quantum discord}
\cite{Zurek-discord}. Considering properly symmetrised difference
between $f(\rho_{AB})$ and $f(\rho_{BA})$ one could in principle
obtain some other measure of asymmetry of quantum
correlations. However it seems that such a measure would not capture
the asymmetry of entanglement content exactly. The functions defined
by means of asymmetric (one-way) class of operations like
$E_D^{\rightarrow}$ seems to bring in a kind of its "own asymmetry"
(the same holds e.g. for one-way {\it distillable key}). Concerning
the correlation functions $f$ such as quantum discord, they seem to
capture the asymmetry of quantum correlations in general rather then
that of entanglement, being nonzero for separable states (the same
holds e.g. for Henderson-Vedral $C_{HV}$ quantity (see \ref{eq:chv})
and one-way distillable {\it common randomness}
\cite{DevetakW03-common}).

The phenomenon of asymmetry has been studied also in other contexts,
such as asymmetry of quantum gate capacities
\cite{HarrowShor2005,LindenSW2005-entpower}, and the possibility of
{\it exchange} of the subsystems of a bipartite state in such a way
that entanglement with a purifying reference system is preserved
\cite{uncom_info}. Although these are very different phenomena, the
possible relation is not excluded.

\section{Entanglement and \secure\ correlations}
\label{sec:Ent_in_QKD}

A fundamental difference between classical and quantum, is that
quantum formalism allows for states of composite systems to be both
{\it pure} and {\it correlated}. While in classical world those two
features never meet in one state, {\it entangled} states can exhibit
them at the same time.

For this reason, entanglement in an astonishing way incorporates basic
ingredients of theory of \secure\ communication. Indeed, to achieve
the latter, the interested persons (Alice and Bob) need a private key:
a string of bits which is {\it i)} perfectly correlated ({\it
  correlations}) and {\it ii)} unknown to any other person ({\it
  security or privacy}) (then they can use it to preform \private\
conversation by use of so-called Vernam cipher \cite{Vernam} ). Now, it is purity
which enforces the second condition, because an eavesdropper who wants
to gain knowledge about a quantum system, will unavoidably disturb it,
randomizing phase via quantum back reaction. In modern terminology we
would say, that if Eve applies CNOT gate, to gain knowledge about a
{\it bit}, at the same time she introduces a {\it phase} error into
the system, which destroys purity, see \cite{Zurek81}.

Let us note that all what we have said can be phrased in terms of
monogamy of entanglement in its strong version: {\it If two quantum
  systems are maximally quantumly correlated, then they are not
  correlated with any other system at all (neither quantumly nor
  classically)} (see \cite{KoashiW-monogamy} in this context). In this
section we shall explore the mutual interaction between entanglement
theory and the concept of private correlations.

\subsection{Quantum key distribution schemes and security proofs
based on distillation of pure entanglement}
\label{subsec:Pure_proofs}

Interestingly, the first protocol to obtain a private key \footnote{We
  call such protocol Quantum Key Distribution (QKD).} - the famous
BB84 \cite{BB84}, did not use the concept of entanglement at
all. Neither uses entanglement another protocol B92 proposed by
Ch. Bennett in 1992 \cite{B92}, and many variations of BB84 like
six-state protocol \cite{Bruss6state}. Indeed, these QKD protocols are
based on sending randomly chosen {\it nonorthogonal quantum
  states}. Alice prepares a random signal state, measures it and send
it to Bob who also measures it immediately after reception.  Such
protocols are called a {\it prepare and measure protocols} (P\& M).

The first entanglement based protocol was discovered by Ekert (see
Sec. \ref{sec:effects}). Interestingly, even Ekert's protocol though
using explicitly entanglement was still not based solely on the
``purity \& correlations'' concept outlined above. He did exploit
correlations, but along with purity argument, used violation of Bell
inequalities. It seems that it was the paper of Bennett, Brassard and
Mermin (BBM) \cite{BBM92} which tilted the later history of
entanglement-based QKD from the ``Bell inequalities'' direction to
``disturbance of entanglement'': upon attack of Eve, the
initially pure entangled state becomes mixed, and this can be detected
by Alice and Bob. Namely they have proposed the following protocol
which is also entanglement based, but is {\it not} based on Bell
inequality. Simply Alice and Bob, when given some (untrusted)
\eprpairs\ check their quality by measuring correlations in
$\{|0\>,|1\>\}$ basis, and the conjugated basis
$|\pm\>={1\over \sqrt2}(|0\>\pm|1\>)$. So
simplified Ekert's protocol is {\it formally} equivalent to the BB84
protocol. Namely the total final state between Alice, Bob and Eve is
the same in both cases\footnote{In BB84, the state consists of
  preparations of Alice, outcomes of Bob's measurement and quantum
  states of Eve. In BBM protocol it is outcomes of Alice and Bob's
  measurement and quantum states of Eve.}. Thus entanglement looks
here quite superfluous, and moreover Bell inequalities appear rather
accidentally: just as indicators of possible disturbance by Eve.

Paradoxically, it turned out recently that the Bell inequalities are
{\it themselves} a good resource for key distribution, and allow to
prove security of private key {\it without} assuming quantum formalism
but basing solely on no-signaling
assumption. \cite{AcinGM-bellqkd,BHK_Bell_key,MW_Bellcrypto}. There is
still an analogy with entanglement: nonlocal correlations are
monogamous \cite{BKP_Bellcrypto}.

Concerning entanglement, Ekert noticed that the equivalence of
entanglement based protocol and BB84 is not complete:\footnote{See
  note added in \cite{BBM92}.} the former has an advantage that Alice
and Bob can postpone measuring \eprpairs\ until they need the key, so
that a burglar breaking into their labs trying to get some
information, would disturb the pairs risking detection, while in BB84,
there is no possibility of storing the key in quantum form.  Thus
entanglement provides {\it potential} key, in a similar way as it
provides potential communication in dense coding (see Sec.
\ref{sec:effects}). However this is not the only advantage of
entanglement: in fact its role turned out to be
indispensable in further development of theory of secure
correlations. Actually, the interaction is bilateral: also the development
of entanglement theory was influenced in essential way by the ideas of
secure correlations, to mention only, that first protocols of
entanglement distillation (fundamental for the whole quantum
communication theory) have been designed by use of methods of
generation of \secure\ key \cite{BBPSSW1996,BDSW1996}.

Another (at least theoretical) advantage of entanglement-based QKD
protocol is that with quantum memory at disposal, one can apply dense
coding scheme and obtain a protocol which has higher capacity than
usual QKD as it was applied by Long and Liu \cite{LongLiu2002} (see
also \cite{Cabello00}).

Moreover, entanglement may help to carry out quantum cryptography
over long distances by use  quantum repeaters  which exploit
entanglement swapping and quantum memory \cite{repeaters}.
We should note that all the above potential advantages of entanglement,
would need quantum memory.

\subsubsection{Entanglement distillation based quantum key distribution protocols.}

Both Ekert's protocol, and its BBM version worked in situation, where
the disturbance comes only from eavesdropper, so if only Alice and Bob
detect his presence, they can abort the protocol. Since in reality one
usually deals with imperfect sources, hence also with imperfect
(noisy) entanglement, it is important to ask if the \secure\ key can
be drawn from noisy \eprpairs. The {\it purification of \eprpairs}
appeared to be crucial idea in this case.  The first scheme of
purification (or {\it distillation}) of entanglement has been
discovered and developed in \cite{BBPSSW1996,BDSW1996} (see
Sec. \ref{sec:distil}). In this scheme Alice and Bob share $n$ copies
of some mixed state, and by means of {\it local quantum operations and
  classical communication} (LOCC) they obtain a smaller amount $k < n$
of states which are very close to the \eprstate.
\begin{equation}
  \xymatrix@1@W=2cm{\rho^{\ot n}
    \ar[r]^{\textrm{\footnotesize Distillation}} &
    |\phi^{+}\>^{\ot k}.}
\end{equation}
The highest asymptotic ratio $k\over n$ on the above diagram is called
{\it distillable entanglement} and denoted as $E_D(\rho)$ (see Sec. \ref{subsec:edec}).  This
concept was adopted in \cite{QPA}, where
distillation process for cryptographical purpose was named {\it
  Quantum Privacy Amplification} (QPA). From $n$ systems in a joint
state $\rho_n$ (which may be in principle supplied by Eve), Alice and
Bob distill \singlets, and finally generate a key via measurement in
computational basis:
\begin{equation}
  \xymatrix@1@W=2cm{\rho_n
    \ar[r]^{\textrm{\footnotesize QPA}} &
    |\phi^{+}\>^{\ot k}.}
\end{equation}

The protocol of QPA assumes that devices used for distillation are
perfect. Moreover, the distillation scheme of Bennett \emph{et al.}
works if the initial state is in tensor product of many copies. The
question of verification, by Alice and Bob whether they indeed share
such state (or whether the final state is indeed the desired
$\phi^+={1\over \sqrt2}|00\>+|11\>$
state) was not solved in \cite{QPA}.

This problem has been tackled by Lo and Chau \cite{Lo-Chau} who have
provided the first both unconditionally secure and fully
entanglement-based scheme\footnote{The first proof of unconditional security of quantum key distribution was provided by Mayers who proved security of BB84 \cite{Mayers84proof}.}.
To cope with imperfections Alice and Bob use {\it fault tolerant quantum
  computing}. In order to obtain \secure\ key, they perform
entanglement distillation protocol of \cite{BDSW1996} to distill
\singlets, and check their quality\footnote{Using the concept of
  another entanglement-based communication scheme --- {\it quantum
    repeaters} \cite{repeatersPRL,repeaters} Lo and Chau established
  quantum key distribution (QKD) over arbitrary long distances.}.

The Lo-Chau proposal has a drawback: one needs a quantum
computer to implement it. On the other hand, the first quantum
cryptographic protocol (BB84) does not need a quantum computer. And
the BB84 was already proved to be secure by Mayers
\cite{Mayers84proof}. Yet, the proof was quite complicated, and
therefore not easy to generalize to other protocols.

\subsubsection{Entanglement based security proofs}

A remarkable step was done by Shor and Preskill \cite{ShorPreskill},
who showed, that one can prove security of BB84 scheme, which is a
\pmp protocol, by considering {\it mentally} a protocol based on
entanglement (a modified Lo-Chau protocol). This was something like
the Bennett, Brassard, Mermin consideration, but in a noisy
scenario. Namely, while using BB84 in presence of noise, Alice and Bob
first obtain a so-called {\it raw key} - a string of bits which is not
perfectly correlated (there are some errors),
and also not perfectly secure (Eve have some knowledge about the key). By looking at
the part of the raw key,  they can estimate
the level of error and the knowledge of Eve.
They then classically process
it, applying procedures of {\it error correction} and {\it privacy
amplification} (the latter aims at diminishing knowledge of Eve).

In related entanglement based scheme, we have {\it coherent} analogues
of those procedures. Without going into details, we can imagine that
in entanglement based scheme, Alice and Bob share pairs in one of four
Bell states (\ref{basis}), which may be seen as the state $\phi^+$
with two kinds of errors: bit error, and phase error.
The error from the previous scheme translates here into
bit error, while knowledge of Eve's into phase error.
Now the task is simply to correct both errors.
Two procedures of a different kind (error correction
and privacy amplification)
are now both of the same type - they correct errors.

After correcting bit error, Alice and Bob are left with Bell states which are all
correlated
\begin{eqnarray}
|\phi^{-}\> = {1\over {\sqrt 2}}(|00\> - |11\>),\nonumber \\
|\phi^{+}\> = {1\over {\sqrt 2}}(|00\> + |11\>).
\label{eq:correlBell}
\end{eqnarray}
Then they apply phase error correcting procedure. I.e. they get to
know which systems are in $\phi^{-}$ and which in $\phi^{+}$ so
that they can rotate each $\phi^-$ into $\phi^{+}$ and finally
obtain a sequence of $\phi^+$ solely. What Shor and Preskill noticed is
that these two quantum procedures are {\it coherent}\footnote{It is worth to observe that formally any QKD
  scheme can be made ``entanglement based'', as according to axioms of
  quantum mechanics any operation is unitary and the only source of
  randomness is the subsystem of an entangled state.  From this point
  of view, even when Alice sends to Bob a randomly chosen signal
  state, as it is in \pmp schemes, according to axioms she sends a
  part of entangled system. Moreover, any operation that Alice and Bob
  would perform on signal states in \pmp scheme, can be done
  reversibly, so that the whole system shared by Alice Bob and Eve is
  in a pure state at each step of the protocol. This principle of
  maintaining purity is usually referred to as {\it coherent}
  processing. Let us note however, that not always the coherent
  application of a protocol that provides a key, must result in the
  distillation of $\phiplus$. In Sec. \ref{subsubsec:private} we will
  see that there is a more general class of states that gives private
  key.} versions of
classical error-correction and privacy amplification
respectively. Thus Shor-Preskill proof can be phrased as follows: {\it
  ``The BB84 protocol is secure because its suitable coherent version
  distills \eprstates''}.

It should be emphasized here, that the equivalence between noisy BB84
protocol and its coherent version does not continue to the very
end. Namely, in distillation, Alice and Bob after finding which pair
is in $|\phi^-\>$ and which in $|\phi^+\>$, rotate $|\phi^-\>$. The
classical procedures performed coherently cannot perform this very
last step (rotation) as no classical action can act as phase gate
after embedding into quantum. However, the key is secure, because
Alice and Bob {\it could have performed the rotation}, but they do not
have to. Indeed, note that if Alice and Bob measure the pairs in basis
$\{|0\>,|1\>\}$ they obtain the same results, independently of whether
they have rotated $|\phi^-\>$ or not. The very possibility of
rotation, means that the key will be secure. Thus the coherent version
of BB84 does not actually give $|\phi^+\>$ itself, but it does if
supplemented with rotations.

The concept of proving security of \pmp protocols by showing that at
the {\it logical level} they are equivalent to distillation of
entanglement, has become very fruitful. In 2003 K. Tamaki, M. Koashi
and N. Imoto \cite{Tamaki03:secB92} showed that B92 is unconditionally
secure, using Shor-Preskill method (see also
\cite{TamakiLutkenhaus04}). They showed that B92 is equivalent to
special entanglement distillation protocol known as {\it filtering}
\cite{Gisin96,HHH1997-distill} (see section
\ref{subsec:two-qubits-distil}). In \cite{lca} the efficient version
of BB84 was proposed, which is still unconditionally secure though the
number of systems that Alice and Bob use to estimate the error rate is
much smaller than in BB84. Again security is proved in Shor-Preskill
style. In \cite{Gottesman-Lo} a \pmp scheme with a {\it two-way}
classical error correction and privacy amplification protocol was
found. It is shown that a protocol with two-way classical
communication can have substantially higher key rate than the one with
the use of one-way classical communication only.
Also security of key distribution using dense coding  \cite{LongLiu2002} was proved
by use of Shor-Preskill techniques \cite{ZhangWenLong2005} (see in this
context \cite{DegiovanniRCRBCC2003-dense-qkd,Wojcik-dense-qkd,RepToCommQDKD}).

Thanks to simplicity of Shor-Preskill approach pure entanglement
remained the best {\it tool for proving unconditional security} of QKD
protocols \cite{Crypt_ent_withent}.  As we will see further this approach can be
generalized by considering {\it mixed} entangled states containing
ideal key.

\subsubsection{Constraints for security from entanglement}

So far we have discussed the role of entanglement in particular protocols of quantum
key distribution. A connection between entanglement and {\it any} QKD protocol has been established by Curty, Lewenstein and L{\"u}tkenhaus \cite{CurtyLewLut}. They have proved that entanglement is necessary precondition of unconditional security. Namely, in case of any QKD protocol Alice and Bob perform some measurements and are left with some classical data from which they want to obtain key. Basing on these data and measurements settings, they must be able to construct a so called {\it entanglement witness} to ensure that the data could not be generated via measurement on some {\it separable} state (see Sec. \ref{subsubsec:Witnesses} ). Let us emphasized, that this holds not only for entanglement based but also for prepare and measure protocols. In the latter case, a kind of ``effective''  entanglement is witnessed (i.e. the one which is actually never shared by Alice and Bob).

It is worth noting, that this approach can be seen as a generalization of
the very first Ekert's approach \cite{E91}. This is because Bell
inequalities can be seen as a special case of entanglement witnesses
(see Sec. \ref{subsec:Bell_witness}). Formally, in any QKD protocol,
Alice and Bob perform repeatedly some POVM's $\{ A_i\}$ and $\{B_i\}$ respectively,
and obtain a probability distribution of the outcomes $P(a,b)$. Now, a necessary condition
for security of the protocol is that it is possible to build out some
entanglement witness $W=\sum_{i,j} c_{ij} A_i\ot B_j$ with real $c_{ij}$, such that
$\sum_{ij}c_{ij}p_{ij} <0$ and ${\tr}W\sigma \geq 0$ for all
separable states $\sigma$. Indeed, otherwise they could make key from
separable states which is impossible \cite{GisinWolf_linking} (see
also discussion in sec \ref{subsec:QKA}). This idea has been studied
in case of high dimensional systems
\cite{NikolopulosAlber05,NikolKhaliqueAlber05} and general upper
bounds on key rates for prepare and measure schemes has been found
\cite{MoroderCL2005-povmintr,symmetric_key_bound}. It is also
connected with optimization of entanglement measures from incomplete
experimental data (see Sec. \ref{subsec:wit_exp_ent_det}).

\subsubsection{Secure key beyond distillability - prelude}

The fact that up to date techniques to prove unconditional security
were based on entanglement {\it purification} i.e.  distilling {\it
pure} entangled states, has supported the belief that possibility of
distilling pure entanglement (singlet) is the only reason for unconditional
security.  For this reason the states which are entangled but not
distillable (i.e. {\it bound entangled}) found in \cite{bound} have
been considered as unlikely to be useful for cryptographical tasks. Moreover
Acin, Masanes and Gisin \cite{ALG_2_equiv}, showed that in the case of
two qubit states (and under individual Eve's attack) one can distill
key by {\it single measurement and classical postprocessing} if and
only if the state contains distillable entanglement. Surprisingly, as
we will see further, it has been recently proved that one can obtain
{\it unconditionally} secure key even from bound entangled states.

The first interesting step towards this direction was due to Aschauer and Briegel,
who showed Lo and Chau's protocol provide key even without the fault tolerant
computing i.e. with realistic noisy apparatuses \cite{AschauerBriegel2002}.

 The crucial property of their approach was
to consider all noise, (which they assumed to act {\it locally} in
Alice and Bob's laboratories) in a {\it coherent} way. Keeping track
of any noise introduced by imperfect devices, they can ensure that the
total state of distilled {\it noisy singlets} together with the state
of their laboratories is in fact a {\it pure} state. This holds
despite the fact that the noisy singlets may have {\it fidelity}, that
is overlap with a state (\ref{eq:singlet_dense}), bounded away from
maximal value 1. This kind of noisy entanglement, which is corrupted
because of local noise, but not due to the eavesdropping, they have
called {\it private entanglement}.


\subsection{Drawing private key from distillable and bound entangled
states of the form $\rho^{\ot n}$ }
\label{subsec:QKA}

Strong interrelation between theory of \secure\ key and entanglement
can already be seen in the scenario, where Alice and Bob share $n$
bipartite systems in {\it the same state} $\rho_{AB}$ and Eve holds
their purification, so that the joint state of Alice, Bob and Eve
systems is a pure state $|\psi_{ABE}\>$. The task of the honest
parties is to obtain by means of {\it Local Operations and Classical
  Communication} (LOCC) the highest possible amount of correlated
bits, that are unknown to Eve (i.e. a \secure\ key). The difficulty of
this task is due to the fact that Eve makes a copy of any classical
message exchanged by Alice and Bob.

The above paradigm, allows to consider a new measure of entanglement:
{\it distillable key} $K_D$, which is similar in spirit to distillable
entanglement as it was discussed in Secs. \ref{subsubsec:em-tr-exact}
and \ref{subsec:edec}. It is given by the number of secure bits of key
that can be obtained (per input pair) from a given state.

Let us discuss in short two extreme cases:
\begin{itemize}
\item All distillable states are key distillable:
  \begin{equation}
    K_D(\rho_{AB})\geq E_D(\rho_{AB}).
  \end{equation}
\item All separable states are key non-distillable:
  \begin{equation}
    K_D(\sigma_{sep})=0.
  \end{equation}
\end{itemize}

To see the first statement, one applies the idea of quantum privacy
amplification described above. Simply, Alice and Bob distill
\singlets\, and measure them locally. Due to ``purity \& correlation''
principle, this gives a secure key.

To see that key cannot be drawn from separable states
\cite{GisinWolf_linking,CurtyLewLut}, note that by definition of
separability, there is a measurement on Eve's subsystem such that
conditionally upon result (say $i$) Alice and Bob share a {\it
  product} state $\rho_A^{(i)}\ot \rho_B^{(i)}$. This means that Alice
and Bob conditionally on Eve have initially no correlations. Of
course, any further communication between Alice and Bob cannot help,
because it is monitored by Eve.

In Sec. \ref{subsubsec:pptkey} it will be shown, that there holds:
\begin{itemize}
\item There are non-distillable states which are key distillable:
  \begin{equation}
    E_D(\rho_{be})=0\quad\&\quad K_D(\rho_{be})>0.
  \end{equation}
\end{itemize}

\subsubsection{Drawing key from distillable states: Devetak-Winter protocol}
\label{subsubsec:cqq}

Here we will present a protocol due to Devetak and Winter, which shows
that from any state, one can draw at least amount of key equal to
coherent information.  This is compatible with the idea, that one can
draw key only from entangled states (states with positive coherent
information are entangled, as shown in \cite{RPH1994}). The coherent
version of the protocol will in turn distill this amount of \singlets\
from the state. In this way Devetak and Winter have for the first time
proved the hashing inequality (\ref{subsec:hasineq}) for {\it
  distillable entanglement}. Thus, again cryptographic techniques
allowed to develop entanglement theory.

Namely, consider a state $\rho_{AB}$, so that the total state
including Eve's system is $\psi_{ABE}$. As said, we assume that they
have $n$ copies of such state. Now Alice performs complete
measurement, which turns the total state into $\rho_{cqq} = \sum_{i}
p_i |i\>\<i|_A\ot \rho^{i}_{BE}$ where subscript $cqq$ reminds that
Alice's system is classically correlated with Bob and Eve
subsystems. The authors considered drawing key from a general cqq
state as a starting point, and showed that one can draw at least the
amount of key equal to
\be
I(A:B) - I(A:E),
\label{eq:Qckbound}
\ee
where $I(X:Y)=S(X)+S(Y)-S(XY)$ is quantum mutual information.

Without going into details, let us note that
the quantity $I(A:B)$ is the common information between
Alice and Bob, hence it says how many correlated bits Alice and Bob
will obtain via error correction. Since $I(A:E)$ is common information
between Eve and Alice, its subtraction means, that in the procedure of
privacy amplification this amount of bits has to be removed from key,
to obtain a smaller key, about which Eve does not know anything.

Now, let us note that in the present case we have
\ben
&&I(A:B)=S(\rho_B) - \sum_ip_i S(\rho^i_B), \nonumber\\
&&I(A:E)=S(\rho_E) - \sum_ip_i S(\rho^i_E),
\een
where $\rho_B^i=\tr_E\rho^i_{BE}$, $\rho_E^i=\tr_B\rho^i_{BE}$,
$\rho_B=\sum_i p_i \rho_i^B=\tr_A\rho_{AB}$, and
$\rho_E=\sum_i p_i \rho_i^E=\tr_{AB}\rho_{AE}$.

Since the measurement of Alice was complete, the states
$\rho_{BE}^i$ are pure, hence $S(\rho^i_E)=S(\rho^i_B)$. Also, since
the total initial state was pure, we have $S(\rho_E)=S(\rho_{AB})$
where $\rho_{AB}$ is initial state shared by Alice and Bob. Thus we
obtain that the amount of key gained in the protocol, is actually
equal to coherent information.

Now, Devetak and Winter have applied the protocol coherently, and
obtained protocol of distillation of singlets proving first general
lower bound for distillable entanglement \begin{equation} E_D(\rho_{AB})\geq S(\rho_B)- S(\rho_{AB}). \end{equation}
This was also used by Devetak to provide the first fully rigorous
proof of quantum Shannon theorem, saying that capacity of quantum
channel is given by coherent information \cite{Devetak2003}.
Thus cryptography was used also to develop entanglement and
quantum communication theory.

\subsubsection{Private states}
\label{subsubsec:private} In the previous section we have invoked a
protocol, which produces a private key, and if applied coherently,
distills \singlets. A fundamental question which naturally arises is
how far this correspondence goes. In particular, one can ask whether
it is possible to distill key only from states from which singlets
can be distilled. Surprisingly,  it has been shown that one can
obtain key even from certain bound entangled states
\cite{pptkey,keyhuge}.
To this end they have characterized
class of states which have subsystem that measured in certain
product basis gives $\log d$ bits of perfect key. The present section is mostly devoted to this class of states which apart from being fundamental for analysis of key distillation constitute a new quality in mixed state entanglement, hence it is also of its own interest.

These states, called {\it private states} or {\it gamma states},
have been proved to be of the following form\footnote{
More precisely, the class of private states is locally equivalent to the one
given by equation (\ref{eq:pdit}), that is contains any state $\gamma^{(d)}$ rotated by local unitary transformation $U_A\ot U_B \ot \id_{A'B'}$.} \be \gamma^{(d)} =
{1\over d} \sum_{i,j=0}^{d-1} |i i\>\<j j|_{AB}\ot
U_{i}\rho_{A'B'}U_{j}^{\dagger} \label{eq:pdit}, \ee where $U_i$
are arbitrary unitary transformations acting on the system $A'B'$.
The whole state resides on two systems with distinguished subsystems $A A'$ and $BB'$ respectively. The $AB$ subsystem, after measuring in
computational basis, gives $\log d$ bits of key, hence it is called
the {\it key part} of the private state. A private state with
$d\times d$-dimensional key part is called shortly a {\it pdit} and
in special case of $d=2$ a {\it pbit}. The subsystem $A'B'$ aims to
{\it protect} the key part, hence it is referred to as a {\it
shield} {}\footnote{Dimensions $d_{A'}$ and $d_{B'}$ are in
principle arbitrary (Without losing generality, we can put them
equal).}. The class of private state has been generalized
to the multipartite case \cite{AH-pditdist}.

\subsubsection{Private states versus singlets}
\label{subsubsec:twisting} A private state can be obtained via
unitary transformation from a simple private state called {\it basic
pdit} \be \gamma^{(d)}_{basic}=|\Phiplus\>\<\Phiplus|_{AB} \ot
\rho_{A'B'}.
\label{eq:basic_pbit}\ee
with\footnote{In this section we will refer to the state $|\Phiplus\>$ as well as $|\phiplus\>={1\over \sqrt{2}}(|00\>+|11\>)$ as to singlet state.
}. $|\Phiplus\>=\sum_{i=0}^{d-1}{1\over \sqrt{d}}|ii\>$.
Namely the following
controlled transformation,  called {\it twisting}, \be U_{\tau} =
\sum_{kl}|k l\>_{AB}\<k l|_{AB}\ot U_{kl}^{A'B'} \label{eq:twisting}
\ee transforms $\gamma_{basic}$ into a private state $\gamma$ given
by (\ref{eq:pdit})
\footnote{The unitaries $U_{i}$ and $U_j^{\dagger}$ in definition of
private state corresponds to $U_{kk}$ and $U_{ll}^{\dagger}$ in definition of twisting.}.
In general the unitary $U_{\tau}$
can be a highly nonlocal operation, in the sense that it cannot be
inverted by means of local operations and classical communication.
Thus the $|\Phiplus\>$ is contained somehow in the private state, but
in a ``twisted'' way: Therefore, unlike in the $|\Phiplus\>$ state,
$\ed$ can be much smaller than $\ec$, as shown by the following
example \cite{pptkey}:
\be
\rho_\gamma =
p|\phi^+\>\<\phi^+|\ot\rho_s + (1-p)|\phi^-\>\<\phi^-|\ot\rho_a
\label{eq:samlldistgama}
\ee
where $|\phi^\pm\>$ are the maximally
entangled states, $\rho_{s,a}$ are normalized projectors onto
symmetric and antisymmetric subspaces and $p={1\over 2}(1 - {1\over d})$. One
computes logarithmic negativity, which bounds $E_D$, to obtain
\be
E_D(\rho_\gamma)\leq \log({1+ {1\over d}}).
\ee
One can show that
in general $K_D\leq \ec$, so that $\ec(\rho_\gamma)\geq 1$.

Although twisting can diminish distillable entanglement, it can not
set it to zero. Indeed it was shown that all private states are distillable
\cite{AH-pditdist}.

Interestingly, private state can be written as $|\Phiplus\>$, where
amplitudes have been replaced by q-numbers: \be \gamma = \Psi
\Psi^{\dagger}, \ee where $\Psi$ is the matrix ({\it not necessary}
a state) similar to a pure state,
\be
\Psi = \sum_{i=0}^{d-1}Y_i^{A'B'}\ot |ii\>_{AB},
\label{eq:pdit_singlet_likeform}
\ee
with $Y_i = U_i\sqrt{\frac{\rho}{d}}$ for some unitary transformation $U_i$
and certain state $\rho$ on $A'B'$. Looking at the state
$|\psi\>=\sum_{j=0}^{d-1} {1\over
\sqrt{d}} e^{i\phi_j}|jj\>$ we see that  $Y_i$ is a noncommutative
version of  $\frac{1}{\sqrt{d}}^{i\phi_j}$. This special form has
not yet been studied thoughtfully. It seems that focusing
on non-commutativity of operators $Y_i$ may lead to better
understanding of the properties of this class of states.

\subsubsection{Purity and correlations: how they are present in p-bit}
In the introductory paragraph to Sec. \ref{sec:Ent_in_QKD} we
have pointed out that the origin of private correlations in quantum
mechanics is the existence of pure state which still exhibits
correlations (unlike in classical world). However, we have
argued then that there are states which exhibit private key, but are
quite mixed. They can have arbitrarily small distillable
entanglement, still offering bits of perfect key. One might wonder if it implies that
we need a more general principle, which explains privacy obtained in quantum mechanics.
There is no simple solution of this problem. Let us first argue,
that although in different context, the principle of "purity \& correlations" still
works here.

To this end, let us note that private state is a unitarly rotated {\it
basic p-bit}. The latter is of the form $\gamma_{basic}=|\Phiplus\>\<\Phiplus|_{AB}\ot
\rho_{A'B'}$ (see Eqs. (\ref{eq:basic_pbit}) and (\ref{eq:twisting})). The unitary $U_{\tau}$ might melt completely the division of $\gamma_{basic}$ into
subsystems $A$, $B$, $A'$ and $B'$. However what is for sure
preserved, is that the private state is a product of some pure state
$\psi$  and $\rho$ across some tensor product. In other words, there
is a \emph{virtual} two-qubit subsystem, which is in \wsinglet\ state
$\psi=\phi^+_{virt}$. (This virtual \wsinglet\ is what we call
``twisted'' version of the original \wsinglet\ state.)  Because of
purity, the outcomes of any measurement performed on this virtual subsystem
are not known to Eve. Thus we already have one ingredient of the
principle --- the role of {\it purity}. Let us now examine the second
ingredient --- {\it correlations}. The pure state is no longer a state of two qubits, one of them on Alice side, and the second on Bob's. The measurements that access individually the virtual qubits are the original local measurements of Alice and Bob, rotated however by the inverse of unitary $U_{\tau}$. Now, it happens that local measurements on key part (AB subsystem) of the private state in standard basis commute with $U_{\tau}$ (\ref{eq:twisting}), for the latter is a unitary controlled just by this basis. Thus at least one correlation measurement on the virtual
qubits is still accessible to Alice and Bob, and can be performed on
the qubits $A$ and $B$.

It is nevertheless natural quite natural to maintain that the  "purity \& correlations" is too strong principle for ensuring security.
Simply, the purity of a virtual subsystem is not the purity of the total state
that Alice and Bob have. In the case of basic p-bit,
it is easy to obtain the whole system in a pure state, simple by removing the mixed subsystem $A'B'$. However, for a generic private state,
the purity is inaccessible: the state of the whole system
is irrevocably mixed.  This is because Alice and Bob have only local access
to the system, hence from some p-bits Alice and Bob may be able to distill
only an amount of purity much smaller than the amount of privacy they can get.
Even more: if we allow arbitrary small inaccuracy,  then
they may not be able to obtain a system in a pure correlated state {\it at all},
while still obtaining secure key.

Thus to get privacy, one does not need to share pure states.
There is a simple explanation how it can be so: A pure state
$\Phiplus$ offers secure key  in {\it any}  correlated basis, while what we need
is just  a single secure basis, because we need only classical
secure key - pair of bits.

\subsubsection{Distillable key as an operational entanglement measure}
\label{subsubsec:kd-pbits}
The concept of private states allows to represent $K_D$ as a
quantity analogous to entanglement distillation:
\begin{equation}
    \xymatrix@1@W=3cm{\rho_{AB}^{\ot n}
      \ar[r]^{\parbox{2cm}{\center\footnotesize LOCC  key distillation }} &
      \gamma^{(d)}_{ABA'B'},}
\end{equation}
where the highest achievable ratio ${\log d\over n}$ in the
asymptotic limit equals the distillable key denoted as
$K_D(\rho_{AB})$. Instead of \singlets\ we distill private states.
Since the class of private states is broader than the class of
maximally entangled states, one can expect that distillable key can
be greater than distillable entanglement. Indeed, this is the case,
and an example is just the private state of Eq.
(\ref{eq:samlldistgama}).

Thus distillable key is a new operational measure of entanglement,
which is distinct from $\ed$. It is also distinct from $\ec$ and it satisfies
\begin{equation}
\ed \leq K_D\leq \ec.
\end{equation}
Moreover it is upper bounded by relative entropy of entanglement \cite{pptkey} and
squashed entanglement \cite{Christandl-PhD}. There is also a bound
involving the best separable approximation \cite{MoroderCL2005-povmintr} which
exploits the fact that admixing separable state can only decrease
$K_D$.  In \cite{symmetric_key_bound} there is also a bound for
one-way distillable key, based on the fact that for a state which
has a symmetric extension, its one-way distillable key must vanish.
Indeed, then Bob and Eve share with Alice the same state, so that
any final key which Alice shares with Bob, she also share with Eve.

\subsubsection{Drawing \secure\ key from bound entanglement.}
\label{subsubsec:pptkey} Bound entanglement is a weak resource,
especially in the bipartite case. For a long time the only useful
task that bipartite BE states were known to perform was activation,
where they acted together with some distillable state. Obtaining
private key from bound entanglement, a process which we will present
now, is the first useful task which bipartite BE states can do {\it
themselves}.

Since distillable entanglement of some private states can be low it
was tempting to admix with small probability some noise in order to
obtain a state which is non-distillable while being still entangled:
\begin{equation} \rho_{total} = (1-p)\gamma + p\rho_{noise} \end{equation}

It happens that for certain private states $\gamma$ the state $\rho_{noise}$ can be adjusted in such a way that the state $\rho_{total}$ is PPT (hence $\ed=0$), and despite this, from many copies of $\rho_{total}$ one can distill key of arbitrarily good quality. That is one can distill private state $\gamma'$ with arbitrary small
admixture of noise.

The first examples of states with positive distillable key and zero
distillable entanglement were found in \cite{pptkey,keyhuge}. We
present here a simple one which has been recently given in
\cite{smallkey}. It is actually a mixture of two private bits (correlated and anticorrelated). The total state has a
matrix form
\begin{eqnarray}
\rho_{ABA'B'}= \frac{1}{2}\left[ \begin{array}{cccc}
p_{1}|X_{1}| & 0 & 0 & p_{1}X_{1} \\
0 & p_{2}|X_{2}|& p_{2} X_{2} & 0 \\
0 & p_{2}X_{2}^{\dagger} & p_{2} |X_{2}| & 0\\
p_{1}X_{1}^{\dagger} & 0 & 0 & p_{1}|X_{1}|\\
\end{array}
\right],
\label{standform}
\end{eqnarray}
with $X_1 = \sum_{i,j=0}^1 u_{ij} |ij\>\<ji|_{A'B'}$ and $X_2 =
\sum_{i,j=0}^{1} u_{ij} |ii\>\<jj|_{A'B'}$ where $u_{ij}$ are the elements
of 1 qubit Hadamard transform and $ p_1 = {\sqrt{2}\over 1 +
\sqrt{2}}$ ($p_2=1-p_1$). This state is invariant under partial
transposition over Bob's subsystem. If we however project its key
part ($AB$ subsystem) onto a computational basis it turns out that
the joint state of Alice, Bob and Eve system is fully classical and of very
simple form: with probability $p_1$ Eve knows that Alice and
Bob are correlated, while with probability $p_2$ that they are
anticorrelated. Thus the mutual information $I(A:E)=0$, and
$I(A:B)=1-H(p_1)$. Thus applying Devetak-Winter
protocol\footnote{Since in  this particular case the state is fully
classical, it would be  enough to use the classical predecessor  of
Devetak-Winter protocol, called Csiszar-K{\"o}rner-Maurer protocol.}
(see formula (\ref{eq:Qckbound})) we obtain a key rate \be
K_D(\rho_{be}) \geq 1 - h(p_1) = 0.0213399 > E_D(\rho_{be})=0. \ee
Basing on this example it is argued \cite{smallkey}, that the volume
of bound entangled key distillable states is non-zero in the set of
states occupying more then 4 qubits. It is however a nontrivial task
to provide new examples. Interestingly, no previously known bound
entangled state has been shown to be key-distillable.

\subsection{Private states --- new insight into
entanglement theory of mixed states}

Investigations concerning distillable key were fruitful to
entanglement theory itself. A new operational measure
of entanglement was obtained, and also a new source of examples of
irreversibility in entanglement distillation was provided.
The private states, or PPT states with nonzero key,
constitute a new zoo of states which are easy to deal with and have
nontrivial properties, in addition to such canonical classes as
Werner, isotropic, Bell diagonal states or maximally correlated
states.
While the simplicity of the latter classes comes from symmetries (e.g. invariance under twirling), simplicity of the class of private states is based just on special asymmetry between the key and the shield part.

 The private bits called {\it flower states} ( given by trace over subsystem E of (\ref{eq:flower-purif})) are the ones for which the squashed entanglement has been
computed. Interestingly, in this case $\esq=\ec$. Moreover basing on this family one can find states with $E_C$ arbitrarily greater than the squashed entanglement and the
latter arbitrarily greater than $E_D$ \cite{Winter-squashed-ent}. The
flower states exhibit locking (see Sec. \ref{subsec:locking}).  There is actually a general link between locking effect and the problem of
drawing key from bound entanglement. Last but not least, the
description of this class of states yields a natural generalization of
pure maximally entangled states to the case of mixed states with
coefficients becoming operators.

\subsection{Quantum key distribution schemes and security proofs
based on distillation of private states - private key beyond purity}

Key distillation described in Sec. \ref{subsec:QKA} relies upon
important assumption. The initial state shared by Alice and Bob
should be tensor product of the same state $\rho_{AB}$. This
assumption is unreal in almost all security applications, since the
eavesdropper can interrupt the communication and entangle copies of
the state. It was then unclear whether one can obtain gap between
distillable key and distillable entanglement (as reported in Sec.
\ref{subsubsec:kd-pbits}) in the general scenario, where Alice and
Bob do not know a priori anything about their states. It hasn't been noticed,
that a positive answer to this question follows from the results on
finite Quantum de Finetti theorem by R. Renner, N. Gisin and B.
Kraus \cite{RGKinfo_sec_proof_short,RGKinfo_sec_proof_long} (see
especially \cite{RR-phd} and \cite{KRdeFinetti}) and the results of
\cite{pptkey} on bound entangled key distillable states. In
the meantime, a more entanglement-based approach has been developed
\cite{bigkey,GeneralUncSec}, which can be seen as a
generalization of Lo and Chau entanglement purification based
approach to the private state distillation one. It has been shown
there, that an unconditionally secure key can be arbitrarily greater
than the amount of entanglement which can be distilled, and even
when the latter is zero. This important result can be rephrased as
follows:
\begin{itemize}
\item There are situations in which one can not send {\it faithfully} any {\it qubit}, but one can send arbitrarily many {\it unconditionally secure  bits}.
\end{itemize}

\subsubsection{``Twisting'' the standard protocol}

The situation is as follows. Alice and Bob have quantum channel
$\Lambda$, which can only produce bound entangled states (this is so
called binding entanglement channel). Clearly, quantum information
cannot be sent faithfully via such channel: otherwise it would be
possible to transmit faithfully  \singlets (see Sec. \ref{sec:communication}). Suppose further that if
Alice sends half of \wsinglet, then, provided Eve has not changed
the channel, they obtain a state $\rho$ of type (\ref{standform}).
(i.e. the state $\rho$ and the channel $\Lambda$ are connected with
each other via Choi-Jamio\l{}kowski isomorphism $(\id\ot
\Lambda)|\Phiplus\>\<\Phiplus|=\rho$). The state has the property that
after measuring the key part in standard basis and processing the
outcomes classically, Alice and Bob obtain secure key.

Since Eve might change the channel, standard approach would be to
check bit error rate and phase error rate. Unfortunately, this would
not work: the channel itself (without any action of Eve) produces
too high errors. Indeed, low error rates would mean that the state is
close to \wsinglet, which is not the case: the state $\rho$ is bound entangled hence its
fidelity with \wsinglet\ is no better than that for separable (i.e. key undistillable) state (see Sec. \ref{subsubsec:Witnesses}).

To overcome this problem, we have to use the fact that  (as discussed in Secs.
\ref{subsubsec:private} and \ref{subsubsec:twisting}) private states
contain perfect \wsinglet\ twisted by some global unitary transformation.
Therefore the present state, which is a noisy version of private
state, contains a twisted {\it noisy} \wsinglet.
Should Alice and Bob have access to the twisted noisy singlet,
they would estimate its quality by measuring on selected number of systems
the observables $\sigma_z\ot \sigma_z$ (bit error) and $\sigma_x\ot \sigma_x$ (phase error).
In the case of twisted singlets, they can still estimate directly bit error by
measuring the observable $\sigma_z\ot \sigma_z$ on the key part
of the shared states. However the observable $\sigma_x\ot \sigma_x$
does not commute with twisting, so that to estimate phase error of a twisted
noisy singlet means to estimate a nonlocal observable.
Fortunately, this can be done, by decomposing the
observable into local ones, and measuring them separately
\footnote{One needs quantum de Finetti theorem to prove that such
separate estimation will report correctly the value of global
observable.}.

The rest of the protocol, is that Alice and Bob measure the key part
of the remaining systems (i.e. not used for error testing),
and performs standard error correction/privacy amplification
procedures based on estimated error level.
How to see that such protocol is secure? One way is to notice, that it does the same to
twisted state, as the standard BB84-like protocol  (such as e.g. \cite{LoChauArdehali2000,lca}) does to the usual state. How
this looks like in terms of entanglement purification? The standard
protocol is secure, because it would produce \singlets. The present
protocol is secure, because, if performed coherently it produces
{\it private state}, i.e. it is twisted purification. Of course, one
can easily convert it into \pmp\ protocol along Shor-Preskill lines.

We have arrived at general principle,  which is actually ``if and only if'':
{\it A protocol produces secure key if and only if its coherent version produces
private states}.

This statement connects entanglement and key distribution in an
ultimate way. It was recently applied by Renes and Smith
\cite{RenesSmith06} who have found an entanglement based proof of the
\pmp protocol with noisy preprocessing of
\cite{RGKinfo_sec_proof_short,RGKinfo_sec_proof_long}. They have
demonstrated its {\it coherent} version which distills private states,
and hence must be secure.

We have quite a paradoxical situation. When it  turned out that one
can draw secure key in the situation where no singlets can be
distilled, it seemed natural that to prove unconditional security,
one cannot use the  techniques based on entanglement purification.
Surprisingly, it is not the case: everything goes through, in a
twisted form.

\subsection{Entanglement in other cryptographic scenarios}
There are many other quantum cryptographic scenarios than quantum key agreement, where entanglement enters. Here we comment briefly on some of them. Interestingly, this time entanglement is important not only because it makes some protocols possible (like QKD, quantum secret sharing, third man quantum cryptography), but because it disallows certain schemes (like quantum bit commitment, quantum one-out-of-two oblivious transfer).

\subsubsection{Impossibility of quantum bit commitment --- when entanglement says no}
Historically, it was claimed that QIT can ensure not only
unconditionally secure key distribution but also a very important
ingredient of classical cryptographic protocols --- a {\it bit
commitment} protocol \cite{Unsec_bitcom}. If so, Alice could {\it
commit} some decision (a bit value) to Bob, so that after committing
she could not change her mind (change the bit value) but Bob also
could not infer her decision before she lets to {\it open} it. Such
a protocol would be one of the most important ingredients in secure
transaction protocols. Unfortunately, it is not the case: Mayers
\cite{Mayers_nobitcom1,Mayers_nobitcom2} and independently Lo and
Chau \cite{LoChau_nobitcom1,LoChau_nobitcom2} have proved under
assumptions plausible in cryptographic context, that quantum bit
commitment is not possible. Paradoxically, it is exactly
entanglement, which assures security of QKD , that is the main
reason for which the quantum bit commitment is not possible. It
shows the following important fact: {\it When the two parties do not
trust each other, entanglement between them may sometimes become the
most unwanted property.}

There were many attempts to perform quantum bit commitment; some of
them invalid as covered by the proof given by Lo and Chau and some
of them being approximated versions of impossible quantum bit
commitment.

While the proof of Lo and Chau is valid, as it was  pointed out by
H. P. Yuen \cite{Yuen_bc} one could weaken assumptions, so that the
Lo-Chau theorem does not apply.  Namely, the initial state of Bob in
this two-party protocol may not be fixed at the beginning. It leads
to considerations similar to a game theoretic situation.
This case is almost covered by the latest result of D'Ariano {\it et
al.} \cite{DAriano_bc}. The latter paper also provides the most recent and
wide review of this topic.

It is important to note, that the same impossibility reasons share other desired protocols such as {\it ideal quantum coin tossing}, {\it one-out-of-two oblivious transfer} and {\it one-sided two party secure computations}
\cite{LoChau_nobitcom2,Lo_noseccomp}.

\subsubsection{Multipartite entanglement in quantum secret sharing}
There are situations in which one does not trust a single man. A
secret should be shared by a few people so that no single person
could know the truth without permission of the others. Classically,
splitting information into pieces among parties, so that each piece
is informationless unless they cooperate, is known as {\it secret
sharing} \cite{Blakley,Shamir}.

Quantum secret sharing protocol, as proposed in \cite{Hillery_secret_sharing}, solves the following related problem, which can be viewed as secure secrete sharing. Charlie wants to perform secret sharing, but is far from Alice and Bob, thus he has to face the additional problem - eavesdropping. He could resolve it if he had two separate secret channels with them,  sending to one person a key, and to the other the encoded message (secret). However there is a more direct way  discovered in \cite{Hillery_secret_sharing}, using $|GHZ\>$ state \be |GHZ\>_{ABC} ={ 1\over{\sqrt{2}}}(|000\> + |111\>) \ee In their protocol all three parties are supplied many copies of the above multipartite state. Then each of them measures randomly either in $\{|+\>,|-\>\}$ basis, or in $\{{
1\over{\sqrt{2}}}(|0\>+i|1\>),{ 1\over{\sqrt{2}}}(|0\>-i|1\>)\}$.
After this step each party announces the basis that was
chosen, but not the result. In half of the cases, Alice and Bob can
infer the result of Charlie's measurement when they meet
and compare the results. Thus the results of Charlie's
measurement, which are random, can serve as a key, and Alice and
Bob can receive its {\it split} copy. Moreover it can be shown, that
this protocol is secure against any eavesdropping and even a
cheating strategy of either Alice or Bob.

We have discussed sharing of {\it classical information} by means of
quantum entanglement. One can also share {\it quantum information},
and entanglement again helps \cite{Hillery_secret_sharing}. This
time Charlie wants to split securely a qubit, so that Alice and Bob
would need to cooperate to recover it. Interestingly, sharing a {\it
quantum secret}, is essentially teleportation of a qubit through a
GHZ state. After Charlie performs teleportation, the qubit
is split between Alice and Bob. To reconstruct the qubit one of the
parties (i.e. Alice) measures his qubit in $\{|+\>,|-\>\}$ basis and
sends the result of his measurement to the other party (i.e. Bob).
Finally the other party applies on of two unitary operations.

This scheme has been further developed within the framework of
quantum error correcting codes in
\cite{Cleve_secret_sharing,Gottesman_secret_sharing}. Interestingly,
the protocol of Hillery at al. can be changed so that it uses
bipartite entanglement only \cite{EPR_secret_sharing}. This
simplification made it possible to implement the scheme in 2001
\cite{Tittelexp_secret_sharing}. For the recent experimental
realization see \cite{Panexp_secret_sharing}, and references
therein.

\subsubsection{Other multipartite scenarios}
\label{subsec:othermulti}
One of the obvious generalization of quantum key distribution is the
so called {\it conference key agreement}. When some $n$ parties who
trust each other want to talk securely, so that each of them could
receive the information, they could do the following: one party
makes $n-1$ QKD ``bipartite'' protocols with all the other parties,
and then using this distributes a single key, which all the parties
will share finally. This however can be achieved in a much simpler
way using a multipartite entangled state. A state which is mostly
used for this task is the $n$ partite GHZ state
\cite{ChenLo_multi_dist}. One can also use for this task multipartite version
of private states \cite{AH-pditdist}.

Another interesting application of multipartite entangled states is
the so called {\it third man quantum cryptography}
\cite{Zuk_third_man_crypto}. This cryptographic scenario involves
traditionally Alice and Bob, and a third person Charlie, who
controls them. Only when Charlie wants, and tells them some
information, they can share a quantum private channel and they
verify, that indeed nobody including Charlie have access to this
channel. This can be easily achieved by employing the $|GHZ\>$ state
and the idea of ``entanglement with assistance'' (see Sec.
\ref{sec:miary}). Any two-qubit subsystem of $|GHZ\>$ state is in
separable state, hence useless for cryptography. However if
Charlie measures a third qubit in $\{|+\>,|-\>\}$ basis, and tells Alice
and Bob holding other two qubits the result of his measurement then
they obtain one of the maximally entangled states. Then Alice and
Bob can verify that they indeed share the entangled states and use
them to launch a QKD protocol.

\subsection{Interrelations between entanglement and classical key agreement}
\label{subsec:inter-ent-CKA}
So far we have discussed the role of entanglement in quantum
cryptography. It is interesting, that entanglement, which is
originally quantum concept, corresponds to \privacy\ in general -
not only in the context of quantum protocols. Here the interaction
between entanglement theory and the domain of classical cryptography
called {\it classical key agreement} (\cka) is presented.

The problem of distilling secret key from correlations shared by
Alice and Bob with presence of an eavesdropper Eve was first studied
by Wyner \cite{Wyner_key_agreement} and Csiszar and Korner
\cite{CsisarKorner_key_agreement}. It was introduced as a classical
key agreement scenario and studied in full generality by Maurer
\cite{Maurer_key_agreement}. According to this scenario, Alice and
Bob have access to $n$ independent realizations of variables $A$ and
$B$ respectively, while the malicious $E$ holds $n$ independent
realizations of a variable $Z$. The variables under consideration
have joint probability distribution $P(A,B,E)$. The task of Alice
and Bob is to obtain via local (classical) operations and public
communication (LOPC) the longest bit-string which is almost
perfectly correlated and about which Eve (who can listen to the
public discussion) knows a negligible amount of
information\footnote{Let us emphasize, that unlike in quantum
cryptography,  it is in principle not possible to bound Eve's
information on classical ground. However, in particular situations,
there may be good practical reasons for assuming this.}.

Here the probability distribution $P(A,B,E)$ is a priori given.
I.e. it is assumed, that Alice and Bob somehow know how Eve is
correlated with their data.

\subsubsection{Classical key agreement --- analogy to distillable entanglement scenario}

Classical key agreement scenario is an elder sibling of the
entanglement distillation-like scenario. This relation was first
found by N. Gisin and S. Wolf
\cite{GisinWolf_QKAvsCKA,GisinWolf_linking}, and subsequently
developed in more generality by Collins and Popescu
\cite{Collins-Popescu}. The analogy has been explored in the last
years and proved to be fruitful for establishing new phenomena in
classical cryptography, and new links between \privacy\ and
entanglement theory. The connections are quite beautiful, however
they still remain not fully understood by now.

The classical key agreement task is described by the following diagram:
\begin{equation}
  \xymatrix@1@W=3cm{{[P(A,B,E)]}^{\otimes n}
    \ar[r]^{\parbox{2cm}{\center\footnotesize classical distillation of key}} &
    {[P(K,K,E')]}^{\otimes k},}
\end{equation}
where $P(K,K,E')$ is perfectly secure distribution satisfying:
\ben
P(K,K)\equiv \{P(i,j)={1\over 2}\delta_{ij}\}   \nonumber \\
P(K,K,E') = P(K,K)P(E') \label{eqs:idealdist} \een where Alice and
Bob hold variable $K$ and $E'$ is some Eve's variable, i.e. Alice
and Bob are perfectly correlated and product with Eve. The optimal
ratio $k\over n$ in asymptotic limit is a (classical) distillable
key rate denoted here as $K(A;B\|E)$ \cite{Maurer_key_agreement}.

Entanglement between two parties (see Sec. \ref{sec:Ent_in_QKD})
reports that nobody else {\it is correlated} with the parties. In
similar way the \privacy\ of the distribution $P(A,B,E)$ means that
nobody knows about  (i.e. is classically correlated with) the
variables $A$ and $B$. In other words, any tripartite joint
distribution with marginal $P(A,B)$ has a {\it product} form
$P(A,B)P(E)$.

Following along these lines one can see the nice correspondence
between maximally entangled state
$|\phi^{+}\>={1\over\sqrt{2}}(|00\>+|11\>)$ and the private
distribution (\ref{eqs:idealdist}), and also the correspondence
between the problem of transformation of the state $\rho_{AB}^{\ot
n}$ into maximally entangled states which is the entanglement
distillation task and the above described task of classical key
agreement.  Actually, the first entanglement distillation schemes
\cite{BBPSSW1996,BDSW1996} have been designed on basis of protocols
of classical key agreement. The feedback from entanglement theory to
classical key agreement was initiated by Gisin and Wolf
\cite{GisinWolf_linking} who asked the question, whether there is
an analogue of bound entanglement, which we discuss in next section.
Subsequently,  in analogy to {\it entanglement cost} which measures
how expensive in terms of \wsinglet\ state is the creation of a
given quantum state $\rho_{AB}$ by means of LOCC operations Renner
and Wolf \cite{renner-wolf-gap} have defined {\it information of
formation} denoted as $I_{form}(A;B|E)$ (sometimes
called ``key cost''). This function quantifies how many \secure\ key
bits (\ref{eqs:idealdist}) the parties have to share so that they
could create given distribution $P(A,B,E)$ by means of LOPC
operations. The axiomatic approach to privacy, resulting
in deriving secrecy monotones (also in multipartite case), has been
studied in  \cite{Cerf-secr-mono,intrinfo}.

\begin{table}
    \centering
    \begin{tabular}{l @{\hskip2mm} | @{\hskip2mm} l}
      \bf entanglement theory & \bf key agreement \\[2mm]
      \hline
      quantum &secret classical \\
      entanglement & correlations\\[2mm]
      quantum & secret classical \\
      communication & communication\\[2mm]
      classical & public classical \\
      communication & communication\\[2mm]
      local actions & local actions\\
        \end{tabular}
        \caption{Here we present relations between basic notions
        of key agreement and entanglement theory following \cite{Collins-Popescu}}
\end{table}

An interesting formal connection between \cka\ and entanglement is
the following \cite{GisinWolf_linking}.
Any classical distribution can be obtained via POVM
measurements on Alice, Bob and Eve's subsystems of a pure quantum
state $|\psi\>_{ABE}$: \be p_{ijk}^{ABE}:={\tr} M_A^{(i)}\ot
M_B^{(j)}\ot M_E^{(k)}|\psi\>\<\psi|_{ABE}
\label{eq:GWprocedure}\ee where
$M_{A,B,E}^{(.)}$ are elements of POVM's of the parties satisfying
$\sum_{l}M_{A,B,E}^{(l)}=\id_{A,B,E}$ with $\id$ the identity operator.

Conversely, with a given tripartite distribution $P(A,B,E)$ one can
associate a quantum state in the following way: \be
P(ABE)=\{{p_{ijk}^{ABE}}\}\longmapsto |\psi_{ABE}\> = \sum_{ijk}
\sqrt{p_{ijk}^{ABE}}|ijk\>_{ABE}. \ee According to this approach,
criteria analogous to those for pure bipartite states transitions
and catalytical transitions known as majorization criteria (see Secs. \ref{subsec:puretrans} and \ref{subsec:catalysis}) can
be found \cite{GisinWolf_linking,Collins-Popescu}. Also
other quantum communication phenomena such as {\it merging}
\cite{SW-nature} and {\it information exchange} \cite{uncom_info} as
well as the no-cloning principle are found to have counterparts in
\cka\ \cite{OppenSpekWint05}.

A simple and important connection between tripartite distributions
containing privacy and entangled quantum states was established in
\cite{GisinAcin_link}. Consider a quantum state $\rho_{AB}$ and its
purification $|\psi_{ABE}\>$ to the subsystem $E$.
\begin{itemize}
\item If a bipartite state $\rho_{AB}$ is {\it entangled} then there exists
a measurement on subsystems $A$ and $B$ such, that for all measurements on subsystem $E$ of purification $|\psi_{ABE}\>$ the resulting probability distribution
$P(A,B,E)$ has {\it nonzero} key cost.
\item If a bipartite state $\rho_{AB}$ is {\it separable}, then for all measurements on subsystems $A$ and $B$ there exists a measurement subsystem $E$ of purification $|\psi_{ABE}\>$ such that the resulting probability distribution $P(A,B,E)$ has {\it zero} key cost.
\end{itemize}

\subsubsection{Is there a bound information?}
In the entanglement distillation scenario there are {\it bound
entangled} states which exhibit the highest irreversibility in
creation-distillation process, as the distillable entanglement is
zero although the entanglement cost is a nonzero quantity(see
Sec. \ref{sec:distil}). One can ask then if the analogous
phenomenon holds in classical key agreement called {\it bound
information} \cite{GisinWolf_linking,renner-wolf-gap}. This question
can be stated as follows:

\begin{itemize}
\item{\it Does there exist a distribution $P(A,B,E)_{bound}$ for which a \secure\ key is needed to create it by LOPC ($I_{form}(A;B|E)>0$), but one cannot distill any key back
from it ($K(A;B\|E)=0$)?}
\end{itemize}

In \cite{GisinWolf_linking} the tripartite distributions obtained
via measurement from bound entangled states were considered as a
possible way of search for the hypothetical ones  with bound
information. To get Eve's variable, one has first to purify a bound
entangled state, and then find a clever measurement to get
tripartite distribution. In this way, there were obtained tripartite
distributions with non-zero key cost. However the no-key
distillability still needs to be proved.

Yet there are serious reasons supporting conjecture that such
distributions exist
\cite{GiReWo02,RenWol02b}. To give an example, in \cite{RenWol02b}
an analogue of the necessary and sufficient condition for
entanglement distillation was found. As in the quantum case the
state is distillable iff there exists a projection (acting on $n$
copies of a state for some $n$) onto 2-qubit subspace which is
entangled \cite{bound}, in the classical case, the key is
distillable iff there exists a binary channel (acting on $n$ copies
of a distribution for some $n$) which outputs Alice's and Bob's
variables, such that the resulting distribution has nonzero key (see
in this context \cite{ALG_2_equiv,AcGisScar03,QKA_CKA_equivalance}).

A strong confirmation supporting hypothesis of {\it bound
information} is the result presented in \cite{renner-wolf-gap},
where examples of distributions which {\it asymptotically} have
bound information were found. Namely there is a family of
distributions $P(A_n,B_n,E_n)$ such that $\lim_{n\rightarrow \infty}
K(A_n;B_n\|E_n)=0$ while $I_{form}(A_n;B_n|E_n)>{1\over 2}$ for all $n$.
Another argument in favor of the existence of bound information in
this {\it bipartite} scenario, is the fact that the {\it
multipartite bound information} has been already proved to exist,
and explicit examples have been constructed \cite{AcinCM-MultiBoundInfo}.
More specifically this time three honest cooperating parties
(Alice, Bob and Clare) and the eavesdropper (Eve) share $n$
realizations of a joint distribution $P(A,B,C,E)_{bound}$ with the
following properties: it has nonzero secret key cost and no pair of
honest parties (even with help of the third one) can distill secret
key from these realizations $P(A,B,C,E)_{bound}$.

Its construction has been done according to a Gisin-Wolf procedure (\ref{eq:GWprocedure}) via measurement of a (multipartite) purification of a bound entangled state in computational basis.

As an example consider a four partite state
\begin{eqnarray}
|\psi_{be}\>=[{1\over{\sqrt{3}}}|GHZ\> +
{1\over{\sqrt{6}}}(|001\>|1\>
\nonumber\\
+|010\>|2\>+|101\>|3\>+|110\>|4\>)]_{ABCE}
\end{eqnarray}
where $|GHZ\>
={1\over{\sqrt{2}}}(|000\>+|111\>)$. It's partial trace over subsystem
$E$ is a multipartite bound entangled state\footnote{
This is one of the states from the family
given in Eq. (\ref{mCirac}), with parameters: $m=3,\, \lambda_0^+={1\over 3},\, \lambda_0^-=\lambda_{10}=0,\, \lambda_{01}=\lambda_{11}={1\over 6}$.} \cite{DCpur,DurC_multi_dist2000}.
If one measures $|\psi_{be}\>$ in computational basis on all four
subsystems, the resulting probability distribution on the labels of
outcomes is the one which contains (tripartite) bound information
\cite{AcinCM-MultiBoundInfo}.

\section{Entanglement and quantum computing}

\subsection{Entanglement in quantum algorithms}

Fast quantum computation is one of the most desired properties of
quantum information theory. There are few quantum algorithms which
outperform their classical counterparts. These are the celebrated
Deutsch-Jozsa, Grover and Shor's algorithm, and their variations.
Since entanglement is one of the cornerstones of quantum information
theory it is natural to expect, that it should be the main
ingredient of quantum algorithms which are better than classical.
This was first pointed out by Jozsa in \cite{Jozsa_ent_qc_first}.
His seminal paper opened a debate on the role of entanglement in
quantum computing. Actually, after more than decade from the
discovery of the first quantum algorithm, there is no common
agreement on the role of entanglement in quantum computation. We
discuss major contributions to this debate. It seems that
entanglement ``assists'' quantum speed up, but is not sufficient for
this phenomenon.

Certainly {\it pure} quantum computation needs some level of
entanglement if it is not to be simulated classically. It was shown
by Jozsa and Linden, that if a quantum computer's state contains
only constant (independent of number of input qubits $n$) amount of
entanglement, then it can be simulated efficiently
\cite{JozsaLinden}.

Next, Vidal showed  that even a polynomial in $n$ amount of
entanglement present in quantum algorithm can also be simulated
classically \cite{Vidal}. The result is phrased in terms of the
number $\chi$ which he defined as the maximal rank of the subsystem
of the qubits that form quantum register of the computer (over all
choices of the subsystem). Any quantum algorithm that maintains
$\chi$ of order $O(poly(n))$ can be efficiently classically
simulated. In other words to give an exponential speedup the quantum
algorithm needs to achieve $\chi$ of exponential order in $n$,
during computation.

This general result was studied by Orus and Latorre
\cite{OrusLatorre} for different algorithms in terms of entropy of
entanglement (von Neumann entropy of subsystem). It is shown among
others that computation of Shor's algorithm generates highly
entangled states (with linear amount of entropy of entanglement
which corresponds to exponential $\chi$). Although it is not known
if the Shor's algorithm provides an exponential speedup over
classical factoring, this analysis suggests that Shor's algorithm
cannot be simulated classically.

Entanglement in Shor's algorithm has been studied in different
contexts
\cite{EkertJozsa,JozsaLinden,ParkerPlenio,ShimoniShapiraBiham}.
Interestingly, as presence of entanglement in quantum algorithm is
widely confirmed (see also
\cite{AdattaV_ent_mixed_comp2006,ADattaFCC_1q_comp2005}), its role is
still not clear, since it seems that amount of it depends on the type
of the input numbers \cite{KendonMunro}.

Note, that the above Jozsa-Linden-Vidal ``no entanglement implies no
quantum advantage on pure states'' result shows the need of
entanglement presence for exponential speed up. Without falling into
contradiction, one can then ask if entanglement must be present for
polynomial speed up when only pure states are involved during
computation (see \cite{KenigsbergMoreRatsaby} and references
therein).

Moreover it was considered to be possible, that a quantum computer
using only {\it mixed}, separable states during computation may
still outperform classical ones \cite{JozsaLinden}. It is shown,
that such phenomenon can hold \cite{Tal1}, however with a tiny speed
up. It is argued that isotropic separable state cannot be entangled
by an algorithm, yet it can prove useful in quantum computing.
Answering to the general question of how big the enhancement
based on separable states may be, needs more algorithm-dependent
approach.

That the presence of entanglement is only {\it necessary} but not
{\it sufficient} for exponential quantum speed up follows from the
famous Knill-Gottesman theorem \cite{GottesmanC1999-Knill,JozsaLinden}.
It states that operations from the so called {\it
Clifford group} composed with Pauli measurement in
computational basis can be efficiently simulated on
classical computer. This class of operations can however produce
highly entangled states. For this reason, and as indicated by other
results cited above, the role of entanglement is still not
clear. As it is pointed out in \cite{JozsaLinden}, it may be that
what is essential for quantum computation is not entanglement but
the fact that the set of states which can occur during computation
can not be described with small a number of parameters
(see also discussion in \cite{Knill-ent-comp2001} and
references therein).

\subsection{Entanglement in quantum architecture}
Although the role of entanglement in algorithms is unclear,
its role in architecture of quantum computers is crucial. First of
all the multipartite cluster states provide a resource for one-way
quantum computation
\cite{RaussendorfBriegelOneWay}. One
prepares such multipartite state, and the computation bases on
subsequent measurements of qubits, which uses up the state.

One can ask what other states can be used to perform universal
one-way quantum computation. In \cite{NestMDB2006-one-way-resource}
it was assumed that universality means possibility
of creating any final state on the part of lattice that
was not subjected to measurements.
It was pointed out that by use of entanglement measures one can
rule out some states. Namely, they introduced an entanglement
measure, {\it entanglement width}, which is defined as the
minimization of bipartite entanglement entropy over some specific
cuts. It turns out that this measure is unbounded for cluster states
(if we increase size of the system). Thus any class of
states, for which this measure is bounded, cannot be resource for
universal computation, as it cannot create arbitrary large cluster
states. For example the \ghz\ state is not universal, since under
any cut the entropy of entanglement is just $1$.
One should note here, that a more natural is weaker notion of universality
where one requires possibility of compute arbitrary classical function. In \cite{GrossE2006-one-way}
it is shown that entanglement width is not a necessary condition
for this type of universality.

An intermediate quantum computing model, between circuit based on
quantum gates, and one-way computing, is teleportation based
computing \cite{GottesmanC1999-Knill}. There two and three
qubit gates are performed by use of teleportation as a basic
primitive. The resource for this model of computation are thus
\eprstates\ and \ghz\ states. Teleportation based computing is of
great importance, as it allows for efficient computation by use of
linear optics \cite{KnillLM2001-lin-opt}, where it is impossible to
perform two-qubit gates
deterministically. Moreover, using it, Knill has significantly
lowered the threshold for fault tolerant computation
\cite{Knill-FT}.

The need of entanglement in quantum architecture has been also studied
from more ``dynamical'' point of view, where instead of entangled
states, one asks about operations which generate entanglement. This
question is important, as due to decoherence, the noise usually limits
entanglement generated during the computation. This effect can make a
quantum device efficiently simulatable and hence useless. To avoid
this, one has to assure that operations which constitute a quantum
device generate sufficient amount of (or a robust kind of)
entanglement.

An interesting connection between entanglement and fault tolerant quantum computation was obtained by Aharonov \cite{dorit-phase}.
She has shown that a properly defined {\it long range entanglement}
\footnote{Another type of long range entanglement was recently defined
\cite{KitaevP2005-top-ent,LevinW2006-top-entr} in the context of topological
order.} vanishes in
thermodynamical limit if the noise is too large. Combining it with the fact that
fault tolerant scheme allows to achieve such entanglement, one obtains a
sort of phase transition (the result is obtained within
phenomenological model of noise).

The level of noise under which quantum computer becomes efficiently
simulatable was first studied in \cite{Aharonov-comp}. It is shown
that quantum computer which operates (globally) on $O(\log n)$ number
of qubits at a time (i.e. with limited entangling capabilities) can be
efficiently simulated for any nonzero level of noise. For the circuit
model (with local gates), the phase transition depending on the noise
level is observed (see also \cite{Aharonov-comp2}).

The same problem was studied further in \cite{HarrowN2003-robust},
however basing directly on the entangling capabilities of the gates
used for computation. It is shown, that so called {\it separable}
gates (the gates which can not entangle any product input) are
classically simulatable. The bound on minimal noise level which turns
a quantum computer to deal only with separable gates is provided
there. This idea has been recently developed in
\cite{VirmaniHP-comp-ent2005}, where certain gates which create only
bipartite entanglement are studied. This class of gates is shown to be
classically simulatable. In consequence, a stronger bound on the
tolerable noise level is found.

\subsection{Byzantine agreement --- useful entanglement for quantum and
classical distributed computation}

As we have already learned the role of entanglement  in
communication networks is uncompromised. We have already described
its role in cryptography (see Sec. \ref{sec:Ent_in_QKD}) and
communication complexity. Here we comment another application -
quantum solution to one of the famous problem in classical
fault-tolerant distributed computing called {\it Byzantine
agreement}. This problem is known to have no solution in classical
computer science. Yet its slightly modified version can be solved
using quantum entangled multipartite state \cite{Gisin01-byzantine}.

One of the goals in distributed computing is to achieve broadcast in
situation when one of the stations can send faulty signals. The
station achieve broadcast if they fulfill conditions which can be
viewed as a natural extension of broadcast to a fault-tolerant
scenario:

\begin{enumerate}
\item if the sending station is not faulty, all stations announces the
  sent value.
\item if the sending station is faulty, then all stations announce the
  same value.
\end{enumerate}

It is proved classically, that if there are $t\geq n/3$ stations
which are out of work and can send unpredictable data, then
broadcast can not be achieved. In quantum scenario one can achieve
the following modification called {\it detectable} broadcast which
can be stated as follows:

\begin{enumerate}
\item if there is no faulty station, then all stations announces the
  received value.
\item if there is faulty station, then all other stations either {\it
    abort} or announces the same value.
\end{enumerate}

This problem was solved in \cite{Gisin01-byzantine} for $n=3$.
We do not describe here the protocol, but comment on the role of
entanglement in this scheme. In the quantum Byzantine agreement there
are two stages: quantum and classical one. The goal of quantum part
is to distribute among the stations some correlations which they
cannot deny (even the one which is faulty). To this end, the
parties perform a protocol which distributes among them
the Aharonov state:
\begin{equation}
\psi={1\over \sqrt6} (|012\>+|201\>+|120\> - |021\> -|102\>-|210\>)
\end{equation}
The trick is that it can be achieved fault-tolerantly, i.e.  even when
one of the stations sends fake signals.

\section*{ACKNOWLEDGMENTS}

We would like to thank J. Horodecka for her continuous help in editing of
this paper.  Many thanks are due to A. Grudka, \L{}. Pankowski,
M. Piani and M. \.Zukowski for their help in editing and
useful discussions.  We would like also to express thanks to many other
colleagues in the field of quantum information for discussions and useful feedback.
We wish also to acknowledge the referees for their fruitful
criticism. This work is supported by EU grants SCALA FP6-2004-IST no.015714 and
QUROPE. K.H. acknowledges support of Foundation for Polish Science. P.H.
is supported by Polish Ministry if Science and Education,
contract No. 1 P03B 095 29. K.H., M.H. and R.H. are supported
by Polish Ministry of Science and Education  under the (solicited) grant
No. PBZ-MIN-008/P03/2003

\bibliographystyle{apsrmp-article-eprint}
\bibliography{rmp6}

\end{document}